%% file: thesis-main.tex
\documentclass[12pt]{book}

\usepackage{afterpage} 

\newcommand\blankpage{%
    \null
    \thispagestyle{empty}%
    \addtocounter{page}{0}%
    \newpage}
    
\usepackage{appendix}
\usepackage{chngcntr}
\usepackage{etoolbox}
\usepackage[font={small}]{caption}   

\AtBeginEnvironment{subappendices}{%
\section*{Appendix}
\addcontentsline{toc}{section}{Appendix}
\counterwithin{figure}{section}
\counterwithin{table}{section}
}

\usepackage{amsmath}
\usepackage{amssymb}
\usepackage{color}
\usepackage{transparent}  
\usepackage[sectionbib]{chapterbib}
\usepackage{fancybox}
\usepackage[colorlinks,bookmarks=false,citecolor=red,linkcolor=blue,urlcolor=green]{hyperref}

\definecolor{dr}{RGB}{150,00,00} 
\definecolor{dg}{RGB}{00,150,00} 
\definecolor{db}{RGB}{00,00,150} 

\usepackage{geometry}		
\usepackage{graphicx}		
\usepackage{verbatim}		
\usepackage{subfig}		

\usepackage{epstopdf}
\usepackage{wrapfig}
\usepackage{bm}			
\usepackage[mathlines]{lineno}
\usepackage{epstopdf}		
\usepackage{pdfpages}		
\usepackage{makeidx}		
\makeindex
\usepackage{nomencl}
\makenomenclature
\usepackage[Bjornstrup]{somufncychap}
\usepackage{layouts} 
\usepackage{fancyhdr}

\usepackage{listings}
\usepackage{minitoc} 

\allowdisplaybreaks
\begin{document}
\include{titlepage/titlepage1}
\include{frontbackmatter/titleback} 
\frontmatter
\cleardoublepage\include{frontbackmatter/declaration} 
\cleardoublepage\include{frontbackmatter/dedication} 
\cleardoublepage\include{frontbackmatter/acknowledgments} %
\cleardoublepage\include{frontbackmatter/publication} 
\begingroup
\hypersetup{linkcolor=black}
\dominitoc 
\tableofcontents
\endgroup
\printnomenclature 
\afterpage{\blankpage}
\mainmatter
\include{chapters/introduction}
\include{chapters/numerical-methods}
\afterpage{\blankpage}
\include{chapters/rabi}
\include{chapters/quantum-ising}
\afterpage{\blankpage}
\include{chapters/falicov-kimball-model}
\printindex
\end{document}

%% file: titlepage/titlepage1.tex

\begin{titlepage}
\begin{center}
\bigskip
\vfill
\bigskip
{\Huge {\bf Quantum Ising Systems,\\[12pt]Edge Modes, and Rabi Lattice}} \\
[50pt]
{\Large{Thesis submitted for the award of the degree of \\ Doctor of Philosophy}} \\
\bigskip
\vfill
\bigskip
{\LARGE Somenath Jalal}
\vfill
\bigskip
\vfill
\hrule
\bigskip
\Large
\begin{tabular} {cc}
\parbox{0.3\textwidth}{\includegraphics[width=2.0cm]{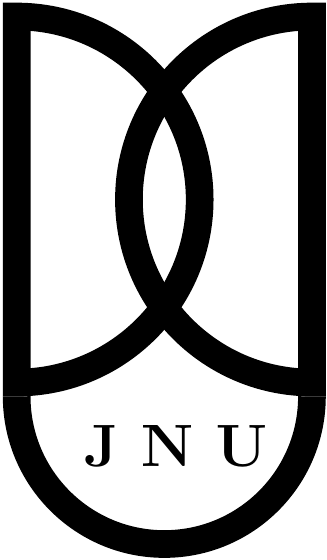}}
&
\parbox{0.5\textwidth}{%
	School of Physical Sciences\\[-5pt]
	Jawaharlal Nehru University\\[-5pt]
	New Delhi, India \\[10pt]
	July 2014
	}
\end{tabular}
\bigskip
\hrule
\end{center}
\end{titlepage}

%% file: frontbackmatter/titleback.tex


\thispagestyle{empty}

\hfill

\vfill


\noindent {\textit{Quantum Ising Systems, Edge Modes, and Rabi Lattice}}\\
{\textcopyright} Somenath Jalal \\ July 2014


\bigskip


\noindent{School of Physical Sciences\\
Jawaharlal Nehru University\\
New Delhi - 110067\\
India
}



%% file: frontbackmatter/declaration.tex

\thispagestyle{empty}

\hrule

\bigskip

\begin{center}
\Huge{Declaration}
\end{center}

\bigskip

\hrule

\bigskip

\noindent I hereby declare that the work carried out in this thesis is entirely original and has been carried out by me in the School of Physical Sciences, Jawaharlal Nehru University, New Delhi,  under the supervision of Dr. Brijesh Kumar. I further declare that it has not formed the basis for the award of any degree, diploma, associateship or similar title of any other university or institution.

\vspace{2cm}

{
Date: July 2014 \hfill Somenath Jalal
}

\vspace{2cm}

{
\flushleft{
\parbox{
0.4\textwidth}{
\begin{center}
Dr. Brijesh Kumar\\
Thesis Supervisor\\
School of Physical Sciences\\
Jawaharlal Nehru University
\end{center}
}
}
}

\vspace{3cm}

\begin{center}
Prof. Subhasis Ghosh\\
Dean\\
School of Physical Sciences\\
Jawaharlal Nehru University
\end{center}

%% file: frontbackmatter/dedication.tex


\thispagestyle{empty}

\vspace*{1.5in}

\begin{center}
{\Large{\emph{I dedicate my thesis to \dots}}} \\[16pt]
my mother {\emph{Jayanti}}, pisimoni {\emph{Sila}}, masimoni {\emph{Malu}}\\
{\&} specially,\\
my loving wife {\emph{Sumi}}.\\
\vspace{0.5in}
My best regards {\&} love to my father {\emph{Shri Ranjit Jalal}} {\&} my brother {\emph{Priyanath}}, for their love {\&} care.
\end{center}

%
%

%% file: frontbackmatter/acknowledgments.tex

\chapter*{Acknowledgements} 
{\small
\noindent Now here I am, writing my Ph.D. thesis, and looking back to my good old days. First I would like to thank my Ph.D. supervisor Dr. Brijesh Kumar for his kind {\&} constant guidance for last six years, till date, in SPS. His writings, lectures and personal conversations have changed me a lot. If there is any  academic achievement, that is due to his blessings.

I would like to acknowledge Council of Scientific and Industrial Research (CSIR), India, for financing me from 2009 January to 2013 December. It helps to get paid for what you love doing.


My first teacher was my father, who taught me how to write. I still remember the classes of Ganesh Sir, Kabul Sir, Mani Babu, Suman Mukherjee and Ashwani Layak (Mathematics), Deb Babu (Physics) from my school days and Debasis Bhattacharya from my B.Sc. days in Burdwan.

My regards to Prof. Shankar Prasad Das, Prof. Subir K. Sarkar, and Prof. Nivedita Deo (DU) who wrote recommendation letters for me. Prof.  S. Patnaik, Prof. D.  Ghosal, Prof. S. Puri, Prof. D. Kumar, Prof. A. Pandey, Prof. H. Bohidar, Prof. S. Ghosh, Prof. A. K. Rastogi, Prof. P. Sen, Prof. S. S. N. Murthy and my guruji, Prof. B. Kumar, all taught me during M.Sc., and I thank them from the bottom of my heart. I like Prof. Brijesh Kumar's and Prof. Sobhan Sen's way of presenting talks and Prof. Pritam Mukhopadhyay for his and his lab's work culture, which I myself try to follow. I had the privilege to talk to Prof. R. Rajaraman during my SPS Friday Journal Club organizer days. I thank him for encouraging us to be involved in extra-curricular activities.


In JNU, I had the opportunity to meet students from all over the India. I have found best friends in Ashwani {\&} Bimla, with whom I shared the Condensed Matter Theory lab.  Special thanks to Dr. Rakesh Kumar, who is and will be a big brother to me. Thank you Pratyay (The Scientist!) and Panchram (Anderson!) for the wonderful times and laughter we shared together. Specially, thanks to two of you for proofreading my thesis.

Thanks to Siddhartha, Dipan and Abhinandan for being good friends of mine. Some M.Sc. friends and seniors will be remembered for the rest of my life. One is surely Ravi Kumar Pujala. Thanks to Amit, Sayantari, Sudhir, Neeta, Prem, Preeda, Avesh, Pawan, Sarita, Priya, Najmul, Avinash.  And from seniors, Sakir, Neeraj, Murari, Gaurav, Budhi, Awadhesh, Nirmal, Deepak (Chem.), Ajay (Chem), Nibedita, Sachin, Bhaskar,  Santosh Kumar (SNU) and Hbar, with whom I collaborated for 4 months. From juniors, I must thank Prasenjit Das, Akhilesh Verma, Elisha Siddiqui, Himadri, Arijit, Kishor, Pankaj, Ravi Prakash, Alok (Math.), Vishal (Parle-G), Ambuj, Abhishek, Manoj (U2), Girish, Manoj (Puri), Him Sweta, Preeti, Sangeeta, Himadri Dhar \dots . I need 5 extra pages here!

Thank you Renny Thomas. Thank you Indraneel Mukherjee for letting me stay in your room in JNU when I had no hostel.


I thank Jagannath ji, Pradip ji, Shipra ji, Tikaram ji, Rahul, Nikhil, Ashis, Anita ji, SPS librarian Rajmani ji, PS to dean Jagdish ji, and A.O. Dharam Pal ji,  for making the office works smooth from day 1.
I thank my Kaveri Hostel's mess workers for serving me meals during my stay in JNU. Thanks to JNU health center's doctors for keeping me fit. Special thanks to Santhosh bhaiya for nice tea and samosas.


I thank Sudipa for being a part of our family. Thank you grandma for waiting to see my thesis. I thank my cousin  Sandip, Santi, Tubai, Chandan, Binuda {\&}  Malay for their love. Love you Dipu. I thank my in-laws and specially my pisemosai, Janakinath Singha for their encouragement. I am lucky to have Bono mama and Mamima.  Thanks to Pratyay's lifepartner Ankita for splendid lunches and cakes.

Last, but not the least, I thank my bad-looking Thinkpad Laptop, `Kanaad' and SPS-Cluster, for working like a beast and finishing most of my research work in time. I also thank Google and arxiv. I thank M.S. Dhoni and the God of Cricket Sachin Tendulkar for gifting us such wonderful cricket matches during my Ph.D. days. The workplace has always been very important part of my life. I really love my desk in SPS, and whole of JNU for her academic and non-academic environment. 

I shall remember you all forever.

}

%% file: frontbackmatter/publication.tex

\chapter*{Publications} 


\bigskip

\begin{enumerate}
\item Quantum Ising dynamics and Majorana-like edge modes in the Rabi lattice model,\\
B.Kumar and {\underline{\bf{S. Jalal}}}, {\href{http://journals.aps.org/pra/abstract/10.1103/PhysRevA.88.011802}{Phys. Rev. A {\bf 88}, 011802(R) (2013)}}[\href{http://arxiv.org/abs/1210.6922}{arXiv:1210.6922}]
\item Edge modes in a frustrated quantum Ising chain, {\underline{\bf{S. Jalal}}} and B. Kumar, {[\href{http://arxiv.org/abs/1407.0201}{arXiv:1407.0201}]}
\end{enumerate}

%% file: chapters/introduction.tex
\def\ahat{\hat{a}}
\def\bhat{\hat{b}}
\def\chat{\hat{c}}
\def\xhat{\hat{x}}
\def\yhat{\hat{y}}
\def\phihat{\hat{\phi}}
\def\psihat{\hat{\psi}}
\def\chihat{\hat{\chi}}
\def\atilde{\tilde{a}}
\def\nhat{\hat{n}}
\def\Dhat{\hat{D}}
\def\Nhat{\hat{N}}
\def\Phat{\hat{P}}
\def\Hhat{\hat{H}}
\def\Uhat{\hat{U}}
\def\k{{\bf k}}
\def\r{{\bf r}}
\def\calU{\mathcal{U}}
\def\calL{\mathcal{L}}
\def\deltavec{\vec{\delta}}
\def\omegatilde{\tilde{\omega}}
\def\ttilde{\tilde{t}}
\def\Deltatilde{\tilde{\Delta}}
\def\a{\hat{a}}
\def\adag{\hat{a}^\dagger}
\def\sighat{\hat{\sigma}}
\def\chihat{\hat{\chi}}
\def\nhat{\hat{n}}
\def\Uhat{\hat{U}}

\chapter{Introduction}

Quantum condensed matter physics is a very active area of research, encompassing broad range of physical problems. It is basically concerned with the understanding of interacting quantum many-particle systems such as electrons inside a solid, atomic gases at low temperatures, frustrated quantum magnets, and so on. From a theoretical perspective, the main objective is to construct theoretical models inspired by real phenomena, to understand their properties, and to make newer predictions, if possible. But generally, these are hard and often intractable problems. While a theorist is often interested in studying the physics of a model over a wide range of parameters, but experimentally it may not even be possible to access all of this. So, the physical content of a model problem remains somewhat poorly understood, unless of course, it is an exactly soluble problem. Recently, however, things have changed for better. Now there is an active effort in simulating these interesting many-body problems using controlled laboratory setups. Thus the `quantum simulators' have come into the picture~{\cite{quantum-simulation}}. It all started with an emulation of the Bose-Hubbard model using ultra-cold atoms in optical lattices~{\cite{BH.ColdAtom,ColdAtoms}. Of course, things have gone beyond the cold-atoms to other interesting possibilities, most notably, to the quantum-cavity (or circuit-QED) arrays~\cite{Tomadin.Fazio.10,CircuitQED.Girvin}. 

Motivated by the recent developments on the cavity-QED arrays~\cite{Greentree.06,Hartmann}, and related systems such as the Josephson junction arrays and circuit-QED problems, we have done some interesting new work on the Rabi lattice model, and through it, on the quantum Ising model and Majorana edge modes therein. Since it forms the bulk of this thesis, in the following sections of this chapter, we give a brief introduction to the quantum Ising model, the atom-photon (quantum cavity) problem, and Rabi lattice. We conclude this introductory chapter by presenting an overview of the problems worked out in the thesis.

\section{Quantum Ising model}{\label{sec:qi}}
The Ising model is a model of great importance in statistical mechanics and condensed matter physics~\cite{pathria,huang}. It basically describes a system of interacting spin-1/2 (two-level) objects, such as a magnet or a binary alloy. While in its usual form, it is a classical problem, the presence of an external field transverse to the Ising axis makes it a quantum mechanical model. The Hamiltonian of such a `quantum' Ising (QI\nomenclature{QI}{Quantum Ising}) model can be written as  


\begin{equation}
H_{QI} = J {\sum_{i=1}^{L-1}} ~ {\sigma^{x}_{i} \sigma^{x}_{i+1}} + h {\sum_{i=1}^{L}}  {\sigma^{z}_{i}} {\label{eqn:qi}}
\end{equation}
where $\sigma_{i}^{x}$, $\sigma_{i}^{z}$ are the Pauli matrices{\index{Pauli Spin matrices}}, $i$ is the site index, and $L$ is the total number of sites on a lattice. The first term of the Hamiltonian is the Ising interaction term, and the second term proportional to $h$ is the transverse field. In the above equation, we have explicitly written it on a one-dimensional lattice with nearest neighbor interactions (and open boundary condition). In general, the geometry of the lattice and the range of the interaction would, of course, be different as per the needs of a physical setting. As the Ising interactions do not commute with the transverse field, the field along $z$-direction necessarily induces quantum fluctuations to the spin states in $x$-directions. Therefore, it is a quantum problem, unlike the usual Ising model for $h=0$. Hence, the name quantum Ising model. It is also called the Transverse Field Ising (TFI{\nomenclature{TFI}{Transverse Field Ising}}) model. Historically, the interest in this model started through a work of de Gennes where it was used to model the proton dynamics in the hydrogen-bonded ferroelectric materials ($e.g.$, KH$_{2}$PO$_{4}$)~\cite{deGennes}. The QI model continues to be a subject of diverse current interests, as clear from a variety of physical contexts in which it occurs~\cite{book.Bikas, randomQI.Fisher, Moessner, QI.Holo.Girvin, Coldea, bermudez-qi-expt,Mila.Square}.

It is minimal spin model that nicely exhibits quantum phase transition from an ordered to a paramagnetic phase. By a quantum phase transition, we mean a change in the physical nature of the ground state due to competing agents in the Hamiltonian~\cite{QPT.Sachdev}. In the QI model, $J$ and $h$ are such competing agents. The possibility of a quantum phase transition in the QI model of Eq.~(\ref{eqn:qi}) can easily be seen by arguing as follows. Suppose $h = 0$, then the spins align themselves according to the Ising interaction. For instance, if $J<0$, then the spins will form an arrangement along in the $+x$ direction (with state $|\rightarrow\rangle$) or in $-x$ direction (with state $|\leftarrow\rangle$), in the ground state. These states are the eigenstates of $\sigma^x$ operator  and written as $|\rightarrow\rangle = \frac{1}{\sqrt{2}}\left(|\uparrow\rangle + |\downarrow\rangle\right)$ and $|\leftarrow\rangle = \frac{1}{\sqrt{2}}\left(|\uparrow\rangle - |\downarrow\rangle\right)$; $|\uparrow\rangle$ and $|\downarrow\rangle$ are the eigenstates of $\sigma^{z}$. This is the ferromagnetic phase with a non-zero expectation value of the magnetic order parameter, p=$\frac{1}{L}\sum_{i}\sigma_{i}^{x}$. However, if $J = 0$ and $h\neq 0$, then all the spins will  point along $+z$ or $-z$ direction, depending upon the sign of $h$.  We can check that $\langle \rightarrow|\sigma_{i}^{z}|\rightarrow\rangle = 0$, and the same in $|\leftarrow\rangle$. Hence, this is called a paramagnetic phase, or more correctly, a quantum paramagnetic phase because of the absence of (Ising) order due to quantum fluctuations in the transverse direction. This suggests that when $h$ and $J$ both are non-zero, then depending upon their relative strengths, the ground state will either behave as ordered or disordered, separated by a critical point. Figure~{\ref{fig:1d-qi}} shows the quantum phase diagram of the QI model of Eq. ({\ref{eqn:qi}}).
\begin{figure}[htbp] 
   \centering
   \includegraphics[width=0.9\textwidth]{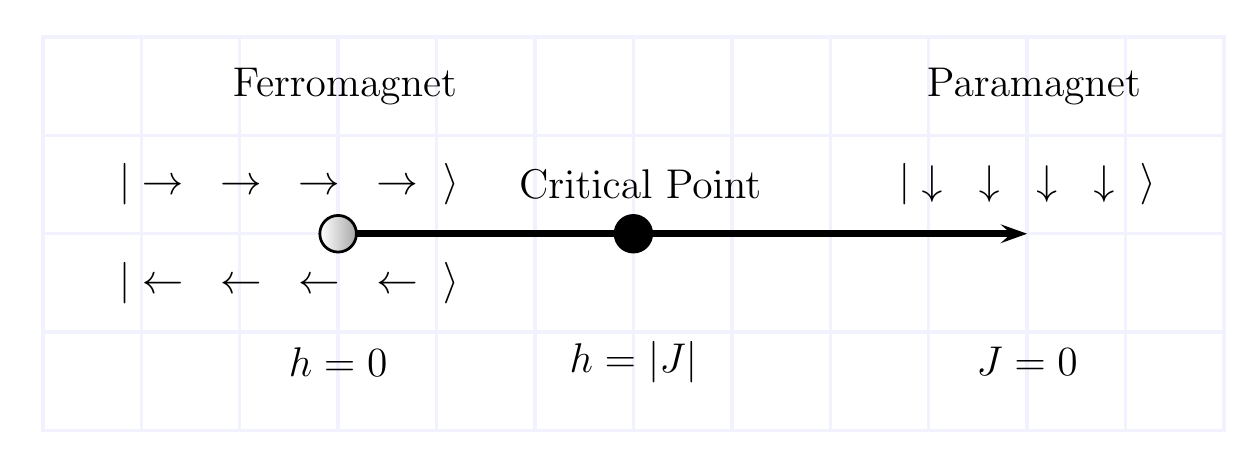} 
   \caption{Quantum phase diagram of the quantum Ising model. Here, $J<0$ and $h>0$. For the QI model in Eq.~(\ref{eqn:qi}), the exact critical point is at $h/|J| = 1$.}
   \label{fig:1d-qi}
\end{figure}

\subsection{Exactly soluble case in one-dimension}
The one-dimensional (1D) spin-1/2 QI model with only nearest-neighbor interaction has been rigorously worked out by Pfeuty using the Jordan-Wigner{\nomenclature{JW}{Jordan-Wigner}}{\index{Jordan-Wigner}} (JW)  fermionization~\cite{QIsing.Pfeuty,LSM}. We briefly summarize it here. The JW transformation is an exact map that relates spin-1/2's into fermions in the following way.
\begin{eqnarray}
&& {\sigma^{-}_{i}} = e^{-i \pi {\sum_{l=1}^{i-1}}\nhat^{}_{l}} {\chat^{}_{i}} \qquad {\sigma^{+}_{i}} = {\chat^{\dagger}_{i}} e^{i \pi {\sum_{l=1}^{i-1}}\nhat^{}_{l}} {\label{eqn:jw}} \\
&& \sigma^z_l = 2 \chat^\dagger_l \chat^{}_l - 1
\end{eqnarray}
Here, $\chat^{}_{i}$'s are the (spinless) fermion operators. The fermionic form of the QI model of Eq. (\ref{eqn:qi}) is given below.
\begin{eqnarray}
H_{QI} = J {\sum_{i=1}^{L-1}} \big[ {( \chat^{\dagger}_{i} \chat^{}_{i+1} + H.C. )} + {  (\chat^{\dagger}_{i} \chat^{\dagger}_{i+1} + H.C. ) } \big]  + h {\sum_{i=1}^L} { ( \chat^{\dagger}_{i} \chat^{}_{i} - \chat^{}_{i} \chat^{\dagger}_{i} ) } {\label{eqn:qi-fermion}}
\end{eqnarray}
One solves this bilinear model of fermion in momentum space (as Pfeuty did), and find the following energy dispersion~\cite{QIsing.Pfeuty}.
\begin{eqnarray}
E(k) = \sqrt{J^2 ~ {\sin^2} {k} + ( J \cos{k} + h )^2 }
\end{eqnarray}

Clearly the dispersion is gapped, except at the critical point $|J|=h$, where the gap is zero. In fact, the energy gap exactly varies as ($h-|J|$). Note that the ordered phase is also gapped due to the discrete (parity) symmetry of the problem. Pfeuty had also  calculated the order parameter, $p$, and spin correlation functions exactly. Notably, \[p=\left(1-\frac{h^{2}}{J^{2}} \right)^{\frac{1}{8}}\] for $|J|>h$ in this soluble problem. For two spins sitting very far from each other inside the bulk of the chain, the spin-spin correlation is $$\lim_{n\rightarrow \infty} \langle \sigma^x_i \sigma^x_{i+n} \rangle \sim p^2$$ in the ordered phase, where $i$ and $i+n$ are two sites in the bulk. In the paramagnetic phase, this spin-spin correlation is zero. Another exact result, that we will use very prominently  in our calculations, is the  end-to-end spin-spin correlation 
\begin{equation}
\rho^{x}_{1,L}=p^{8}+\mathcal{O}(1/L)
\label{eqn:end2end}
\end{equation} on an open chain. Importantly, we realized that this behavior of the end-to-end correlation is actually a signature of the Majorana edge modes that arise in the ordered phase of the QI chain, and we used it effectively to investigate the edge modes in a few different models discussed in this thesis.

\subsection{Majorana edge modes{\index{Majorana Fermion}}}
There has always been a great interest in finding the Majorana fermions in nature~\cite{marcel-race4majorana}. The developments in quantum computation and condensed matter, in particular the proposed role of Majorana fermions in the error-free (topological) quantum computation~\cite{Kitaev.FreeMajorana,Nayak.RMP,Topo_QC_Kitaev}, have further invigorated their search~\cite{das-nature2012,Mourik.Majorana}. Notably, for the 1d quantum Ising model in the fermionized form [Eq.~\eqref{eqn:qi-fermion}],  Kitaev made a remarkable observation that the two ends of an open chain carry a Majorana fermion each~\cite{Kitaev.FreeMajorana}. A Majorana fermion is its own Hermitian (that is, an anti-particle of itself)~\cite{Kitaev.FreeMajorana, Wilczek.Majorana, note.Majorana}, and appears as a zero energy mode in the Kitaev/QI problem.

To see the occurrence of Majorana edge modes in the QI chain, let us rewrite the fermion operators in terms of the Majorana fermions. Please note that a pair of Majorana fermion operators, $\psi_{l}$ and $\phi_{l}$, can describe the spinless fermion on the $l^{th}$ site in the following manner. 
\begin{eqnarray}
&& \phi_{l} = \chat^{\dagger}_{l} + \chat_{l} \qquad i \psi_{l} = \chat^{\dagger}_{l} - \chat_{l} \\
{\mbox{or, conversely}} \nonumber
&& \chat^{\dagger}_{l} = {\frac{1}{2}} ( \phi_{l} + i \psi_{l}) \qquad \chat^{}_{l} = {\frac{1}{2}} ( \phi_{l} - i \psi_{l}) {\label{eq:majo}}
\end{eqnarray}
A Majorana fermion can be viewed as a linear superposition of a particle and its hole with equal weights. The $\phi_{l}$ and $\psi_{l}$ are all Hermitian, and they anti-commute among themselves and with all the other fermions. Moreover,  $\phi_{l}^{2}=\psi_{l}^{2}=1$. 
\begin{figure}[htbp] 
\centering
\includegraphics[width=0.8\textwidth]{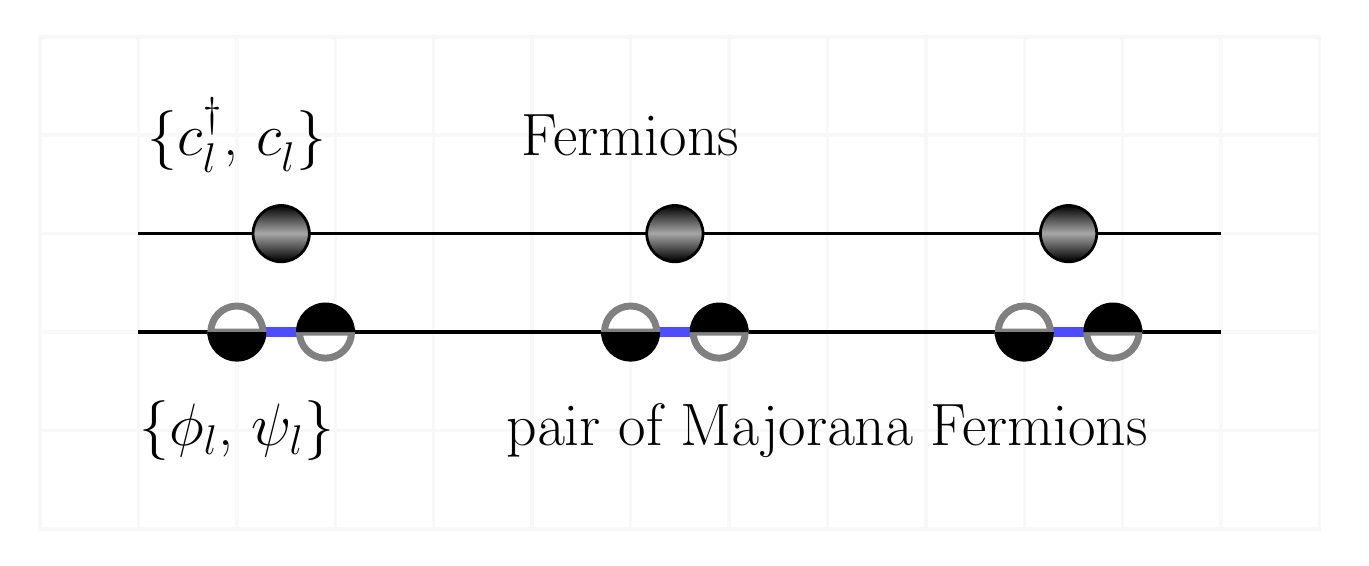} 
\caption{A fermion is consists of two Majorana fermions. Here, the filled circles denote fermions, while the half-filled circles schematically represent Majorana fermions. It is to indicate that a Majorana fermion is linear combination of fermionic creation and annihilation operators, or  that it is `half' of a full fermion.}
\label{fig:majorana}
\end{figure}

The nearest neighbor QI model (see Eqs.~(\ref{eqn:qi}) and (\ref{eqn:qi-fermion})) on an open chain takes the following form in Majorana fermion language. 
\begin{eqnarray}
H_{QI} &=& J_{} ~ {\sum_{l=1}^{L-1}} i \psi_{l} \phi_{l+1} + h ~ {\sum_{l=1}^{L}} i\psi_{l} \phi_{l}
\end{eqnarray}
Now, for a moment, consider $h=0$ case. In this case, the Hamiltonian in the Majorana fermion language clearly shows that $\psi_{1}$ is connected to $\phi_{2}$, $\psi_{2}$ is connected to $\phi_{3}$ etc. But $\phi_{1}$ and $\psi_{L}$ are not connected to any other Majorana operators (or to each other, as its an open chain).  Thus, there are two Majorana modes at the free ends of the chain. However, even when $h\neq 0$, Kitaev (and Pfeuty) still found the edge modes in the ordered phase, but with a finite decay into the bulk.



\section{Cavity-QED systems}
The term cavity-QED{\index{cavity-QED}}, or quantum cavity, generically refers to the interacting matter-radiation problems inside an enclosure (cavity) with quantized radiation. Since an important part of this thesis is concerned with studying the Rabi lattice (a quantum cavity array), let us briefly go through the basics of the Rabi model. 

\begin{figure}[htbp] 
\centering
\includegraphics[width=0.8\textwidth]{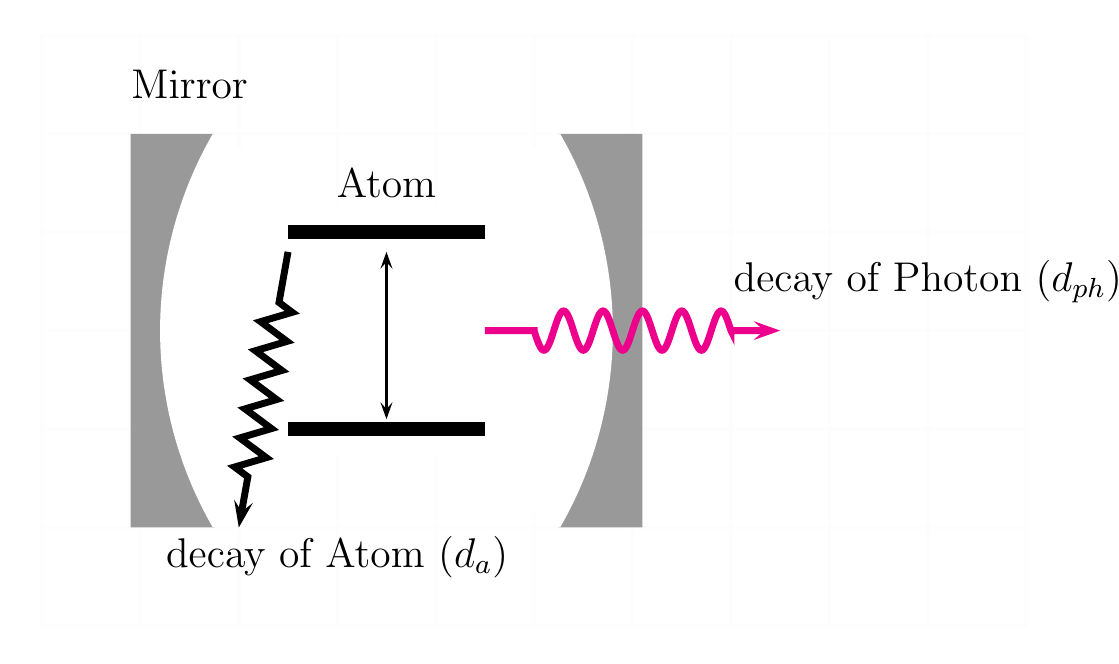}
\caption{Schematics of an cavity containing a two-level atom. 
The quality factor of a cavity is defined as $ Q = \frac{\omega}{d_{ph}}$, and the cooperative factor{\index{Cooperative factor}} as $\xi = \frac{{\gamma^{2}}}{{2 d_{a} d_{ph} }}$, where $\gamma$ is the atom-photon interaction and $\omega$ is the photon frequency.}
\label{fig:cavity}
\end{figure}

\subsection{Minimal atom-photon problem}
\def\E{\vec{E}}
\def\B{\vec{B}}
\def\a{\hat{a}}
\def\nhat{\hat{n}}
\def\chat{\hat{c}}
\subsubsection{Two-level atom}

Let us consider a `simplified' atom with only two energy levels, namely ground state $|g\big>$ with energy $\epsilon_{1}$ and excited state $|e\big>$ with energy $\epsilon_{2}$ as in Fig. {\ref{fig:atom-photon-model}}. The Hamiltonian of such an isolated two-level atom can be written as follows.
\begin{equation}
H_{atom} = \epsilon_{1} |g\big> \big< g| + \epsilon_{2} |  e \big> \big< e | {\label{h-atom}}
\end{equation}
Let us define the Pauli operators in the two-dimensional Hilbert space (HS){\index{Hilbert-Space}}{\nomenclature{HS}{Hilbert Space}} as: $\sigma^{+} = |e\rangle\langle g|$, $\sigma^{-} = |g\rangle\langle e|$ and $\sigma^z=|e\rangle\langle e| - |g\rangle\langle g|$. Moreover, $ I = |g\rangle\langle g| + |e\rangle\langle e|$ is the identity operator. Now the $H_{atom}$ can be rewritten as 
\begin{eqnarray}
H_{atom} &=& \epsilon_{1} \frac{I - \sigma^{z}_{}}{2} + \epsilon_{2} {\frac{I + \sigma^{z} }{2}} I  = {\frac{\epsilon_{2} - \epsilon_{1}}{2}} {\sigma^{z}_{}} + {\frac{\epsilon_{1} + \epsilon_{2}}{2}} I \cr
&=& {\frac{\epsilon}{2}} {\sigma^{z}} + {\mbox{constant}}, 
\end{eqnarray}
where $\epsilon = \epsilon_{2} - \epsilon_{1}$, is the energy gap between the two levels of the atom. This is also called the transition energy. The constant term does not take part in the dynamics of the system. So, it can in general be ignored.



\subsubsection{Quantized radiation}
\begin{figure}[ht] 
   \centering
   \includegraphics[width=0.8\textwidth]{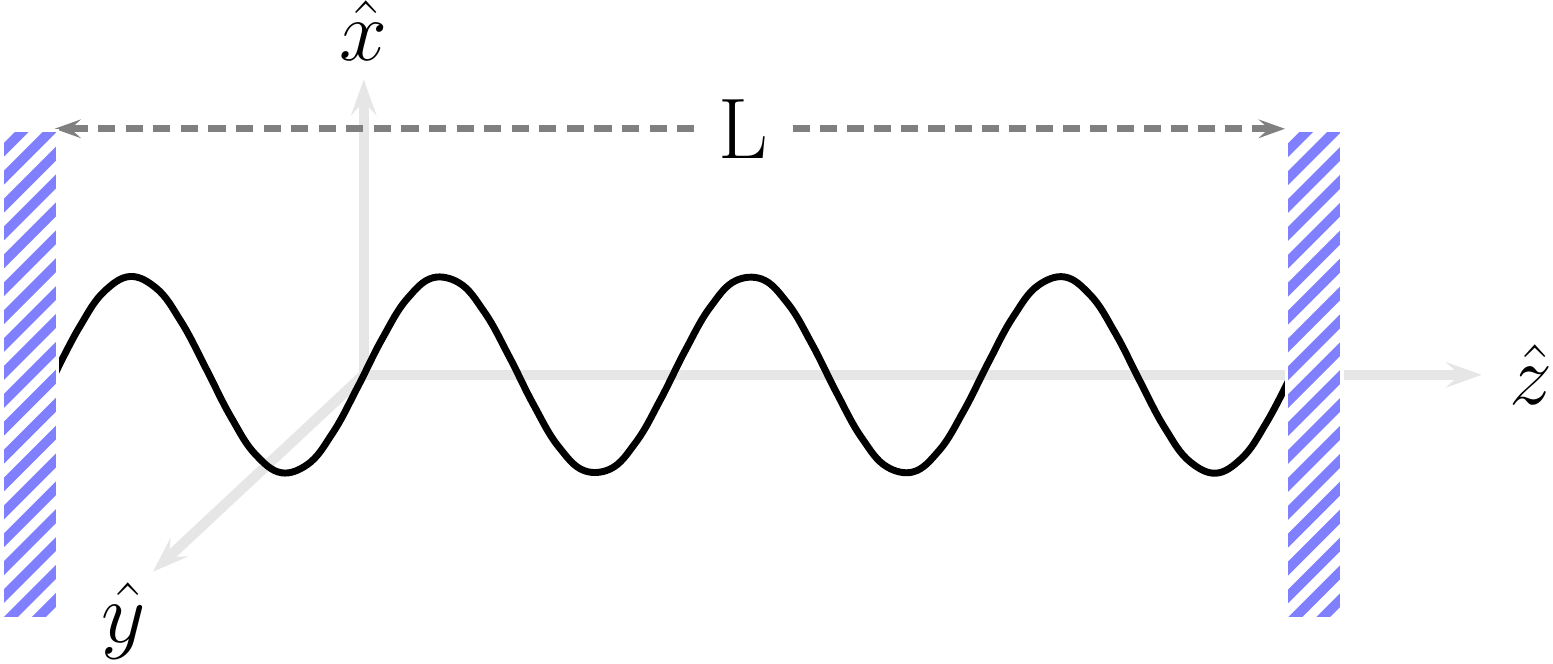} 
   \caption{Electromagnetic wave inside a cavity (resonator) of length $L$ moving along $+z$ axis and polarized along $x$ axis.}
   \label{fig:light}
\end{figure}
Here, we discuss the quantization of electromagnetic radiation in free space~\cite{Scully.Zubairy}. We know the radiation consists of electric and magnetic field. The strength of electric field is $c$ times stronger than the magnetic field, where $c$ is the velocity of the radiation. From the Maxwell's equations in free space one gets the wave equation for both electric and magnetic field. For electric field we have the wave equation
\begin{eqnarray}
\nabla^{2} {\E} = {1 \over c^{2}} {\frac{\partial^{2} \E }{\partial t^{2}}} {\label{wave-eqn}}
\end{eqnarray}
Let us solve this equation inside the cavity shown in Fig. {\ref{fig:light}}. Without loss of generality, let us consider the wave is moving along $+z$ direction and polarized along $+x$ direction. The solution of Eq. ({\ref{wave-eqn}}) then reads as
\begin{eqnarray}
{E_{x}}(z,t) = {\sum_{j}} A_{j} {q_{j}}(t) \sin(k_{j} z)
\end{eqnarray}
where $q_{j}$ is the normal mode amplitude with the dimension of length. Both $A_{j}$ and $k_{j}$ will depend on the boundary conditions. $k_{j} = {\frac{j \pi}{L}}$, with $j = 1,2,3 \dots$ and $A_{j} = \left(\frac{2 \omega_{j}^{2} m_{j}}{V \epsilon_{0}} \right)^{1/2}$ with $\omega_{j} = j \pi c/L$, the frequency of  photon. At the boundaries of the resonator electric field is zero. Thats why the solution is a sine function of $z$. Similarly, the solution for magnetic field will be
\begin{eqnarray}
{H_{y}}(z,t) = {\sum_{j}} A_{j} {\frac{\dot{q_{j}} (t) \epsilon_{0}}{k_{j}}} \cos(k_{j} z)
\end{eqnarray}
where $ \dot{q_{j}} $ is generalized velocity. We define new `oscillator' operators $\a^{\dagger}_{j}$ and $\a^{}_{j}$ where 
\begin{eqnarray}
q_{j} (t) = {\frac{1}{\sqrt{2 m_{j} \omega_{j}}}} \left( \a^{}_{j} { e^{- i \omega_{j} t}} + \a^{\dagger}_{j} { e^{ i \omega_{j} t}}  \right); \qquad {\mbox{consider $\hbar$ = 1}}
\end{eqnarray}
These canonical bosonic operators $\a^{\dagger}_{}$ and $\a^{}_{}$  are the photon creation and annihilation operator, respectively. Now, in terms of these photon operators, the electromagnetic field Hamiltonian take the following form.
\begin{eqnarray}
{\cal{H}} &=& {\frac{1}{2}} {\int_{v}} d \tau (\epsilon_{0} E_{x}^{2} + \mu_{0} H_{y}^{2}) = {\frac{1}{2}} {\sum_{j}} \left( m_{j} \omega_{j}^{2} q_{j}^{2} + \frac{p_{j}^{2}}{m_{j}} \right) \cr
&=& {\sum_{j}} \omega_{j} \left( \a^{\dagger}_{j} \a^{}_{j} + 1/2 \right)
\end{eqnarray}
Since, in experiments, one is typically concerned with a particular resonant mode of the cavity, the minimal radiation problem is just that of a single mode given by
\begin{eqnarray}
H_{photon} = \omega_{} \left( \a^{\dagger}_{} \a^{}_{} + 1/2 \right).
\label{eqn:singlemodeH}
\end{eqnarray}
For a single mode, the electric field operator can be written as 
\begin{eqnarray}
E_{x} (z,t) = {{\cal{E}}_{j}} \left( \a^{}_{j} { e^{- i \omega_{j} t}} + \a^{\dagger}_{j} { e^{ i \omega_{j} t}} \right) \sin(k_{j} z) {\label{eqn:etime}}, 
\end{eqnarray}
where ${\cal{E}}_{j} = \left( \frac{\omega_{j}}{\epsilon_{0} V} \right)^{1/2}$ has the dimension of electric field. Here, the time dependence is explicitly appears. But it naturally arises through the Heisenberg's equations of motion with respect to Eq.~(\ref{eqn:singlemodeH}). Hence, the essential time-independent form of the electric field operator is $E \sim (\a^{}_{} + \a^{\dagger}_{})$.


\subsubsection{Atom-Photon interaction{\label{sec:atom-photon}}}
\begin{figure}[htbp] 
   \centering
   \includegraphics[width=0.7\textwidth]{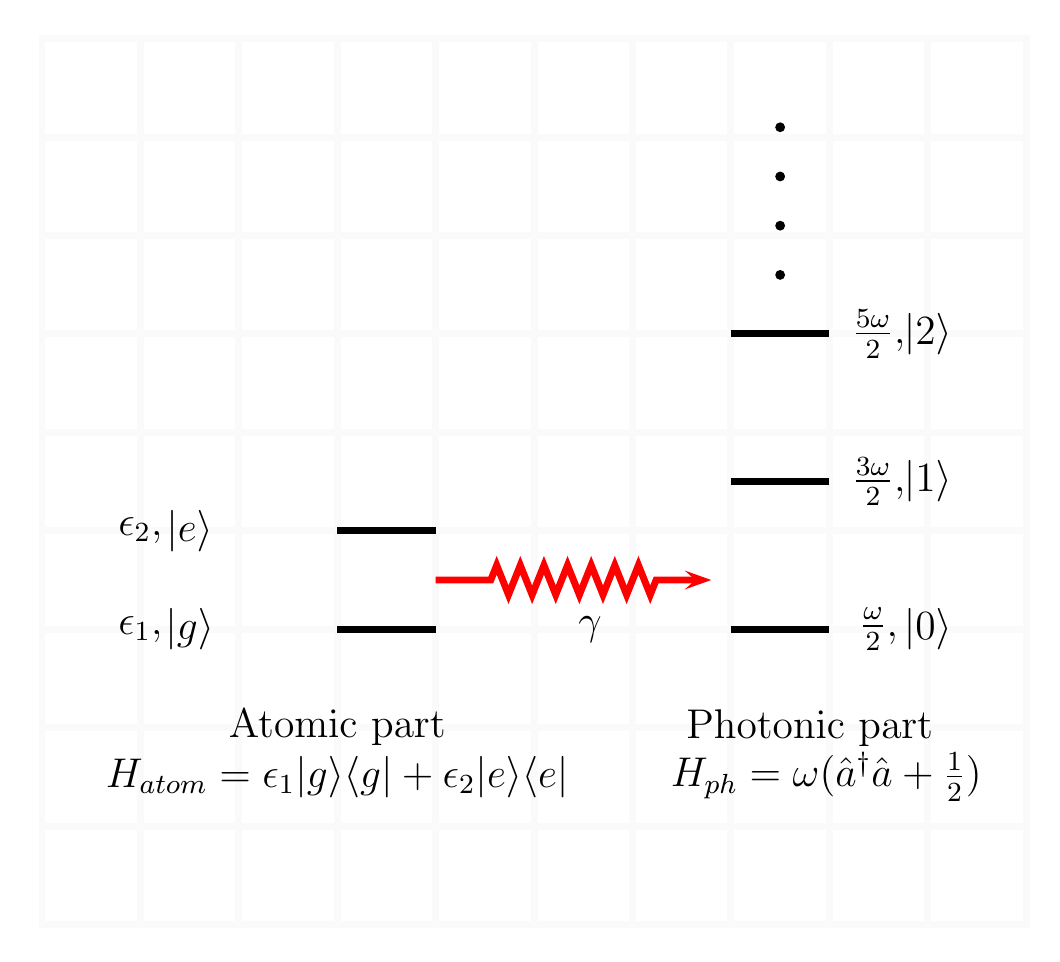} 
   \caption{Energy-levels of a two-level atom in the left side. Equally spaced  energy levels of photon in the right side. They are interacting with interaction strength $\gamma$. $\epsilon_1$ and $\epsilon_2$ are the energy of atomic levels and $\omega$ is photon energy.}
   \label{fig:atom-photon-model}
\end{figure}
In a typical quantum-optical problem, the wave length of light, say for red color, is $\sim 700$ nm, is much larger than the size of an atom (about $0.05$ nm). When an atom is exposed to such an electromagnetic radiation, the atom doesn't feel its spatial variation, and the field appears position independent ($\vec{k} \cdot \vec{r} \ll 1$). In this long-wavelength approximation, the atom-photon interaction can be written as
\begin{eqnarray}
H_{int} = e {\vec{r}} \cdot {\E} = {\vec{p}} \cdot {\E} =\gamma \sigma^{x}_{} (\a^{}_{} + \a^{\dagger}_{}).
\end{eqnarray}
This is the dipole interaction, similar to the classical dipole in an electric field. Here, the atomic dipole operator $\vec{p} \propto \sigma^{x}$, where $\sigma^{x}= |e \rangle \langle g| + |g\rangle \langle e |$. 

Now we can write the Hamiltonian of the  minimal atom-photon problem as
\begin{eqnarray}
H_{Rabi} &=& H_{atom} + H_{photon} + H_{int} \cr
&=&  {\frac{\epsilon}{2}} {\sigma^{z}} + \omega_{} \left( \a^{\dagger}_{} \a^{}_{} + 1/2 \right) + \gamma \sigma^{x}_{} (\a^{}_{} + \a^{\dagger}_{}) {\label{eqn:rabi}}
\end{eqnarray}
This is often called the Rabi model{\index{Rabi Model}}, or the Rabi quantum-cavity . The interaction term $\gamma \sigma^{x}_{} (\a^{}_{} + \a^{\dagger}_{})$ is simple looking, but the Rabi Hamiltonian has not been exactly solved yet. An easily soluble variant of this model is known as Jaynes-Cummings model (JCM)~{\cite{Jaynes.Cummings}}{\index{Jaynes-Cummings model}}, which we shall describe now. 

Consider the dipole interaction. 
\begin{eqnarray}
\sigma^{x}_{} (\a^{}_{} + \a^{\dagger}_{}) = (\sigma^{+} \a^{\dagger} + \sigma^{-} \a) + (\sigma^{+} \a + \sigma^{-} \a^{\dagger}) {\label{eqn:sxa}}
\end{eqnarray}
Here, $\sigma^{+} \a^{\dagger}$ is the process where atom goes from ground state to excited state and simultaneously creating a photon. Similarly $\sigma^{-} \a^{}$ is the process where atom comes to ground state from excited state  by absorbing one photon. These two situations are not conserving energy. Whereas the physical processes are atom goes to excited state to ground state by absorbing one photon, thus the term $\sigma^{+} \a^{}$, or emits one photon and comes to ground state from excited state, $\sigma^{-} \a^{\dagger}$. In the rotating wave approximation {\index{RWA}}(RWA{\nomenclature{RWA}{Rotating Wave Approximation}}), one can ignore the energy non-conserving terms, which simplifies the problem. 

To understand the RWA, consider the time dependence of dipole interaction with respect to the non-interacting part of $H_{Rabi}$. We know the time dependence of any operator is $${\hat{A}}(t) = e^{i H_0 t/\hbar} {\hat{A}} e^{- i H_0 t / \hbar}$$ where $H_{0}$ is unperturbed Hamiltonian. For atomic and photonic operators
\begin{eqnarray}
\frac{\partial \sigma^{+} (t)}{\partial t} = i [ H_{atom} , \sigma^{+} (t)] \qquad \frac{\partial \a^{\dagger} (t)}{\partial t} = i [ H_{photon} , \a^{\dagger} (t)]
\end{eqnarray}
Then we have $\a^{\dagger}(t) = \a^{\dagger} e^{i \omega t}$ and $\sigma^+ (t) = \sigma^+ e^{i \epsilon t}$ etc. From Eq. (\ref{eqn:sxa}) one can write the interaction term as 
\begin{eqnarray}
 H_{int} &=& \gamma [ ( \sigma^- \a^{\dagger} e^{i (\omega - \epsilon) t}  + \sigma^+ \a e^{- i (\omega - \epsilon) t )}  \cr && + ( \sigma^- \a  e^{- i (\omega+\epsilon) t} + \sigma^+ \a^{\dagger}  e^{i (\omega+\epsilon) t} ) ] 
\end{eqnarray}
When atomic transition energy $\epsilon$ is equal to the photon frequency $\omega$, then the system is at resonance. So close to resonance ($\epsilon \sim \omega $), the terms $  \sigma^- \a  e^{- i (\omega+\epsilon) t} + \sigma^+ \a^{\dagger}  e^{i (\omega+\epsilon) t}  $ rotates faster than $  \sigma^- \a^{\dagger} e^{i (\omega - \epsilon) t}  + \sigma^{+} \a e^{- i (\omega - \epsilon) t}$. And faster terms average out to be zero. This is called RWA. 
\begin{eqnarray}
H^{RWA}_{int} = \gamma (\sigma^{+} \a^{}_{} + \sigma^{-}\a^{\dagger}_{})
\end{eqnarray}
Under RWA, the Rabi Hamiltonian takes the following form.
\begin{eqnarray}
H_{JC} = {\frac{\epsilon}{2}} {\sigma^{z}} + \omega_{} \left( \a^{\dagger}_{} \a^{}_{} + 1/2 \right) + \gamma (\sigma^{+} \a^{}_{} + \sigma^{-}\a^{\dagger}_{}) {\label{eqn:jc}}
\end{eqnarray}
Hamiltonian in Eq. ({\ref{eqn:jc}}) is called Jaynes-Cummings model (JCM){\nomenclature{JCM}{Jaynes-Cummings Model}}{\index{Jaynes-Cummings model}}. This model is exactly solvable and have discrete eigenspectrum. One important thing to define  here is the polariton{\index{Polariton}} number which is the excitation for JCM. It is defined  as  
\begin{eqnarray}
n_{p} = \a^{\dagger} \a^{} + \sigma^{+}\sigma^{-} {\label{eqn:np}}.
\end{eqnarray}
The $n_p$ is conserved for JCM but not conserved for Rabi model. i.e, $ [ n_{p}, H_{JC}] = 0$, and  $ [ n_{p},H_{Rabi} ] \neq 0$

\subsection{{\label{subsec:qed}}Cavity QED array: Emulators of many-body physics}
\subsubsection{Jaynes-Cummings model viz. Bose-Hubbard model}


The lattice models of JC quantum cavities have received much attention, recently, for emulating the quantum phase diagram of Bose-Hubbard (BH) model by exhibiting Mott-insulator (MI){\nomenclature{MI}{Mott-Insulator}} to superfluid (SF){\nomenclature{SF}{Superfluid}} type quantum phase transition (QPT){\nomenclature{QPT}{Quantum Phase Transition}}{\index{quantum phase transition}} for light~\cite{Tomadin.Fazio.10,Hartmann,Greentree.06,Littlewood,Blatter}. It is hoped that the lattice models of such quantum cavities may be realized by constructing an array of cavities in a photonic band-gap material~\cite{Greentree.06,QD.Photonic}, or as a circuit-QED{\nomenclature{QED}{Quantum Electro-dynamics}} system~\cite{CircuitQED.Girvin,CircuitQED.Fink,Houck.Koch}. Such laboratory made systems will be useful in studying interesting quantum many-body problems, much like the cold atoms in optical lattices~\cite{ColdAtoms}.

\begin{figure}[ht] 
   \centering
   \includegraphics[width=0.8\textwidth]{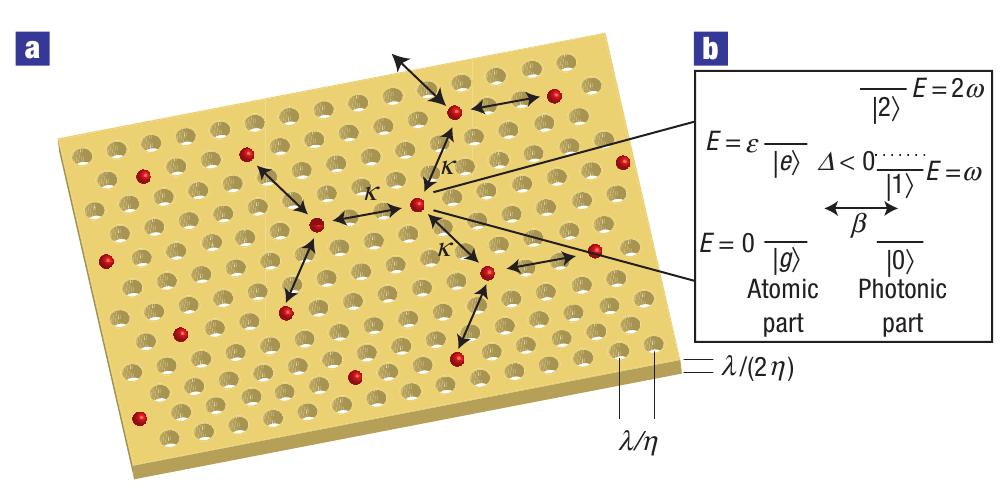} 
   \caption{A proposed cavity-QED array. Figure taken from Ref.~{\cite{Greentree.06}}. Dots are the sites where two level atom is interacting with a photon.}
   \label{fig:cavity-array}
\end{figure}

The qualitative phase diagram of Jaynes-Cummings lattice model (JCLM) and  BH model is shown in Fig.~{\ref{fig:jc-phase-bh-phase}}. Here the quantum cavities are coupled via coherent photon hopping to form a lattice. The photon disperse from cavity to cavity while the atom stay inside the particular cavity where it belongs. The Hamiltonian of JCLM is
\begin{eqnarray}
H_{JCLM} &=& \omega {\sum_l} \left( {\adag_l \a_l + {1 \over 2}} \right) + {\epsilon \over 2} {\sum_l} \sighat^{z}_l +  \gamma {\sum_l} {\left( { {\sighat^{-}_l} \adag_l + {\sighat^{+}_l} \a^{}_l} \right)} \cr && - {\frac{t}{2}} {\sum_{l, \delta}} {\left( {\adag_l \a_{l+\delta} + h.c } \right)}
\end{eqnarray}

\noindent where $\omega$ is photon frequency, $\epsilon$ atomic transition energy, $\gamma$ is atom-photon interaction strength and $t$ is photon hopping amplitude. $l$ is site index, and the $\delta$ summation is over all nearest neighbor (defined by site index  $l+\delta$). The first three terms in the above Hamiltonian are local~[see Section {\ref{sec:atom-photon}}]. The  hopping term is the only non-local term. As total polariton number, $N^{total}_p = {\sum_l} n^{}_{p,l} = {\sum_l} (\a^\dagger_l \a^{}_l + \sigma^+_l \sigma^-_l)$, is conserved quantity for this system (as $[N^{total}_p , H_{JCLM}] = 0$), one introduces chemical potential $\mu$ to control the number of polariton inside each cavity. So the grand-canonical Hamiltonian becomes $H = H_{JCLM} - \mu N^{total}_p$ and we have the quantum phase diagram for $H$ as in Fig.~{\ref{fig:jclm}}. The $y$-axis is $\tilde{\omega} = \frac{(\omega - \mu)}{\gamma}$ and $x$-axis is $\tilde{t} = \frac{t z}{\gamma}$. The lobes represent Mott insulating phase given by constant polariton number density ($n_p$). For larger hopping amplitude there is superfluid phase.  To draw this figure we have set $\epsilon = \omega$, i.e. resonance condition. If one is away from resonance, the phase diagram is qualitatively same with resized lobes. 

\begin{figure}[ht]
\centering
\subfloat[Quantum phase diagram of JCLM. The lobes are Mott-Insulating regions with fixed polariton number density ($n_{p}$). Outside the lobe is called Superfluid region (shaded). Note that the hopping amplitude ($\ttilde$) axis is in logarithmic scale.]{
\label{fig:jclm}
 \includegraphics[width=0.45\textwidth]{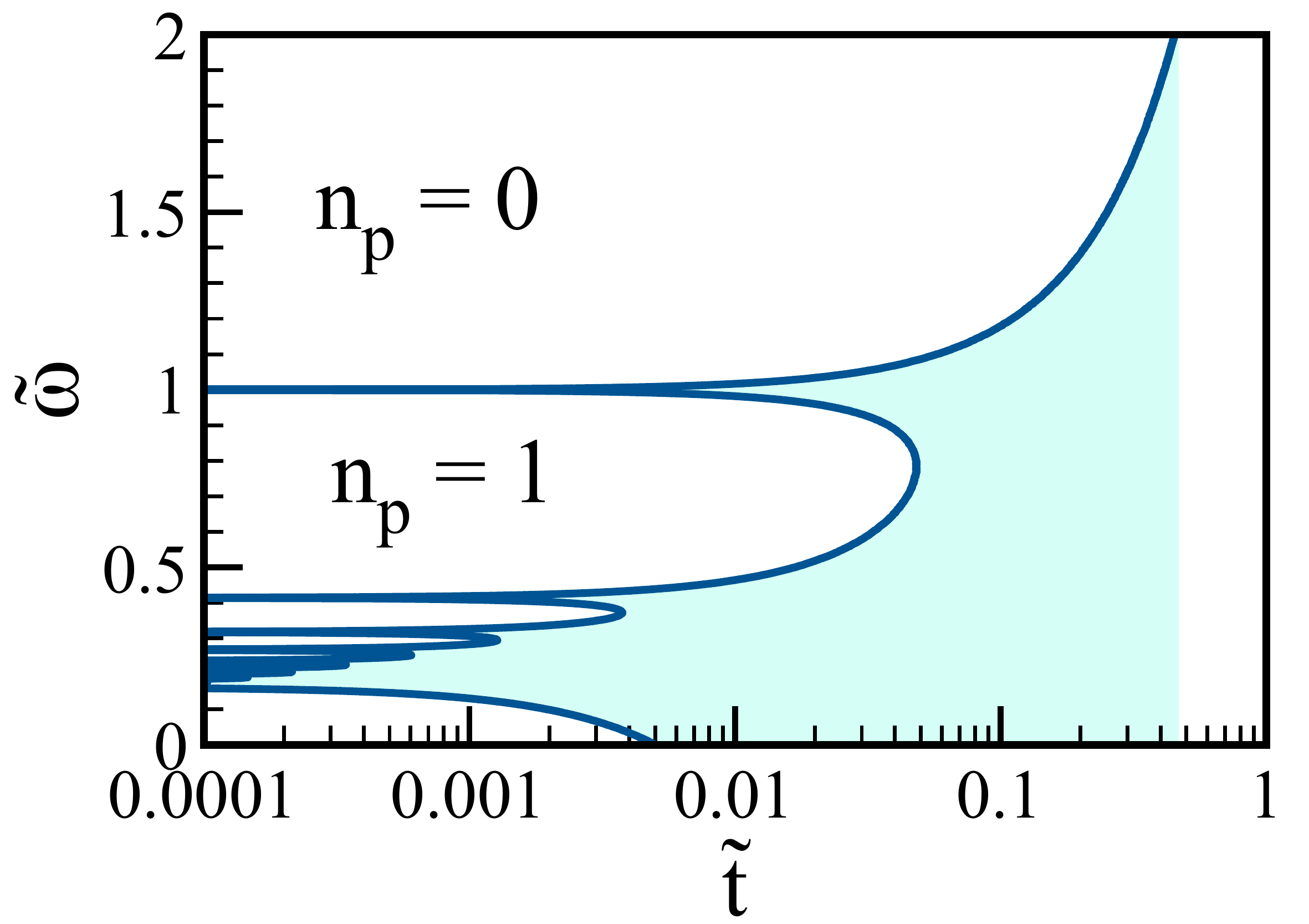} 
} \qquad 
\subfloat[Quantum phase diagram of Bose-Hubbard model. The Mott-Insulating lobes (shaded) are for constant boson number density ($n$) and outside it is the Superfluid region. The hopping amplitude axis is linear.]{
\label{fig:bh}
\includegraphics[width=0.45\textwidth]{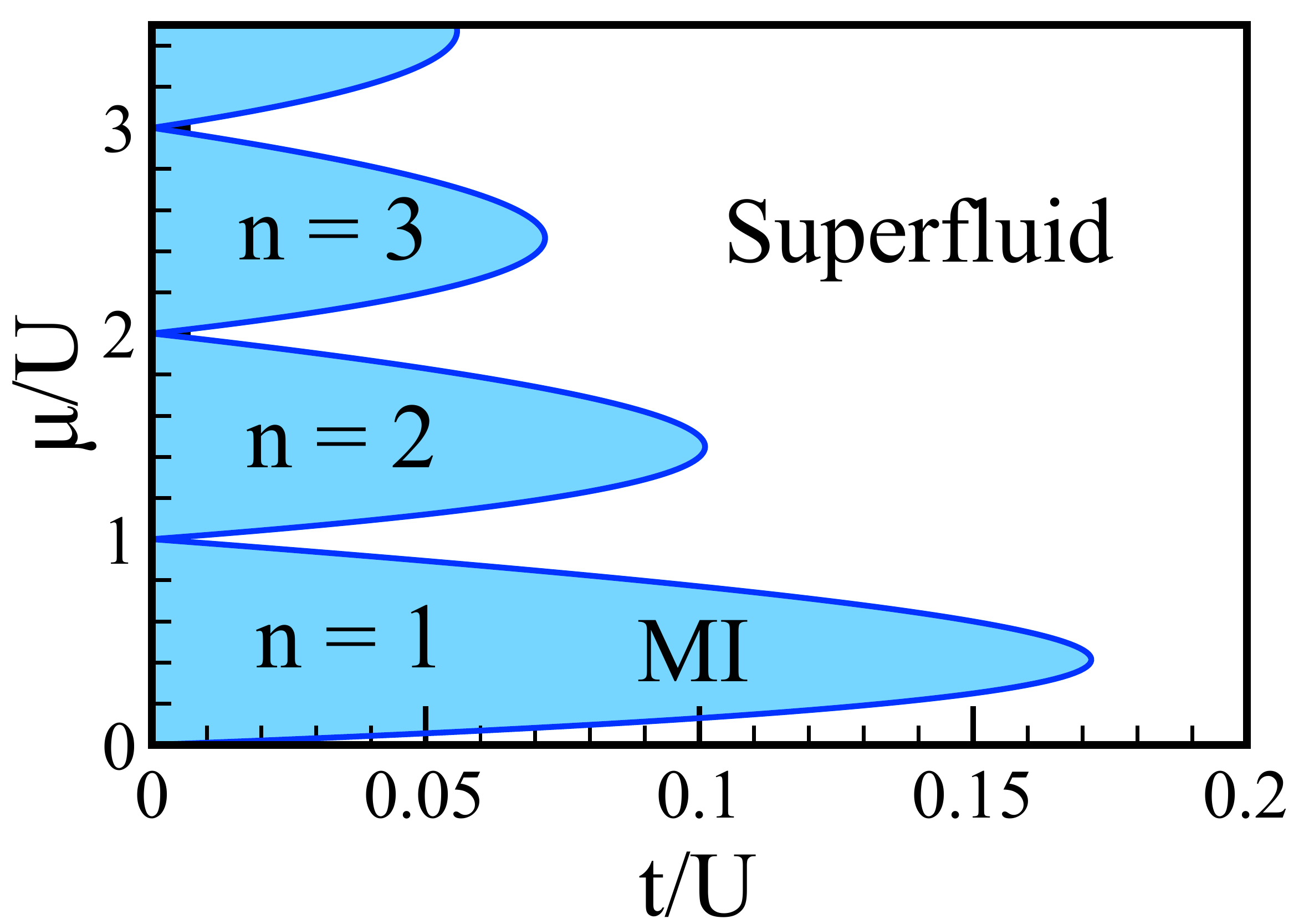}
}
\caption{Quantum phase diagrams of Jaynes-Cummings lattice and Bose-Hubbard model.}
\label{fig:jc-phase-bh-phase}
\end{figure}%

There is striking similarity of its phase diagram with that of Bose-Hubbard model. One can view the JC lattice as a problem of competition between the photon hopping and the effective photon-photon interaction generated by atom-photon interaction, similar to the the Bose-Hubbard model given by
$$H_{BH} = - t {\sum_{\langle i,j\rangle}} ( \ahat^\dagger_i \ahat^{}_{j} +H.C) + U {\sum_i} {\nhat_i (\nhat_i -1)} - \mu {\sum_i} \nhat_i,$$ where
\noindent $i$ is site index, $-t$ is hopping amplitude and the summation is over nearest-neighbor sites. $U$ is onsite repulsion strength, $\mu$ is chemical potential and $n_i$ is number of bosons on $i^{th}$ site. The phase diagram is shown in Fig.~{\ref{fig:bh}} on $\frac{t}{U}$-$\frac{\mu}{U}$ plane. Here also we get lobe like feature, called Mott-Insulating lobes, for constant boson number per site and a superfluid phase for high enough hopping ($t$) strength, where the bosons are being delocalized from one-site to another. The quantum phase transition is because of the competing interaction between the boson hopping and onsite repulsion term.


\section{Overview of the thesis}

After having introduced the basics of quantum Ising model and cavity-QED systems, we now present a brief overview of the work presented in this thesis.

In Chapter-2, we discuss some methods of calculations used for the work presented in the following chapters. In particular, we describe DMRG (density matrix renormalization group), Exact diagonalization and classical Monte Carlo methods. These are well known numerical techniques, and very useful to do simulation work in theoretical condensed matter physics. We also briefly describe a mean-field approach based on the Schwinger boson representation of spins. 

In Chapter-3, we present our calculations and findings on the Rabi lattice model. There, we systematically show that the quantum phase transition in Rabi lattice follows quantum Ising dynamics. This statement is rigorously shown in the limit of strong atom-photon coupling. After this, we neatly discuss the occurrence of Majorana-like edge modes in one-dimensional Rabi lattice. In this context, we find the relation between the end-to-end dipole correlation and the polarisation [see Eq.~(\ref{eqn:end2end})] to be very useful. Thus, in this thesis, we propose that the quantum Ising model can be simulated on Rabi lattice, and through this, one can also realise Majorana edge modes in a one-dimensional Rabi lattice with open boundaries~\cite{bkumar.somenath}.  

There is a general worry that the Majorana fermion modes in a quantum Ising system are very sensitive to local physical perturbations. This is unlike the Majorana fermions in an actual fermi system (Kitaev chain). For example, a longitudinal field acts unfavourably upon the Majorana edge modes in a QI chain. Motivated by these concerns, in Chapter-3, we also study the effects of the longitudinal field (random as well as uniform) on the edge modes in the nearest-neighbour quantum Ising chain, and discuss the conditions which seem to help the edge modes under this adverse perturbation~\cite{bkumar.somenath}. We find that, despite no topological protection, these edge modes do have a chance of survival on energetic grounds.

Carrying on with our studies on the stability of the Majorana modes in QI chain against longitudinal operators, we further investigate the edge modes in a frustrated QI chain in Chapter-4. There, we numerically study the one-dimensional $J_{1}$-$J_{2}$ quantum Ising model~\cite{J1J2QI}, using DMRG, cluster mean-field theory and fermionic mean-field theory. Here, $J_{1}$ is the nearest and $J_{2}$ is the next-nearest neighbour interaction. Its quantum phase diagram is discussed in the $J_{1}$-$J_{2}$ plane. It has two ordered regions, one of which has simple ferromagnetic or N\'eel order depending upon the sign of $J_{1}$, and the other is the double-staggered antiferromagnetic phase. These ordered phases are separated by a quantum disordered region, consisting of a gapped and a gapless (critical) quantum paramagnetic phases. The occurrence of the Majorana-like edge-modes in the ordered phases is investigated by calculating the end-to-end spin-spin correlations on open chains. We find that the ordered phases indeed support edge modes despite frustration. While the ferromagnetic/N\'eel phase support two edge modes, the double-staggered phase supports four edge modes. The fermion mean-field theory, however, seem to suggest that out of these four edge modes, two may be the zero energy Majorana type.

In the last chapter of this thesis, that is Chapter-5, we present some calculations from our ongoing work on the Falicov-Kimbal model for electrons~\cite{fkmodel}. It is a model in which the electrons of one spin state ($\uparrow$ or $\downarrow$) hop with amplitude $t$, while the others with opposite spin state don't hop. In addition to this spin-dependent hopping, it also has a local Hubbard interaction, $U$. This is also a quantum Ising problem in the sense that while in the limit $U/t\gg1$ at half-filling, this model behaves as the classical Ising model, but for any finite $U$, it always has quantum mechanical charge fluctuations (instead of spin fluctuations due to transverse field). Our objective is to understand the effect of charge fluctuations on the standard Ising thermodynamics and other properties. We present some results from the Monte Carlo simulations on square lattice, and note the deviations from the known results for the classical Ising model.


\clearpage
\bibliographystyle{unsrt}
\bibliography{chapters/ref-all}

%% file: chapters/numerical-methods.tex
\def\adag{{\hat{a}}^{\dagger}}
\def\a{\hat{a}}
\def\Hhat{\hat{H}}
\def\U{\hat{U}}
\def\Udag{{\hat{U}}^\dagger}
\def\s{\sigma}
\def\r{\rho}
\def\ahat{\hat{a}}
\def\nhat{\hat{n}}
\def\chat{\hat{c}}
\def\fhat{\hat{f}}
\def\dhat{\hat{d}}
\def\psihat{\hat{\psi}}
\def\phihat{\hat{\phi}}
\def\Qhat{\hat{Q}}
\def\ntilde{\tilde{n}}
\def\shat{\hat{\sigma}}
\def\htilde{\tilde{h}}
\def\atilde{\tilde{A}}

\definecolor{somu1}{RGB}{15,16,57}
\definecolor{somu2}{RGB}{65,120,250}
\definecolor{somu3}{RGB}{234,122,23}
\definecolor{somu4}{RGB}{17,172,38}
\definecolor{somubg}{RGB}{255,255,230}

\lstset{language=Fortran, 
keywordstyle=\color{somu2},
stringstyle=\color{somu3}, 
commentstyle=\color{somu4}, 
backgroundcolor=\color{somubg},
}




\chapter[Methods of calculations]{\label{sec:dmrg-ed} Methods of calculations} 
\begin{center}
\parbox{
0.8\textwidth}{\footnotesize
{\bf{\small About this chapter}}
\\[10pt]
We briefly describe different useful numerical and analytical calculations which have been used to study the problems described in this thesis. The numerical tools are density matrix renormalization group (DMRG) method, exact diagonalization (ED) techniques for spin systems, and classical Monte Carlo simulation. We also describe a mean-field theory based on Schwinger boson representation for spins.
\\[15pt]
}

\end{center}

\minitoc

\section{{\label{sec:dmrg}}Density Matrix Renormalization Group}

The Density Matrix Renormalization Group{\index{DMRG}}, aka DMRG, is the best method to study the ground state, or a few low energy excited states, of the one-dimensional quantum systems~{\cite{srw-noack-prl1992,srw-prl1992,srw-prb1993}}. It helps in accessing large system sizes. Here, we discuss the basic idea of DMRG calculations for any general one dimensional problem. In our present work, we have implemented this algorithm to a quantum Ising chain with nearest and next-nearest neighbor interaction, and also on Rabi lattice model which consists of spins-$1/2$ objects (two-level atom) and bosons (photon). The DMRG techniques has become one of the best methods to study low dimensional systems and there are lots of recent articles~{\cite{Schollwock2005,chiara, noack}} which help us to learn  the DMRG.

Below we point out the main ideas and algorithmic steps for DMRG
  
\begin{enumerate}
\item A general Hamiltonian of the following form is taken 
\begin{eqnarray}
H = {\sum^{L}_{l=1}} h_l + {\sum_{l=1}^{L-1}} A_l A_{l+1}
\end{eqnarray}
The first term is local term and the second term is interaction between nearest neighbor sites. $L$ is total number of sites on a one-dimensional open chain. 
We have to find the eigenvalue and ground state eigenvector of this large matrix to find any physical quantities of the system. 
\item To diagonalize the $H$ matrix of size $n^L \times n^{L}$, where $n$ is the dimension of local operator, we need $n^{2L}$ bits of computer memory. For example Heisenberg model of spin-$1/2$ particles of total 20 sites will take $2^{20} \times 2^{20}$ bits of computer RAM which is approximately 128 GB.

\item Let us fix the convention to define a matrix. $[A_{l}]_{m_{1} \times m_{2}}$ 
means the matrix form of $l^{th}$ site operator $A_{l}$ has $m_{1}$ number of rows and $m_{2}$ number of columns. i.e., $A_{l}$ is a $m_{1} \times m_{2}$ matrix.
\begin{figure}[htbp] 
   \centering
   \includegraphics[width=0.6\textwidth]{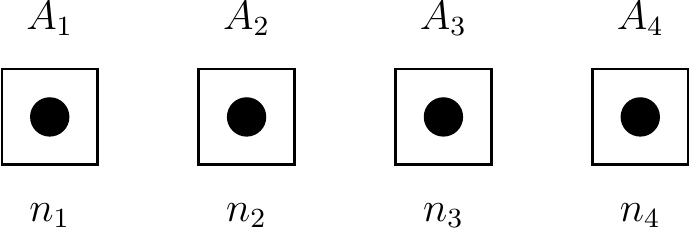} 
   \label{fig:4sites}
\end{figure}
\item {\label{item:makesb}}Let us take 4 sites at first. The model Hamiltonian becomes 
\begin{eqnarray}
H = h_1 + h_2 + h_3 + h_4 + A_1 A_2 + A_2 A_3 + A_3 A_4 {\label{eqn:h4site}}
\end{eqnarray}
This $4$ sites make the superblock. $h_{l}$ is the local Hamiltonian terms for the $l^{th}$ site, and $A_{l}$ is the local operator. Matrix dimension of each site is $n_{1} \times n_{1}$, $n_{2} \times n_{2}$, $n_{3} \times n_{3}$, and $n_{4} \times n_{4}$. That sets the linear dimension of superblock $n_{super} = n_{1} n_{2} n_{3} n_{4} $.

\item In matrix form this should be written as 
\begin{eqnarray}
H_{\mbox{superblock}} = h_1 \otimes I_{2} \otimes I_{3} \otimes I_{4}   +  I_{1} \otimes h_{2} \otimes I_{3} \otimes I_{4} \cr 
+ I_{1} \otimes I_{2} \otimes h_{3} \otimes I_{4} + I_{1} \otimes I_{2} \otimes I_{3} \otimes h_{4} \cr
+ A_{1} \otimes A_{2} \otimes I_{3} \otimes I_{4} + I_{1} \otimes A_{2} \otimes A_{3} \otimes I_{4} \cr
+ I_{1} \otimes I_{2} \otimes A_{3} \otimes A_{4} 
\end{eqnarray}
Above $I_{l}$ is identity matrix of dimension $n_{l} \times n_{l}$
\item $H_{superblock}$ is a matrix of dimension $n_{super} \times n_{super}$
\item We solve the full Hamiltonian or superblock of Eq.~(\ref{eqn:h4site}), which contains $4$ sites, and find the ground state ($|\psi_0\rangle$) only. This process takes time and memory. So we have used Davidson (or one can use Lanczose algorithm also. These methods are iterative diagonalization methods, which takes less memory and time) algorithm to find only the ground state, i.e., lowest energy eigenstate. We did not use standard brute force Linear Algebra PACKage (LAPACK) subroutines to diagonalize the superblock, as this is much slower compare to Davisdon~{\cite{saad,davidson}}. But Davidson subroutine make use of the LAPACK{\nomenclature{LAPACK}{Linear Algebra PACKage}} subroutines internally.

\item So we have found the ground state ($|\psi_0\rangle$) of the superblock of 4 sites. This $|\psi_0\rangle$ is a one dimension array or column vector with dimension $n_{super} \times 1$. Construct the density matrix from this ground state. Full density matrix is defined by
\begin{eqnarray}
\rho = |\psi_0\rangle \langle \psi_0 | {\label{eqn:rhofull}}
\end{eqnarray}
which means the system is precisely at the ground state with probability 1.

\item $\rho$ is the full density matrix of the superblock (4 sites here) and this is a square matrix,  $\rho = [\rho]_{n_{super} \times n_{super}}$.

\item Now let us make two blocks (shaded part in the figure below) out of the 4 sites. Namely system block, which is consists of 1st and 2nd site. Linear dimension of the block is $n_{sys} = n_{1} n_{2} $. And environment block, taking 3rd and 4th site, with linear dimension $n_{env} = n_{3} n_{4}$.
\begin{figure}[htbp] 
   \centering
   \includegraphics[width=0.6\textwidth]{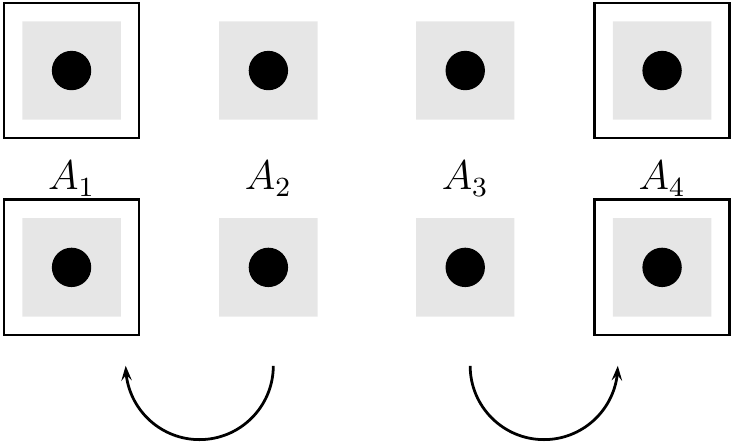} 
   \includegraphics[width=0.6\textwidth]{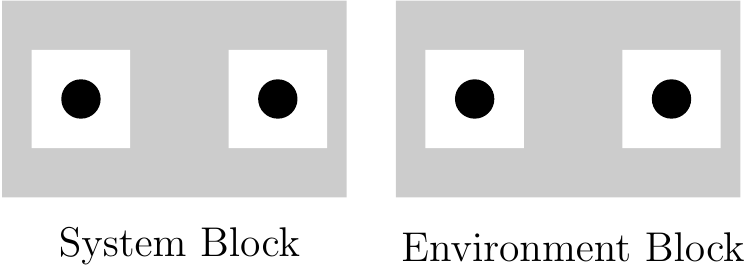}
   \label{fig:4sites}
\end{figure}

\item Any matrix in the system block will be a matrix of dimension $n_{1} n_{2} \times n_{1} n_{2}$ or $n_{sys} \times n_{sys}$. 

\item For example, Hamiltonian for system block ($1^{st}$ and $2^{nd}$ site) is 
$$ [ H_{sys} ]_{n_{sys} \times n_{sys}} = h_{1} \otimes I_{2} + I_{1} \otimes h_{2} + A_{1} A_{2}$$

\item Now find the density matrix for the system block. For this we trace out the environment part from the full density matrix $\rho$ of Eq. (\ref{eqn:rhofull}) So
\begin{eqnarray}
\rho_{sys} = Tr_{env} ~ [ \rho ]{\label{eqn:rhosys}}
\end{eqnarray}
This is the algorithm to directly find $\rho_{sys}$ from $\rho$


{\color{blue}
\begin{verbatim}
	allocate(rho_sys(ns,ns))
	do is=1,ns
		do js=1,ns
			sumrhos=0.0d0
			do ie=1,ne
!----------------------------------------------
!	 as we are tracing over environment, je=ie
!-------------------------------------------
				je=ie
				i=ie+ne*(is-1)
				j=je+ne*(js-1)
				sumrhos=sumrhos+g0(i)*g0(j)
!-------------------------------------
!	the wave function is real; so g0(i)*g0(j)^* = g0(i)*g0(j)
!---------------------------------------
			end do
			rho_sys(is,js)=sumrhos
		end do
	end do
\end{verbatim}
}

\item Similar thing is for finding $\rho_{env}$, by tracing out the system part from the full density matrix.

\item Now comes the most important part of the DMRG method, which is the renormalization part. $\rho_{sys}$ is the density matrix of the system part of  superblock and we want to extract maximum information from it. The eigenvalues of density matrix of system contains the maximum information about the system. We diagonalize $\rho_{sys}$ matrix and keep, let's say, $m$ eigenstates corresponding highest $m$ eigenvalues in descending order, thus keeping most probable states from the system matrix. We use LAPACK~{\cite{lapack}} subroutines to diagonalize $\rho_{sys}$.

\item One calculates truncation error to set the limit of convergence. The truncation error is calculated as $${\rm Error} = 1.0 - {\sum_{i=1}^{m}} {\epsilon^{\rm sys}_{i}}$$ where ${\epsilon^{\rm sys}_{i}}$ is the eigenvalues of system's density matrix $\rho_{sys}$. Obviously we increase the values of $m$ to minimize the error, let's say we get error $\sim 10^{-9}$.


\item Unitary operator $U_{sys}$ is created by arranging those $m$ eigenvectors columnwise and $U_{sys} = [ U_{sys} ]_{n_{sys} \times m}$.

\item Similarly we diagonalise $\rho_{env}$ to collect $m$ highest eigenvalues and their eigenvectors. Arranging the eigenvector columnwise, we construct $U_{env} = [ U_{env} ]_{n_{env} \times m}$ 

\item The system block, which included $1^{st}$ and $2^{nd}$ site, is diffused to create a new renormalized  $1^{st}$ site. 
\begin{figure}[htbp] 
   \centering
   \includegraphics[width=0.5\textwidth]{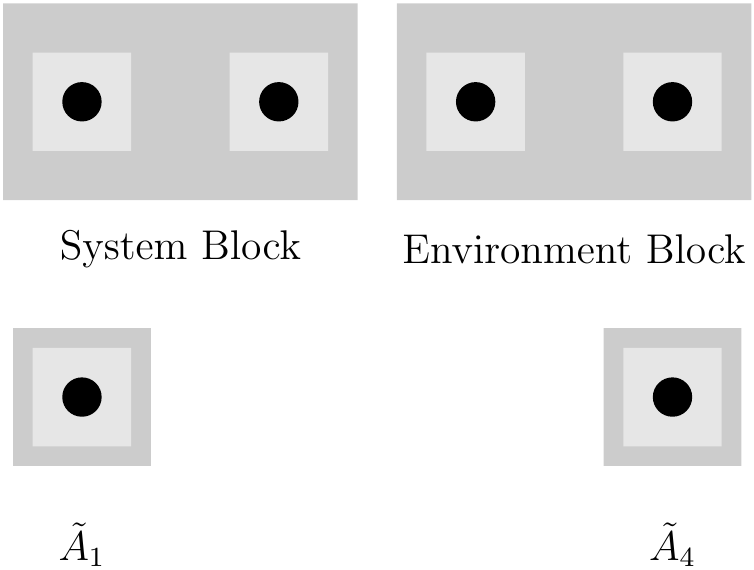} 
   \label{fig:4sites}
\end{figure}

\begin{eqnarray}
&& {\tilde{h}}_1 = {U_{sys}}^{\dagger} (H_{sys}) {U_{sys}}^{} \\
&& {\tilde{A}}_1 = {U_{sys}}^{\dagger} (I_2 \otimes A_2) {U_{sys}}^{}
\end{eqnarray}

\item Similarly $3^{rd}$ and $4^{th}$ site will make the renormalized $4^{th}$ site.

\begin{eqnarray}
&& {\tilde{h}}_4 = {U_{env}}^{\dagger} (H_{env}) {U_{env}}^{} \\
&& {\tilde{A}}_4 = {U_{env}}^{\dagger} (A_3 \otimes I_4) {U_{env}}^{}
\end{eqnarray}

\item We are left with a renormalized $1^{st}$ and $4^{th}$ sites. Bring the original un-normalized $A_2$ and $A_3$ between them. Again the lattice has now 4 operators, the new tilde operators at the end of the chain and two sites in the middle. But actually its now $6$ sites as tilde operators contain $2$ sites each.

\item Remember, ${\tilde{A}}_1$ is the created from the fact that $A_2$ is included in the system and this is ${\tilde{A}}_1$, which will now interact with newly added $A_2$. 

\begin{figure}[h] 
   \centering
   \includegraphics[width=0.5\textwidth]{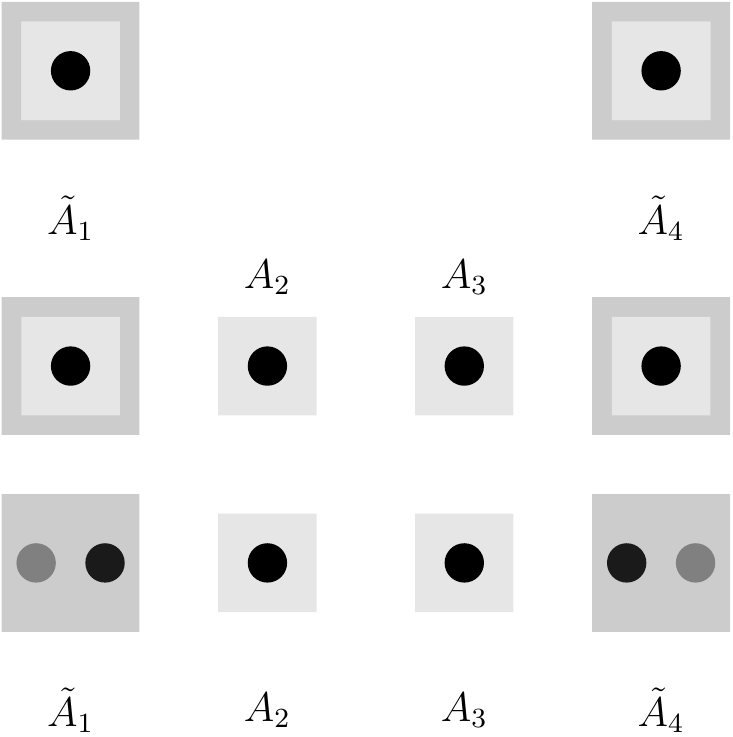}
   \label{fig:4sites}
\end{figure}

\item  Its understood that new renormalized ${\tilde{A}}_1 $ contains both the original $A_1$ and $A_2$, which are shown in the figure as light shaded and dark shaded dots.

\item The old $A_1$ is the light shaded dot inside the first block, along with dark dot ${\tilde{A}}_1 $ to the right of the block. At this stage there are 6 sites in total.

\item The size of ${\tilde{A}}_1  = [ {\tilde{A}}_1 ]_{m \times m}$, $A_2 = [ A_2 ]_{n_2 \times n_2}$, $A_3 = [A_3]_{n_3 \times n_3}$ and ${\tilde{A}}_4  = [ {\tilde{A}}_4 ]_{m \times m}$

\item This is end point of our first DMRG step. Where the superblock size has the linear dimension $n_{super} = m \times n_{2} \times  n_{3} \times m $

\item Again we go to Step. (\ref{item:makesb}). But from now on the superblock Hamiltonian will look like 

\begin{eqnarray}
H_{\mbox{new superblock}} = {\tilde{h}}_1 \otimes I_{2} \otimes I_{3} \otimes I_{4}   +  I_{1} \otimes h_{2} \otimes I_{3} \otimes I_{4} \cr 
+ I_{1} \otimes I_{2} \otimes h_{3} \otimes I_{4} + I_{1} \otimes I_{2} \otimes I_{3} \otimes \htilde_{4} \cr
+ \atilde_{1} \otimes A_{2} \otimes I_{3} \otimes I_{4} + I_{1} \otimes A_{2} \otimes A_{3} \otimes I_{4} \cr
+ I_{1} \otimes I_{2} \otimes A_{3} \otimes \atilde_{4} 
\end{eqnarray}

\item Eventually we have to solve for the ground state of this superblock Hamiltonian. One can set the size of the superblock by setting the value of $m$, i.e., number of states kept from the system's (or environment's) density matrix. 

\item We find expectation value of some physical quantities and see whether that changes with the value of $m$ being increased. There we converge it. 

\item After keeping $m$ fixed, we add system size by one site at a time in each DMRG step. Thus total number of sites increased by two sites at each DMRG step. We continue addition of sites until the above convergence criteria is satisfied.

\item This method is also called infinite site DMRG, because there is no limit of adding two sites to the chain. But eventually we settle at some large but finite length of the chain when our results (e.g. average energy per sites, average expectation value of a physical quantity etc.) converges.

\begin{figure}[htbp] 
   \centering
   \includegraphics[width=0.5\textwidth]{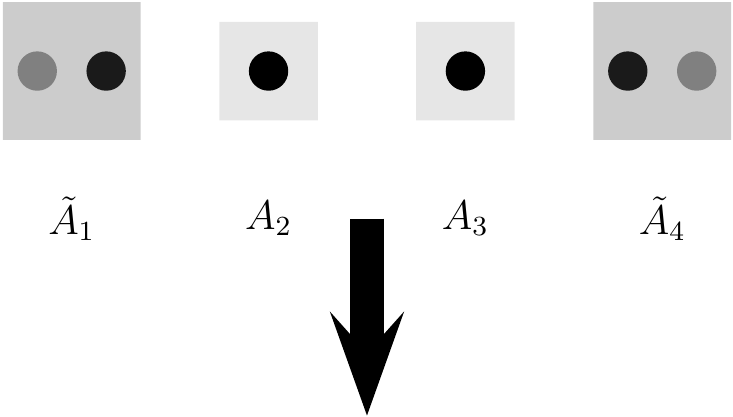}

   \includegraphics[width=0.5\textwidth]{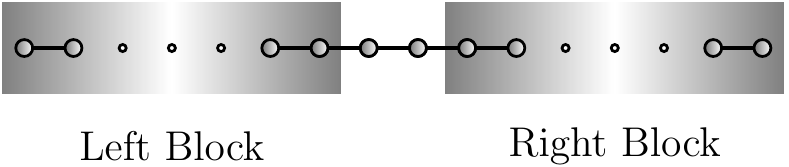}
   \label{fig:4sites}
\end{figure}

\item The final stage of the lattice will be like the above. Only renormalized local Hamiltonians $\htilde_1, h_2, h_3$ and $\htilde_4$ and operators  $\atilde_1, A_2, A_3$ and $\atilde_4$  are there for the Hamiltonian to be constructed. All other information has gone inside the block after renormalization.

\end{enumerate}


\clearpage

\def\sig{\vec{S}}
\def\ua{\uparrow}
\def\da{\downarrow}

\section{Exact Diagonalization}
Let us describe the steps for numerical exact diagonalization (ED){\index{Exact Diagonalization}} of spin system. Here, we shall discuss only the spin-1/2 Heisenberg model in one-dimension (1D). The algorithm can be used for 2D or 3D also. However, it can only be used on small clusters, as the computer memory requirement grows exponentially with the number of sites the cluster. Many well written documents are available on the web~{\cite{sandvik,lin-prb-ed}}. We also found a talk by A. Lauchli helpful~{\cite{lauchli-ed-talk}}. 

The Hamiltonian of Heisenberg model for spin-$1/2$ particles is written below.
\begin{eqnarray}
H = {\sum_{\langle i,j \rangle}} {\sig}_{i} \cdot {\sig}_{j}
\end{eqnarray}
Here, the interaction strength is taken as unity, and $\langle i,j \rangle$ means the summation is on nearest neighbor spins only. One can do ED for general interaction problem and large number of particles depending on the computational resources available. Its all about writing the basis of your problem and constructing the Hamiltonian matrix tactically to be solved by the computer. Below we shall describe the ED algorithm stepwise.

\begin{enumerate}

\item We have taken $4$ spins on an open chain. Hamiltonian for this is
\begin{eqnarray}
H_{4 - spins} = \sig_{1} \cdot \sig_{2} + \sig_{2} \cdot \sig_{3} + \sig_{3} \cdot \sig_{4} {\label{eqn:heisen}}
\end{eqnarray}

\item The dimension of the Hilbert-Space (HS) is $2^{4} = 16$. The basis states can be written as 
\begin{eqnarray}
0 - | \ua \ua \ua \ua \rangle \nonumber \\
1 - | \ua \ua \ua \da \rangle \nonumber \\
2 - | \ua \ua \da \ua \rangle \nonumber \\
3 - | \ua \ua \da \da \rangle \nonumber \\
4 - | \ua \da \ua \ua \rangle \nonumber \\
5 - | \ua \da \ua \da \rangle \nonumber \\
6 - | \ua \da \da \ua \rangle \nonumber \\
7 - | \ua \da \da \da \rangle \nonumber \\
8 - | \da \ua \ua \ua \rangle \nonumber \\
9 - | \da \ua \ua \da \rangle \nonumber \\
10 - | \da \ua \da \ua \rangle \nonumber \\
11 - | \da \ua \da \da \rangle \nonumber \\
12 - | \da \da \ua \ua \rangle \nonumber \\
13 - | \da \da \ua \da \rangle \nonumber \\
14 - | \da \da \da \ua \rangle \nonumber \\
15 - | \da \da \da \da \rangle \nonumber
\end{eqnarray}

\item Define the $\ua$ as binary $0$ and $\da$ as binary $1$. Then we can write these states as binary representation. E.g. $ \ua \da \ua \da $ can be seen as $0 ~1 ~ 0 ~ 1 $

\item We label this states as $0, 1, 2, \dots , 15~(2^{4}-1)$ by just transforming the binary to decimal number of those states.

\item Hamiltonian of Eq.~(\ref{eqn:heisen}) should be applied on these states and this will produce linear combination of states inside this HS as for example
\begin{eqnarray}
H_{4 - spins}   | \ua \da \ua \ua \rangle  = | \ua \da \ua \ua \rangle  + {\frac{1}{2}}   | \da \ua \ua \ua \rangle + {\frac{1}{2}}  | \ua \ua \da \ua \rangle
\end{eqnarray}

\item In our example of $4$ spins on an open chain, the Hamiltonian is applied on the all 16 states stated above. Now we shall keep track of the states created by application of the Hamiltonian, and thus we construct the Hamiltonian matrix of size $16 \times 16$.

\item To find expectation value of any operator, we have to write that operator in the above specified basis, and this will also be a $16 \times 16$ matrix  on this HS. 

\item One solves the Hamiltonian matrix for its eigenvalues and eigenvectors and finds the ground state. Then we can calculate the expectation value of any physical quantities in that ground state.  

\item One is mostly interested in the ground state of the Hamiltonian. And if the Hamiltonian has any conserved quantity, we can use that fact to construct the basis of the HS, to reduce the dimension of it. As for example, in the Heisenberg model, total spin along $z$ axis ($S^{z}_{\rm total}$) is a conserved quantity $[S^{z}_{\rm total}, H] = 0$. So we write the basis in $S^{z}_{\rm total} = 0$ sector of the HS. 

\item Why $S^{z}_{\rm total} = 0$ sector? For four spin-$1/2$ objects, $S_{\rm total}$ is $0, 1, 2$ and all of them has $S^{z} = 0$ part. So $S^{z}_{\rm total} = 0$ has all the information from the higher $S_{\rm total}$ sectors also.

\item The above statement is very important. That symmetry reduces the dimensions of HS from $16$ to $6$ for our example of 4-sites Heisenberg model of spin-$1/2$ objects.

\item The basis states can be written as 
\begin{eqnarray}
3 - | \ua \ua \da \da \rangle \nonumber \\
5 - | \ua \da \ua \da \rangle \nonumber \\
6 - | \ua \da \da \ua \rangle \nonumber \\
9 - | \da \ua \ua \da \rangle \nonumber \\
10 - | \da \ua \da \ua \rangle \nonumber \\
12 - | \da \da \ua \ua \rangle \nonumber \\
\end{eqnarray}
and here half of the spins are $\ua$ and other half are $\da$.

\item Next the Hamiltonian matrix is constructed by tracking the states created by the Hamiltonian after application on the above $6$ states. Likewise we construct any operator on this basis states.

\item One can see the matrices created are sparse in nature. That mean, there are few non-zero elements in the matrix and most of the elements are zero. One must use suitable methods for matrix diagonalization, to find the eigenstates of the sparse Hamiltonian. 

\item To find the expectation values, one needs fast matrix multiplication subroutines for multiplying sparse matrices with vectors.

\end{enumerate}



\section{Classical Monte Carlo Simulation}
\subsection*{Classical Ising model in two dimension}
The classical Monte Carlo {\index{Monte Carlo simulation}} (MC) is one of the oldest numerical techniques that has been ever useful in studying different physical problems. In this Section we shall write about MC simulation for a two-dimensional classical Ising model~{\cite{krauth,binder}}. We have taken two-dimensional (2D) lattice with periodic boundary condition (PBC). The model is described by the following Hamiltonian without any external magnetic field.

\begin{eqnarray}
H = - {\frac{J}{2}} {\sum_{\langle i,j \rangle}} \sigma_{i} \sigma_{j}
\end{eqnarray}
where $\sigma_{i}$ can take values $+1$ or $-1$, as they are classical spins. The $j$ index summation goes over all the nearest neighbors spin of $i^{th}$ site. Their interaction strength is taken to be $J = 1$ and interaction is only on the nearest neighbor spins. We will study the finite temperature physics of this model by means of classical Monte Carlo (MC) simulation and find the temperature dependence of some physical quantities, e.g. specific heat, susceptibility and magnetization. 2D classical Ising model undergoes a second order phase transition at a critical temperature $T_c$. The critical temperature can be determined  by looking at the discontinuity of specific heat and susceptibility for that particular temperature. To find $T_c$ more systematically  we shall calculate the Binder Cumulant as a function of temperature for different lattice size. The main motivation of this part of calculation is to discuss the classical MC simulation for 2D classical Ising model and reproduce well-known results, 
and find the critical exponents. One important comment: The critical exponents for 2D classical Ising model are same as the critical exponents for 1D quantum Ising model.

\subsection*{\em Algorithm of Classical Monte-Carlo simulation}
\subsection*{Create 2D lattice with PBC}
\begin{figure}[htbp] 
  \centering
  \includegraphics[width=0.7\textwidth]{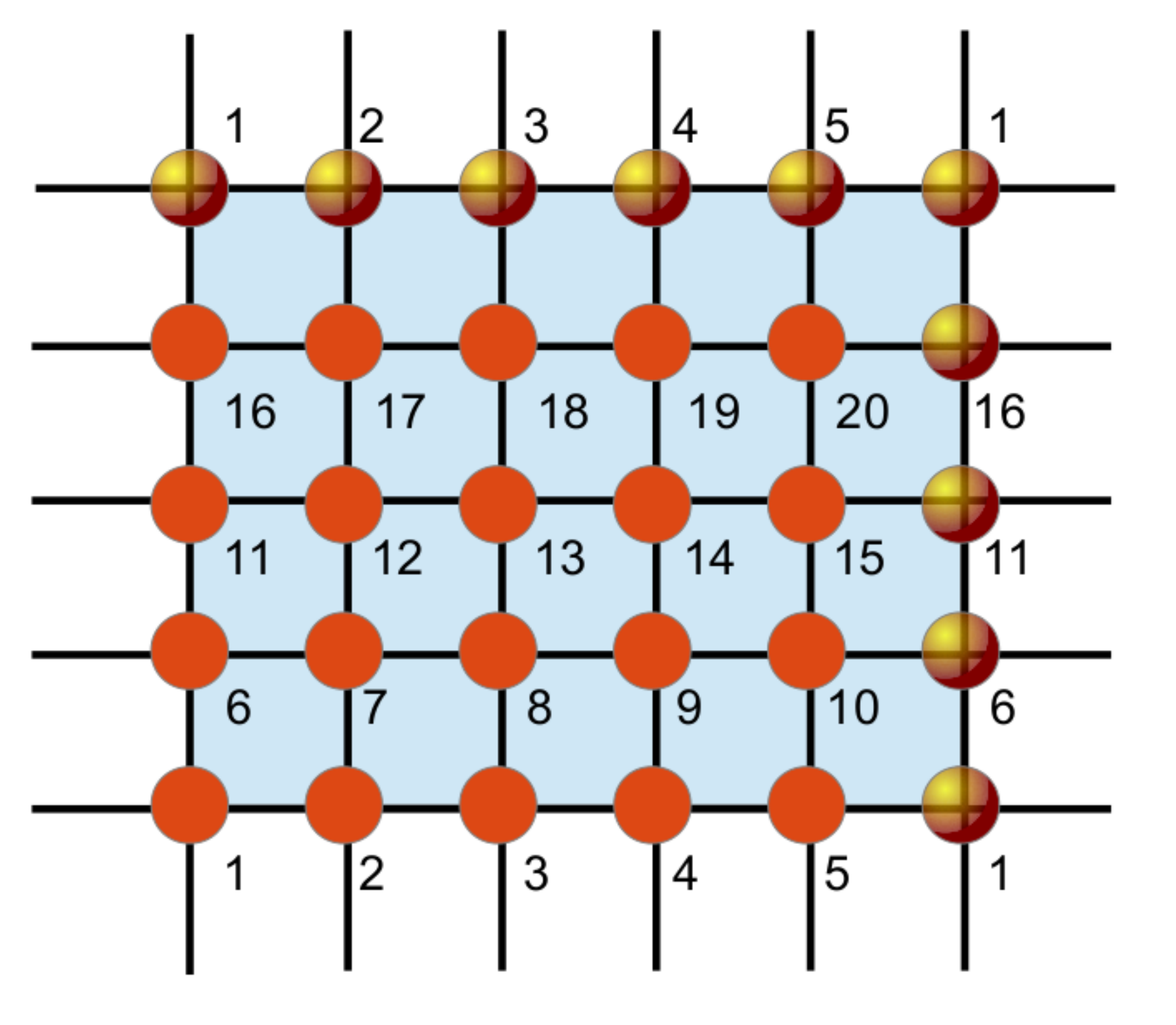} 
  \caption{A 2D $5 \times 4$ lattice with periodic boundary condition.}
  \label{fig:example}
\end{figure}
Here is the algorithm to create the 2D lattice with initial energy and magnetization conditions.
{\color{dg}
\begin{verbatim}
! E = total energy of the system
! m = sum of all spins
! spin(x,y) = value of spin at (x,y) Cartesian coordinate
! up    = (x,y+1) site of the lattice
! right = (x+1,y) site of the lattice
! Lx = number of sites in x direction
! Ly = number of sites in y direction
! spin(1:Lx,1:Ly)

	E=0.0d0
	m=0.0d0
	spin = 1.0d0
	
	do y=1,Ly
		if (y == Ly)then
			up = 1
		else
			up = y + 1
		end if
		do x=1,Lx
			if (x == Lx)then
				right = 1
			else
				right = x + 1
			end if
			neighbor_sum = spin(x,up) + spin(right,y)
			E = E - spin(x,y)*neighbor_sum
			m = m + spin(x,y)
		end do
	end do
\end{verbatim}
}
\begin{figure}[hp] 
   \centering
   \includegraphics[width=0.98\textwidth]{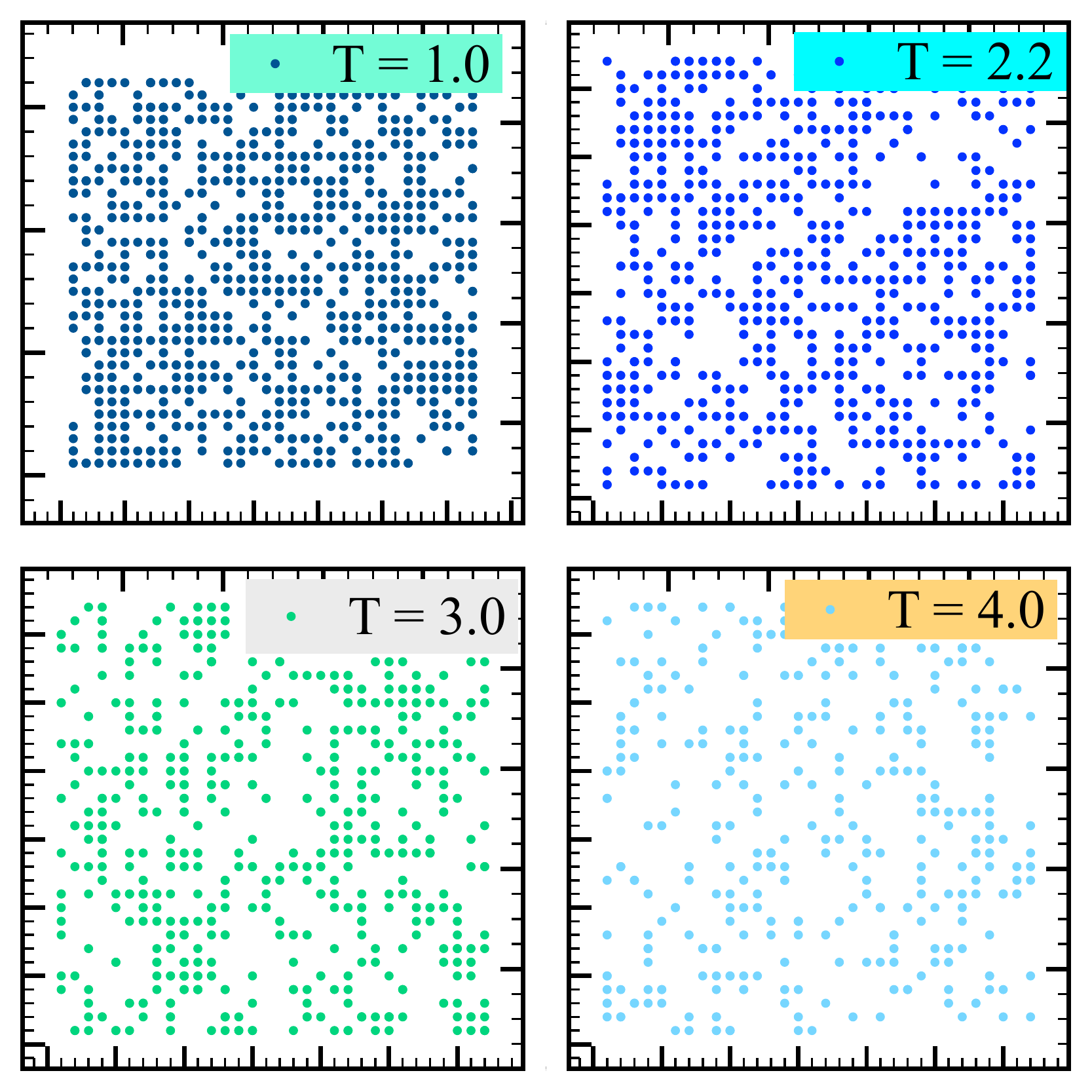} 
   \caption{This is the snapshot of the spin configurations for different temperature value. The lattice size is $32 \times 32$ with $10,000$ Monte-Carlo steps. The dots represents the value of the spin $\sigma_i$ to be $+1$, white space is for $-1$. The initial configuration of the spins was all $+1$. }
   \label{fig:example}
\end{figure}

\subsection*{Steps to follow}

\begin{enumerate}
\item Set up all the spins to have the value $+1$ or $-1$ uniformly. Set a small temperature, let's say $T = 0.1$
\item Calculate total energy ($E$) and total magnetization ($m$) of the system.
\item Start the MC loop
\item Again start a loop for spin-flipping processes. This should be $N$ times. For one MC step we try to flip $N$ number of random spins, so that on an average every spin gets a chance to flip. Whether the flip is accepted or nor will be decided by the following rule,
\item {\label{step:random-site}}Pick up a random site inside the 2D lattice. Flip the spin of that site. ($\sigma_i  \rightarrow -\sigma_i$)

\item Find its all neighbor's spin configurations.
 \item Calculate change in total energy and total magnetization due to this single spin flip. Here is the algorithm where the change in energy and magnetization is calculated (not the total energy and magnetization) by just looking at the values of neighboring spin values
{\color{db}
\begin{verbatim}
!	ileft, iright, iup, idown are  respectively the
!	left, right, up and down sites of 
!	of the i^th site

	if (x==1)then
	left=spin(Lx,y)
	else
	left=spin(x-1,y)
	end if
	if (x==Lx) then
	iright=spin(1,y)
	else
	iright = spin(x+1,y)
	end if
	if (y==1)then
	idown = spin(x,Ly)
	else
	idown = spin(x,y-1)
	end if
	if (y==Ly)then
	iup=spin(x,1)
	else
	iup=spin(x,y+1)
	end if
	isum = ileft + iright + iup + idown
	
	de = - 2.d0*spin(x,y)*(ileft + iright + iup + idown)
	dm = 2.d0*spin(x,y)
\end{verbatim}
}

\item {\label{step:metropolis}}If change in energy is less than zero, we accept the flip and go to step ({\ref{step:random-site}}) with reduced (updated) energy and magnetization, otherwise find a new random number, and compare that new random number  with the value of  $e^{-de/T}$. If the random number is less than $e^{-de/T}$, then we accept the flip, otherwise completely reject the flip without changing the energy and magnetization, and go to step ({\ref{step:random-site}}) 

\item One repeats these steps (\ref{step:random-site}) to (\ref{step:metropolis}) minimum $N$ times as said above. And cumulatively updating the energy and momentum changes in each flipping processes. 
\item Thus, one MC step is complete.

\item For each temperature we perform approximate $10^{5}$ MC steps and find the total cumulative energy and  magnetization. Then divide by ($N$ $\times$ {Number of MC steps}) to get the average quantities.

\end{enumerate}

\subsection*{\em Simulation results:}
The following quantities, magnetization, specific heat and susceptibility are calculated on a square lattice of  size $64 \times 64$, with periodic boundary condition. Total Monte-Carlo steps are $10^5$ and we have taken an ensembles of $100$ such copies. In each ensemble, first 1000 Monte-Carlo steps were not taken in the averaging process. 
\subsection*{Magnetization}

Average magnetization is defined as sum of all the spins divided by total number of spins. Mathematically 
\begin{eqnarray}
\langle m \rangle = {\frac{1}{N}} {\sum_{i}} \sigma_{i}
\end{eqnarray}
$N$ is total number of spins or sites. $N = L \times L$ for a square lattice. Below (but close to) critical temperature the magnetization has the universal behavior given by the equation
\begin{figure}[htbp] 
   \centering
   \includegraphics[width=0.6\textwidth]{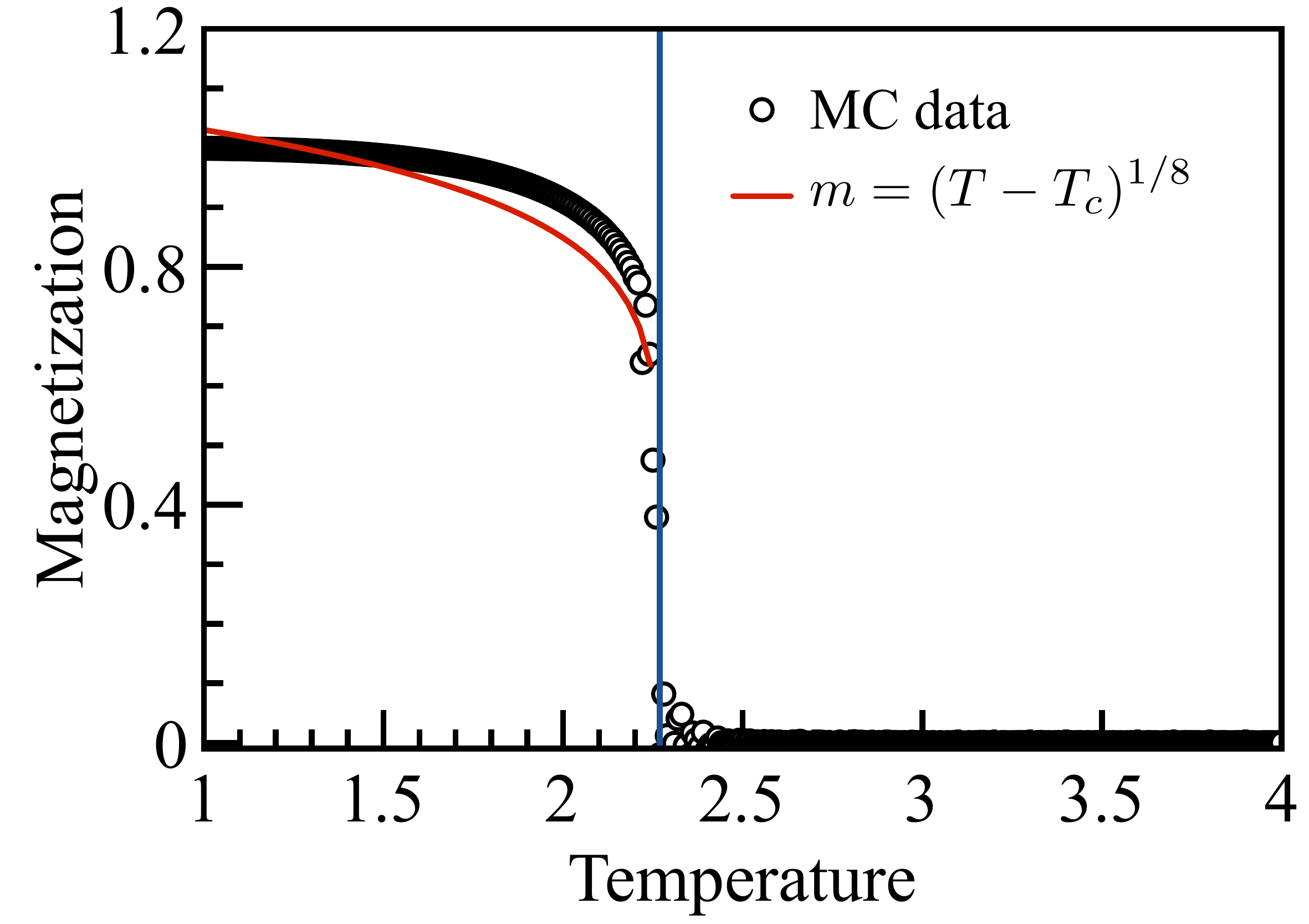} 
   \caption{Average magnetization as a function of temperature. The critical exponent $\beta = 1/8$.}
   \label{opi}
\end{figure}
\begin{eqnarray}
\langle m \rangle \sim ( T_c  - T )^\beta \qquad {\mbox{for}}~~T < T_c
\end{eqnarray}
here $\beta = {1 \over 8}$ is a critical exponent. The critical temperature, as a function of Ising  interaction strength $J$, is given by the formula for 2D Ising model as 
\begin{eqnarray}
T_c = {\frac{2 J}{{\rm ln}(1.0+\sqrt{2})}}  \approx 2.269 {\mbox{~ ~ (for J = 1)}}
\end{eqnarray}
See Fig.~{\ref{opi}}. Average magnetization is plotted as a function of temperature. The solid vertical line is at $T = T_{c} = 2.269$. The analytical curve near $T_c$ is $m \sim ( T_c  - T )^{1/8}$ for $T < T_c$, and we go near $T_c$ from below. For $T > T_c$, $ m = 0$. This is a continuous jump, and 2D classical Ising model undergoes a second order phase transition at $T_c$

\subsection*{Specific Heat {\&} Susceptibility}
\begin{figure}[hp]
\centering
\subfloat[ Specific heat as a function of temperature. The solid vertical line is at $T = T_{c} = 2.269$ ]{
\label{fig:cv}
\includegraphics[width=0.45\textwidth]{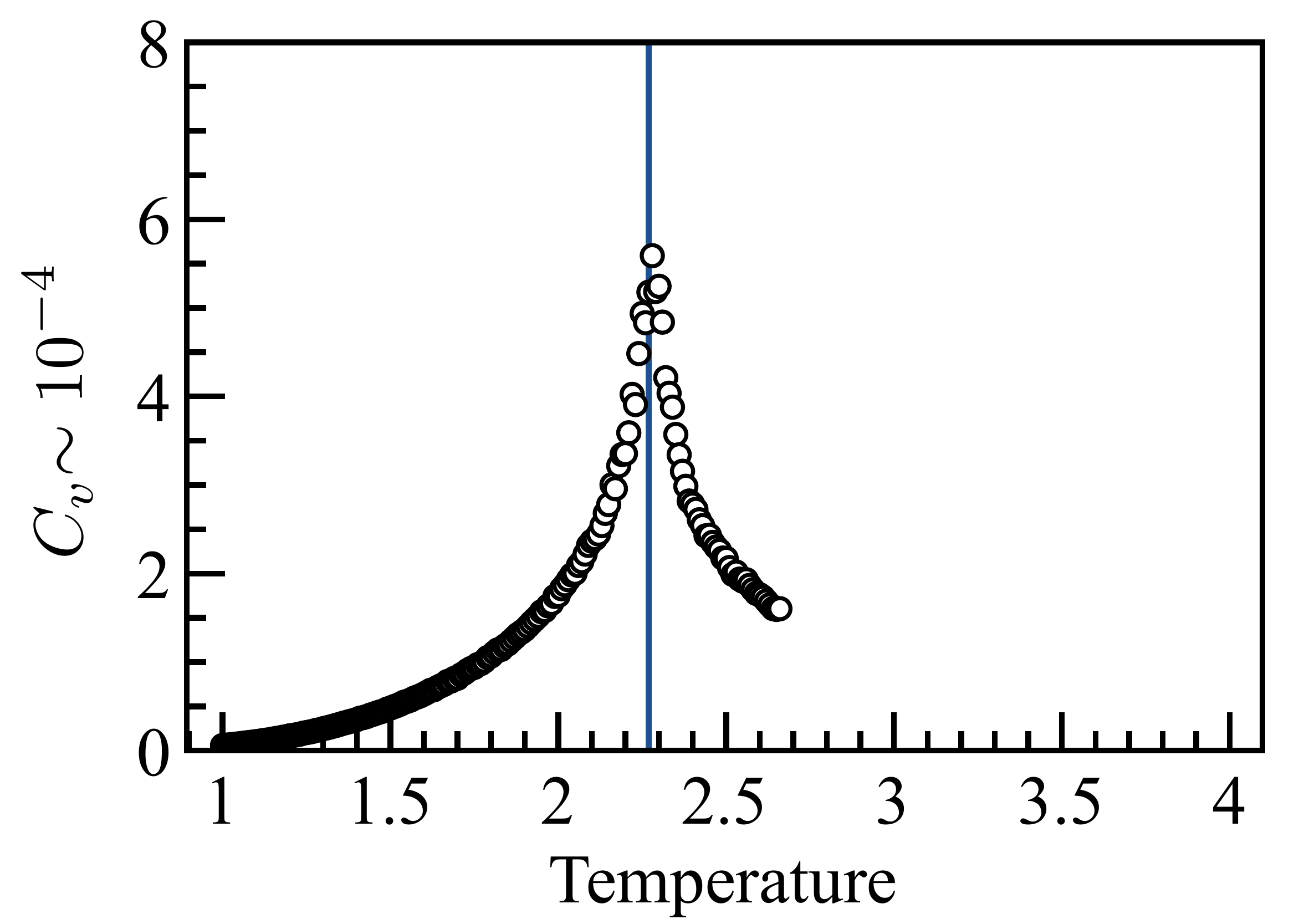} 
} \qquad 
\subfloat[ Magnetic susceptibility as a function of temperature. The vertical line is at $T = T_{c} = 2.269$. The analytical curve is $\chi \sim |T - T_c|^{-\gamma}$, where $\gamma = 7/4$.]{
\label{fig:xi}
\includegraphics[width=0.45\textwidth]{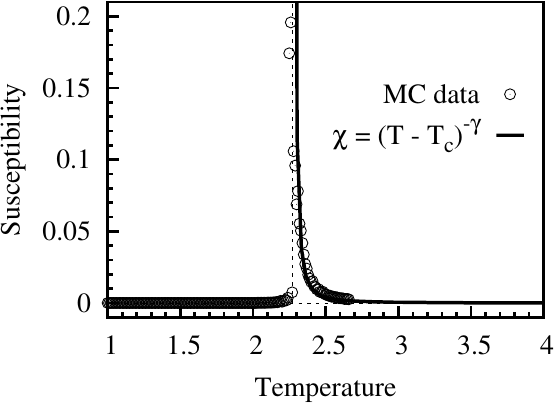} 
}
\caption{Specific heat and Susceptibility as a function of temperature, comparing MC simulation and analytical result.}
\label{fig:cv-xi-vs-temp}
\end{figure}

Specific heat is defined as $C_{v} = {\frac{ {\langle E^{2} \rangle} - { \langle E \rangle }^{2} }{T^{2}}} $ and susceptibility is defined as $ \chi = {\frac{ {\langle m^{2} \rangle} - { \langle m \rangle }^{2} }{T}}$. Susceptibility has analytical form as  $\chi \sim |T - T_c|^{-\gamma}$, where the critical exponent $\gamma = 7/4$. See Fig.~{\ref{fig:cv-xi-vs-temp}} for simulation result. 


\subsection*{Binder Cumulant}
\begin{figure}[htp] 
   \centering
   \includegraphics[width=0.7\textwidth]{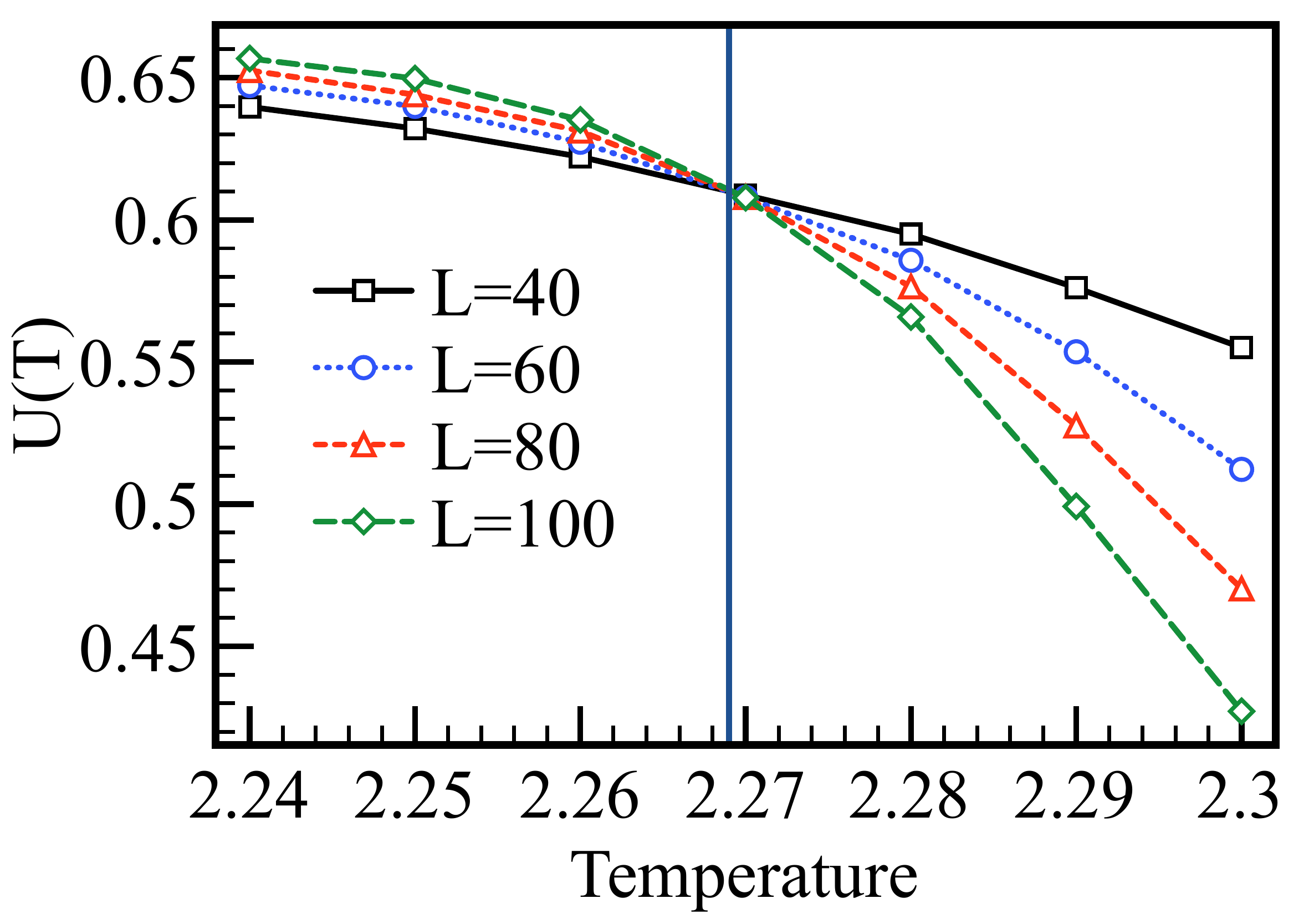} 
   \caption{Binder cumulant as a function of temperature for different lattice sizes with periodic boundary condition.}
   \label{fig:bc}
\end{figure}
Binder Cumulant/ratio is defined as
\begin{eqnarray}
U(T) = 1 - \frac{\langle m^4 \rangle }{3{\langle m^2 \rangle}^2}
\end{eqnarray}
In Fig.~{\ref{fig:bc}} Binder cumulant is plotted as a function of temperature for different lattice size with periodic boundary condition. $L$ is the linear dimension of square lattice. So the total number of sites are $L \times L$. All the curves intersect at $T=T_c$. The simulation is done for $10^5$ Monte-Carlo steps for each ensemble, and we have taken $100$ of such ensembles, and taken the average over all the ensembles to get the final data points. The solid vertical line is at $T = T_{c} = 2.269$


\def\ttilde{\tilde{t}}
\def\cbar{\bar{c}}
\def\bbar{\bar{b}}

\def\Deltatilde{\tilde{\Delta}}
\def\omegatilde{\tilde{\omega}}
\def\lambdatilde{\tilde{\lambda}}
\def\epsilontilde{\tilde{\epsilon}}

\def\adag{{\hat{a}}^\dagger}
\def\a{{\hat{a}}}
\def\bdag{{\hat{b}}^\dagger}
\def\b{{\hat{b}}}
\def\cdag{{\hat{c}}^\dagger}
\def\c{{\hat{c}}}

\def\akdag{{\hat{a}}^\dagger_\k}
\def\ak{{\hat{a}}^{}_\k}
\def\bkdag{{\hat{b}}^\dagger_\k}
\def\bk{{\hat{b}}^{}_\k}
\def\xhat{\hat{x}}
\def\phat{\hat{p}}

\def\yhat{\hat{y}}

\def\amkdag{{\hat{a}}^\dagger_{-\k}}
\def\amk{{\hat{a}}^{}_{-\k}}
\def\bmkdag{{\hat{b}}^\dagger_{-\k}}
\def\bmk{{\hat{b}}^{}_{-\k}}
\def\xk{{\hat{x}}_\k}
\def\pk{{\hat{p}}_\k}

\def\a{\hat{a}}
\def\adag{\hat{a}^\dagger}
\def\sighat{\hat{\sigma}}
\def\chihat{\hat{\chi}}
\def\nhat{\hat{n}}
\def\Uhat{\hat{U}}

\section{\label{sec:Bose-MFT}Bosonic Mean-field theory}


In the ordered phases of quantum spin systems, the spins arrange themselves in a particular order, and they all have a non-zero uniform expectation value. Now small fluctuations of the spins at a very low temperature around their mean orientations can have interesting effects on the ordered phase itself. And this is the heart of Spin-Wave theory{\index{Spin-Wave theory}}. In this approach, the spins are written in terms of bosonic operators in order to discuss the relevant physics. The magnitude of spin defines the dimension of Hilbert-space (HS). It is ($2 S +1$) dimensional HS for a spin values $S$, which is a finite. But the Bosonic Hilbert-space in infinite dimensional. We can map spin operators to two bosonic operators using Schwinger-Boson (SB){\nomenclature{SB}{Schwinger-Boson}} representations~{\cite{assa-1988,assa-book}}, and a local constraint projects the infinite dimension to the finite dimensional  sub-space.




%
The SB{\index{Schwinger Boson representation}} representation is for general spin-$S$. However, here we consider it only for spin-$1/2$ operators (because that is what we use it for later). But the ideas presented are applicable for any $S$. The Pauli spin matrices are written in terms of two bosonic operators $\hat{b}$ and $\hat{c}$ as
\begin{eqnarray}
&& \sigma^z_l = \bdag_l \b_l - \cdag_l \c_l \label{sigz} \\
&& \sigma^+_l = \bdag_l \c_l  \\
&& \sigma^-_l = \cdag_l \b_l \\
&& \rm{I} = \bdag_l \b_l + \cdag_l \c_l \qquad  \mbox{local constraint}
\end{eqnarray}

We use a physically meaningful mean-field approximation here. Suppose in the ground state is $\langle \sigma^z \rangle < 0$. Now we can think of the $\b$ type of bosons which will disperse through an average field, or background, of $\c$ type of bosons. So the excitation is only with the $\b$ bosons. We replace $\hat{c}$ by a number $\cbar$ and create the fluctuation due to $\hat{b}$ operator on $\cbar^2$ value.  So when $\langle \sigma^z \rangle$ is $-1$, $\cbar^2 = 1$ and $\langle {\bdag \b \rangle}$ is $0$. The excitation $\langle {\bdag \b \rangle}$ will reduce $\cbar^2$. When $\langle \sigma^z \rangle = 0$, $\cbar^2 = 0.5$. In this mean-field approximation, 
\begin{equation}
\sigma^z_l = \bdag_l \b_l - \cbar^2, \qquad \sigma^+_l = \bdag_l \cbar, \qquad \sigma^-_l = \b_l \cbar 
\end{equation}
The constraint $\bdag_l \b_l + \cbar^2 = 1$ will be satisfied on the average through a Lagrange multiplier. This approach has been used to do a calculation on Rabi lattice. That is presented in Appendix~\ref{subsec:Bose-MFT}.

\clearpage

%

\bibliographystyle{unsrt}
\bibliography{chapters/ref-all}


%% file: chapters/rabi.tex
\def\ahat{\hat{a}}
\def\bhat{\hat{b}}
\def\chat{\hat{c}}
\def\xhat{\hat{x}}
\def\yhat{\hat{y}}
\def\phihat{\hat{\phi}}
\def\psihat{\hat{\psi}}
\def\chihat{\hat{\chi}}
\def\atilde{\tilde{a}}
\def\nhat{\hat{n}}
\def\Dhat{\hat{D}}
\def\Nhat{\hat{N}}
\def\Phat{\hat{P}}
\def\Hhat{\hat{H}}
\def\Uhat{\hat{U}}
\def\k{{\bf k}}
\def\r{{\bf r}}
\def\calU{\mathcal{U}}
\def\calL{\mathcal{L}}
\def\deltavec{\vec{\delta}}
\def\omegatilde{\tilde{\omega}}
\def\ttilde{\tilde{t}}
\def\Deltatilde{\tilde{\Delta}}
\def\a{\hat{a}}
\def\adag{\hat{a}^\dagger}
\def\sighat{\hat{\sigma}}
\def\chihat{\hat{\chi}}
\def\nhat{\hat{n}}
\def\Uhat{\hat{U}}




\chapter[Quantum Ising dynamics and Edge modes in Rabi lattice]{Quantum Ising dynamics and Edge modes in \\ Rabi lattice model}{\label{chap:rabi}}

\begin{center}
\parbox{
0.8\textwidth}{\footnotesize
{\bf{\small About this chapter}}
\\[10pt]
The properties of the lattice model of Rabi quantum cavities are investigated. It is rigorously shown that the atomic dipoles in the Rabi lattice (of arbitrary geometry) exhibit quantum Ising dynamics in the limit of strong atom-photon interaction. This quantum Ising dynamics governs the para- to ferro-electric phase transition in the ground state. It also implies the existence of two Majorana-like edge modes in the ferro-electric phase on a one-dimensional Rabi lattice with open boundaries. The relation $\rho^x_{1L}=p^8$ between the end-to-end dipole correlation, $\rho^x_{1L}$, and the polarization, $p$, is proposed as an observable signature of these edge modes. The DMRG calculations on the one-dimensional Rabi lattice model clearly support the strong coupling quantum Ising behavior, and correctly yield the proposed end-to-end dipole correlation. The stability of these edge modes against the longitudinal field coupled to the dipoles is also studied. The conditions which protect the edge modes against such adverse perturbations are identified. Miscellaneous calculations that generally help in the basic understanding of the Rabi lattice problem are also done and presented in the appendix.
\\[15pt]
}
\end{center}

\minitoc

\section{Introduction}

The lattice models of quantum cavities have received much attention, recently, for emulating  the Bose-Hubbard model by exhibiting Mott-insulator to superfluid type quantum phase transition for photons~\cite{Tomadin.Fazio.10,Hartmann,Greentree.06,Littlewood,Blatter}. A quantum cavity (also called a cavity-QED system) refers to an interacting matter-radiation system inside a high-Q resonator (cavity). The simplest of it can be modeled as a two-level atom interacting with a single mode of quantized radiation. It is hoped that the lattice models of such quantum cavities may eventually be realized, for example, by engraving an array of cavities in a photonic band-gap material~\cite{Greentree.06,QD.Photonic}, or as a circuit-QED system~\cite{CircuitQED.Girvin,CircuitQED.Fink,Houck.Koch}. Such engineered systems will indeed be useful in studying interesting quantum many-body problems, much like the cold atoms in optical lattices~\cite{ColdAtoms}.

In complex physical systems, one is often concerned with investigating the quantum phase transitions. The superfluid to insulator transition in the ground state of the Bose-Hubbard (BH) model is an example of a quantum phase transition driven by the competition between kinetic energy and local repulsion~\cite{BH_Fisher, BH_Sheshadri}. See Subsection~{\ref{subsec:qed}} for discussion on this. Another famous case which beautifully typifies quantum phase transitions is the quantum Ising (QI) model~\cite{QIsing.deGennes,QIsing.Katsura,QIsing.Pfeuty}. See Sec.~{\ref{sec:qi}}. It describes an Ising system in a field transverse to the Ising axis. The QI model has not only been used to study a variety of physical problems~\cite{QIsing.deGennes,Columbite1,quench.QI,Moessner.QI}, but it has also served as an important point of reference in the general understanding of  quantum phase transitions~\cite{QPT.Sachdev}.
 
The one-dimensional (1d) spin-1/2 QI model with only nearest-neighbor interaction is solvable, and has been rigorously worked out by Pfeuty using the Jordan-Wigner fermionization~\cite{QIsing.Pfeuty,LSM}. For the 1d QI model in the fermionized form, Kitaev made a remarkable observation that the two ends of an open chain carry a Majorana fermion each~\cite{Kitaev.FreeMajorana}. A fermion which is Hermitian (that is, an anti-particle of itself) is called a Majorana fermion~\cite{Kitaev.FreeMajorana, Wilczek.Majorana, note.Majorana}. There has always been a great interest in finding the Majorana fermions in nature. 
The developments in quantum computation and condensed matter, in particular the proposed role of Majorana fermions in the error-free (topological) quantum computation~\cite{Kitaev.FreeMajorana,Nayak.RMP,Topo_QC_Kitaev}, have further invigorated their search~\cite{Mourik.Majorana}. 

In this Chapter, we have systematically generated the quantum phase diagram of Rabi lattice model (RLM){\index{Rabi lattice model}} in Section~{\ref{sec:rabi-qpd}} with the help of numerical calculations and analytics. By Rabi lattice we mean an array of the Rabi quantum cavities where each cavity has a two-level atom (or a spin-1/2) interacting via dipolar interaction with a single photon mode, and the inter-cavity coupling leads to photon hopping. We present a case for the realization of Majorana fermion modes, through the QI dynamics, in the Rabi lattice model~\cite{bkumar.somenath}. As it is shown that the Rabi lattice model in the limit of strong atom-photon interaction rigorously tends to the QI model. It automatically implies that the 1d Rabi lattice also has two Majorana fermion modes. This is in contrast to the lattice models formed by the Jaynes-Cummings type cavities~\cite{Tomadin.Fazio.10}, which show superfluid-insulator transition for photons, but no QI behavior. We identify and propose $\rho^x_{1L}=p^8$ as a signature of these Majorana modes, where $p$ is the spontaneous polarization and $\rho^x_{1L}$ denotes the end-to-end correlation of atomic dipoles in the ground state on an open chain of length $L$. We also identify practical conditions which save the QI Majorana modes from the detrimental longitudinal field, both uniform and random in Section~{\ref{sec:edge}}. Below we systematically discuss the emergence of QI dynamics and Majorana fermions in the Rabi lattice model, and support it by the density matrix renormalization group (DMRG) calculations  in one dimension.

\section{{\label{sec:rabi-qpd}}Rabi lattice model}

A Rabi quantum cavity is a minimal atom-photon problem described by the following Hamiltonian, known as  Rabi model, inside $l^{th}$ cavity
\begin{eqnarray}
\Hhat_{R,l} = \omega (\nhat_l +\frac{1}{2}) +\frac{\epsilon}{2}\sigma^z_l +\gamma\sigma^x_l (\ahat^\dag_l+\ahat^{ }_l) {\label{eqn:rabi-ham}}
\end{eqnarray}
where $\omega$ is the photon energy and $\epsilon$ is the atomic transition energy. The dipole interaction, $\vec{p}\cdot\vec{E}$, is written as $\gamma\sigma^x (\ahat^\dag+\ahat)$, where $\gamma$ denotes the atom-photon coupling, $\ahat$ ($\ahat^\dag$) is the photon annihilation (creation) operator, and the Pauli operator, $\sigma^x = |e\rangle\langle g| + |g\rangle\langle e|$, measures the atomic dipole, $\vec{p}$~\cite{Scully.Zubairy}. Moreover, $\sigma^z=|e\rangle\langle e| - |g\rangle\langle g|$. The kets $|g\rangle$ and $|e\rangle$ denote the atomic ground and excited levels, respectively. The $\Hhat_{R,l}$ is simple looking but there is no exact solutions yet, although it has been studied extensively~\cite{RabiModel.Graham, RabiModel.Reik, RabiModel.Braak}. Polariton number ($n_p = \ahat^\dagger_{l} \ahat^{}_{l} + \sigma^+_l \sigma^-_l $) is not conserved for Rabi model but the local parity operator $P_{l} = -\chi_{l} \sigma^{z}_{l}$ is. Here $\chi_{l} = (-1)^{n_{l}}$ and $n_{l} = \ahat^{\dagger}_{l} \ahat$. So the Rabi model has discrete parity symmetry. In the lattice models of quantum cavities, the photon can jump from one cavity to another (let's consider without dissipation) while the atom stays inside the particular cavity where it belongs. The Hamiltonian is
\begin{eqnarray}
\Hhat &=& \omega {\sum_l} \left( {\adag_l \a_l + {1 \over 2}} \right) + {\epsilon \over 2} {\sum_l} \sighat_l +  \gamma {\sum_l} {\sighat^x_l}{\left( {\adag_l + \a^{}_l} \right)} \cr && - t {\sum_{l, \delta}} {\left( {\adag_l \a_{l+\delta} + h.c } \right)} {\label{eqn:rlm}}
\end{eqnarray}
$t$ is nearest-neighbor photon hopping amplitude. For RLM the local parity operator is not conserved as the photon hopping is constantly changing the photon number inside each cavity. But the global parity $\Pi_l ( -\chi_{l} \sigma^{z}_{l})$ is conserved for the RLM. RWA has changed the symmetry properties of JCLM and RLM, and that is the reason there are qualitatively two different physics in this two different models. The breaking of this discrete symmetry is followed by a quantum phase transition. As it only has a discrete symmetry (global parity), a continuous phase transition in the Rabi lattice model would occur by breaking it, and a gapped excitation will occur in the `ordered' phase. This point has been noted recently in Ref.~\cite{RabiLattice.ZHeng}. In fact, we expect  the Rabi lattice to exhibit quantum Ising transition, as hinted at by the fluctuating local parity picture presented above. Below, we neatly establish this by transforming the $\Hhat$ to a form wherein the quantum Ising dynamics becomes evident.

\subsection{Quantum Ising dynamics}
Let's solve Rabi Hamiltonian of Eq.~(\ref{eqn:rabi-ham}). We use unitary operator which will decouple the atom part ($\sigma^{x}_{l}$) and the photon part ($\ahat^{\dagger}_{l} + \ahat^{}_{l}$). Since the atomic dipole operator $\sigma^x_l$ has eigenvalues $\pm 1$, we can absorb it into the electric field, $(\ahat^\dag_l+\ahat^{ }_l)$, by the unitary transformation, 
\begin{eqnarray}
\Uhat=\prod_l (\Phat^+_l+\Phat^-_l\chihat_l)
\end{eqnarray}
where 
\begin{eqnarray}
\Phat^\pm_l=(1\pm\sigma^x_l)/2
\end{eqnarray}
Under this transformation, 
\begin{eqnarray}
\Uhat^\dag \ahat_l \Uhat &=& \sigma^x_l\ahat_l \\
\Uhat^\dag \nhat_l \Uhat &=& \nhat_l \\
\Uhat^\dag \sigma^z_l \Uhat &=& \chihat_l\sigma^z_l \\
\Uhat^\dag \sigma^x_l \Uhat &=& \sigma^x_l
\end{eqnarray}
Thus, in the transformed version of the Rabi cavity, the dipole interaction turns into the displacement field, $\gamma (\ahat^\dag_l+\ahat^{ }_l)$, for photons. The $\sigma^z_l$ operator is to be understood as the local parity, and $\sigma^x_l$ continues to be the atomic dipole as it was before applying $\Uhat$.

Now consider the RLM, $\Hhat$ of Eq.~{\eqref{eqn:rlm}}. Under $\Uhat$, it transforms to the following form.
\begin{eqnarray}
&& \Uhat^\dag \Hhat \Uhat = \Hhat_1 \nonumber = \sum_l \left[\omega\left(\nhat_l+\frac{1}{2}\right)+\gamma\left(\ahat^\dag_l+\ahat^{ }_l\right) \right] \nonumber\\
&&+\frac{\epsilon}{2}\sum_l\chihat_l\sigma^z_l -t\sum_{l,\delta}\sigma^x_{l}\sigma^x_{l+\delta} \left( \ahat^\dag_l\ahat^{ }_{l+\delta} + \ahat^\dag_{l+\delta}\ahat^{ }_l \right) {\label{eqn:H1}}
\end{eqnarray}
Let us look at the form of $\Hhat$ and $\Hhat_{1}$ for comparison. The $1^{st}$ term is as it was, the photon energy term. The $2^{nd}$ term of the Hamiltonian, $\gamma\left(\ahat^\dag_l+\ahat^{ }_l\right)$, is a displacement field which was the atom-photon interaction. This term guarantees that $\langle \ahat_l\rangle \neq 0$. Thus, a static `electric' field, proportional to $\gamma$, is ever present. The $3^{rd}$ term was energy contribution from the atom. Now, this atomic transition energy, $\epsilon$ (renormalized by the expectation of $\chihat_l$), will act as a `transverse' field on the parity, $\sigma^z_l$. At last, there was the photon hopping term in Eq. (\ref{eqn:rlm}), $ - t {\sum_{l, \delta}} {\left( {\adag_l \a_{l+\delta} + h.c } \right)} $, which would generate an `Ising' interaction between the atomic dipoles, $\sigma^x_l$ (the last term in Eq.~{\eqref{eqn:H1}} which involves hopping). Thus emerges the quantum Ising dynamics in the Rabi lattice model. Although the photon fluctuations above the static field would also be present, one may neglect them in the strong $\gamma$ limit. We shall show that in the next part. As mentioned before strong-coupling limit is when the atom-photon interaction energy is larger than any other energy scales of the system, and where we can say the decay is very small compare to the atom-photon interaction energy. So in our case $\gamma \gg \omega, \epsilon$ is the limit where we can say, the system is in strong-coupling limit.

We `displace' $\Hhat_1$ such that the $\nhat_l$ absorbs $\gamma(\ahat^\dag_l+\ahat_l)$. This is done by the displacement operator
\begin{eqnarray}
\Dhat=\prod_l e^{-\frac{\gamma}{\omega}(\ahat^\dag_l-\ahat^{ }_l)}
\end{eqnarray}
Since
\begin{eqnarray}
\Dhat^\dag \ahat_l \Dhat = \ahat_l-\frac{\gamma}{\omega} ~ ; \qquad \Dhat^\dag \chihat_l\Dhat = \chihat_l \, e^{-\frac{2\gamma}{\omega}(\ahat^\dag_l-\ahat^{ }_l)}
\end{eqnarray}
$\Hhat_1$ of Eq.~{\eqref{eqn:H1}} is transformed by applying the unitary operator $\Dhat$ as follows,
\begin{eqnarray}
&& \Dhat^\dag\Hhat_1\Dhat = \Hhat_2 = \sum_l\left[ \omega\left(\nhat_l+\frac{1}{2}\right) -\frac{\gamma^2}{\omega} \right] + \nonumber\\
&& \frac{\epsilon}{2}e^{-\frac{2\gamma^2}{\omega^2}}\sum_l \chihat_l \, e^{-\frac{2\gamma}{\omega} \ahat^\dag_l}e^{\frac{2\gamma}{\omega} \ahat_l}\,\sigma^z_l -2t\frac{\gamma^2}{\omega^2}\sum_{l,\delta}\sigma^x_l\sigma^x_{l+\delta} \nonumber \\
&& -t\sum_{l,\delta}\sigma^x_l\sigma^x_{l+\delta} \left\{\left[\ahat^\dag_l\ahat_{l+\delta} -\frac{\gamma}{\omega}\left(\ahat^\dag_l+\ahat_{l+\delta}\right)\right] + h.c. \right\} \label{eqn:H2}
\end{eqnarray} 
This is the Rabi lattice model explicitly in terms of the static field, $\gamma/\omega$, and the photon fluctuations. Now, bring the strong coupling limit, $\gamma/\omega \gg 1$, to work. We now do two things here. Neglect the photon fluctuations in $\Hhat_2$. And look for the lowest energy eigenstate which is for $\nhat_{l} = 0$. Let's analyze the effects of these two considerations on the Hamiltonian $\Hhat_{2}$ of Eq. (\ref{eqn:H2}) term by term. For $\nhat^{}_{l} = 0$ the first term is a constant energy term. The value of the constant energy may be large for $\gamma/\omega \gg 1$, but this does not affect system property as its a constant shift of energy. The second term involving the atomic transition energy is proportional to $ \frac{\epsilon}{2}e^{-\frac{2\gamma^2}{\omega^2}}$ as the $\chihat_l$ expectation is $1$ in the ground state and exponential part does not contribute. Similarly the last term is zero. We are left with the minimal Hamiltonian which is 
\begin{eqnarray}
\Hhat_{QI} =  \mathcal{E}_0 + \frac{\epsilon}{2}e^{-\frac{2\gamma^2}{\omega^2}}\sum_l\sigma^z_l -2t\frac{\gamma^2}{\omega^2}\sum_{l,\delta}\sigma^x_l\sigma^x_{l+\delta} \label{eq:HQI}
\end{eqnarray} 
we get the structure of quantum Ising model, wherein for different photon occupancies $|\{n_l\}\rangle $, the dipoles and parities follow QI dynamics with different reference energies and weakly differing transverse fields. Here, the Ising interaction,
\begin{eqnarray}
J=-2t\frac{\gamma^2}{\omega^2} {\label{eqn:j}}
\end{eqnarray}
between atomic dipoles is ferro-electric, and the transverse field acting on the local parities, 
\begin{eqnarray}
h=\frac{\epsilon}{2} \exp{(-\frac{2\gamma^2}{\omega^2})} {\label{eqn:h}}
\end{eqnarray}
which is in exponential scale.   $\mathcal{E}_0=\frac{\omega^2-2\gamma^2}{2\omega}L$ in $\Hhat_{QI}$, although it has no role in the dynamics. Notably, even when $t$ is only moderately small ($t\gamma^2/\omega^2\lesssim \omega$), the ground state of the Rabi lattice model is still  described reasonably by the $\Hhat_{QI}$. We have checked this, and the above proposition, about $\Hhat_{QI}$ by doing exact numerics on small clusters. For large $L$, the DMRG calculations presented below clearly confirm the strong coupling QI dynamics.

\subsection{Ground state properties}

Since the QI model is a well-studied problem, having $\Hhat_{QI}$ as the limiting Hamiltonian immensely helps in understanding the ground state of the Rabi lattice model. It predicts two distinct phases that undergo quantum Ising transition between them.

Effective interaction strength $J$ in Eq.~(\ref{eqn:j}) is a function of $\omega$ and $\gamma$ and $t$ whereas the effective transverse field $h$ in Eq.~(\ref{eqn:h}) is function of $\omega$, $\gamma$ and  $\epsilon$. So keeping $\omega$ and $\epsilon$ fixed (we have already fixed $\gamma = 1$), we can change hopping strength $t$ to change the effective dipole-dipole interaction strength and move across the quantum phase transition point. 

For $|J| \gtrsim h$, the Rabi lattice would exhibit spontaneous polarization. In this ferro-electric{\index{ferro-electric}} (FE){\nomenclature{FE}{Ferro-Electric}} ground state, the order parameter,
\begin{eqnarray}
p = \frac{1}{L}\sum_l\langle\sigma^x_l\rangle \neq 0
\end{eqnarray}
For $|J| \lesssim h$, each Rabi cavity in the lattice would behave as independent, and $p=0$ [para-electric{\index{para-electric}} (PE){\nomenclature{PE}{{Para-Electric}}} phase]. And they are connected by a gapless point. 

\begin{figure}[htbp] 
   \centering
   \includegraphics[width=0.49\textwidth]{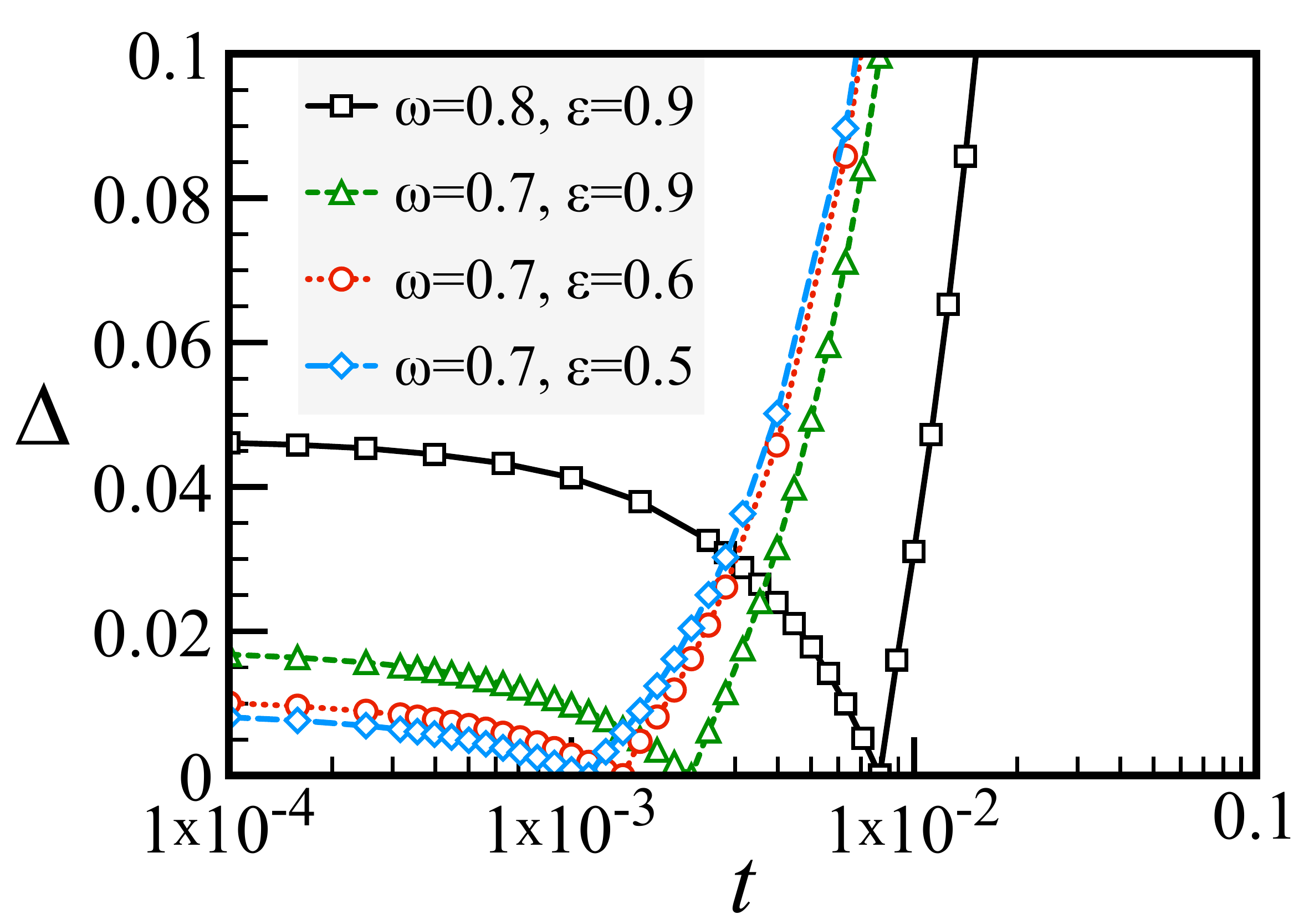} 
   \includegraphics[width=0.49\textwidth]{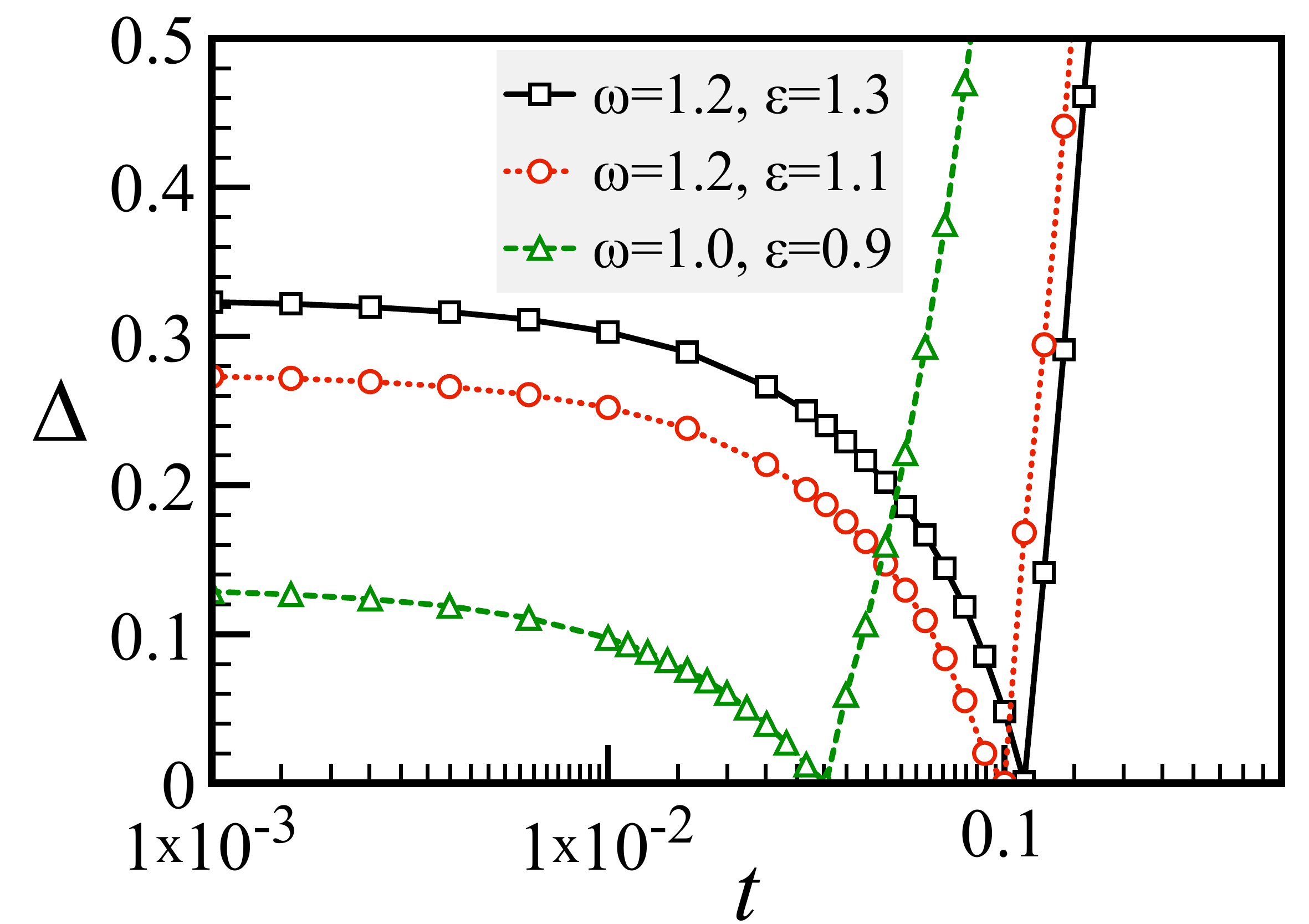} 
   \caption{Gap $\Delta$ (difference between lowest two energies) are plotted as a function of hopping amplitude $t$ for different values of $\omega, \epsilon$.}
   \label{fig:gap-t}
\end{figure}

This can also be restated in terms of photons, as often done in the literature on cavity lattice models. The photon order parameter, $\psi=\langle\ahat_l\rangle$, after having applied $\Uhat$ and $\Dhat$, becomes $\psi=\langle \sigma^x_l(\ahat_l-\frac{\gamma}{\omega})\rangle \approx -p\frac{\gamma}{\omega}$. Thus, in the FE phase, the photons exhibit `superfluidity'. The PE phase may likewise be called a Mott insulating phase. However, it not quite the same because a Mott phase is characterized by an integer polariton number per cavity, which is not true for the Rabi cavity. Moreover, the FE phase is not the usual superfluid described by a complex $U(1)$ order parameter with gapless modes. Instead, it has an Ising ($Z_2$) order parameter with gapped excitations. Notably the excitations in both the phases 
are gapped, except at the transition point where the gap closes. See Fig. {\ref{fig:gap-t}}


\begin{figure}[htbp] 
   \centering
\includegraphics[width=0.6\textwidth]{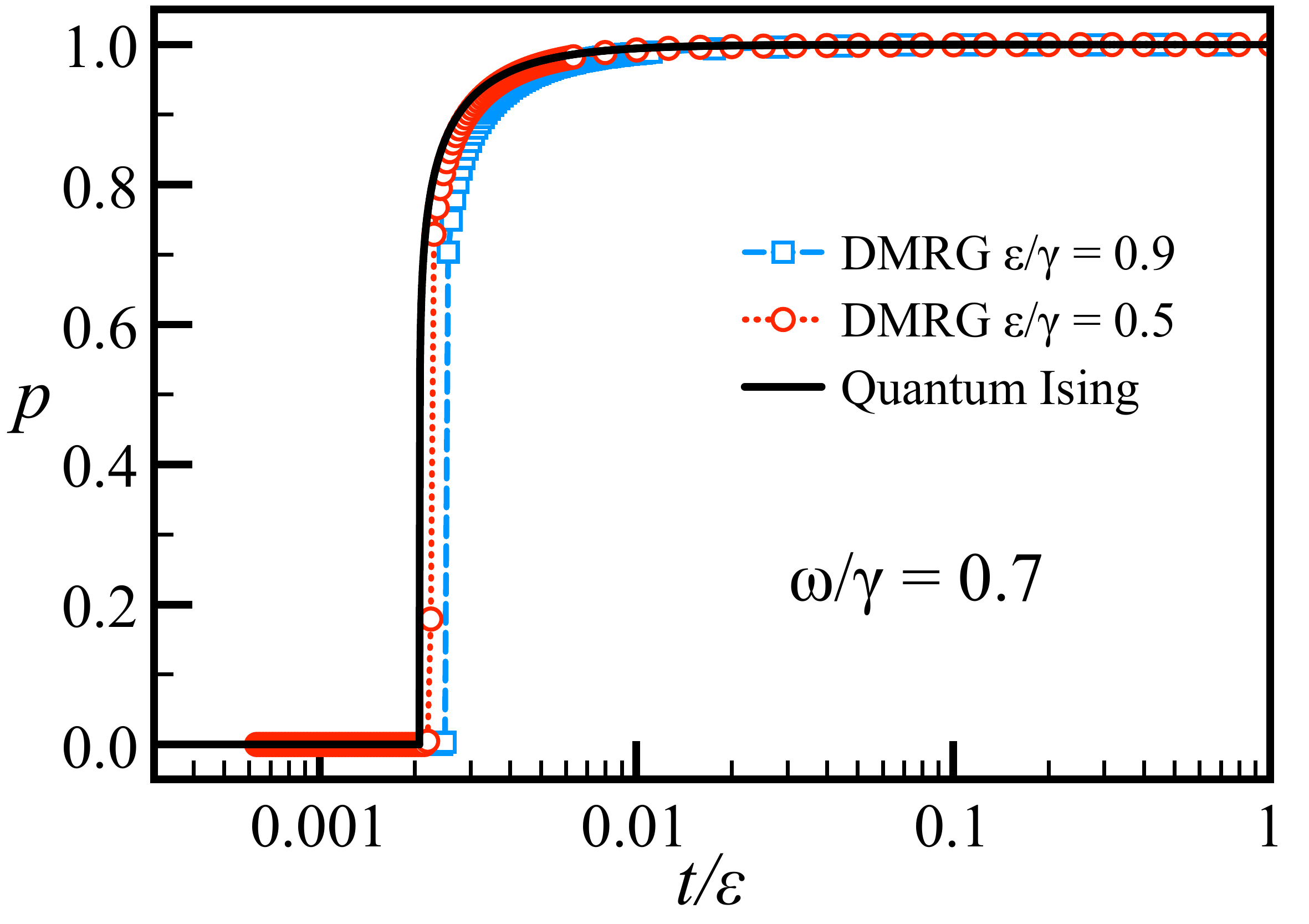} 
\caption{Polarization, $p$, vs. hopping, $t$, in the 1d Rabi lattice for different values 0f $\epsilon$. The black line is the exact strong coupling prediction.}
\label{fig:mx_vs_t}
\end{figure}

The factor, $\exp{(-\frac{2\gamma^2}{\omega^2})}$, in the transverse field, $h$, too has some important consequences. For $\frac{\gamma}{\omega}\gg 1$, it renders the variations in $\epsilon$ unimportant, as $h$ would invariably be small regardless of the value of $\epsilon$. Thus, it is immaterial whether $\epsilon$ is near resonance ($\sim \omega$) or far. It also implies that the critical hopping, $t_c$, for the transition from PE to FE phase would vary on exponentially small scale.

To explicitly validate the strong coupling behavior, and to see deviations from it, we numerically investigate the 1d Rabi lattice using DMRG 
(a nice method for studying 1d quantum problems~\cite{srw-prb-1993,Schollwock2005}). Since the $\Hhat_{QI}$ in 1d is exactly solvable, it allows us to make proper comparisons with the DMRG results. For $\Hhat_{QI}$ in 1d, the exact polarization in the FE phase (in the thermodynamic limit) is~\cite{QIsing.Pfeuty}  where
\begin{eqnarray}
p = \left(1-\frac{1}{\lambda^2} \right)^{\frac{1}{8}} \qquad {\mbox{and}} \qquad \lambda=|J|/h = \frac{4t}{\epsilon} \frac{\gamma^2}{\omega^2} \exp{(\frac{2\gamma^2}{\omega^2})}
\end{eqnarray}

Using DMRG, we compute $p$ as a function of $t$ in the ground state of the Rabi lattice model (in the $\Hhat_2$ form). It clearly exhibits the PE to FE phase transition, and agrees nicely with the strong coupling QI form, as shown in Fig.~\ref{fig:mx_vs_t}. The DMRG calculations are done for chain lengths upto 600, and by taking four photon states ($n_l=0,1,2,3$) per Rabi cavity.

\subsection{{\label{sec:rabi-qpd}}Quantum phase diagram}

Now we describe the nature of the phase transition for Rabi lattice model. Since the exact quantum critical point for the PE-FE transition in $\Hhat_{QI}$ in 1d is $\lambda=1$~\cite{QIsing.Pfeuty}, the strong coupling critical hopping is 

\begin{eqnarray}
t_c= \frac{\epsilon}{4}\frac{\omega^2}{\gamma^2}\exp{(-\frac{2\gamma^2}{\omega^2})}
\end{eqnarray}

\begin{figure}[htbp] 
   \centering
\includegraphics[width=0.6\textwidth]{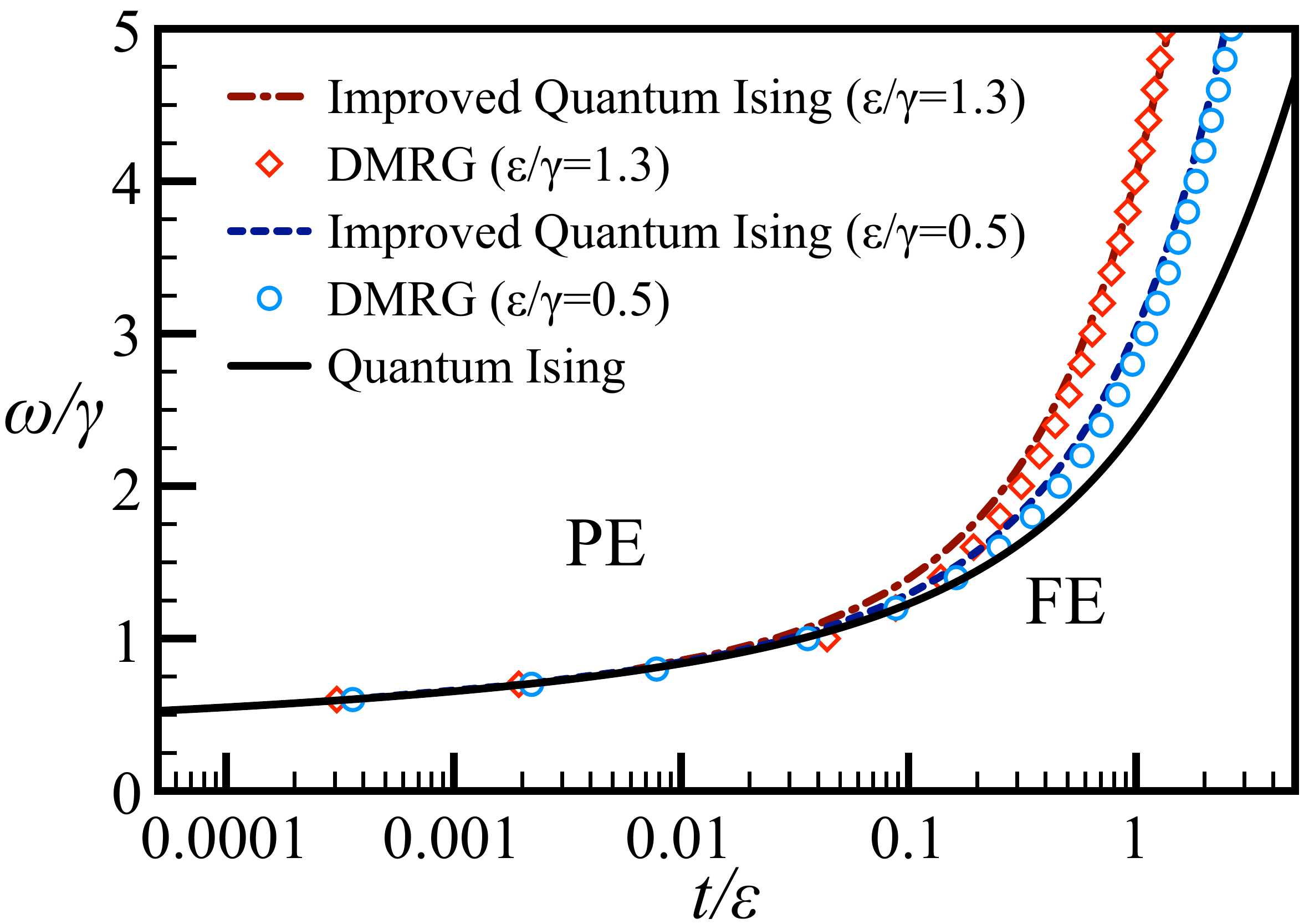} 
\caption{Quantum phase diagram for the Rabi lattice model in 1d. The PE (FE) denotes the para- (ferro-) electric phase.}
\label{fig:rabiQPD}
\end{figure}

The $t_c$ for $\epsilon=\omega$ that has been derived recently in Ref.~\cite{RabiLattice.Schiro} is just a particular case of the present result. For a given $\frac{\gamma}{\omega}$, the strong coupling $t_c$  scales as $\epsilon$. Thus, in the plane of $\frac{t}{\epsilon}$ and $\frac{\omega}{\gamma}$, the strong coupling phase boundary would be a universal curve, independent of $\epsilon$. It is true in any dimension. In Fig. {\ref{fig:rabiQPD}}, the DMRG calculated phase boundary is compared with the strong coupling prediction. Clearly, for $\frac{\gamma}{\omega} \gtrsim 1$, the numerical data approaches the universal ${t_c}/{\epsilon}$.

For $\frac{\gamma}{\omega}\lesssim 1$, the deviations from the strong coupling QI behavior are approximately accounted for by an improved $\Hhat_{QI}$, wherein $\omega$ in Eq.~(\ref{eq:HQI}) is replaced by $\omega-zt\rho^x_1$. It is basically correcting $\omega$ for the photon dispersion. Here, $z$ is the nearest-neighbor coordination, and $\rho^x_1 = \langle \sigma^x_l\sigma^x_{l+\delta}\rangle$ is the nearest-neighbor dipole correlation. This is effected by applying a more general displacement, $\prod_l e^{-\alpha(\ahat^\dag_l-\ahat_l)}$, on $\Hhat_1$, and  approximating $t\alpha\sum_{l,\delta}\sigma^x_l\sigma^x_{l+\delta} (\ahat_l+\ahat_{l+\delta}+h.c.)$ (arising from the hopping after the displacement) by $zt\alpha\rho^x_1\sum_l(\ahat^\dag_l+\ahat_l)$. Putting the linear photon terms to zero gives $\alpha=\gamma/(\omega-zt\rho^x_1)$, and ignoring the photon fluctuations gives the improved $\Hhat_{QI}$. The improved QI phase boundaries in Fig.~\ref{fig:QPD} are in good agreement with DMRG. For $\frac{\gamma}{\omega}\ll1$, however, it will not be as good. 

\section{{\label{sec:edge}}Edge modes}

{\index{Edge modes}}On an open chain, the strong coupling $\Hhat_{QI}$ guarantees that the Rabi lattice has two Majorana fermions in the FE phase. While the basic degrees of freedom in the QI (or Rabi lattice) model are not fermions, the exact quasiparticles in the 1D QI model are fermions, of which, one corresponds to having two Majorana modes localized at the ends of the chain in the ordered phase~\cite{QIsing.Pfeuty,Kitaev.FreeMajorana}. We discuss two points of practical interest about these Majorana modes: 
\begin{enumerate}
\item An observable signature
\item The conditions under which the edge-modes survive a constant longitudinal perturbation $\eta\sum_l\sigma^x_l$ or a random field fluctuation like $\sum_l {\eta_l \sigma^x_l}$ 
\end{enumerate}

\begin{figure}[htbp] 
\centering
\includegraphics[width=0.6\textwidth]{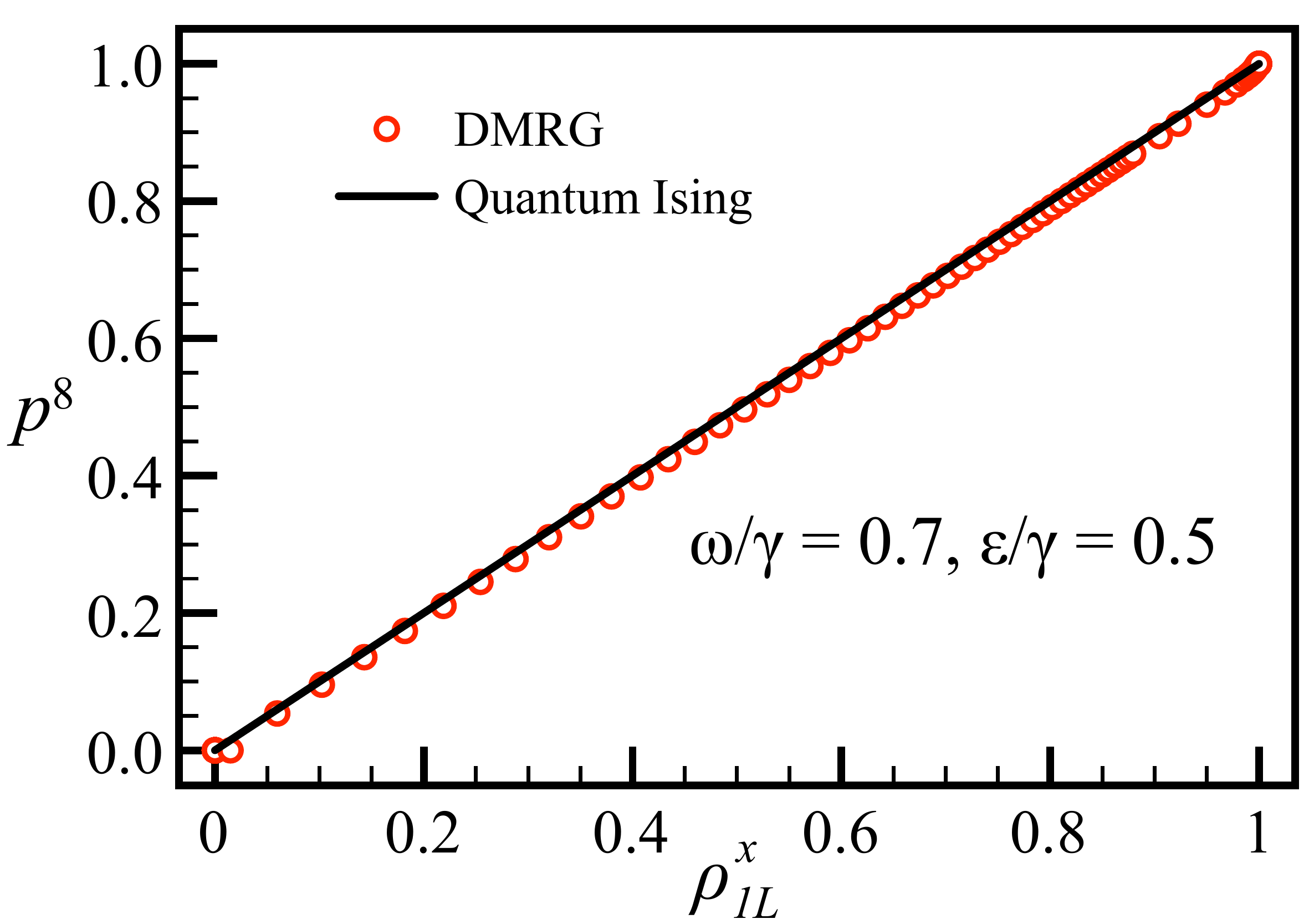} 
\caption{End-to-end correlation, $\rho^x_{1L}$, vs. $p^8$ in the ground state of the Rabi lattice model for $L=600$.}
\label{fig:end2end}
\end{figure}

A careful reading of Ref.~\cite{QIsing.Pfeuty} reveals that the end-to-end correlation, $\rho^x_{1L}=\langle\sigma^x_1\sigma^x_L\rangle$, in the ground state of the open chain QI model has a special exact relation with the spontaneous polarization, $p$. It is: 
\begin{eqnarray}
\rho^x_{1L} = p^8+\mathcal{O}(1/L)
\end{eqnarray}
This is clearly at odds with the $p^2$ behavior of the long-ranged dipole correlation in the bulk. We identify that this striking relation between $\rho^x_{1L}$ and $p$ is a direct consequence of these Majorana modes~\cite{LSM}. Thus, we propose $\rho^x_{1L}=p^8$ as a signature thereof. It compares two observables only, and doesn't explicitly involve any model parameters. We indeed find this behavior in the 1d Rabi lattice (see Fig.~\ref{fig:end2end}, which is generated by varying $t$, i.e., same as varying interaction strength $J$ for QI model).

The twofold ground state degeneracy of the QI model for $\lambda>1$, in the thermodynamic limit, is due to the zero energy of Majorana modes. For a finite $L$, this degeneracy is lifted by an amount, $\lambda^{-L}$~\cite{QIsing.Pfeuty}. Since it is very small compared to the bulk quasiparticle gap for any finite but large enough $L$, the Majorana modes remain localized within a length of order $\frac{1}{\ln{\lambda}}$ at the two ends~\cite{LSM,Topo_QC_Kitaev}. Thus, the Majorana modes retain their identity as long as the splitting of ground state degeneracy is weak compared to the bulk gap. In view of this, we believe that the perturbation, $\eta \sum_l\sigma^x_l$, will not be detrimental to the Majorana modes if $\eta = \frac{v}{L}$ (or weaker) for a $v\sim\mathcal{O}(1)$ and $\lesssim(\lambda-1)$. Deep inside the ordered phase ($\lambda\gg 1$), where these modes are highly localized, we expect them to better survive against a non-zero $v$. Thus, $\eta \sim \frac{1}{L}$ is the `safe' strength of a longitudinal field (even if non-uniform) on a finite $L$ array. In practical terms, $\eta=\frac{v}{L}$ could be 
an electric field  due to a potential drop $v$ along the array. 

\begin{figure}[htbp] 
\centering
\includegraphics[width=0.6\textwidth]{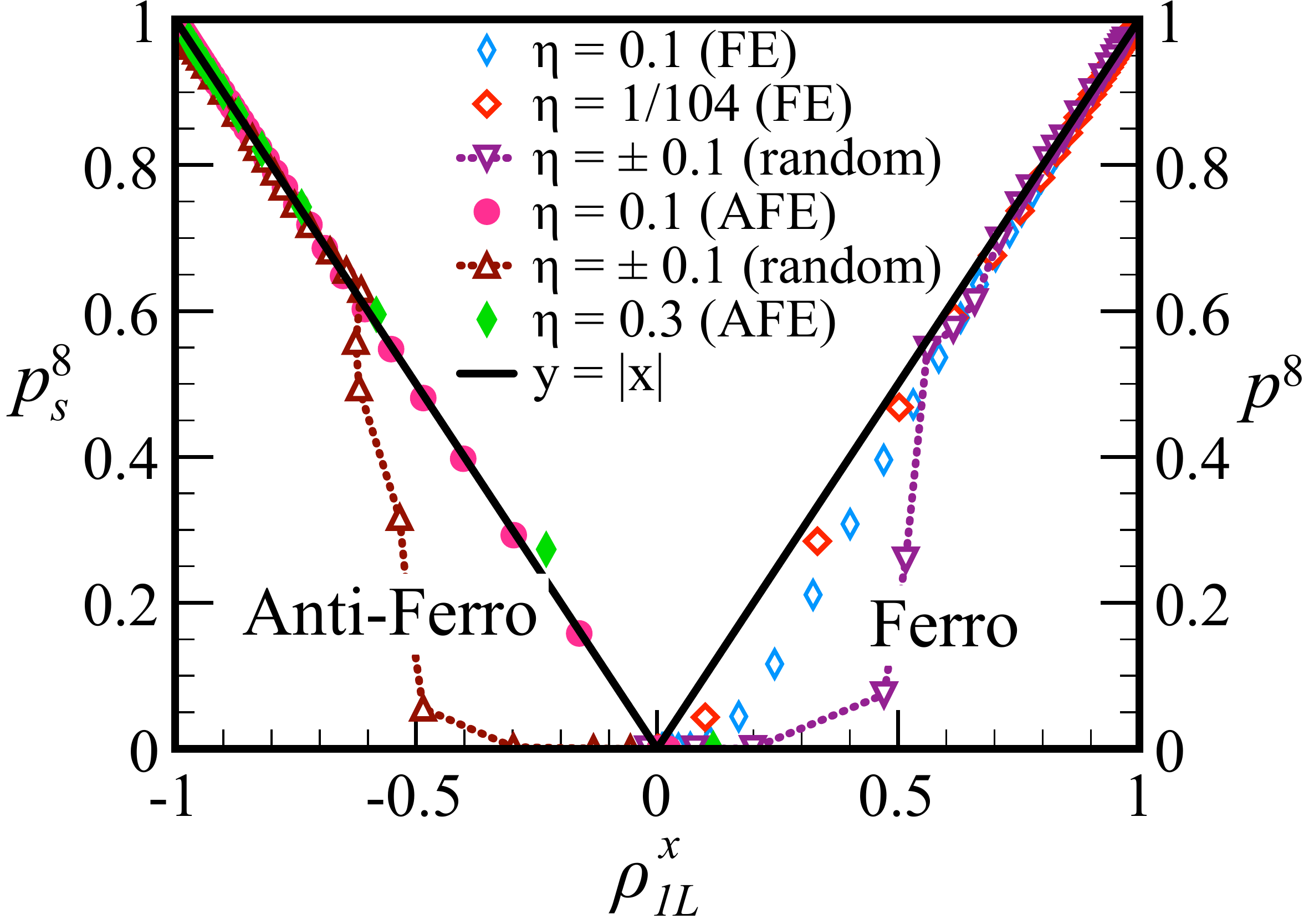} 
\caption{End-to-end correlation in the QI chain with $\eta\sum_l\sigma^x_l$. Except for $\eta= \frac{1}{104}$ ($L=104$, FE), all others have $L=400$. In the random case, $\eta_l$ (on every site) is randomly $\pm 0.1$, and the data is averaged over 100 ensembles. Here, $p_s$ denotes the staggered order parameter for the anti-ferro $J$, and $h=1$.}
\label{fig:end2endeta}
\end{figure}

An interesting alternative to protect these Majorana modes from an $\eta$ stronger than $\frac{1}{L}$ is to have an anti-ferro $J$ (the opposite sign of $t$). Since the ground state is anti-FE ordered for the nearest-neighbor case, its degeneracy is immune to a uniform $\eta\sim \mathcal{O}(1)$, provided $\eta$ is reasonably smaller than the gap. Moreover, to have a staggered $\eta^{ }_l\sim (-1)^l$ which can lift the degeneracy of the anti-FE ground state is not trivial like having a uniform $\eta$, and is probabilistically very unlikely (if $\eta_l$'s are randomly $\pm$). In Fig.~\ref{fig:end2endeta}, the DMRG data on the QI chain shows good agreement with our observations. Notably for strong $J$ (deep inside the ordered phase), the different cases of $\eta$ (including the random $\eta_l=\pm0.1$) fall on the exact line. Moreover, the anti-ferro case exhibits better protection for the Majorana modes (for uniform $\eta$). Thus, the Majorana fermion modes arising due to the QI dynamics are observable objects, and not as fragile as one believed.

\section{Summary}

We have studied a lattice mode of the Rabi quantum cavities, namely the Rabi lattice model (RLM). The  quantum phase diagram of the RLM is generated by means of analytical and numerical calculations. The emergence of QI dynamics in the Rabi lattice model crucially depends upon not enforcing RWA on the dipole interaction. It is shown the RLM exhibits quantum Ising dynamics in strong atom-photon coupling limit. And thus a para-electric (PE){\index{para-electric}} to ferro-electric (FE){\index{ferro-electric}} phase transition is seen. Through the strong coupling QI dynamics, the 1d Rabi lattice with open boundaries realizes two Majorana modes in the ordered phase. These edge modes are observable via the relation, $\rho^x_{1L}=p^8$, between the end-to-end dipole correlation and the order parameter. Thus, we predict the occurrence of Majorana-like edge modes in Rabi lattice. These modes do have a chance to survive against the longitudinal field. The anti-ferro order is particularly good at it. This work has been published~\cite{bkumar.somenath}. 
In view of a recent proposal~\cite{Solano}, it appears that the strong-coupling regime of the Rabi problem may now become accessible.

Below, in the Appendix, we quickly present some more calculations that we did on Rabi lattice, for a general understanding of this problem.



\clearpage


\def\ahat{\hat{a}}
\def\bhat{\hat{b}}
\def\atilde{\tilde{a}}
\def\nhat{\hat{n}}
\def\Nhat{\hat{N}}
\def\Phat{\hat{P}}
\def\k{{\bf k}}
\def\r{{\bf r}}
\def\deltavec{\vec{\delta}}
\def\omegatilde{\tilde{\omega}}
\def\ttilde{\tilde{t}}
\def\Deltatilde{\tilde{\Delta}}

\def\ttilde{\tilde{t}}
\def\cbar{\bar{c}}
\def\bbar{\bar{b}}

\def\Deltatilde{\tilde{\Delta}}
\def\omegatilde{\tilde{\omega}}
\def\lambdatilde{\tilde{\lambda}}
\def\epsilontilde{\tilde{\epsilon}}

\def\adag{{\hat{a}}^\dagger}
\def\a{{\hat{a}}}
\def\bdag{{\hat{b}}^\dagger}
\def\b{{\hat{b}}}
\def\cdag{{\hat{c}}^\dagger}
\def\c{{\hat{c}}}

\def\akdag{{\hat{a}}^\dagger_\k}
\def\ak{{\hat{a}}^{}_\k}
\def\bkdag{{\hat{b}}^\dagger_\k}
\def\bk{{\hat{b}}^{}_\k}
\def\xhat{\hat{x}}
\def\phat{\hat{p}}

\def\yhat{\hat{y}}

\def\amkdag{{\hat{a}}^\dagger_{-\k}}
\def\amk{{\hat{a}}^{}_{-\k}}
\def\bmkdag{{\hat{b}}^\dagger_{-\k}}
\def\bmk{{\hat{b}}^{}_{-\k}}
\def\xk{{\hat{x}}_\k}
\def\pk{{\hat{p}}_\k}

\def\a{\hat{a}}
\def\adag{\hat{a}^\dagger}
\def\sighat{\hat{\sigma}}
\def\chihat{\hat{\chi}}
\def\nhat{\hat{n}}
\def\Uhat{\hat{U}}


%
\begin{subappendices}
%

      
In this following Sections, we study the following generalized version of the Rabi lattice model.  
\begin{equation}
H_{\eta} = H_{ph} + H_{at} + V_\eta
\label{eq:H_eta}
\end{equation}
where 

\begin{subequations}
\begin{eqnarray}
H_{ph} &=& \omega\sum_{l}\left(\ahat^\dag_l\ahat^{ }_l +\frac{1}{2}\right)- \frac{t}{2} \sum_{l,\deltavec} (\ahat^\dag_l\ahat_{l+\deltavec} + h.c.) \label{eq:H_ph}\\
H_{at} &=& \frac{\epsilon}{2}\sum_l \sigma^z_l \label{eq:H_at}\\
V_\eta &=& \gamma\sum_l\left[(\sigma^+_l\ahat^{ }_l + \sigma^-_l\ahat^\dag_l)+\eta(\sigma^+_l\ahat^\dag_l+\sigma^-_l\ahat^{ }_l)\right] \label{eq:V_eta}
\end{eqnarray}
\end{subequations}
Here, $\eta\in [0,1]$ is a mathematical parameter interpolating the Hamiltonian between the physically interesting cases of $\eta=0$, the JC lattice model ($H_0$), and $\eta=1$, the Rabi problem ($H_1$). 

From an alternative perspective, $H_1$ appears to be special and different from the rest of $H_\eta$. Define the new photon operators, $\atilde_l= (\ahat^{ }_l + \eta\ahat^\dag_l)/{\sqrt{1-\eta^2}}$. This is a valid Bogoliubov transformation~\cite{Bogoliubov}. It transforms the interaction to $V_\eta=\gamma\sqrt{1-\eta^2}\sum_l(\sigma^+_l\atilde^{ }_l + h.c.)$, that is interestingly the JC interaction. In addition, there also appear in $H_{ph}$ the terms of the pairing type, $\atilde^\dag_l\atilde^\dag_{l^\prime}$, whose strength is proportional to $\frac{\eta}{1-\eta^2}$. Since this transformation is singular at $\eta=1$, it is not applicable to the Rabi model. This makes the $H_\eta$ for $\eta<1$ to have an apparent closeness to the JC model, while the $H_1$ stands alone.

Clearly, the $H_0$ and $H_1$ seem to represent two distinct ends of the family of $H_\eta$. How much the properties of the two differ, and how these properties evolve with $\eta$, is not obvious a priori. As mentioned earlier, the $H_0$ is a well studied model. Most notably, it is known to exhibit a superfluid to insulator like quantum phase transition. Here, the superfluid phase is a coherent state of photons on the lattice, characterized by $\psi = \langle \ahat^\dag_l\rangle \neq 0$. The insulating phase, with $\psi=0$, refers to a state of localized polaritons (entangled atom-photon states). The multi-lobed quantum phase diagram of $H_0$, where each lobe is an insulating phase labeled by the polariton number (an integer), is remarkably similar to that of the Bose-Hubbard model. We would similarly like to know the ground state phase diagram of $H_1$ which hasn't been investigated much.  At the very outset, let us state what we find from our calculations. It is that the quantum phase diagram of $H_1$ is different from that of $H_0$, with no such multi-lobed features of the Bose-Hubbard type. Of course, there is more to it than briefly stated, and will be discussed at length in the following sections.

We employ two different mean-field approaches to study $H_\eta$. In the first approach, one self-consistently studies a local problem (a single quantum cavity, or a cluster thereof) immersed in a mean-field bath of the photon condensate.  This idea was pioneered by Sheshadri {\it et al.}~\cite{Sheshadri.93} more than twenty years back, for studying the superfluid and insulating phases in the Bose-Hubbard model. Recently this method has been successfully applied to study the JC lattice model~\cite{Greentree.06}. This is discussed in Section~{\ref{subsec:Cluster-MFT}}. It is the simplest method to investigate the polariton-insulator to photon-superfluid quantum phase transition. We have called it Photonic MFT. The second approach is a bosonic mean-field theory described in Section~\ref{subsec:Bose-MFT}. It is based on the Schwinger boson representation of the atomic operators, and then apply small fluctuation in the spins from its the stable  configurations. The  results of the two methods are found consistent with each other. In the last Section~{\ref{sec:ent}} we have calculated the entanglement entropy for two QED cavities.


\section{\label{subsec:Cluster-MFT} Cluster mean-field theory}

{\index{Cluster mean-field theory}}Since the quantum cavities in the lattice model are coupled through the photon hopping, a simple mean-field scheme would be to decouple the hopping. It converts the lattice problem into a local problem of a single quantum cavity coupled (through $t$) to the photon condensate given by the order parameter, $\psi$, which is to be determined self-consistently by solving the local problem exactly. We do the same for $H_\eta$. Actually, we study $\tilde{H}_\eta = H_\eta -\mu \Nhat_{pol}$, that is the grand-canonical version of  $H_\eta$. The `chemical potential' $\mu$ fixes the average polariton number. 

The single cavity mean-field Hamiltonian can be written as follows.  
\begin{eqnarray}
H_{\rm mf,1} &= (\omega-\mu)\left(\ahat^\dag\ahat +\frac{1}{2}\right) +\frac{(\epsilon-\mu)}{2}\sigma^z -  zt\psi (\ahat^\dag+\ahat) \nonumber \\ &  + \gamma\left[(\sigma^+\ahat + \sigma^-\ahat^\dag) + \eta(\sigma^+\ahat^\dag+\sigma^-\ahat)\right] \label{eq:Hmf1}
\end{eqnarray}


Here, the term proportional to $zt\psi$ arises as a result of the mean-field approximation of the photon hopping. It couples the quantum cavity to the photon condensate through the mean-field order parameter $\psi$, where the self-consistent equation, $\psi=\frac{1}{2}\langle\ahat +\ahat^\dag\rangle$, determines $\psi$. Similar calculations has also been done by Schiro {\it et al.}~{\cite{RabiLattice.Schiro}}.

A natural extension of the above scheme is to consider a cluster of quantum cavities self-consistently coupled to the photon condensate. The basic idea is to treat the intra-cluster interactions and couplings exactly, while the boundary of the cluster is assumed to couple to the photon condensate though $t\psi$. 
In the present case, we also make calculations on a two-site cluster, whose Hamiltonian can be written as follows.
\begin{equation}
H_{\rm mf,2} = H^{(L=2)}_\eta -(z-1)t\psi \sum_{l=1}^2 (\ahat^\dag_l+\ahat^{ }_l) \label{eq:Hmf2}
\end{equation}
Here, $H_\eta^{(L=2)}$ is the $H_\eta$ [of Eq.~(\ref{eq:H_eta})] for $L=2$ (that is, for two quantum cavities). The second term in $H_{\rm mf,2}$ is the mean-field term that couples the two-site cluster to the photon condensate. Notice the factor `$(z-1)t\psi$' in $H_{\rm mf,2}$ in comparison with `$zt\psi$' in $H_{\rm mf,1}$. This difference arises because the photon hopping between the two cavities of the cluster has been taken exactly in $H_\eta^{(L=2)}$. Therefore, on a lattice with coordination number $z$, only the coupling to remaining $(z-1)$ `external' cavities is treated within the mean-field approximation. The self-consistent equation for $\psi$ in two cavity case can be written as: $\psi =\frac{1}{4}\langle \ahat_1 +\ahat_2 + h.c. \rangle$. Although one would like to study larger clusters to get more accurate data, we do calculations only on $H_{\rm mf,1}$ and $H_{\rm mf,2}$. The results of the two are in qualitative agreement with each other, and provide useful understanding of the phase transition in the ground state of $H_\eta$. 

We compute $\psi$ by solving for the ground state of $H_{\rm mf,1}$ (and $H_{\rm mf,2}$) self-consistently. The interpretation of the results is straightforward. For $\psi\neq 0$, the ground state is a photon condensate, which in physical terms would imply a macroscopic electric field (or polarization) in the system. For $\psi=0$, it is a polariton insulator.

In our calculations, we have set $\gamma$ as the unit of energy. The dimensionless parameters, in terms of which we discuss the results, are $\omegatilde=\frac{\omega-\mu}{\gamma}$, $\ttilde=\frac{zt}{\gamma}$, $\Deltatilde=\frac{\omega - \epsilon}{\gamma}$, and $\eta$. We have discussed the quantum phase diagram in the plane of $\omegatilde$ and $\ttilde$ for $\Deltatilde=0, 1$ for both single site and two site case.. 


\begin{figure}[ht] 
   \centering
   \includegraphics[width=0.46\textwidth]{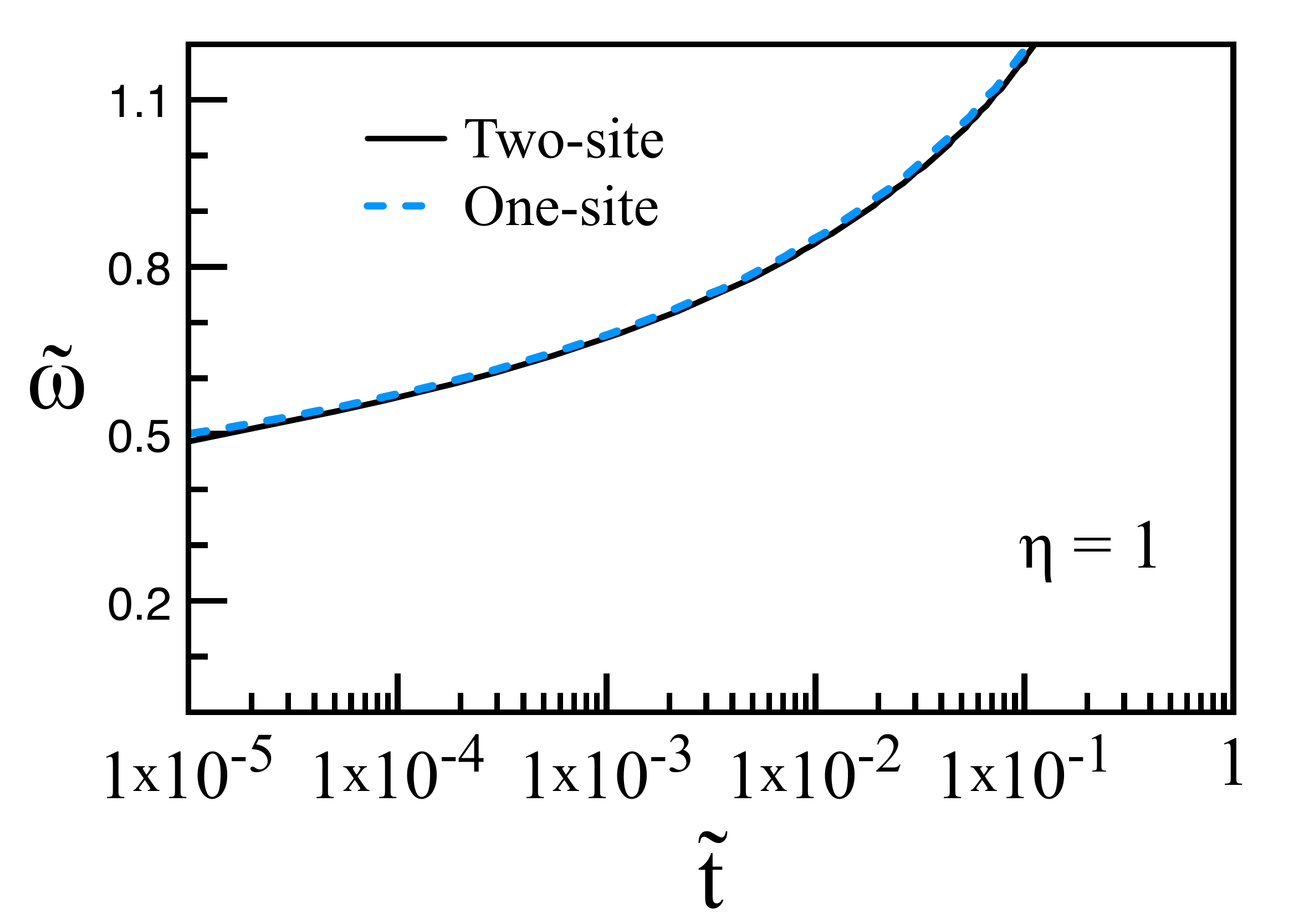}
   \includegraphics[width=0.46\textwidth]{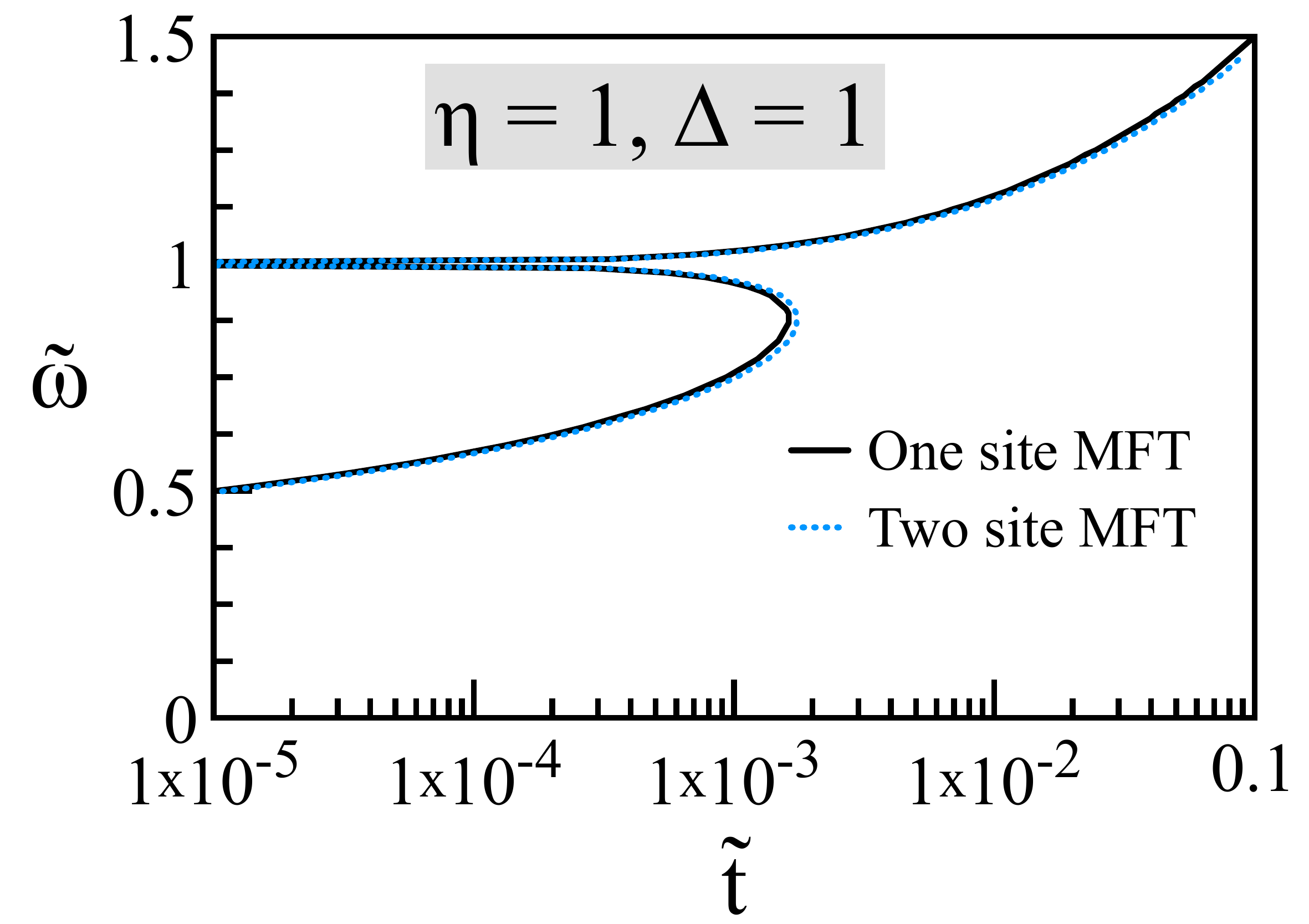}
   \caption{Quantum phase diagram of  $H_{\rm mf,1}$ and $H_{\rm mf,2}$ on $\ttilde$ - $\omegatilde$ plane. This is our take on the full Rabi problem i.e., for $\eta = 1$. Left figure is for $\eta = 1.0, \Delta = 0.0$ and the right figure is for $\eta =1.0, \Delta = 1.0$ For negative $\Delta$ there is no lobe like feature even for $\eta = 1.0$.}
   \label{fig:tw-eta1-delta0-n20}
\end{figure}


\section{\label{subsec:Bose-MFT}Bosonic mean-field theory}



In this appendix, we study the Rabi lattice model using a mean-field approach described in Section~\ref{sec:Bose-MFT}. The basic idea is to do a spin-wave like mean-field analysis with respect to the ground state of the atoms. 




%
In the Schwinger boson representation, the Pauli spin matrices in terms of two bosonic operators $\hat{b}$ and $\hat{c}$ are written as
\begin{eqnarray}
&& \sigma^z_l = \bdag_l \b_l - \cdag_l \c_l \label{sigz} \\
&& \sigma^+_l = \bdag_l \c_l  \\
&& \sigma^-_l = \cdag_l \b_l \\
&& \rm{I} = \bdag_l \b_l + \cdag_l \c_l \qquad  \mbox{local constraint}
\end{eqnarray}

We use a physically meaningful mean-field approximation here. The atomic ground state is achieved when $\langle \sigma^z \rangle$  is $-1$ as $\epsilon > 0$. Now we can think of the $\b$ type of bosons which will disperse through an average field or background of $\c$ type of bosons. So the excitation is only with the $\b$ bosons. We replace $\hat{c}$ with a complex number $\cbar$ and create the fluctuation due to $\hat{b}$ operator on $\cbar^2$ value.  So when $\langle \sigma^z \rangle$ is $-1$, $\cbar^2 = 1$ and $\langle {\bdag \b \rangle}$ is $0$. On excitation $\langle {\bdag \b \rangle}$ will have some small positive value when $\cbar^2$ will be greater than $-1$ towards $0$ (zero) keeping the constrain maintained. When the atom is fully exited $\langle \sigma^z \rangle$ is $1$, $\cbar^2 = 0$. $\langle \sigma^z \rangle = 0$ gives $\cbar^2 = 0.5$. This simple mean-field approximation helps us to write the interacting part of the atom-photon problem in a fashion which is exactly diagonalizable.
\begin{eqnarray}
&& \sigma^z_l = \bdag_l \b_l - \cbar^2 \qquad \sigma^+_l = \bdag_l \cbar \qquad \sigma^-_l = \b_l \cbar 
\end{eqnarray}
And of course the constrain $\bdag_l \b_l + \cbar^2 = 1$ which will be connected to the Hamiltonian through a Lagrange multiplier. The atom-photon interaction Hamiltonian is written with constrain, included using Lagrange multiplier $\lambda$.

\begin{eqnarray}
H &=& \omega {\sum_l} {\left( {\a^\dagger_l \a_l + {1 \over 2}} \right)} + {\epsilon \over 2} {\sum_l}{\sigma^z_l} + \gamma {\sum_l} {\left( {\a^\dagger_l + \a_l} \right)} {\sigma^x_l} - t {\sum_{l, \delta}} {\left( {\a^\dagger_l \a_{l, \delta} + h.c} \right)} \cr && + \lambda {\sum_l} {\left( {1 - ( \b^\dagger_l  \b^{}_l + {\bar{c}}^2 )} \right)}
\end{eqnarray}
Where we have taken $\vec{\delta}$ summation is over $+ \hat{x}$ and $+ \hat{y}$ i.e., half of the nearest neighbor (nn) coordinates are summed over. To be specific, here we take square lattice. The $\omega$ is photon frequency, $\epsilon$ is atomic transition energy and $t$ is photon hopping amplitude. All the parameters are defined in the units of atom-photon coupling strength, $\gamma$. Detuning parameter $\Delta$ is defined as $\Delta = \omega - \epsilon$. In experiments, its the $\Delta$ knob, which can be adjusted to make the system near or far from resonance. $\Delta =0$ is resonance condition. We replace the spin matrices ($\sigma$'s) with the bosonic operators ($\b$)  and the mean-field parameter $\cbar$. As the Hamiltonian is of bilinear in form, we can do Fourier transform to make the problem a local one. And the Hamiltonian in $\k$ space looks like

\begin{eqnarray}
H = L {\epsilon_0} + {\sum_\k}{\omega_\k} { {{\a^\dagger}_\k \a^{}_\k } } + {\alpha} {\sum_{\k}} {\bkdag \bk} +  {\gamma}{\cbar} {\sum_{\k}} \left( { \akdag + \amk } \right) \left( { \bmkdag + \bk} \right) 
\end{eqnarray}
where $\epsilon_0 = {\omega \over 2} - {\alpha} ( 1 - \cbar^2 ) + {\epsilon}({{1 \over 2} - \cbar^2})$, $\omega_\k = \omega - 2 t {\sum_{\delta}} \cos {(\k \cdot \delta)}$ and $\alpha = {\epsilon \over 2} - \lambda$ are the redefined quantities. We have to diagonalize this Hamiltonian which is the full Rabi lattice problem. We use the simple harmonic oscillator technique to solve this problem, by defining displacement and momentum operators for two oscillators as  $\xhat_{\k}$ - $\phat_{x, \k}$ as $\xhat_\k = {\frac{1}{\sqrt{2 \omega_\k}}} \left( {\amkdag + \ak } \right) $ and $\phat_{x, \k} = i {\sqrt{\frac{\omega_k}{2}}} \left( {\amkdag - \ak } \right) $ and $\yhat_{\k}$ - $\phat_{y, \k}$ as $\yhat_\k = {\frac{1}{\sqrt{2 \omega_\k}}} \left( {\bmkdag + \bk } \right) $ and $\phat_{y, \k} = i {\sqrt{\frac{\omega_k}{2}}} \left( {\bmkdag - \bk } \right) $. They follow the commutation relation $\left[ {\xhat_{\k_1}}, \phat_{x, {\k_2}} \right] = i \delta (\k_1 + \k_2)$ etc. This allows us to write the Hamiltonian as

\begin{eqnarray}
H &=& L {\epsilon_1} +  {\sum_\k} { \omega_\k} {\left( { \xhat_{-\k} \xhat_\k + \phat_{x, -\k} \phat_{x, \k}} \right)} 
+ {\alpha \over 2} {\sum_{\k}} {\left( { \yhat_{-\k} \yhat_\k + \phat_{y, -\k} \phat_{y, \k}} \right)} \nonumber \\
&+&  2 {\gamma}{\cbar} {\sum_{\k}} {\xhat_{-\k}} \yhat_\k
\end{eqnarray}
This form of Hamiltonian is of two simple harmonic oscillator. $\epsilon_1 = {\epsilon}({{1 \over 2} - \cbar^2}) - {\alpha} ( {3 \over 2} - \cbar^2 )$. We write equation of motion for $\xhat_\k, \yhat_\k, \phat_{x,\k}$ and $\phat_{y,\k}$ and find the normal modes and frequencies after diagonalizing the Hamiltonian. That mean, we find the suitable unitary transformation to diagonalize this Hamiltonian. The diagonal Hamiltonian is given by
\begin{eqnarray}
H_{diagonal} &=& L \epsilon_1 + {\sum_\k} {\omega_{+, \k}} \left( \adag_{u, \k} \a_{u, \k} + {1 \over 2} \right) + 
 {\sum_\k} {\omega_{-, \k}} ( \adag_{v, \k} \a_{v, \k} + {1 \over 2} )
\end{eqnarray}
So the ground state energy is given by $e_g = {\epsilon_1} + {1 \over {2 L}} {\sum_\k} ({\omega_{+, \k}} + {\omega_{-, \k}} ) $, where $ \omega^2_{\pm , \k} = {1 \over 2} {\left[ {\omega^2_\k + \alpha^2 } \right]} {\pm}  \epsilontilde_\k $ and $\epsilon^2_\k = {1 \over 4} {\left[ {\omega^2_\k - \alpha^2 } \right]}^2 + 4 {\gamma^2} {{\bar{c}}^2} {\omega_\k} \alpha $. We minimize this ground state energy with respect to $\alpha$ and $\cbar^2$ to get two equations which are to be solved self-consistently. 
\begin{eqnarray}
\alpha &=& \epsilon - {\gamma}{\frac{\alpha}{2 L}} {\sum_\k} {\frac{\omega_\k}{\epsilon_\k}} {\left( { {\frac{1}{\omega_{+ , \k}}} - {\frac{1}{\omega_{- , \k}}} } \right)}  \label{alpha} \\
{\bar{c}}^2 &=& {3 \over 2} - {\frac{\alpha}{4 L}} {\sum_\k} {\left( { {\frac{1}{\omega_{+ , \k}}} - {\frac{1}{\omega_{- , \k}}} } \right)} \nonumber \\
&+& {1 \over {8 L}} {\sum_\k} {\frac{ (\omega_\k^2 - \alpha^2 ) \alpha - 4 {\gamma^2} {\bar{c}}^2 {\omega_\k} }{\epsilon_k}} {\left( { {\frac{1}{\omega_{+ , \k}}} - {\frac{1}{\omega_{- , \k}}} } \right)} ~~ ~ ~ {\label{cs}}
\end{eqnarray}
\begin{figure}[htbp] 
   \centering
   \includegraphics[width=0.6\textwidth]{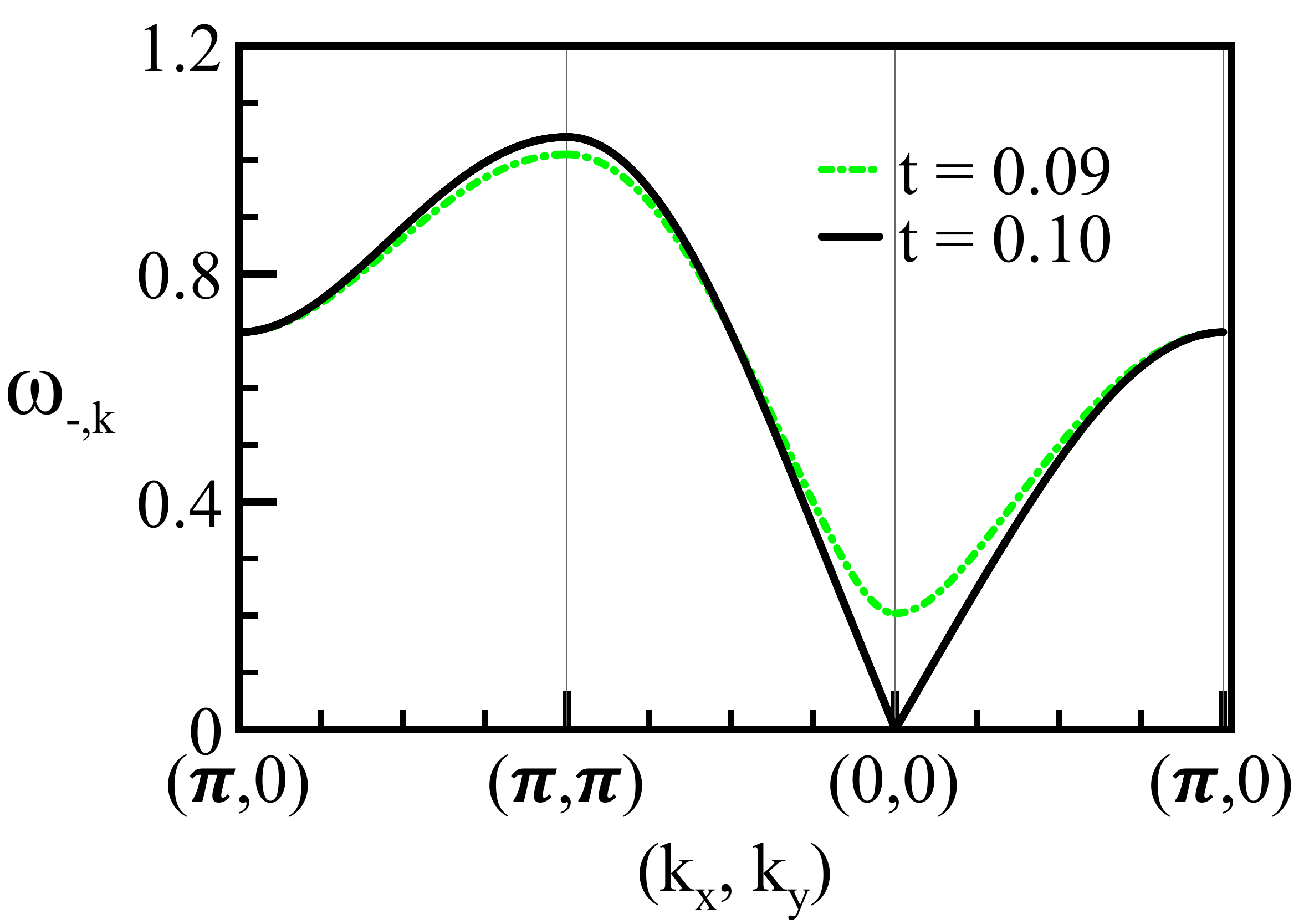} 
   \caption{The dispersion relation of $\omega_{-, \k}$ is plotted for two different values of $t$ namely,  0.1 (the black solid curve) and 0.09 (the green dotted line) for same values of $\omega = 1.704224, \c^2 =  0.800532$ and $\alpha = 2.4552$. This shows how the gap becomes zero as we approach to the phase boundary.}
   \label{fig:dis-diff-t}
\end{figure}
These two equations will give us the quantum phase diagram on $t - \omega$ plane. To show that there is  transition between order (gapped) and disordered (gapless) phases we plot the gap, which is given by  $\omega_{-, \k = 0,0}$ as a function of $t$ and $\omega$.
\begin{figure}[htbp] 
   \centering
  \includegraphics[width=0.95\textwidth]{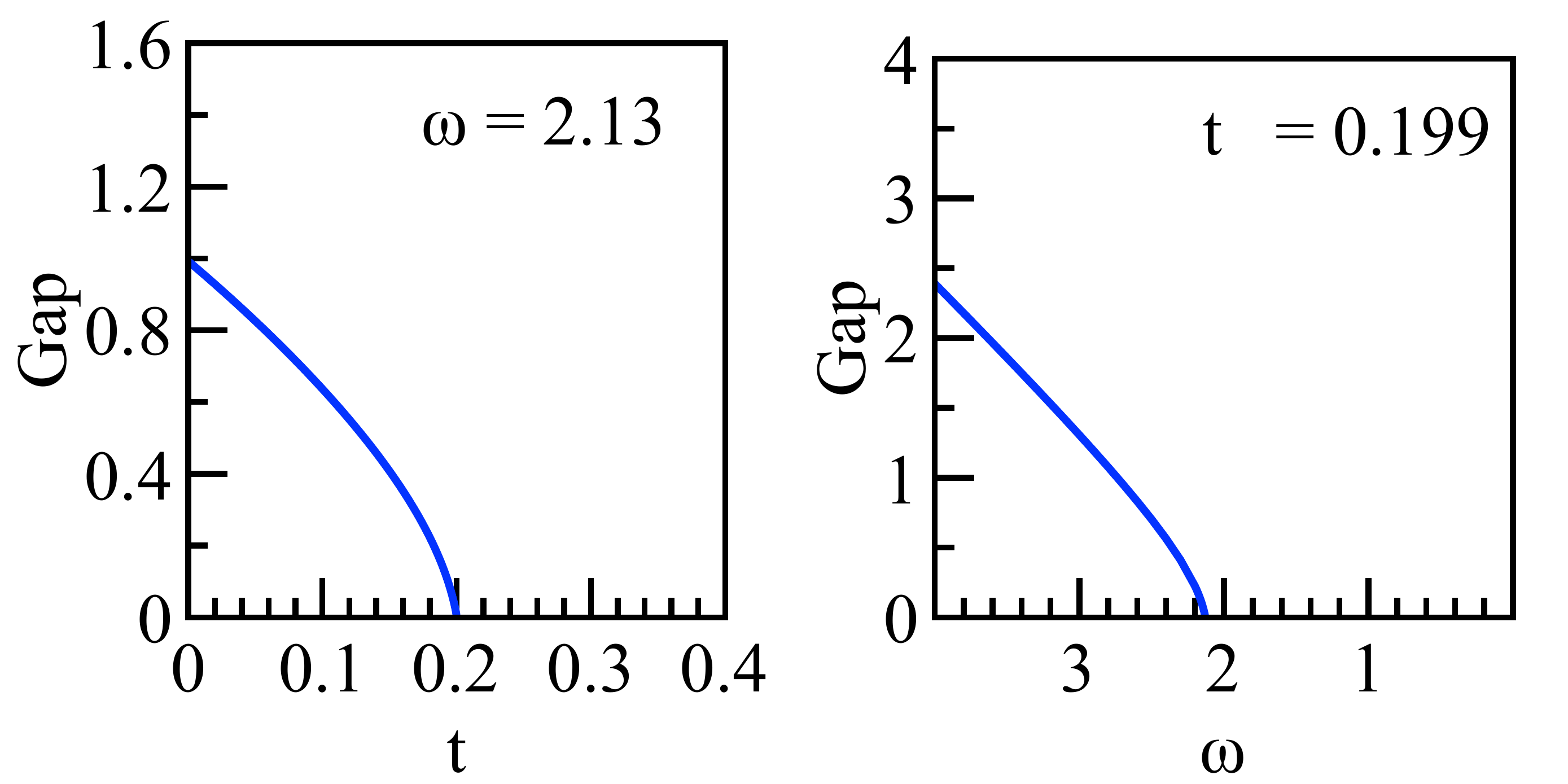} 
   \caption{ The gap ($\omega_{-, \k = 0,0}$) is plotted as a function of $t$ and $\omega$ across the phase diagram for a fixed value of $c^{2} = 0.87, \alpha = 2.63$. This is for the resonance case i.e., $\Delta = 0$.}
   \label{fig:mft-tw}
\end{figure}
To get the boundary of the phase on $t - \omega$ plane we set $\omega_{-, \k} = 0$ and this gives us the condition on $\alpha$ which is
\begin{equation}
\alpha = 4 {\cbar^2}/ {( \omega - 4 t )} \label{alp}
\end{equation}
This expression of $\alpha$ is used to redefine the Eq.~(\ref{alpha}) as
\begin{equation}
\omega = 4 t + {\frac{4 \cbar^2}{\epsilon}} {\left[ {1 + {1 \over {2 L}} {\sum_\k} {\frac{\omega_\k}{\epsilon_\k}} \left( { {\frac{1}{\omega_{+ , \k}}} - {\frac{1}{\omega_{- , \k}}} } \right)} \right]} \label{omi}
\end{equation}
This equation along with Eq.~(\ref{cs}) is now our two self-consistent equations which are to be solved to get the phase diagram. First we take some initial value of $\cbar^2, \omega, t$ and calculate $\alpha$. We put this $\alpha$ in Eq.~(\ref{cs}) and get a new $\cbar^2$. This value of $\cbar^2$ and $\alpha$ is used in Eq.~(\ref{omi}) to get a new $\omega$.  And again with these values of $\omega$ and $\cbar^2$ we find $\alpha$ from Eq.~(\ref{alp})Thus for a fixed $t$ we iterate Eq.~(\ref{omi}) and Eq.~(\ref{cs}) self-consistently to get a $\omega$ which corresponds to a particular $t - \omega$ point on the phase diagram. Thus we generate the complete phase diagram. This is to remind the reader that Rabi lattice model discussed here is same as our initial 1 site and 2-site MFT for $\eta = 1$ case.
\begin{figure}[htbp] 
\centering
\includegraphics[width=0.6\textwidth]{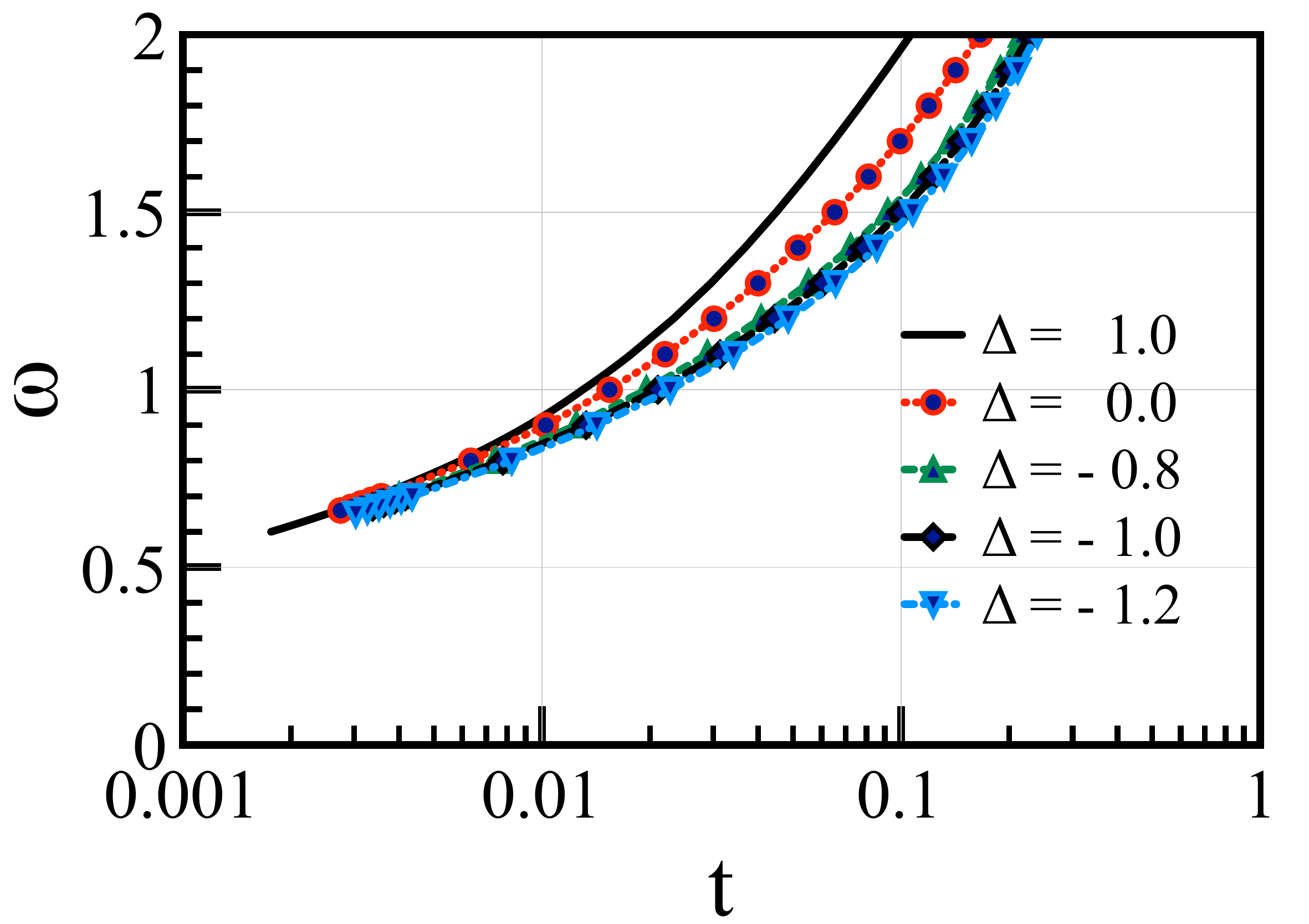} 
\caption{The quantum phase diagram generated from the SB mean-field approximation for the Rabi Lattice model for different detuning $\Delta$ parameter.}
\label{fig:mft-tw}
\end{figure}



\section{{\label{sec:ent}}Entanglement between two optical cavities}
In classical systems the phase transition occurs with a macroscopic order below a certain temperature. In quantum mechanics, zero temperature phase transition happens because of competing interaction between the agents inside the Hamiltonian. In both the cases, at phase transition point the correlation length becomes infinite. Entanglement is measure of correlations strictly for quantum systems. Thus we are interested in checking what is the behavior of `Entanglement' between two QED cavities near the phase transition point. 

The ground state of the two-site cavity problem can be written as
\begin{eqnarray}
{| {\Psi} \big >} &=& {\sum_{n_1 , n_2 }} {\Bigg\{} { {C_{n_1, n_2, \uparrow, \uparrow}} {|n_1 , n_2\big>} \otimes {| \uparrow, \uparrow \big>} } + { {C_{n_1, n_2, \uparrow, \downarrow}} {|n_1 , n_2\big>} \otimes {| \uparrow, \downarrow \big>} } \nonumber \\
&+& { {C_{n_1, n_2, \downarrow, \uparrow}} {|n_1 , n_2\big>} \otimes {| \downarrow, \uparrow \big>} } + { {C_{n_1, n_2, \downarrow, \downarrow}} {|n_1 , n_2\big>} \otimes {| \downarrow, \downarrow \big>} }  {\Bigg\}} \nonumber \\
&=& {\sum_{\sigma_1, \sigma_2}} {\big| \Psi_{\sigma_1, \sigma_2}\big>} \otimes {\big| {\sigma_1, \sigma_2} \big>}
\end{eqnarray}
where both $\sigma_{1,2}$ can be $\uparrow, \downarrow$ and ${\big| \Psi_{\sigma_1, \sigma_2}\big>}$ is defined as
\begin{equation}
{\big| \Psi_{\sigma_1, \sigma_2}\big>} = {\sum_{n_1, n_2}} C_{n_1, n_2, \sigma_1, \sigma_2} {\big| n_1, n_2 \big>}
\end{equation}
To compute entanglement in $|\Psi\rangle$, we first construct its density matrix.
The density matrix  for the state $| \Psi \big>$ of two-coupled cavities is  defined as,
\begin{eqnarray}
\rho &=& |\Psi \big> \big< \Psi | \nonumber \\
&=& {\sum} {C_{{n_1, n_2, \sigma_1, \sigma_2}}} {C_{{m_1, m_2, \tau_1, \tau_2}}^{\star}} {| n_1, n_2 \big> \big< m_1, m_2 |} \otimes {| \sigma_1, \sigma_2 \big> \big< \tau_1, \tau_2 |}
\end{eqnarray}
To be explicit and clear, the single sum is over eight sums and they are defined as
${\sum_{n_1}}$, ${\sum_{n_2}}$, ${\sum_{m_1}}$, ${\sum_{m_2}}$, ${\sum_{\sigma_1}}$, ${\sum_{\sigma_2}}$, ${\sum_{\tau_1}}$ and ${\sum_{\tau_2}}$, where $n$ and $m$ are the boson number index and $\sigma$, $\tau$ are spin index. We take the trace over the first cavity and define~{\cite{bk-ent}} inter-cavity bipartite entanglement (entropy) of a pure state as
\begin{eqnarray}
{\rho_{2}} &=& {Tr_{n_1}} {Tr_{\sigma_1}}{(}\hat{\rho}{)} \qquad \\
{\rho_{1}} &=& {Tr_{n_2}} {Tr_{\sigma_2}}(\hat{\rho}) \\
E &=& 1 - Tr ({{\rho_2}^2}) = 1 - Tr ({{\rho_1}^2})
\end{eqnarray}
This is our formula of calculating entanglement entropy. $\rho_2$ can be calculated as
\begin{equation}
\rho_2 = {\sum_{\sigma_1}} {\big< \sigma_1 | {\Bigg (} { \sum_{n_1} \big< n_1|\rho|n_1 \big>} {\Bigg )} | \sigma_1 \big>} 
\end{equation}
So when we trace over one photon the expression looks like below.
\begin{eqnarray}
\big< n | \hat{\rho} | n \big> = {\sum_{n_2, m_2, \sigma_1, \sigma_2, \tau_1, \tau_2}} {C_{n, n_2, \sigma_1, \sigma_2}} {C_{n, m_2, \tau_1, \tau_2}^\star} {| n_2 \big> \big< m_2 |} \otimes {| \sigma_1, \sigma_2 \big> \big< \tau_1, \tau_2 |}
\end{eqnarray}
And tracing over one of the atomic degree of freedom the final density matrix looks like
\begin{eqnarray}
{\big< \sigma |{\big< n | \hat{\rho} | n \big>} | \sigma\big> } &=& {\sum_{n_2, m_2, \sigma_2, \tau_2}} {C_{n, n_2, \sigma, \sigma_2}} {C_{n, m_2, \sigma, \tau_2}^\star} {| n_2 \big> \big< m_2 |} \otimes {| \sigma_2 \big> \big< \tau_2 |} \nonumber \\
&=& \rho_{n, \sigma} \nonumber
\end{eqnarray}
\begin{equation}
\rho_2 = \sum_{\sigma = \uparrow, \downarrow} \sum_{n} {\rho_{n, \sigma}}
\end{equation}
We have two parameters $\ttilde = z t /\gamma$ and $\omegatilde = (\omega-\mu)/\gamma$ on which entanglement between two cavities depends. We plot entanglement as a function of $\omegatilde$ using $\Deltatilde =(\Delta-\mu)/\gamma$ and $\eta$ as a parameter in the figures given below.
Clearly, the entanglement shows peaked-singularities at the point of quantum phase transition.

\begin{figure}[pt]
\centering
\subfloat[$\ttilde-\omegatilde$ phase and Entanglement vs. $\tilde{\omega}$ for different $\ttilde$ values. $\eta = 0.85$, $\Delta = 1.34$]{
\label{fig:ent-w-eta0p85-delta1p34}
\includegraphics[width=0.45\textwidth]{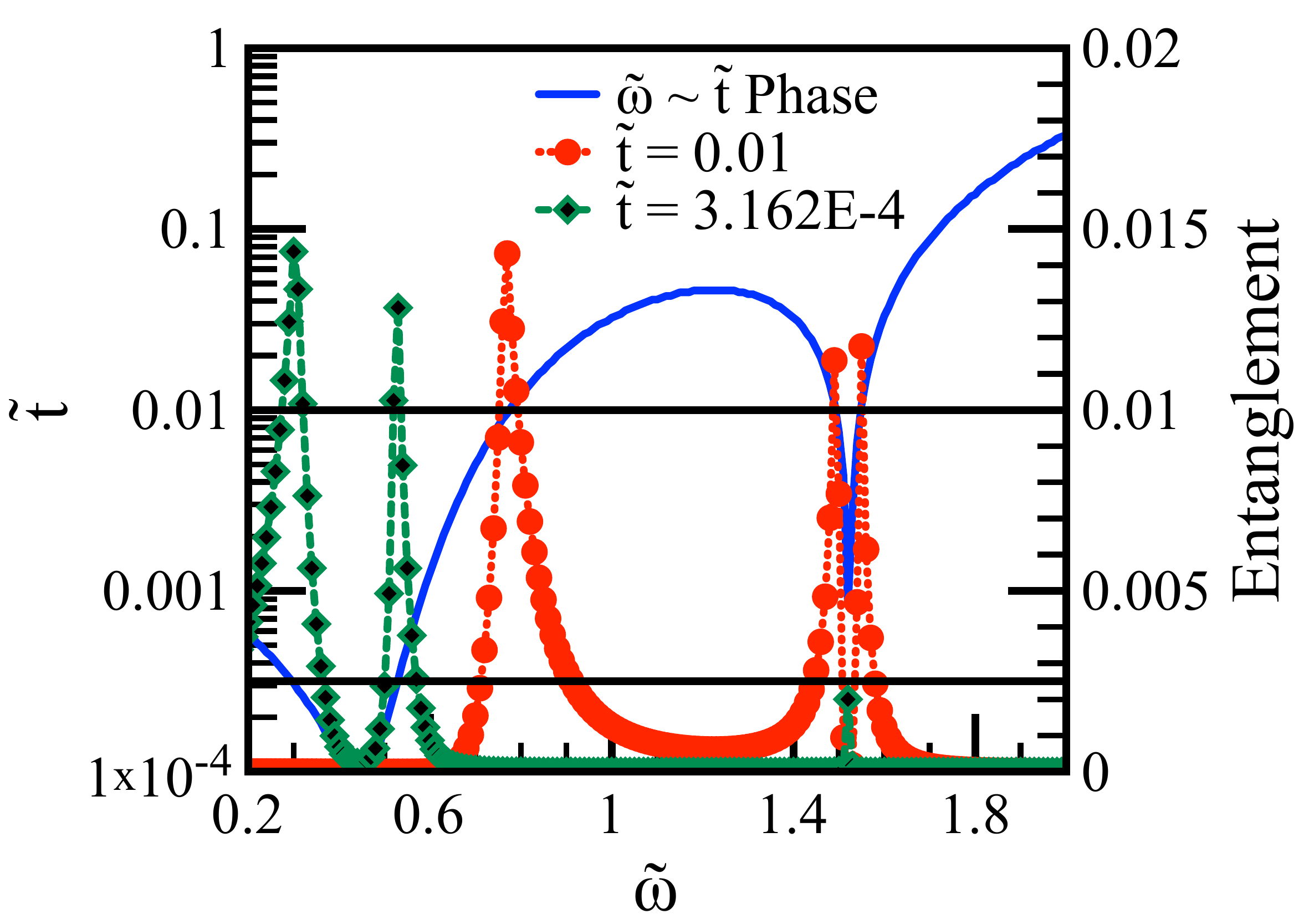} 
}\qquad
\subfloat[$\ttilde-\omegatilde$ Phase and Entanglement vs. $\tilde{\omega}$ for different $\ttilde$ values. $\eta = 1.0$, $\Delta = 1.0$]{
   \label{fig:ent-w-eta1-delta1}
   \includegraphics[width=0.45\textwidth]{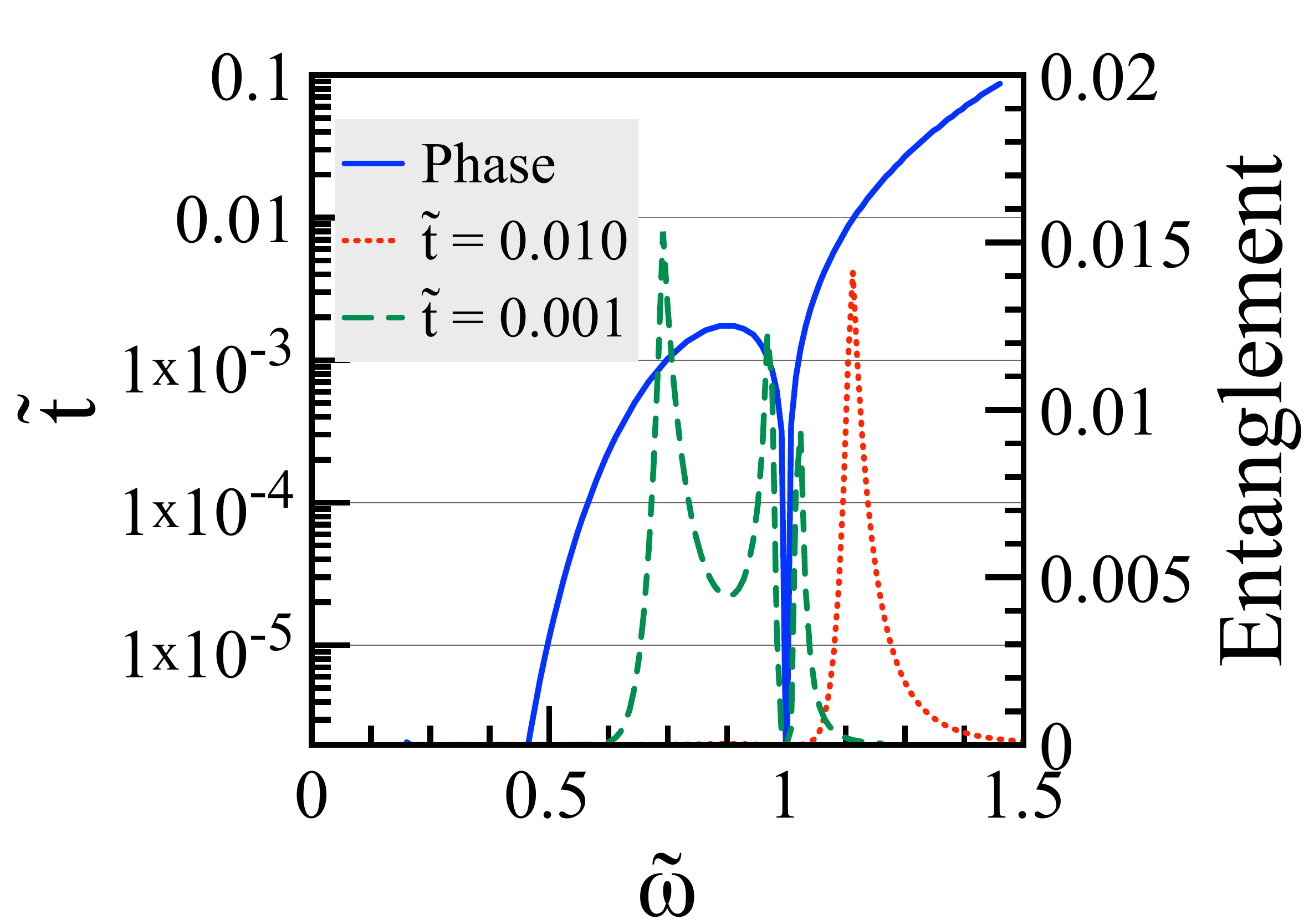} 
}
\caption{ Fig.~{\ref{fig:ent-w-eta0p85-delta1p34}} is a plot of both the phase diagram and entanglement in the same plot. Y1 axis is $\tilde{t}$ and Y2 axis is entanglement. The common $x$ axis is $\omegatilde$. $\Delta = 1.34$, $\eta = 0.85$; Fig.~{\ref{fig:ent-w-eta1-delta1}} is for $\eta = 1$, $\Delta = 1$. For $\ttilde = 0.01$ if we move along $\omegatilde$ axis we cross the phase diagram twice and so we see exactly at those two points the entanglement is maximum and otherwise its zero. The entanglement is plotted on Y2 axis both for $\ttilde = 0.01$ and $0.001$
}
\label{all-entangle}
\end{figure}

%
\end{subappendices}

\clearpage
\bibliographystyle{unsrt}
\bibliography{chapters/ref-all}


%% file: chapters/quantum-ising.tex
\def\adag{{\hat{a}}^{\dagger}}
\def\a{\hat{a}}
\def\Hhat{\hat{H}}
\def\U{\hat{U}}
\def\Udag{{\hat{U}}^\dagger}
\def\s{\sigma}
\def\r{\rho}
\def\ahat{\hat{a}}
\def\nhat{\hat{n}}
\def\chat{\hat{c}}
\def\fhat{\hat{f}}
\def\dhat{\hat{d}}
\def\psihat{\hat{\psi}}
\def\phihat{\hat{\phi}}
\def\Qhat{\hat{Q}}
\def\ntilde{\tilde{n}}
\def\shat{\hat{\sigma}}

\chapter{Edge-modes in a frustrated quantum Ising chain}

\begin{center}
\parbox{0.8\textwidth}{
\footnotesize
{\bf {\small About this chapter}}
\\[10pt]
The ground state properties of an Ising chain with nearest ($J_{1}$) and next-nearest neighbor ($J_{2}$) interactions in a transverse field ($h$) are investigated using DMRG, cluster mean-field theory and fermionic mean-field theory. Its quantum phase diagram is discussed in the $J_{1}$-$J_{2}$ plane, in units of $h$. It has two ordered regions, one of which has simple ferromagnetic or N\'eel order depending upon the sign of $J_{1}$, and the other is the double-staggered antiferromagnetic phase. These ordered phases are separated by a quantum disordered region, consisting of a gapped and a gapless (critical) quantum paramagnetic phases. 
The occurrence of the Majorana-like edge-modes in the ordered phases is investigated by calculating the end-to-end spin-spin correlations on open chains. It is found that there occur four edge modes in the double-staggered phase, while the ferromagnetic and N\'eel phases support two edge modes, except very near $J_{1}=0$ where it seems they have four edge modes like the exact case at $J_{1}=0$ itself.
\\[15pt]
}
\end{center}

\minitoc

\section{\label{sec:intro} Introduction}
The Ising model in a transverse field is an important physical problem.  It is also referred to as the quantum Ising (QI) model because the presence of a transverse field causes quantum fluctuations to the Ising spins. Historically, the interest in this model started through a work of de Gennes where it was used to model the proton dynamics in the hydrogen-bonded ferroelectric materials ($e.g.$, KH$_{2}$PO$_{4}$)~\cite{deGennes}. The QI model continues to be a subject of diverse current interests, as clear from a variety of physical contexts in which it occurs~\cite{book.Bikas, randomQI.Fisher, Moessner, QI.Holo.Girvin, Coldea, bermudez-qi-expt,Mila.Square}. A notable recent case, for instance, is that of the Rabi lattice model which, in the strong coupling limit, exhibits QI dynamics~\cite{bkumar.somenath}.

The one-dimensional (1D) spin-1/2 QI model with only nearest-neighbor interaction, $J_1$, is exactly soluble under JW fermionization~\cite{pfeuty}. It undergoes a quantum phase transition from a disordered to a doubly degenerate ordered ground state as the strength of $J_1$ increases relative to the transverse field, $h$. In the fermionic form, it becomes Kitaev's superconducting quantum wire that harbors two Majorana fermion modes at the free ends of an open chain~\cite{Kitaev.QWire}. These Majorana-like edge-modes arise only in the ordered phase, and do not exist in the disordered phase. There is much current interest in realizing the Majorana modes for quantum computation~\cite{Nayak.RMP, jason-majorana}, and a possible way of achieving this could be through Kitaev's quantum wire, viz., the 1D QI model. 

The Majorana edge modes in Kitaev's quantum wire are topologically protected, as no local perturbation can couple these modes sitting at two opposite ends of the wire. But in the QI chain where the basic physical variables are spins, not fermions, the longitudinal field acts unfavorably upon them. However, through energetic (if not topological) considerations, the edge-modes in a QI chain still stand a chance of survival against such detrimental perturbations, as discussed  in chapter {\ref{chap:rabi}} and also in reference ~\cite{bkumar.somenath}. There, we used the relation, $\rho^x_{1L} = p^8$ (derived by Pfeuty in Ref.~\cite{pfeuty}), between the end-to-end spin-spin correlation, $\rho^x_{1L}$, and the order parameter, $p$,  as a  signature of these edge-modes, and found that it is satisfied even in the presence of the longitudinal fields (uniform as well as random) for strong enough $J_1$ under suitable conditions. Continuing with our studies of the edge modes in the presence of longitudinal operators, we further like to understand the effects of the Ising interactions beyond nearest-neighbor. 

Our basic motivation is to investigate the occurrence of the edge modes in a frustrated quantum Ising problem. Therefore, in this chapter, we study the $J_1$-$J_2$ QI model on an open chain. As described in Eq.~(\ref{eq:model}) of Sec.~\ref{sec:model}, this model has a next-nearest neighbor Ising interaction, $J_{2}$, in addition to $J_{1}$ and $h$. It is a minimal QI problem that has frustration. It is also known in the literature as quantum ANNNI (anisotropic or axial next-nearest neighbor Ising) model, because it is the quantum equivalent of the two-dimensional classical ANNNI model (through the transfer matrix in statistical mechanics)~\cite{QIj1j2.Barber,ANNNI.Selke}. Although it is a well-studied problem~{\cite{dirk}}, the edge modes, it seems, have never been investigated in this model. Since this problem is not exactly soluble, we study it numerically by employing DMRG (density matrix renormalisation group) and cluster-mean-field theory{\index{Cluster mean-field theory}} (CMFT) methods. The calculations and their results are discussed in Secs.~\ref{subsec:QPD} and~\ref{sec:EM}. Through these calculations, we particularly look for the signatures of the edge-modes in the ordered phases of the $J_{1}$-$J_{2}$ QI chain by computing the end-to-end spin-spin correlations. We find that the regions dominated by $J_{2}$ has four edge-modes, while the rest of it support two edge-modes. The main results are summarized in Fig.~\ref{fig:summary} in Sec.~\ref{sec:sum}.

\section{\label{sec:model} The $J_1$-$J_2$ quantum Ising model}
\begin{figure}[htbp] 
\centering
\includegraphics[width=0.8\textwidth]{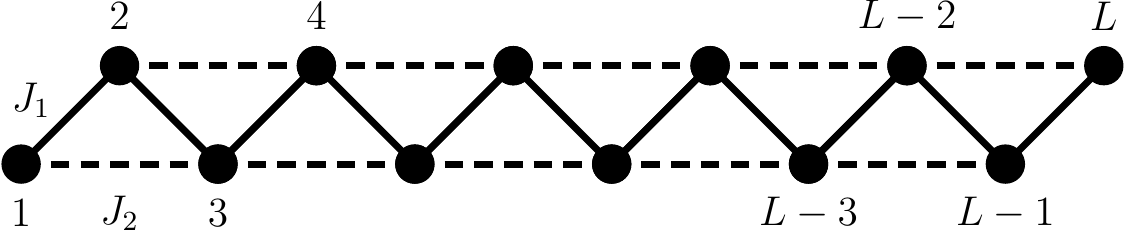} 
\caption{The $J_1$-$J_2 $ quantum Ising chain of $L$ spins. The solid lines denote $J_{1}$ interaction, and the dashed lines $J_{2}$.}
\label{fig:model}
\end{figure}

On a chain with open boundaries, the Ising model with nearest and next-nearest neighbor interactions in a transverse field can be written as follows.
\begin{eqnarray}
\Hhat = J_1 {\sum_{i=1}^{L-1}} {\sigma^x_{i} \sigma^x_{i+1}} + J_2 {\sum_{i=1}^{{L-2}}} {\sigma^x_i \sigma^x_{i+2}} + h {\sum_{i=1}^{L}} {\sigma^{z}_{i}} {\label{eq:model}}
\end{eqnarray}
Here, $\sigma_i^x$ and $\sigma_i^z$ are the Pauli operators, and $L$ is the total number of spins. It is also depicted in Fig.~\ref{fig:model}.

\begin{figure}[ht]
\centering
\includegraphics[width=0.6\textwidth]{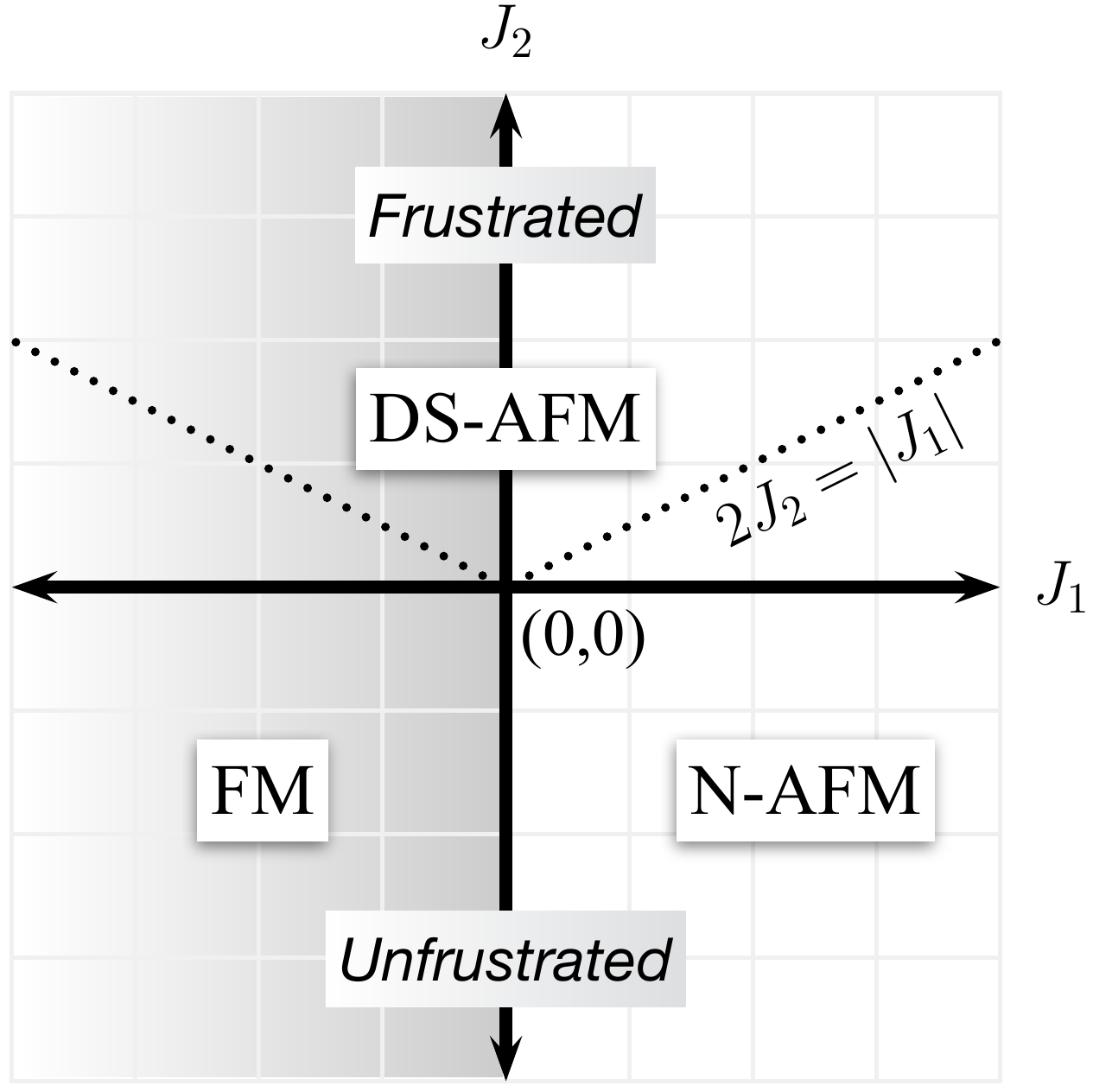} 
\caption{The frustrated and unfrustrated regions of $\Hhat$ in the $J_1$-$J_2$ plane. The dotted line, $2J_{2}=|J_{1}|$, is the line of maximum frustration, which together with $J_{1}=0$ line for negative $J_{2}$ form the phase boundaries between three phases, FM (ferromagnet), N-AFM (N\'eel antiferromagnet) and DS-AFM (double-staggered antiferromagnet), of the classical $J_{1}$-$J_{2}$ Ising chain.
}
\label{fig:j1j2plane}
\end{figure}

The $J_1$ and $J_2$ terms in $\Hhat$ compete for setting the order in the ground state. Whether the competing interactions frustrate the spins, or not, is decided by their signs and relative magnitudes. Of the four quadrants in the $J_1$-$J_2$ plane, as shown in Fig.~\ref{fig:j1j2plane}, the two in the lower half-plane for $J_2<0$ correspond to the unfrustrated cases. This is because a negative $J_{2}$ does not act against $J_{1}$. For instance, a positive $J_{1}$ favors N\'eel antiferromagnetic (N-AFM) order in which the nearest neighbor spins are anti-parallel, and therefore, the second neighbor spins are parallel to each other. This simultaneously satisfies a negative $J_{2}$ and a positive $J_{1}$. Hence, no frustration.  Likewise for $J_{1}<0$, which favors ferromagnetic (FM) order. On the other hand, a positive $J_{2}$, which favors the anti-parallel alignment of the second-neighbor spins, always competes against the $J_{1}$ of either sign. Hence, $J_2>0$ is the frustrated case. 

A property of the $\Hhat$ of Eq~(\ref{eq:model}) is that $J_{1}$ exactly maps to $-J_{1}$, without affecting $J_{2}$ and $h$, under the transformation $\{\sigma^{x}_{i}\} \rightarrow -\{\sigma^{x}_{i}\}$ on any one of the two sub-lattices (even or odd sub-chain in Fig.~\ref{fig:model}). It would suffice,  therefore, to study $\Hhat$ either for negative or positive $J_{1}$ only, as the physics of one case is an image of the other. For instance, the FM order becomes N-AFM under this mapping. In this chapter, we do calculations only for $J_{1}<0$, the shaded left-half of the $J_{1}$-$J_{2}$ plane in 
Fig.~\ref{fig:j1j2plane}. 

While the FM or N-AFM ordering in the ground state is supported naturally by $J_{2}<0$, these phases  also extend energetically into the frustrated upper-half plane. For the classical case, that is $h=0$, one finds that the FM/N-AF phases extend up to $J_{2} < |J_{1}|/2$. At precisely, $J_{2}=|J_{1}|/2$, the dotted MG (Majumdar-Ghosh~\footnote{The Majumdar-Ghosh model is a historic 1D problem of frustrated quantum (Heisenberg) spin-1/2's~\cite{MG}. Since it precisely corresponds to $J_2=J_1/2$, it has become common to refer to this relation between $J_1$ and $J_2$ as MG point or line (depending upon the context).}) line in Fig.~\ref{fig:j1j2plane}, the frustration is maximum, with macroscopic degeneracy and no unique order in the classical ground state. Above this line, the $J_{2}$ dominates, and the classical ground state exhibits double-staggered antiferromagnetic (DS-AFM) order, that is, $|\cdots ++--++-- \cdots\rangle$~\footnote{The double-staggered AFM here is same as the `antiphase' in the conventional ANNNI literature.}. Here, $|+\rangle$ and $|-\rangle$ are the eigenstates of $\sigma^{x}$, such that $\sigma^{x}|\pm\rangle = \pm|\pm\rangle$.

The transverse field, assisted by strong frustration around the MG line, is expected to produce an extended quantum disordered region on both sides of the MG line. But sufficiently away from the MG line, the Ising interactions would overcome the transverse field to generate the expected classical orders. Below, we present the quantum phase diagram of the $J_{1}$-$J_{2}$ QI chain. Although it has been studied variously for its quantum phase 
diagram~\cite{QIj1j2.Barber,QIj1j2.Bikas,QIj1j2.Brazil,QIj1j2.Subinay,QIj1j2.DMRG,QIj1j2.Adam,review.dutta}, we compute it anyway because we will need it later for our investigations of the edge modes. Besides, we plot it differently.  As in Fig.~\ref{fig:j1j2plane}, we treat $J_{1}$ and $J_{2}$ as free parameters, and put $h=1$ for the rest of the discussion. Since $\Hhat$ is not exactly solvable, except for $J_{2}$ or $J_{1}=0$, the calculations presented here are numerical in nature. But through them, we gain a fair understanding of the ground state properties of $\Hhat$.


\subsection{\label{subsec:QPD} Quantum Phase Diagram} 
To generate the quantum phase diagram of $\Hhat$, we compute spin-spin correlation, energy-gap, order parameters and transverse polarization in the ground state of $\Hhat$. We do calculations only for $J_{1}<0$, as it contains complete information about positive $J_{1}$. Since DMRG is a nice method for studying 1D quantum systems, we too use it here for our problem. It is a numerical method that iteratively truncates the Hilbert space by keeping only the most probable contributions to the ground state, and thus allows access to large system sizes~\cite{dmrg1,dmrg2}. We also do cluster-MFT calculations on small chains.  

\begin{figure}[htp] 
   \centering
   \includegraphics[width=0.6\textwidth]{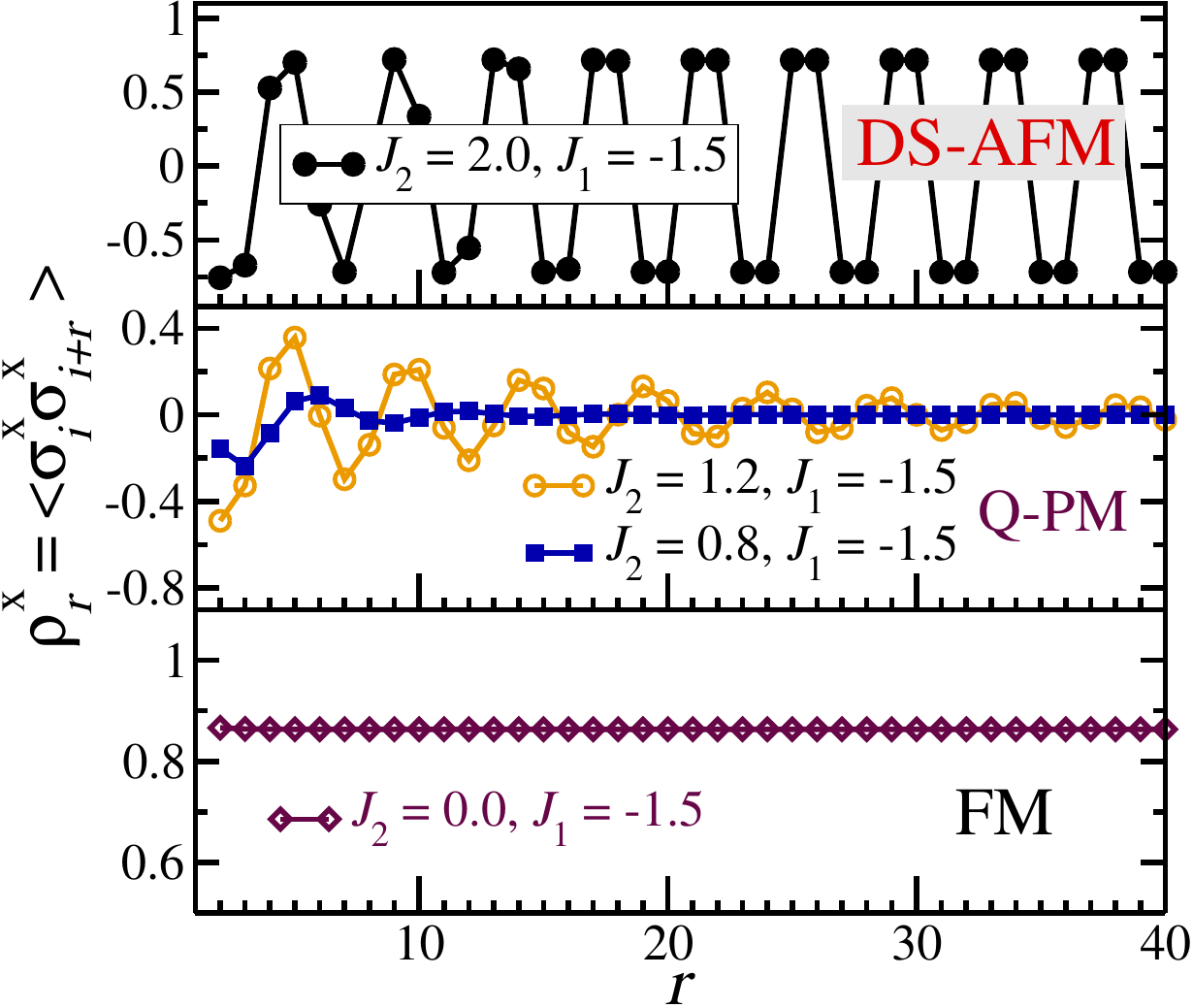} 
   \caption{The spin correlation functions in different phases for $J_{1} = -1.5$, $h = 1.0$.}
   \label{fig:xx-corr}
\end{figure}

\begin{figure}[htp] 
   \centering
   \includegraphics[width=0.6\textwidth]{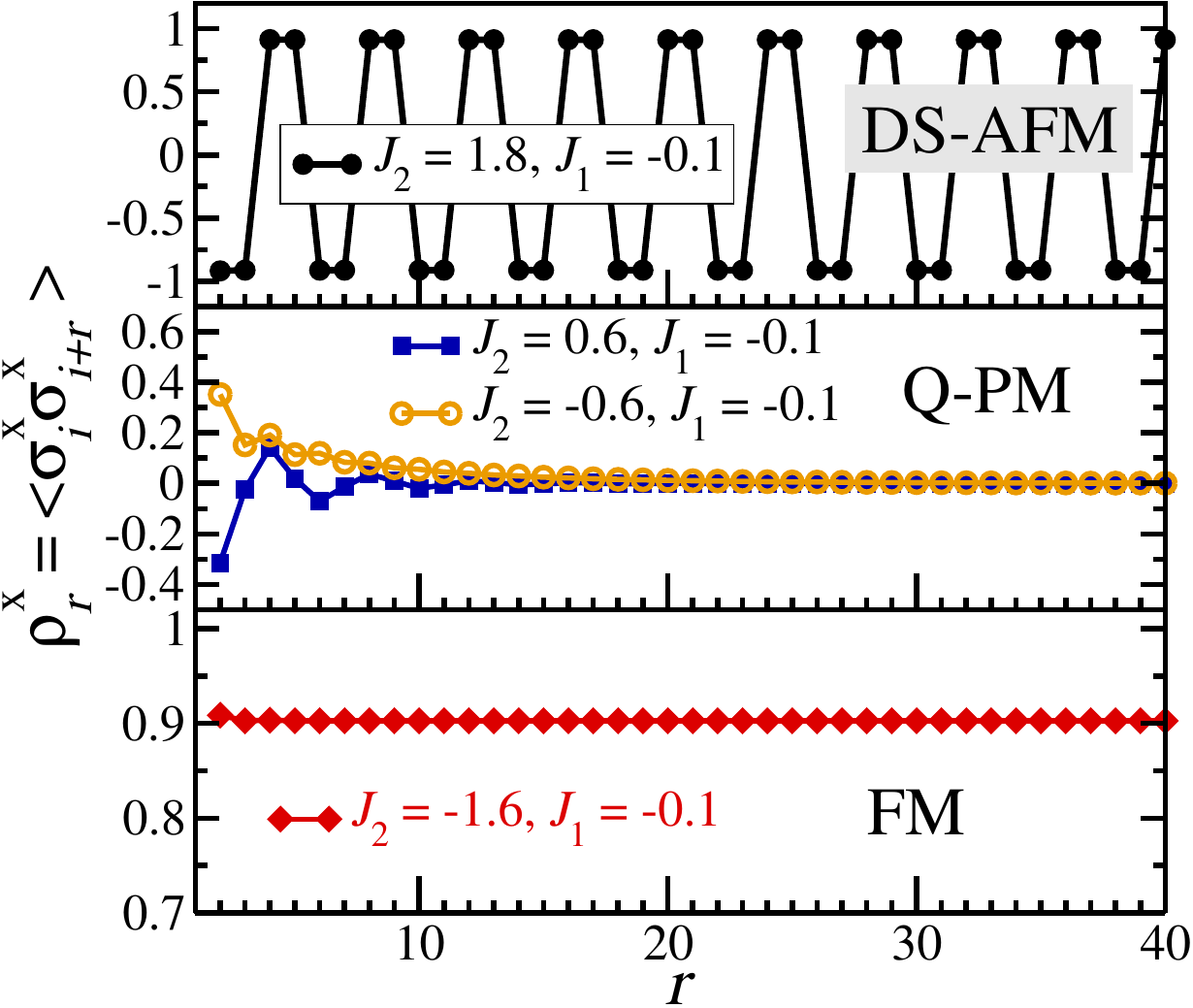} 
   \caption{The spin-spin correlation functions in different phases (written inside each panel) for $J_{1} = -0.1$. Different $J_{2}$ value suggests the position on the quantum phase diagram. See Fig.~{\ref{fig:QPD}} for guidance.}
   \label{fig:xx-corr-j1-0p1}
\end{figure}

The spin correlation function, $\rho^{x}_{r} = \langle \sigma^{x}_{i}\sigma^{x}_{i+r}\rangle$, in the ground state shows three different types of behavior for different interaction strengths, as plotted in Fig.~\ref{fig:xx-corr}. For a fixed $J_{1}$, we calculate $\rho^{x}_{r}$ for $J_{2}$ sufficiently below and above the MG line, and in-between around it. Expectedly, the ground state of $\Hhat$ shows long-ranged FM and double-staggered AFM correlations for $J_{2}$ well below and above the MG line, respectively. Near MG line, on both sides, where the transverse field helped by frustration has the best chance to kill the order, we find that $\rho^{x}_{r}$ decays to zero in two noticeably different ways. Look at the middle panel in Fig.~\ref{fig:xx-corr} carefully. For a $J_{2}$ a bit farther above the MG line, $\rho^{x}_{r}$ decays visibly slowly compared to elsewhere in the disordered region, where it decays to zero very rapidly. For comparison see the middle panel of Fig.~{\ref{fig:xx-corr-j1-0p1}} where the correlation vanishes rapidly in the Q-PM region. This is consistent with the known algebraic and exponential decay behaviors~\cite{QIj1j2.DMRG,QIj1j2.Adam}.

\begin{figure}[htp] 
   \centering
   \includegraphics[width=0.7\textwidth]{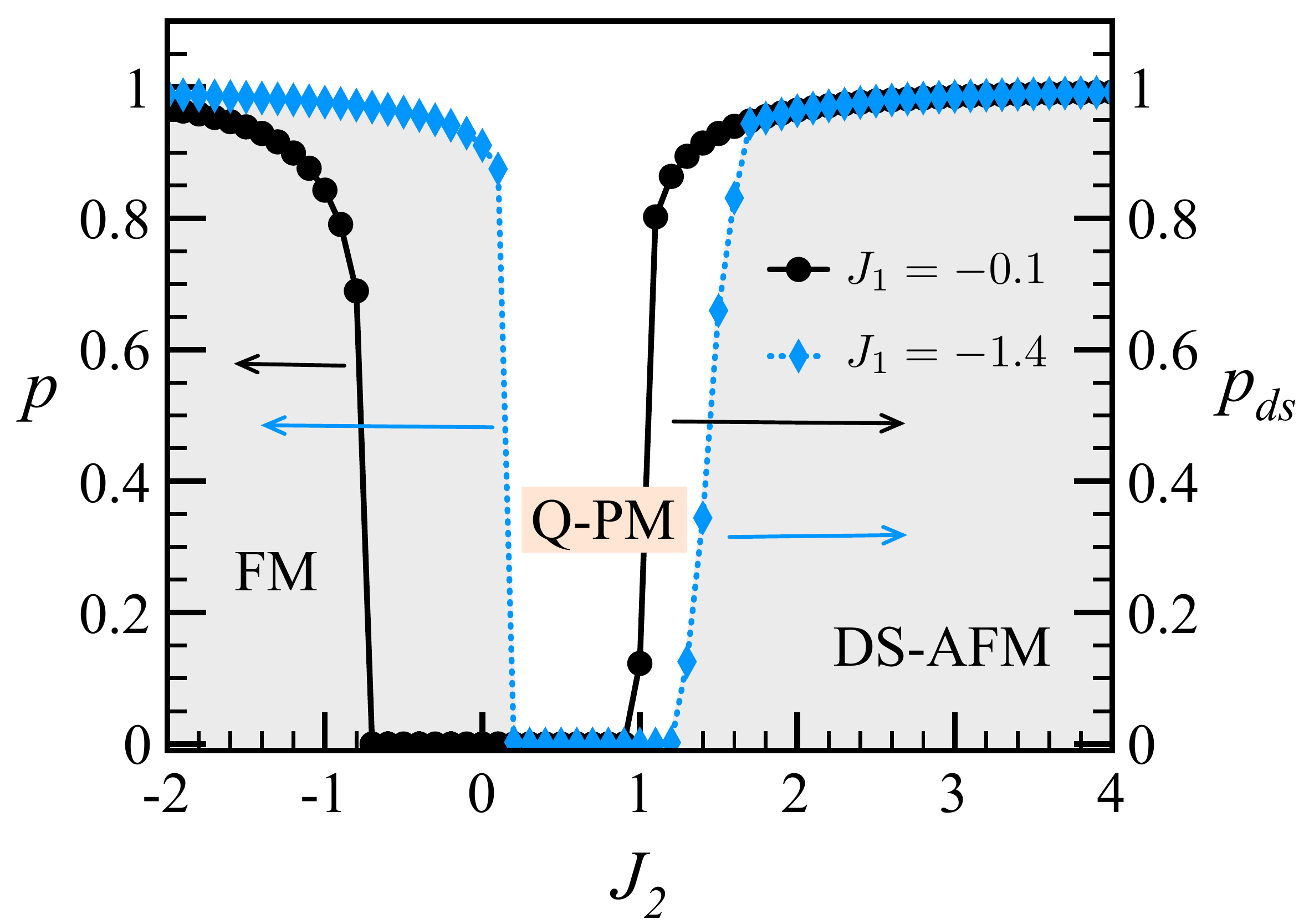}
   \caption{The order parameters vs. $J_{2}$ for fixed values of $J_{1}$. The shaded region is for $J_{1} = -1.4$, where the left side is FM (ferromagnetic) region and the right side is DS-AFM (double-staggered antiferromagnetic) region and the middle white part is Q-PM (quantum-paramegnet).}
   \label{fig:p-pds}
\end{figure}

The FM order parameter, $p$, is the ground state expectation of the uniform spin polarisation, $p=\frac{1}{L}\sum_{i}\langle\sigma^{x}_{i}\rangle$. Moreover, the long-range FM order implies $\rho^{x}_{r\rightarrow\infty} = p^{2}$. We use these two definitions to compute $p$,  both of which give consistent results. In the DS-AFM phase, the order parameter, $p^{}_{ds}$, is defined as $ p_{ds}^{2}=|\rho^{x}_{r \rightarrow\infty} | $, or as $p^{ }_{ds}=\frac{1}{L}\sum_{n=1}^{L/2}(-)^{n}\langle\sigma^{x}_{2n-1} +\sigma^{x}_{2n}\rangle$. Figure~\ref{fig:p-pds} presents the variation of $p$ and $p^{ }_{ds}$ along the $J_{2}$ axis for fixed values of $J_{1}$.  It reveals two quantum phase transitions, separately characterised by the vanishing of $p$ and $p_{ds}$.
 
\begin{figure}[htbp]
\centering
\subfloat[Energy-gap as a function of $J_{2}$ for different $J_{1}$ values]{
\label{fig:gap-vs-j2-diff-j1}
\includegraphics[width=0.47\textwidth]{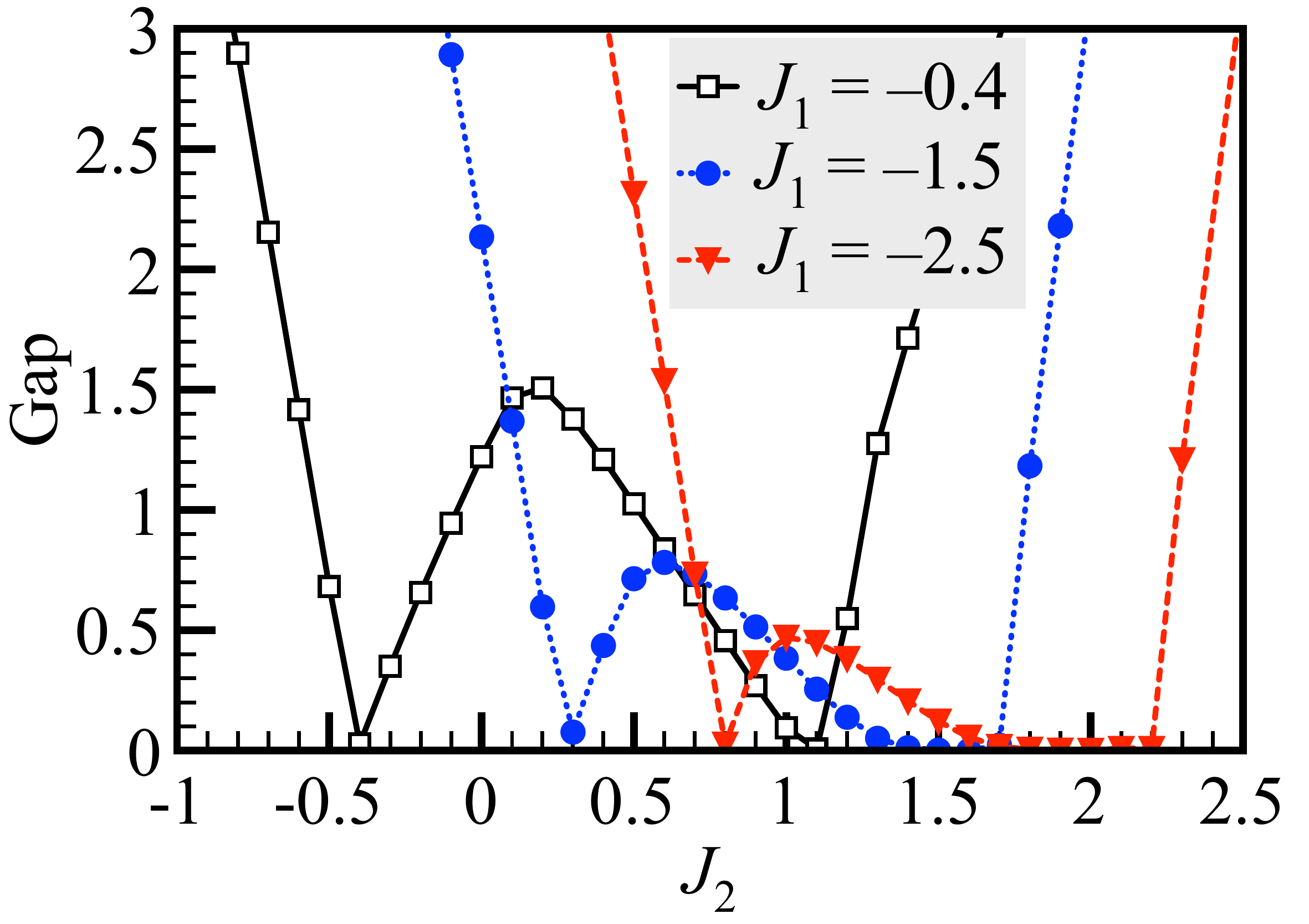} 
} \qquad
\subfloat[Three different phases from gap vs $J_{2}$ data]{
\label{fig:gap-vs-j2-j1-1p5}
\includegraphics[width=0.4\textwidth]{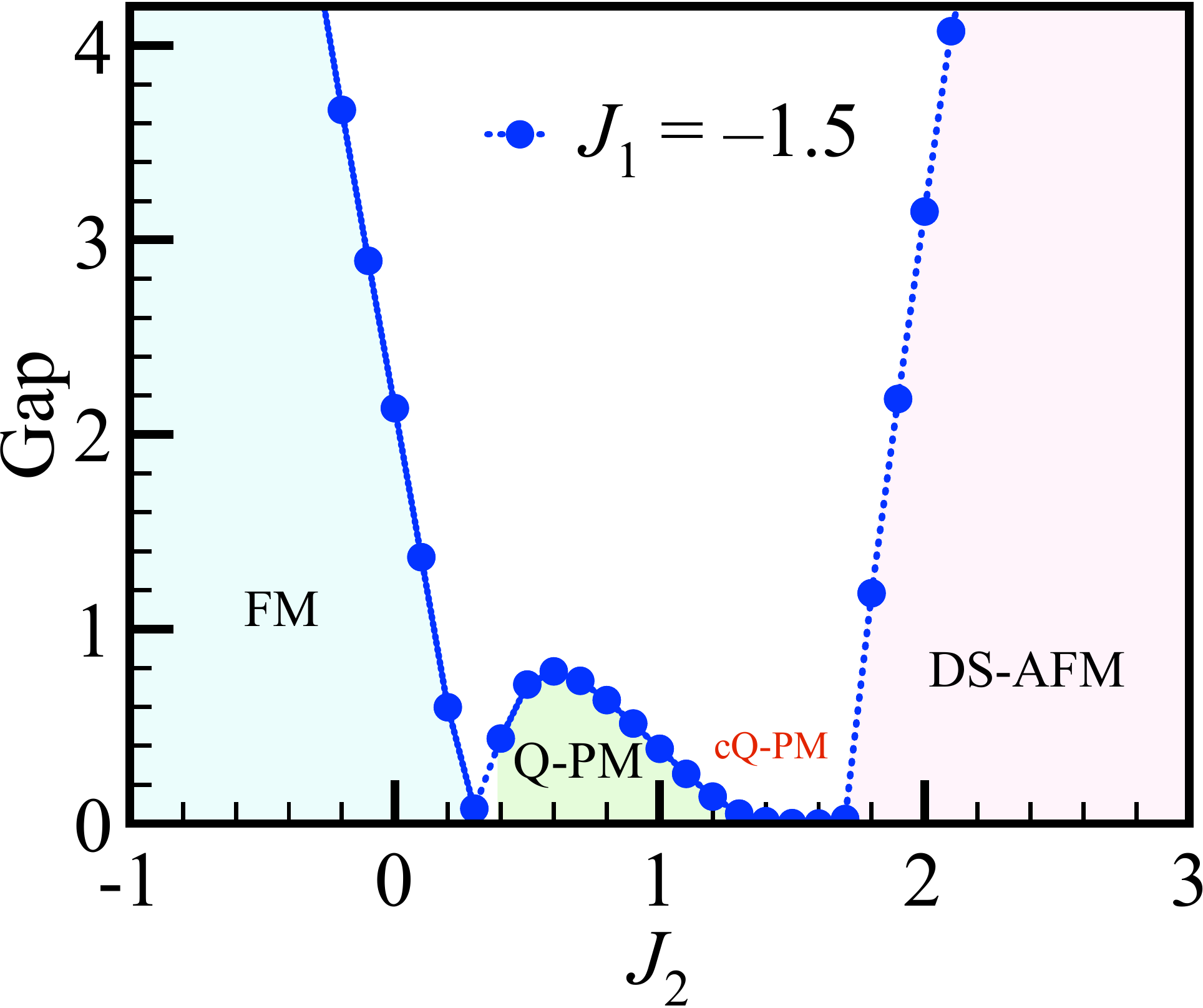}
} 
\caption{DMRG result for a Ising chain of length 400 and 64 states kept during system and environment density matrix truncation.}
\label{fig:gap}
\end{figure}

As $\Hhat$ only has a discrete (parity) symmetry, we find both the ordered phases to be gapped, as shown in Fig.~\ref{fig:gap}. The two gaps are zero at the respective critical points, and grow continuously to non-zero values in the ordered phases. On the disordered side of the FM critical point, the gap is again non-zero and continuously varying. It is the same region of parameters in which $\rho^{x}_{r}$ decays exponentially. We term this phase as quantum paramagnetic (Q-PM). The quantum disordered state adjoining the DS-AFM phase shows gaplessness in a small range of $J_{2}$ for a fixed $J_{1}$ (beyond which it becomes the gapped Q-PM phase). This is as if the critical point has expanded into a finite region of criticality. See the tail like feature of the gap data. The gap is almost zero for a finite rage of $J_{2}$. This gapless critical phase is the one in which one finds the algebraic spin-spin correlation. We call it a critical quantum paramagnetic (cQ-PM) phase~\footnote{In the ANNNI literature, this phase is called the `floating' phase, due to the associated physical picture in the classical context}. 


\begin{figure}[ht] 
   \centering
   \includegraphics[width=0.5\textwidth]{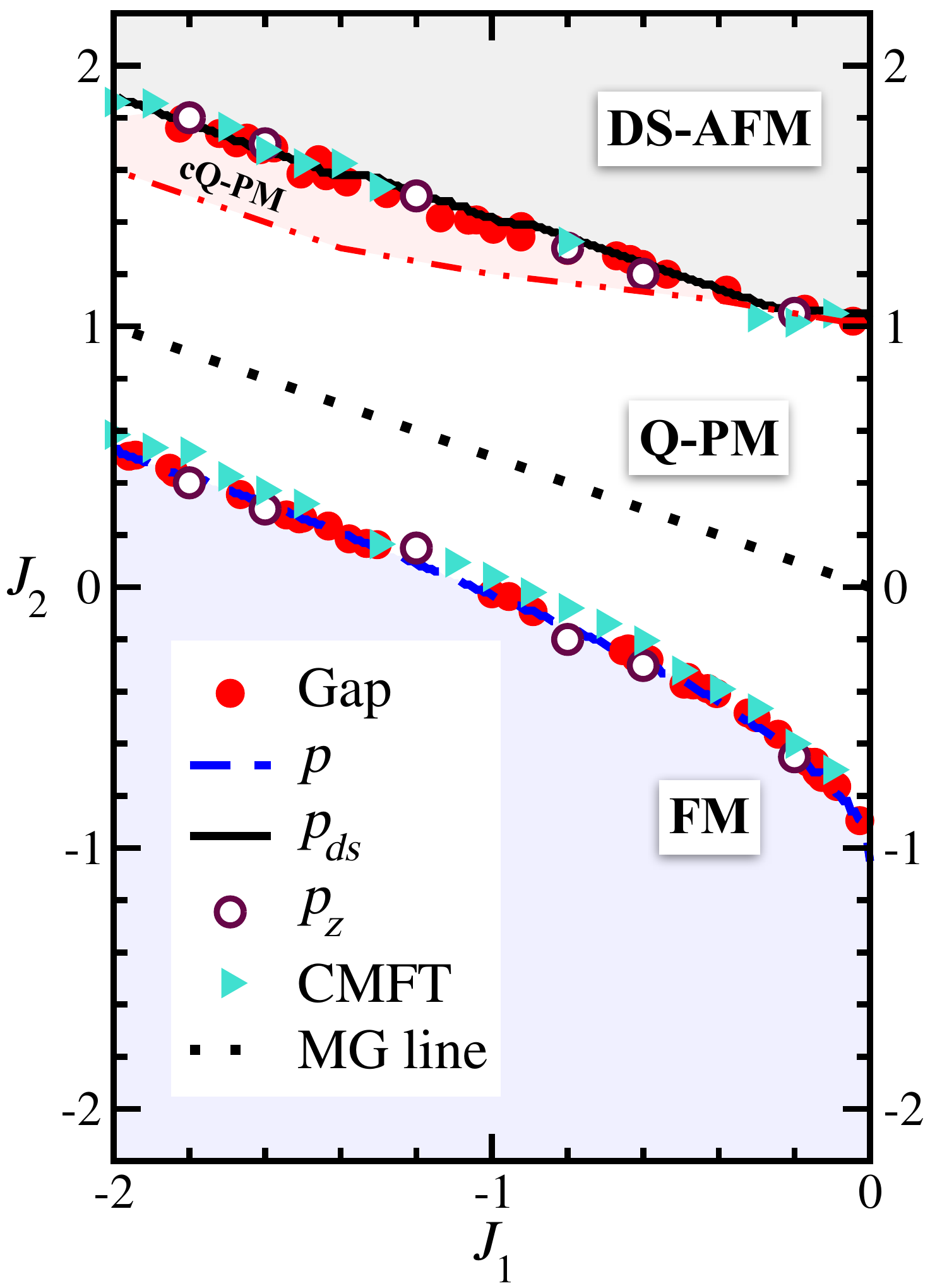}
  \caption{Quantum phase diagram of the one-dimensional $J_{1}$-$J_{2}$ quantum Ising model from DMRG and CMFT calculations.}
   \label{fig:QPD}
\end{figure}

By scanning through the energy-gaps and the order parameters in the $J_{1}$-$J_{2}$ plane, we identify the regions of FM, Q-PM, cQ-PM and DS-AFM phases. The resulting quantum phase diagram of $\Hhat$ is given in Fig.~\ref{fig:QPD}. Its basic topology is similar to the corresponding classical case shown in Fig.~\ref{fig:j1j2plane}, except that the disordered MG line has now turned into a quantum disordered region (consisting of Q-PM and cQ-PM phases) bounded by the critical lines of quantum phase transition to DS-AFM phase on the upper side and to FM phase on the lower side. The phase boundaries generated by tracking the gaps and the order parameters, from DMRG calculations on the chains of lengths up to 600 sites, are consistent with each other. We also compute, $p_{z}=\frac{1}{L}\sum_{i=1}^{L}\langle\sigma^{z}_{i}\rangle$, the transverse polarisation in the ground state. For the exactly soluble QI problem, $p_{z}$ is known to have a kink at the critical point, while varying continuously across it. Likewise, we also find the kink-points of $p_{z}$ to fall on the critical lines found from the gap and order parameter calculations. 

\begin{figure}[htbp] 
   \centering
   \includegraphics[width=0.7\textwidth]{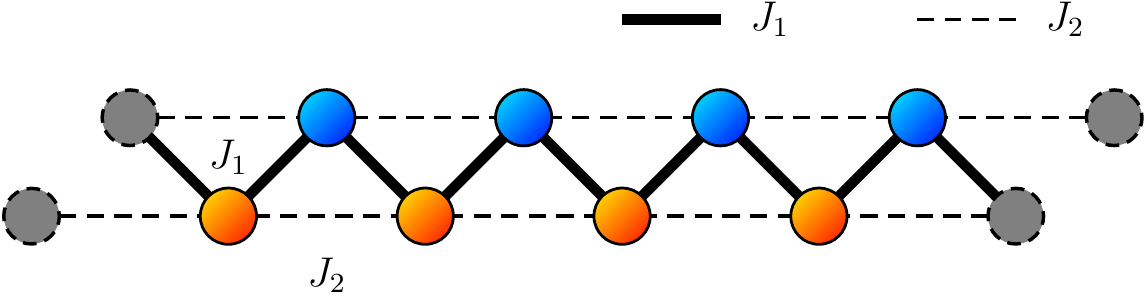} 
   \caption{CMFT: Exact cluster with mean-field on both side.}
   \label{fig:cmft-bothside}
\end{figure}

The CMFT data in Fig.~\ref{fig:QPD} is $1/L \rightarrow 0$ extrapolation of the finite $L$ critical points from the cluster-mean-field calculations on the chains of length 4, 8 and 12. We numerically diagonalize $\Hhat$ coupled to a mean-field order parameter through boundary spins, and determine the order parameter self-consistently. The CMFT models we studied are: $ \Hhat +  p[(J_1 + J_2)(\sigma^{x}_{1}+\sigma^x_L) +  J_2(\sigma^{x}_{2}+\sigma^x_{L-1})]$ for the FM phase, and $ \Hhat +  p^{ }_{ds}[(J_1 + J_2)(\sigma^{x}_{1}-\sigma^x_L) +  J_2(\sigma^{x}_{2}-\sigma^x_{L-1})]$ for the DS-AFM phase. The model is schematically shown in Fig.~{\ref{fig:cmft-bothside}}.
The critical points thus calculated also agree with the DMRG data. 

These numerical phase boundaries correctly start from the exact critical points, $(J_{1},J_{2})=(0,\pm 1$), of two decoupled $J_{2}$-only QI chains. Moreover, the lower one correctly goes through the exact critical point $(-1,0)$. They also become parallel to the MG line for large $J_{1}$ and $J_{2}$.

\section{\label{sec:EM} Edge modes in the ordered phases}
The nearest-neighbor QI model on an open chain is famously known to have two exact Majorana edge modes in the ordered phase~\cite{Kitaev.QWire}. To see this, apply Jordan-Wigner transformation on $\Hhat$, and look at the trivial case of $J_{2}=h=0$. The JW transformation is a canonical and invertible map that relates the Pauli operators to the spinless fermions. It can be defined as  $\sigma^{+}_{i} = {\chat^{\dagger}_{i}} \prod_{l=1}^{i-1}\Qhat_{l}$ and $\sigma^{z}_{l}  = - \Qhat_{l} $, where,  $\chat^{\dag}_{l}$'s are the fermion creation operators, $\Qhat_{l}=e^{i \pi \nhat_{l}} = 1 - 2 \nhat_{l}$, and $\nhat_{l}=\chat^{\dag}_{l}\chat^{ }_{l}$. Moreover, $\chat^{\dag}_{l} = \frac{1}{2}(\phihat_{l} + i\psihat_{l})$, where $\phihat_{l}$ and $\psihat_{l}$ are two Majorana (Hermitian) fermions that anticommute mutually and with other fermions, and $\phihat^{2}_{l}=\psihat^{2}_{l}=1$. Under this transformation, the $\Hhat$ of Eq.~(\ref{eq:model}) takes the following form.
 \begin{equation}
 \Hhat_{1} = J_{1} {\sum_{l=1}^{L-1}}  i \psihat_{l} \phihat_{l+1} + J_{2} {\sum_{l=1}^{L-2}}  i \psihat_{l} \Qhat_{l+1} \phihat_{l+2} + h {\sum_{l=1}^{L}} i \psihat_{l} \phihat_{l}
 \label{eq:model-JW}
 \end{equation}

In the simplest case of $J_{2}=h=0$, the Majorana operators, $\phihat_{1}$ and $\psihat_{L}$, at the ends of the chain do not figure in $\Hhat_{1}$. Thus, it has a zero energy eigen-mode described by two Majorana fermions localized at the opposite edges. Even when $h\neq 0$, but $J_{2}=0$, this problem can be solved exactly. This is what Pfeuty and Kitaev did, and found that in the ordered phase these edge-modes occur with an amplitude that is not strictly localized at the edges but decays with a finite spread into the bulk. However, when $J_{2}$ is also non-zero, things become difficult. A simplified version of this problem, in which  $i \psihat_{l} \Qhat_{l+1} \phihat_{l+2}$ in the $J_{2}$ term is replaced by  $i \psihat_{l} \phihat_{l+2}$ (by dropping $\Qhat_{l+1}$), has been given some attention recently, because it is a bilinear fermion problem amenable to exact solutions~\cite{niu2012prb}. But this is  not same as studying $\Hhat$. 
Since it is not quite studied what happens to the edge modes in the $J_{1}$-$J_{2}$ QI chain, we attempt to answer it here in this section. 

Consider the case of $h=0$, and $J_{1}$ and $J_{2}\neq 0$. Here too, $\phihat_{1}$ and $\psihat_{L}$ are absent in $\Hhat_{1}$. That is, the edge modes do exist for any $J_{2}$, at least in the absence of the transverse field. In fact, by this observation, they would exist even for longer range Ising interactions, as long as $h=0$. This clearly suggests that two Majorana-like edge modes may occur in the ground state of the $J_{1}$-$J_{2}$ model even when $h\neq 0$. Another case to note is that of $J_{1}=0$, and $J_{2}$ and $h$ non-zero. In this case, we have a problem of two independent exactly solvable QI chains (see Fig.~\ref{fig:model}), which realise `four' edge modes, two for each sub-chain. Encouraged by these observations, we now look for the edge-modes in the full $J_{1}$-$J_{2}$-$h$ problem. With no immediate help from analytics, we focus on numerics using DMRG and CMFT. It turns that our simple-minded CMFT calculations also prove quite helpful in a clear analysis of the edge modes.

Since the basic variables in $\Hhat$ are spins, not fermions, we need to devise suitable means to infer the presence of Majorana-like edge modes directly in terms of spins. In this context, Pfeuty's relation, $\rho^{x}_{1L} = \langle \sigma^{x}_{1} \sigma^{x}_{L} \rangle = p^{8}$, in the ordered phase of the nearest-neighbor QI chain with open boundaries, becomes particularly important to us. 
Just contrast it with the correlation between any two far-away spins in the bulk behaving as $p^{2}$. This $p^{8}$ behaviour of the end-to-end spin-spin correlation is an exact indicator of the Majorana-like edge modes in the nearest neighbour QI chain~\cite{pfeuty}.  While no such relation is known for the $J_{1}$-$J_{2}$ problem, we still like to use it {\em empirically} as a signature of the edge-modes. Our past experience has been encouraging in this regard, as we have used this relation to draw meaningful inferences on the edge-modes in the 1D Rabi lattice, and in the QI chain with longitudinal field~\cite{bkumar.somenath}.

We set two simple rules for the analysis of the edge modes. The first rule states that `if any long-ranged spin-spin correlation goes as $p^{2}$ (or $p_{ds}^{2}$), it can only be a bulk correlation'. That is, for the edges to be special, their spin-spin correlation {\em must not} behave as $p^{2}$. The second rule is that `if the correlation between the spins on the edges approaches $p^{8}$ (or $p_{ds}^{8}$), as one goes deeper into the ordered phase, then it confirms the occurrence of the edge modes therein'. Our second rule is obviously inspired by Pfeuty's relation. 

Let us also take note of two limiting views on the $J_{1}$-$J_{2}$ chain. If $J_{1}$ is strong compared to $J_{2}$, then it is obviously a single chain with the sites $1$ and $L$ as the edges, and $\rho^{x}_{1,L}$ as the end-to-end correlation. It can have only two edge modes. But if $J_{1}$ is quite small compared to $J_{2}$, it can be viewed as a problem of two weakly-coupled chains (see Fig.~\ref{fig:model}). Accordingly, the sites $2$ and $L$ could behave as the edges of the sub-chain of even-numbered sites, while $1$ and $L-1$ would be the ends of the odd sub-chain. In this case, $\rho^{x}_{2,L}=\langle\sigma^{x}_{2}\sigma^{x}_{L}\rangle$ and $\rho^{x}_{1,L-1}=\langle\sigma^{x}_{1}\sigma^{x}_{L-1}\rangle$ may also exhibit $p^{8}$ behavior, which would imply the occurrence of four edge-modes (which is exact for $J_{1}=0$). 

\subsection{FM Phase}

We now calculate the end-to-end and other {\em near-end} spin-spin correlations. Since the exponent of the order parameter $p$ is the object of our study here, we plot the $\log$ of different correlations against $\log{p}$, and compare with the physically motivated $y=mx$ lines, where $x$ is $\log{p}$ and the slope $m$ is 2 for the bulk correlation and 8 for the edge modes.

\begin{figure}[htbp]
\centering
\includegraphics[width=0.6\textwidth]{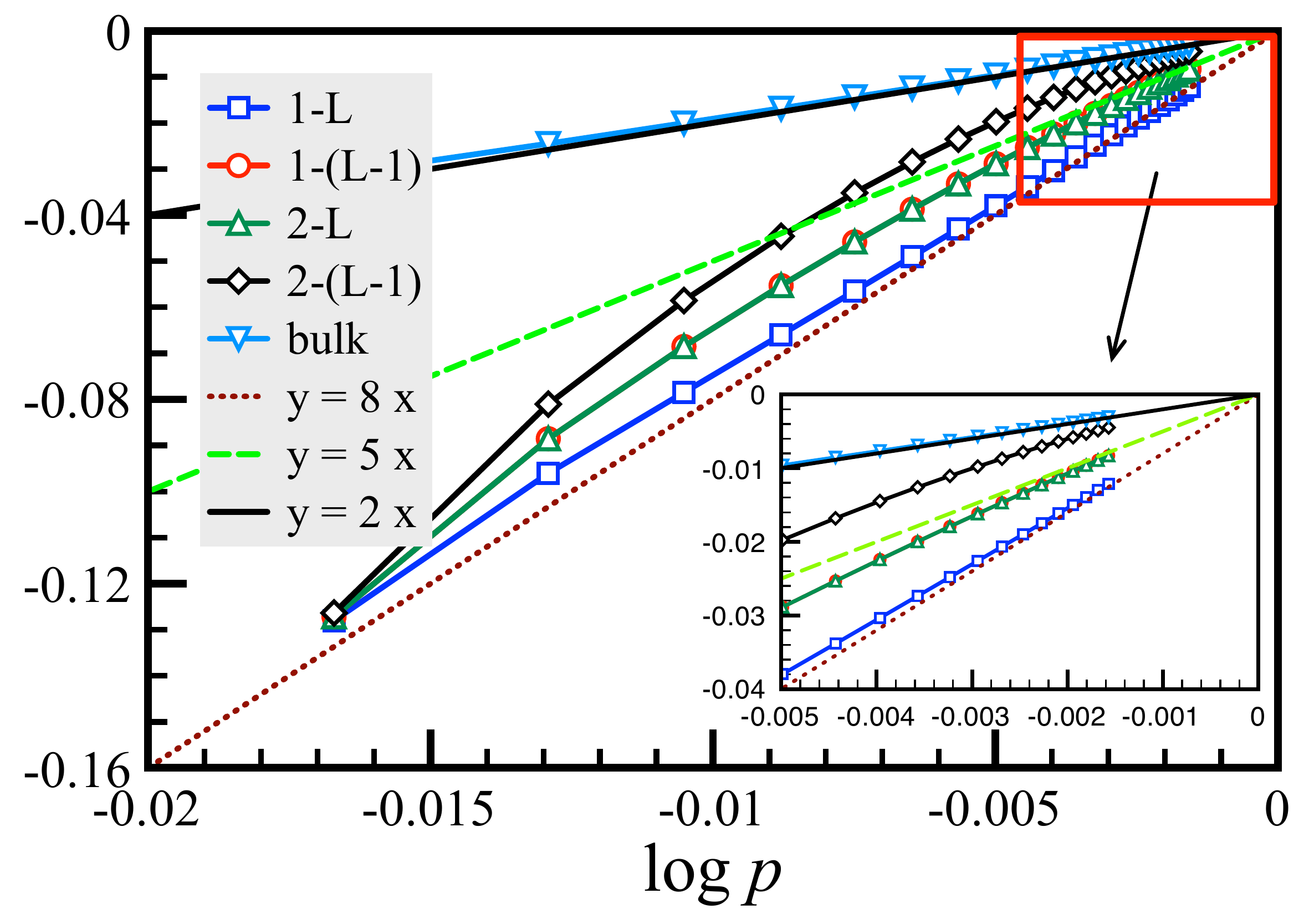} 
\caption{The $\log$ of spin correlations for $J_{2} = -2.0$, generated by varying $J_{1}$, from DMRG calculations for $L=200$. Here, 1-L in the plot legends denotes $\langle \sigma^{x}_{1}\sigma^{x}_{L}\rangle$, and likewise for other near-end correlations. The `bulk' is a correlation between two far away spins that are also far away from the ends. The zoomed inset is at large negative $J_{1}$.}
\label{fig:corr-dmrg-fm-j2-2p0}
\end{figure}

The data from a DMRG calculation is shown in Fig.~\ref{fig:corr-dmrg-fm-j2-2p0}, where, in addition to $\rho^{x}_{1,L}$, we also plot $\rho^{x}_{1,L-1}$, $\rho^{x}_{2,L}$, $\rho^{x}_{2,L-1}$, and the correlation between two spins deep inside the bulk. This data is parametrically generated by varying $J_{1}$ (from 0 to $-4$) for a fixed $J_{2}$ ($=-2$). While the bulk spin-spin correlation correctly falls on $y=2x$ line, the $\rho^{x}_{1,L}$ data follows $y=8x$. Clearly, the edges behave differently from the bulk. As $\rho^{x}_{1,L}$ follows $p^{8}$ rule, it implies the existence of two edge-modes in the FM phase.


The other notable features in Fig.~\ref{fig:corr-dmrg-fm-j2-2p0} are the behaviours of $\rho^{x}_{1,L-1}$, $\rho^{x}_{2,L}$ and $\rho^{x}_{2,L-1}$. They all start from $y=8x$ line at $J_{1}=0$, and as $J_{1}$ grows more and more negative, they neatly approach $y=2x$ or $5x$ lines. See the inset of Fig.~\ref{fig:corr-dmrg-fm-j2-2p0} for clarity. The spins at sites 2 and $L-1$ are clearly part of the bulk, as $\rho^{x}_{2,L-1}$ tends to $p^{2}$ for strong enough $J_{1}$. Very close to $J_{1} = 0$, the $\rho^{x}_{1,L-1}$ and $\rho^{x}_{2,L}$ behave like $p^{8}$, as expected for the end-to-end spin correlations for the odd and even sub-chains. But when $J_{1}$ grows stronger, they approach $p^{5}$. This suggests that $\langle \sigma^{x}_{1}\rangle=\langle\sigma^{x}_{L}\rangle=p^{4}$, and $\langle \sigma^{x}_{2}\rangle=\langle\sigma^{x}_{L-1}\rangle=p^{4}$ or $p$ if they were to respectively behave as the edges or the bulk. This numerical observation for the local expectations of spins is also consistent with the behaviour of $\rho^{x}_{2,L-1}$ and $\rho^{x}_{1,L}$. 

\begin{figure}[ht] 
   \centering
   \includegraphics[width=0.6\textwidth]{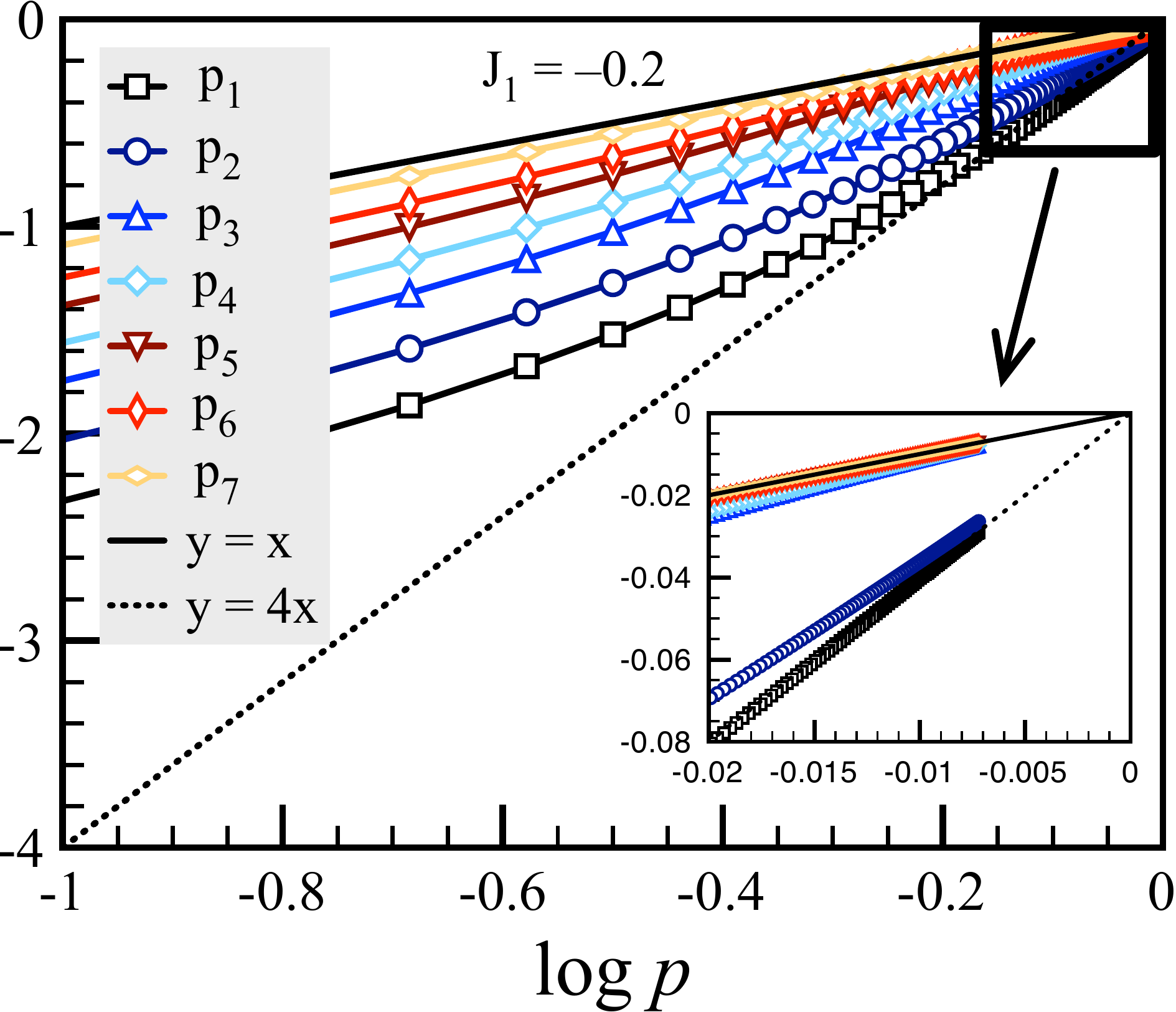}
   \caption{Cluster-MFT in the FM phase. Here, $p_{i}=\langle \sigma^{x}_{i}\rangle$, and $p$ is the FM order parameter. The zoomed inset is at large negative $J_{2}$. It shows only $p_{1}$ and $p_{2}$ are edge spin for large $J_{2}$ and all other spins are inside bulk.}
   \label{fig:cmft-fm-ed}
\end{figure}

\begin{figure}[htbp] 
   \centering
   \includegraphics[width=0.7\textwidth]{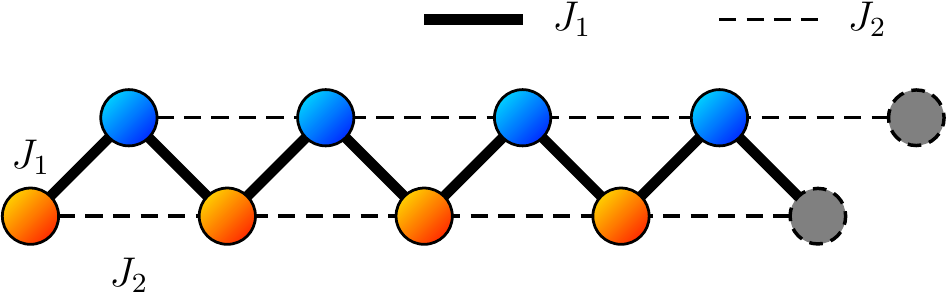} 
   \caption{Exact cluster with mean-field on one side of the chain. Two sites in the right are the mean-field ($p$ for FM side or $p_{\rm ds}$ for DS-AFM side) spins}
   \label{fig:cmft-oneside}
\end{figure}

Motivated by these observations, we also do CMFT{\nomenclature{CMFT}{Cluster Mean-Field Theory}}{\index{Cluster mean-field theory}} calculations. The mean-field model studied for the FM phase is $\Hhat_{FM} = \Hhat +  p[(J_1 + J_2)\sigma^x_L  +  J_2\sigma^x_{L-1}]$, where one end of the cluster is kept free, while the other couples to the `bulk' mean-field, $p$, to be calculated self-consistently. See Fig.~{\ref{fig:cmft-oneside}} for the schematic diagram of the cluster. This allows us to look into the local behavior of the spins near the free-end and compare it with $p$. We do exact numerical diagonalization of $\Hhat_{FM}$ on the chains of lengths up to $16$, and calculate $p_{i}=\langle\sigma^{x}_{i}\rangle$ for the first few spins on the free-end side, that is, $i=1$, 2, 3 etc. Fig.~{\ref{fig:cmft-fm-ed}} is plot of $\log {p_{i}}$ vs $\log {p}$ for $J_{1} = -0.2$. We have collected the data as a function of $J_{2}$ starting from the Q-PM phase boundary to deep inside the FM region downwards. This clearly shows when $J_{2}$ is small compared to $J_{1}$ all the spin expectations behaves as $p^{}$ (as if its a single chain) but as the effect of $J_{2}$ comes into play (see the inset), the $1^{st}$ spin and $2^{nd}$ spin of the free-end behaves as end/edge spins of the chain and follow the $y = 4 x$ line, whereas rest of the spins (from $3^{rd}$ spin onwards) behave as bulk spin and follow the $y = x$ line.

\begin{figure}[htbp] 
   \centering
   \includegraphics[width=0.6\textwidth]{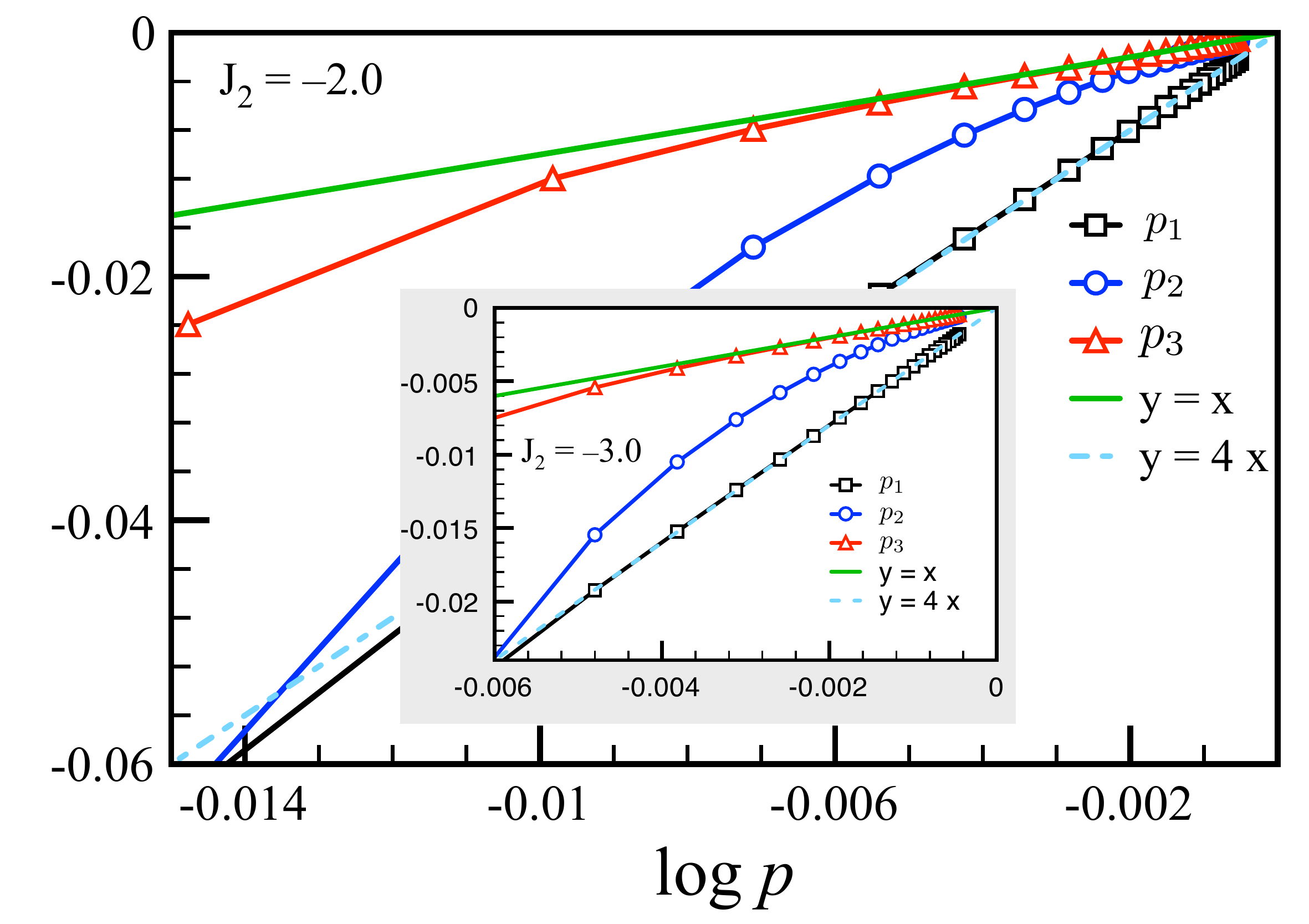}
   \caption{Cluster-MFT combined with DMRG in the FM phase. Here, $p_{i}=\langle \sigma^{x}_{i}\rangle$, and $p$ is the FM order parameter. Inset is same data for $J_{1} = -3.0$.}
   \label{fig:cmft-fm}
\end{figure}

In order to access larger chain lengths, we also combine CMFT with DMRG to grow the cluster size. Both approaches give us consistent results. In Fig.~\ref{fig:cmft-fm}, we present one such data. Notably, the $p_{1}$ goes as $p^{4}$, while $p_{2}$ starts from $p^{4}$ (for $J_{1}=0$) and approaches $p$ for strong $J_{1}$ (for different fixed values of $J_{2}$). This is exactly like what we inferred from the DMRG data. Hence, the simple-minded CMFT is consistent with DMRG, and both these calculations provide clear evidence for the occurrence of two edge-modes in the FM phase. The behavior of $p_{2}$ and $\rho^{x}_{2,L-1}$ also indicates that, very close to $J_{1}=0$, there may occur four edge modes (as in two decoupled sub-chains).

\subsection{DS-AFM Phase}
\begin{figure}[htbp] 
   \centering
   \includegraphics[width=0.6\textwidth]{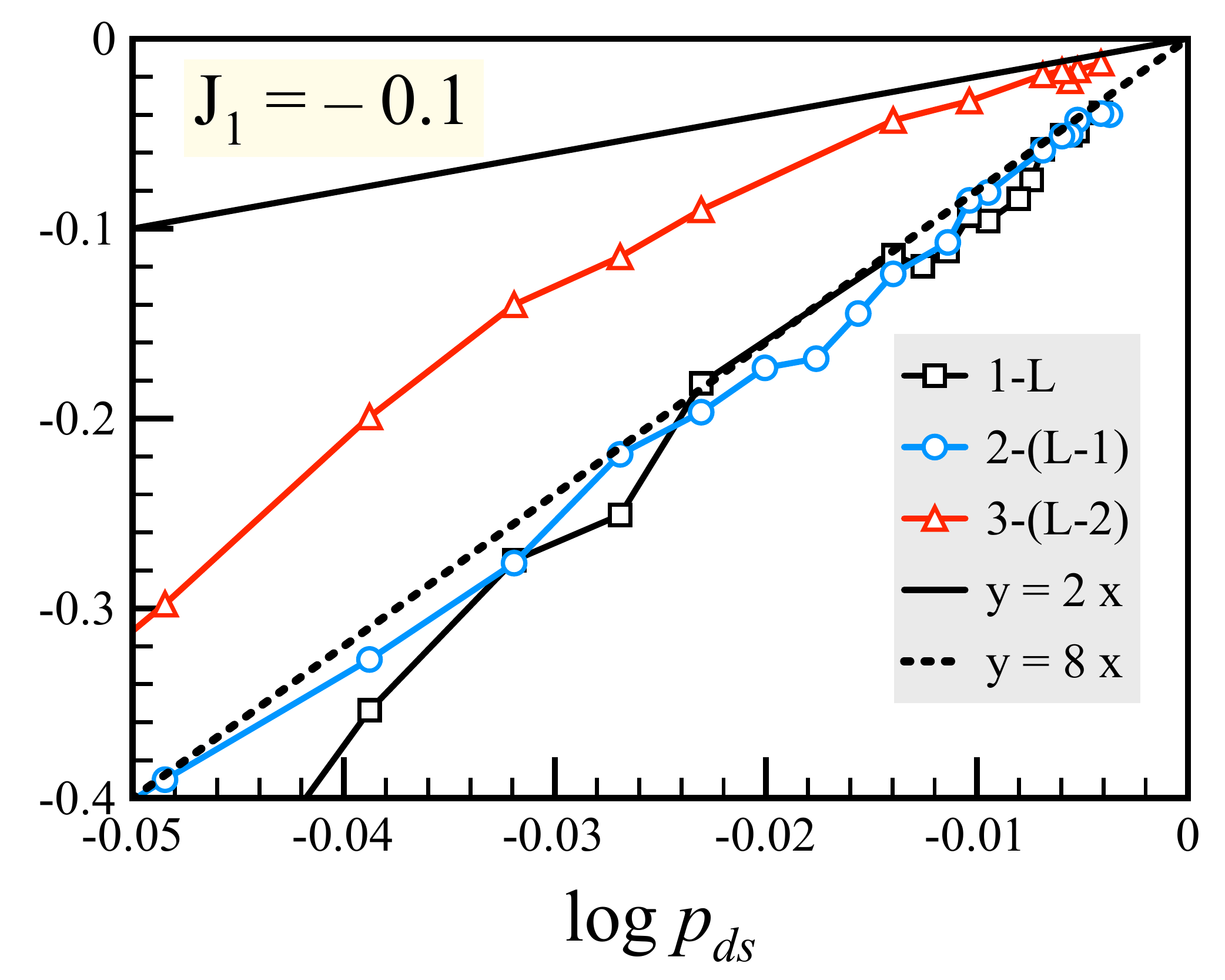} 
   \caption{Different near end-to-end correlation from DMRG data for $J_{1} = -0.1$.}
   \label{fig:dmrg-ds-afm}
\end{figure}

\begin{figure}[ht] 
   \centering
   \includegraphics[width=0.7\textwidth]{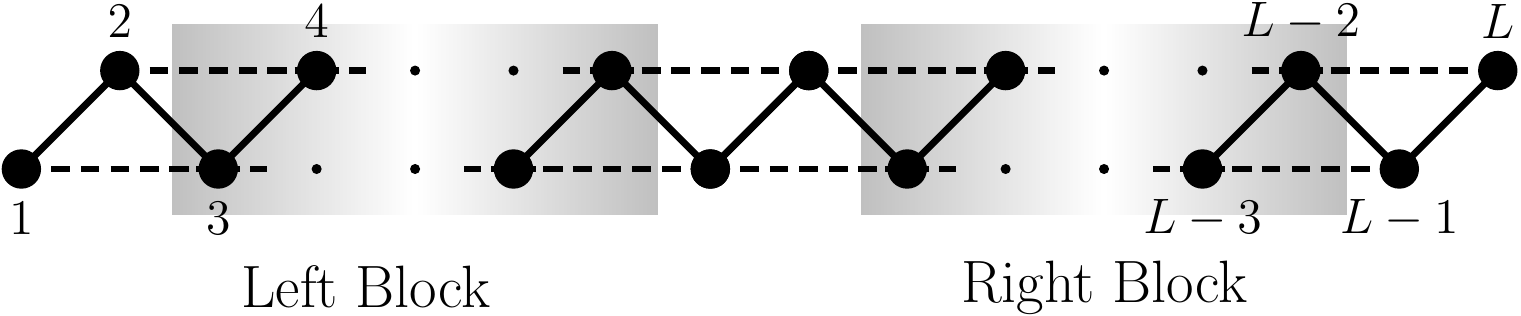} 
   \caption{Free-ends DMRG. While the left and right (shaded) blocks are iteratively renormalised, the two boundary spins (on both sides) are kept un-renormalised.}
   \label{fig:free-end-dmrg}
\end{figure}

We do the same analysis in the DS-AFM phase, except that now we compare different end-to-end spin correlations with different powers of $p^{ }_{ds}$. Since this phase is more frustrated (and fourfold degenerate), we find quite a bit of scatter (see Fig.~{\ref{fig:dmrg-ds-afm}}) in the end-to-end correlations in our simple implementation of DMRG (without using parity symmetry). The data is more scattered for large negative $J_{1}$. We did not face this difficulty in the FM phase. To improve it, we set up DMRG slightly differently by keeping the spins 1 and 2 at the left-end, and $L-1$ and $L$ at the right-end, free (as in Fig.~\ref{fig:free-end-dmrg}). That is, unlike in the usual DMRG approach wherein all the spins of the left and right blocks are updated, we now keep the last two spins at the left and the right ends un-renormalized, while everything in between is being updated. It is found to give better results. 

\begin{figure}[htp]
\centering
\includegraphics[width=0.6\textwidth]{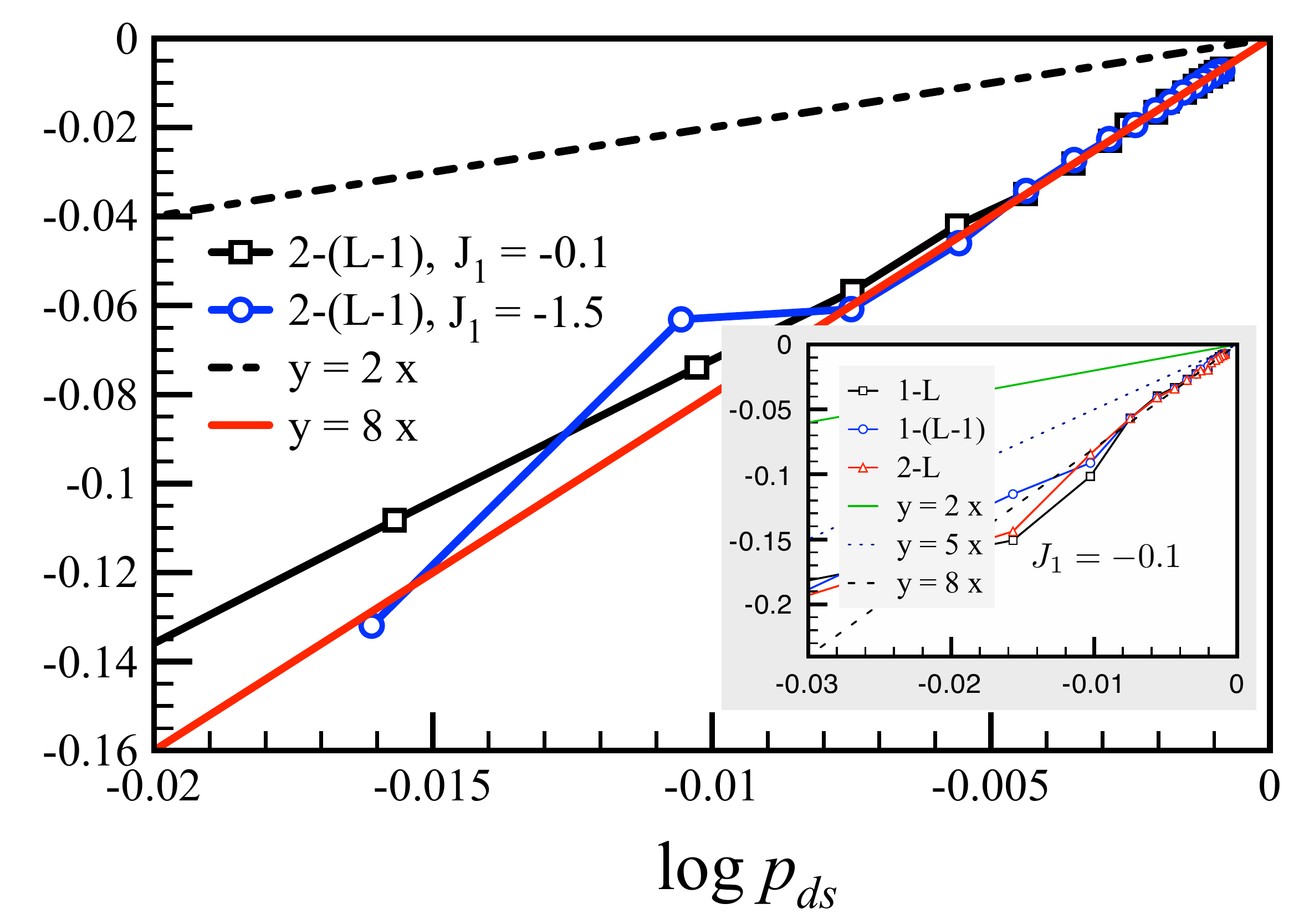} 
\caption{The $\log{\rho^{x}_{2,L-1}}$ from DMRG calculation in the DS-AFM phase for two different $J_{1}$. Here, $J_{2}$ varies from the upper critical line to some larger positive values. (Inset) The $\log$ of $\rho^{x}_{1,L}$, $\rho^{x}_{1,L-1}$ and $\rho^{x}_{2,L}$ vs. $\log{p^{ }_{ds}}$ in the same phase.}
\label{fig:free-end-dmrg-afm}
\end{figure}

The data for different end-to-end correlations from a free-ends DMRG calculation is presented in Fig.~\ref{fig:free-end-dmrg-afm}. This is clearly in accordance with the view that when $J_{2}$ dominates (which is so in the DS-AFM phase), the two sub-chains tend to behave as two. A nice check of this comes from $\rho^{x}_{2,L-1}$. While 1 and $L$ are the actual free-ends of the chain, $2$ and $L-1$ are not. Therefore, if $2$ and $L-1$ were to behave like the free ends (of the respective sub-chains), then $\langle \rho^{x}_{2}\rangle$ and $\langle \rho^{x}_{L-1}\rangle$ should each behave as $p_{ds}^{4}$, or in other words, $\rho^{x}_{2,L-1}$ should behave as $p^{8}_{ds}$. Interestingly, this expectation is precisely met by our numerical data. In Fig.~\ref{fig:free-end-dmrg-afm}, we plot $\log{\rho^{x}_{2,L-1}}$ vs. $\log{p_{ds}}$ for $J_{1}=-0.1$ and $-1.5$. This data is generated by varying $J_{2}$ from the points on the upper critical line to $J_{2}=8$, and it follows $y=8x$ line all along. Moreover, at no point it shows the tendency to go towards $y=2x$ line, that is, of showing the bulk behavior. We have checked it for different values of $J_{1}$ in the DS-AMF phase. This is unlike the FM phase (see Fig.~\ref{fig:corr-dmrg-fm-j2-2p0}), where $\rho^{x}_{2,L-1}$ can behave as both, depending upon the strength of $J_{1}$ relative to $J_{2}$. Furthermore, $\rho^{x}_{1,L-1}$ and $\rho^{x}_{2,L}$ expectedly approach $p^{8}_{ds}$ behavior (see the inset). This data suggests that the DM-AFM phase always supports four edge modes.

\begin{figure}[htbp]
   \centering
   \includegraphics[width=0.6\textwidth]{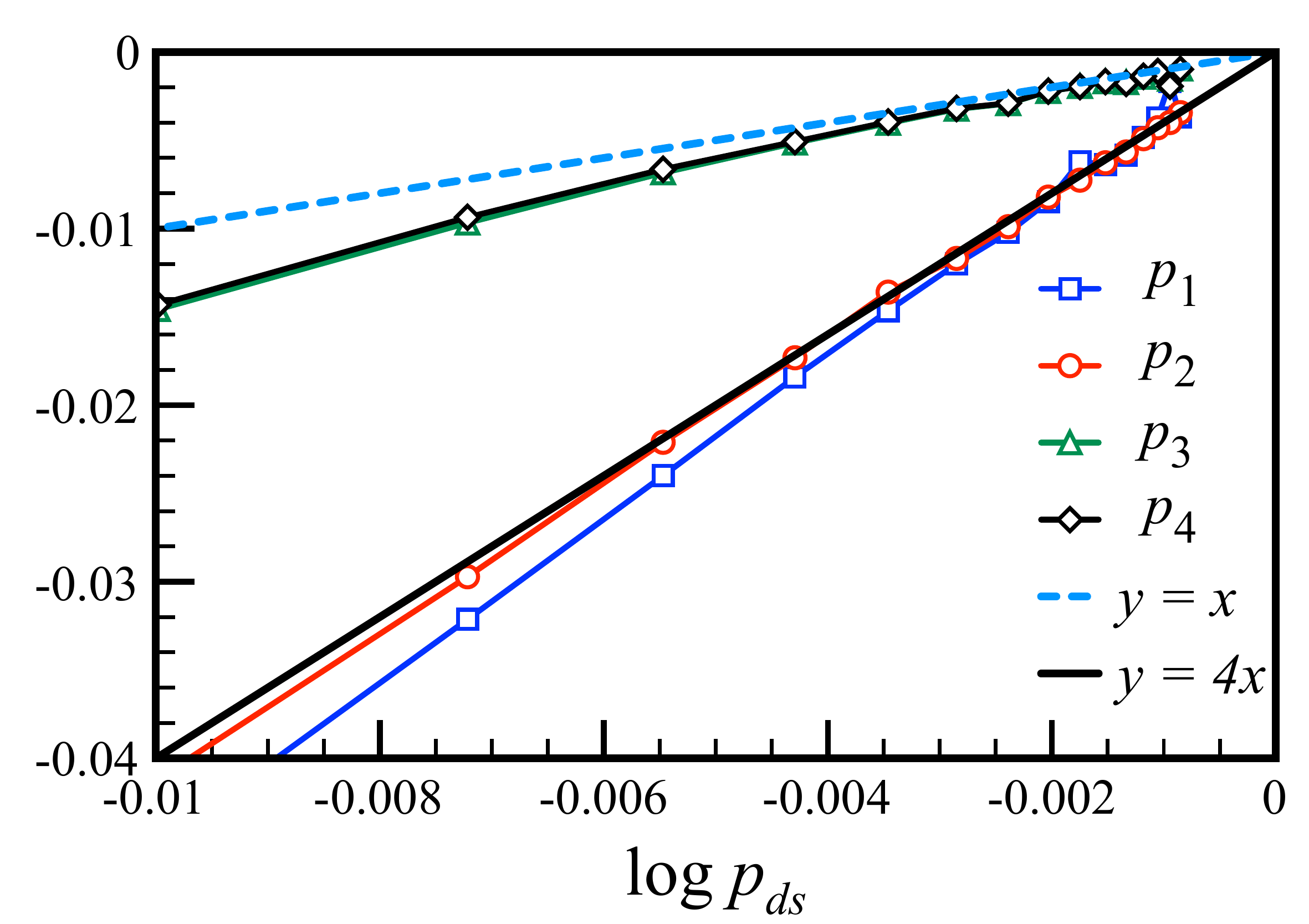}
   \caption{$\log{p_{i}}$ vs. $\log{p_{ds}}$ from CMFT calculations ($L = 200$).}
   \label{fig:cmft-dmrg-afm}
\end{figure}

\begin{figure}[ht] 
   \centering
   \includegraphics[width=0.6\textwidth]{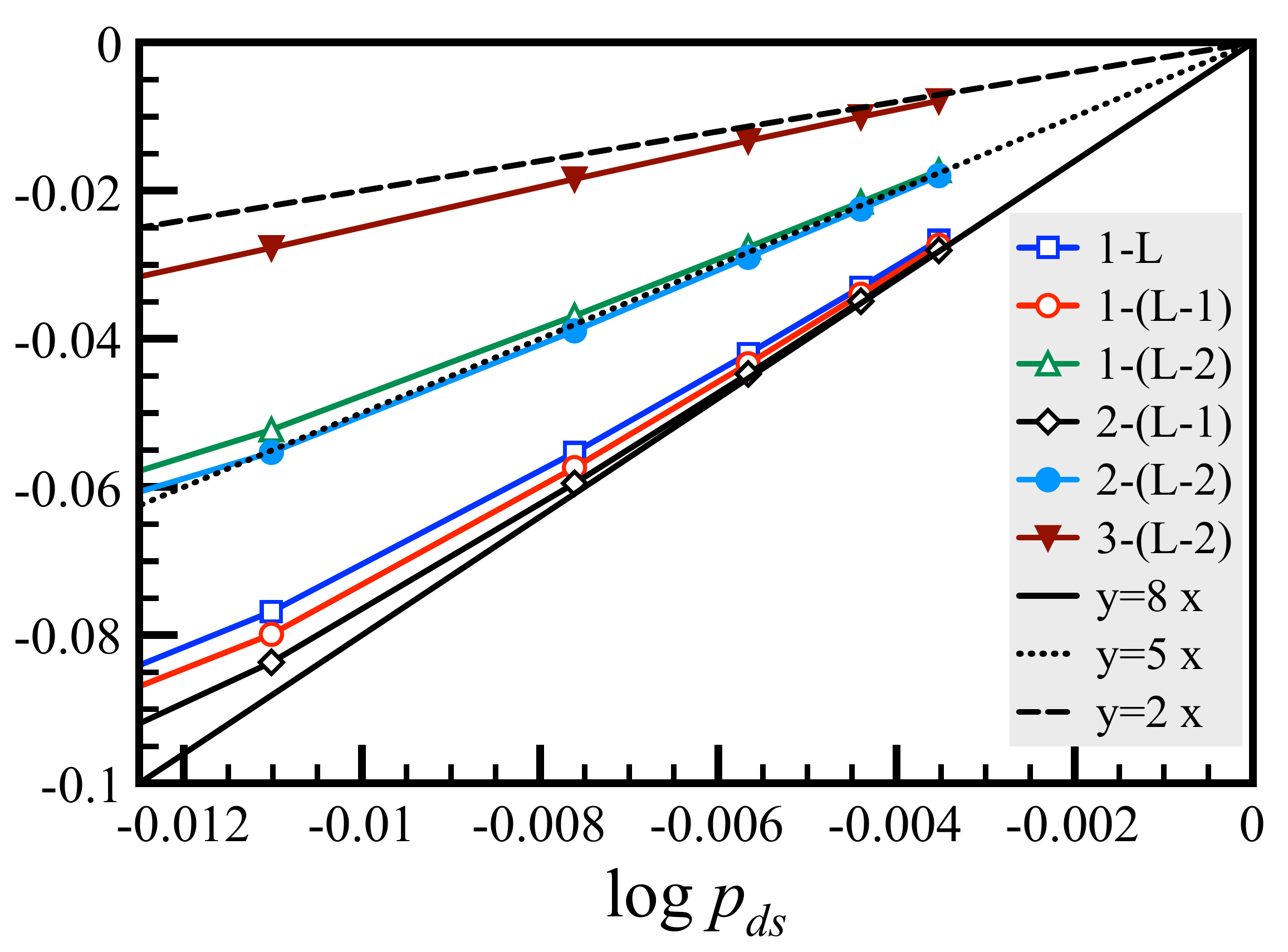} 
   \caption{Exact numerical diagonalization results for $\log{p_{i}}$ in the ground state of $\Hhat$ for 16 spins. Here, $J_{1} = -0.1$ and data is collected by changing $J_{2}$ in the positive direction in the DS-AFM phase.}
   \label{fig:ed16} 
\end{figure}

In Fig.~\ref{fig:cmft-dmrg-afm}, we present the data for the local spin expectation values, $\{p_{i}$\}, near the free end from a CMFT calculation (combined with DMRG). The mean-field model in this case is $\Hhat_{DS-AFM} = \Hhat + p_{ds}[\pm (J_1 + J_2)\sigma^x_L  \pm  J_2\sigma^x_{L-1}]$, where four possible combinations of the signs correspond to four double-staggered mean-field orders, and they all give the same result. The order parameter $p_{ds}$ is determined self-consistently. See Fig.~{\ref{fig:cmft-oneside}} for schematic diagram of the cluster. While $p_{3}$ and $p_{4}$ approach $p_{ds}$ (as expected for the spins in the bulk), the spin expectations at two sub-chain edges, $p_{1}$ and $p_{2}$, exhibit $p_{ds}^{4}$ behaviour, consistent with the DMRG data in Fig.~\ref{fig:free-end-dmrg-afm}. We also do an exact numerical diagonalization (without mean-field) calculation of $\Hhat$ with 16 spins. Technically $p_{ds}$ is zero here. But we take the square root of the modulus of the correlation two spins away from the edges and from each other as $p_{ds}$, and compare it with different end-to-end correlations (see Fig.~\ref{fig:ed16}). Even this simple calculation reveals a behaviour that is consistent with DMRG and CMFT. All of these calculations clearly ascertain that the frustrated DS-AFM phase of the $J_{1}$-$J_{2}$ QI chain indeed supports four edge modes.



\section{\label{sec:fermionic-mft} Fermionic mean-field theory}

One dimensional quantum Ising model can be transformed into a superconducting problem using Jordan-Wigner transformation. It is known that the nearest neighbor Ising term $\sigma^{x}_{l} \sigma^{x}_{l+1}$ can be written in terms of Majorana operator as $i \psihat_{l} \phihat_{l+1}$. So the two Majorana fermions, namely $\phihat_{1}$ and $\psihat_{L}$ do not take part in the energy spectrum of the fermionic Hamiltonian, and there comes the  zero energy mode corresponding to doubly degenerate ground state of Majorana fermion~{\cite{Kitaev.QWire,diptiman-talk,volovik}}. For the nearest neighbor quantum Ising chain the ferromagnetic phase ($J/h > 1$) has two Majorana modes whereas the paramagnetic disordered state ($J/h < 1$) has no Majorana zero modes. The problem we are going to address in the following subsections is to find Majorana zero modes in the presence of next-nearest neighbor interaction term $J_{2}$, if at all it survives. As discussed previously, we shall look for regions of two, four and no Majorana zero modes in the fermionic mean-field theory framework.  Similar fermionic mean-field calculations has been done before~{\cite{QIj1j2.Bikas}}, but not with the same motivations as ours. Recently, $3$-spin interaction model has also been studied~{\cite{niu2012prb}}, which happens to be exactly solvable because of the bilinear nature of the problem in the fermionic language. Our goal is to study Eq.~(\ref{eq:model}) as it is.


Let's take $\Hhat_{1}$ from Eq.~(\ref{eq:model-JW})
\begin{eqnarray}
\Hhat_{2} = && {\sum_{l=1}^{L-1}} J_{1} i \psihat_{l} \phihat_{l+1} + {\sum_{l=1}^{L-2}} J_{2} i \psihat_{l} \phihat_{l+1} i \psihat_{l+1} \phihat_{l+2} + h {\sum_{l=1}^{L}} i \psihat_{l} \phihat_{l} {\label{h-majorana}}
\end{eqnarray}
This problem is not bilinear in fermionic operator and not solvable. For this reason we decouple the terms~{\cite{LSM}} into two Majorana fermion operators. After making it bilinear in fermion, we diagonalize this in Nambu basis. The dispersion relation is calculated and dispersion minima is taken to be the boundary line of the phase diagram. The detail calculations are describes in the following section.

\subsection{Uniform mean-field theory}
The most general decoupling of the second term (involving $J_{2}$) of Eq.~{\eqref{h-majorana}} looks like 
\begin{eqnarray}
i \psihat_{l} \phihat_{l+1} i \psihat_{l+1} \phihat_{l+2} &\simeq&  m_{0} i \psihat_{l+1} \phihat_{l+2}  + m_{0} i \psi_{l} \phi_{l+1} - m^{2}_{0} - (m_{2} - m_{1}) i \phihat_{l+1} \phihat_{l+2} \cr &&
- (- m_{2} - m_{1}) i \psihat_{l} \psihat_{l+1} + (m^{2}_{1} - m^{2}_{2})  -m_{3} i \psihat_{l+1} \phihat_{l+1} \cr &&
- m_{z} i \psihat_{l}\phihat_{l+2} + m_{3} m_{z}
\end{eqnarray}
We are calling this ``uniform mean-field theory" because we have taken the mean-field parameters same for every interactions or sites. The mean-field parameters are defined as 
\begin{eqnarray}
m_{0} &=&  {\frac{1}{L}} {\sum_{l}} \left< i\psi_{l} \phi_{l+1} \right>  = {\frac{1}{L}} {\sum_{l}} \left< i \psi_{l+1} \phi_{l+2} \right> \\
m_{2} - m_{1} & = & {\frac{1}{L}} {\sum_{l}} \left< i \psihat_{l} \psihat_{l+1} \right> \\ 
- m_{2} - m_{1} &=& {\frac{1}{L}} {\sum_{l}} \left< i \phihat_{l+1} \phihat_{l+2} \right> \\
m_3 &=& {\frac{1}{L}} {\sum_{l}} \left< {i \psi_{l} \phi_{l+2}} \right> \\
m_{z} &=& {\frac{1}{L}} {\sum_{l}} \biggl< i \psi_{l+1} \phi_{l+1} \biggr> 
\end{eqnarray}
After this mean-field decoupling of $J_{2}$ term Eq.~(\ref{h-majorana}) becomes
\begin{eqnarray}
&& H = E_{0} + {\sum_{l}} \biggl[ J_{1} i \psihat_{l} \phihat_{l+1} + J_{2} \biggl( m_{0} i \psihat_{l+1} \phihat_{l+2} +  m_{0} i \psi_{l} \phi_{l+1} \cr && + (m_{1} - m_{2}) i \phihat_{l+1} \phihat_{l+2} + ( m_{1} + m_{2}) i \psihat_{l} \psihat_{l+1} - m_{3} i \psihat_{l+1} \phihat_{l+1} \cr && - m_{z} i \psihat_{l}\phihat_{l+2} \biggr) + h i \psihat_{l} \phihat_{l} \biggr] {\label{eq:j1j2majo}}
\end{eqnarray}
Where the constant energy term is $E_{0} =  J_{2} L \bigl( - m^{2}_{0} + m^{2}_{1} - m^{2}_{2} + m_{3} m_{z} \bigr) $.
All mean-field parameters are real constants and the expectation values are taken in the ground state of the Hamiltonian of Eq.~{\eqref{eq:j1j2majo}}. We again transform the Hamiltonian into fermionic language (in terms of $\chat$ operators). After doing this decoupling, the problem becomes bilinear in fermions and can be diagonalized by means of Bogoliubov transformation. In real space Eq.~{\eqref{eq:j1j2majo}} is 
\begin{eqnarray}
&& H = E_{0} + {\sum_{l}} \biggl[ J_{1} \biggl\{ \left( \chat^{\dagger}_{l} \chat^{\dagger}_{l+1} + \chat^{}_{l+1} \chat^{}_{l} \right) \label{eqn:hc} + \left( \chat^{\dagger}_{l} \chat^{}_{l+1} + \chat^{\dagger}_{l+1} \chat^{}_{l} \right) \biggr\} \cr && + J_{2} \biggl\{ m_{0} \biggl\{ \left( \chat^{\dagger}_{l+1} \chat^{\dagger}_{l+2} + \chat^{}_{l+2} \chat^{}_{l+1} \right) + \left( \chat^{\dagger}_{l+1} \chat^{}_{l+2} + \chat^{\dagger}_{l+2} \chat^{}_{l+1} \right) \biggr\} \cr && + m_{0} \biggl\{ \left( \chat^{\dagger}_{l} \chat^{\dagger}_{l+1} + \chat^{}_{l+1} \chat^{}_{l} \right) + \left( \chat^{\dagger}_{l} \chat^{}_{l+1} + \chat^{\dagger}_{l+1} \chat^{}_{l} \right) \biggr\} \cr && + i (m_{1} - m_{2}) \biggl\{ \left( \chat^{\dagger}_{l+1} \chat^{\dagger}_{l+2} - \chat^{}_{l+2} \chat^{}_{l+1} \right) + \left( \chat^{\dagger}_{l+1} \chat^{}_{l+2} - \chat^{\dagger}_{l+2} \chat^{}_{l+1} \right) \biggr\} \cr &&  - i ( m_{1} + m_{2}) \biggl\{ \left( \chat^{\dagger}_{l} \chat^{\dagger}_{l+1} - \chat^{}_{l+1} \chat^{}_{l} \right) - \left( \chat^{\dagger}_{l} \chat^{}_{l+1} - \chat^{\dagger}_{l+1} \chat^{}_{l} \right) \biggr\} \cr && - m_{3} \biggl( {\chat^{\dagger}_{l+1}}{\chat^{}_{l+1}} - {\chat^{}_{l+1}} {\chat^{\dagger}_{l+1}} \biggr)  - m_{z} \biggl\{ \biggl( {\chat^{\dagger}_{l}}{\chat^{\dagger}_{l+2}} + {\chat^{}_{l+2}} {\chat^{}_{l}} \biggr) + \biggl( {\chat^{\dagger}_{l}}{\chat^{}_{l+2}} + {\chat^{\dagger}_{l+2}} {\chat^{}_{l}} \biggr)  \biggr\} \biggr\} \cr && + h \left( \chat^{\dagger}_{l}\chat^{}_{l} - \chat^{}_{l}\chat^{\dagger}_{l}  \right) \biggr] \nonumber 
\end{eqnarray}
Taking Fourier transformation of $\chat_{l}$ and $\chat^{\dagger}_{l}$ the Hamiltonian in momentum space looks like
\begin{eqnarray}
&& H = E_{0} + {\sum_{k}} \biggl[ i \alpha_{k} \left( {\chat^{\dagger}_{k}} {\chat^{\dagger}_{-k}} - {\chat^{}_{-k}} {\chat^{}_{k}} \right) + {\beta_{k}} \left( {\chat^{\dagger}_{k}} {\chat^{}_{k}} - {\chat^{}_{-k}} {\chat^{\dagger}_{{-k}}} \right)  \cr && + \gamma_{k} \left( {\chat^{\dagger}_{k}} {\chat^{\dagger}_{-k}} + {\chat^{}_{-k}} {\chat^{}_{k}} \right) - {\delta_{k}} \left( {\chat^{\dagger}_{k}} {\chat^{}_{k}} + {\chat^{}_{-k}} {\chat^{\dagger}_{{-k}}} \right) \biggr] \label{eqn:ham}
\end{eqnarray}
\begin{figure}[htbp] 
   \centering
   \includegraphics[width=0.6\textwidth]{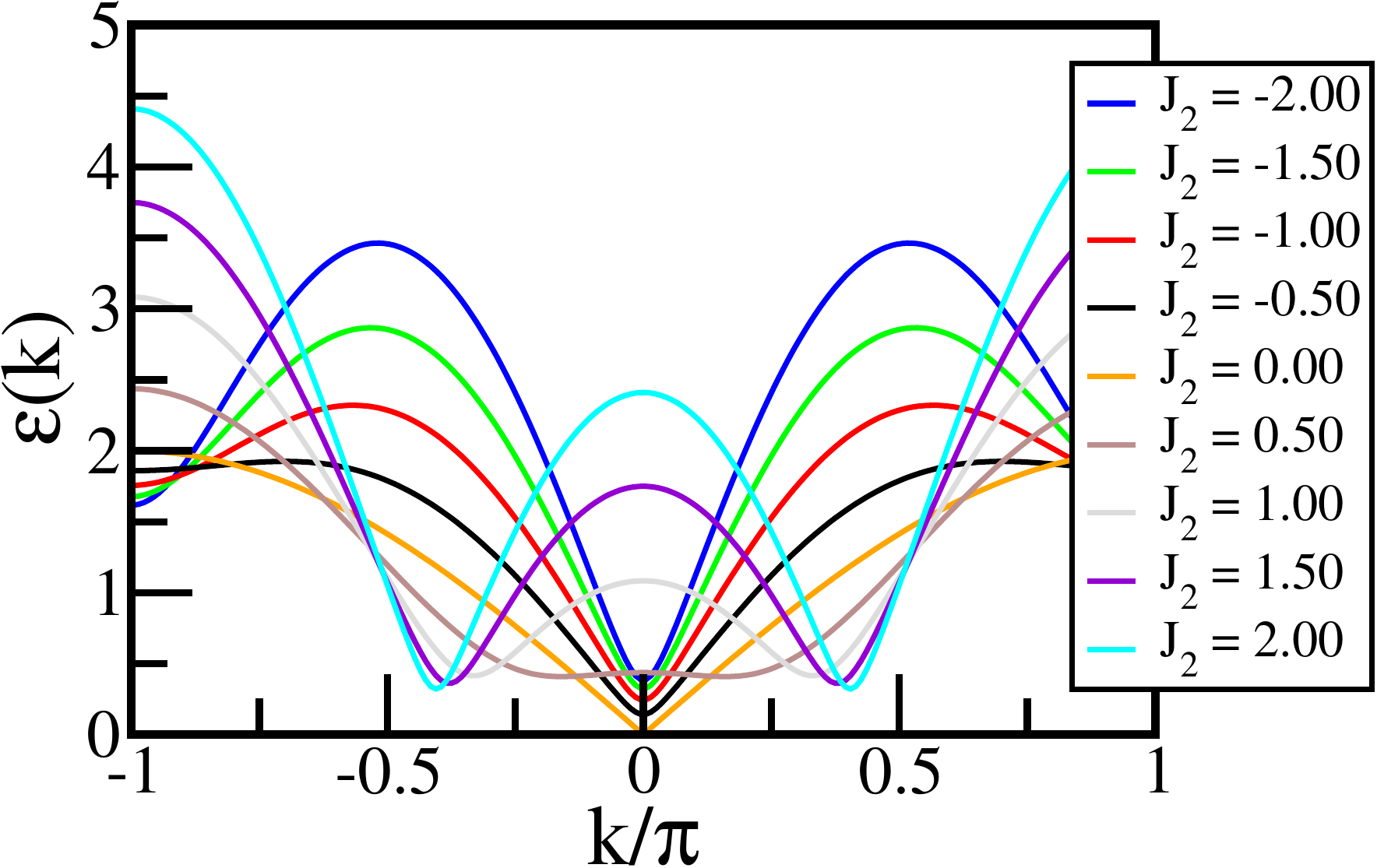} 
   \caption{Dispersion for $J_{1} = -1$ and different $J_{2}$ values.}
   \label{fig:dis}
\end{figure}
We diagonalize Eq.~({\ref{eqn:ham}}) by the Bogoliubov transformation
\begin{eqnarray}
{\chat^{}_{k}} =  \cos{\theta_{k} \over 2}  {\fhat^{}_{k}}  - {\fhat^{\dagger}_{-k}} \sin{\theta_{k} \over 2} e^{i \phi_{k}} \\
{\chat^{\dagger}_{-k}} = {\fhat^{}_{k}} \sin{\theta_{k} \over 2 } e^{ - i \phi_{k}} + {\fhat^{\dagger}_{-k}} \cos{\theta_{k} \over 2 }
\end{eqnarray}
and the diagonalized Hamiltonian becomes 
\begin{eqnarray}
H = \sum_{k} \biggl[ \epsilon_{a,k} {\fhat^{\dagger}_{k}} {\fhat^{}_{k}} + \epsilon_{b,k} {\fhat^{\dagger}_{-k}} {\fhat^{}_{-k}} \biggr] + E_{0} - {\sum_{k}} \omega_{k} \label{eqn:hdiag}
\end{eqnarray}
where
\begin{eqnarray}
\epsilon_{a,k} &=&  \omega_{k} - \delta_{k} \\
\epsilon_{b,k} &=&  \omega_{k} + \delta_{k} 
\end{eqnarray}
are the dispersion relations.
\begin{eqnarray}
\omega_{k} &=& \sqrt{ { \gamma^{2}_{k} + \alpha^{2}_{k} } +  \beta^{2}_{k} } \\
\tan{\theta_{k}} &=& \frac{ \sqrt{ \gamma^{2}_{k} + \alpha^{2}_{k} } }{\beta_{k}} \\
\tan{\phi_{k}}  &=& \frac{\alpha_{k}}{ \gamma_{k} } \\
\alpha_{k} &=& \bigl( J_{1}\sin{k} + 2 J_{2} m_{0} \sin{k} - J_{2} m_{z}\sin{2 k} \bigr) \\
\beta_{k} &=& \bigl( J_{1} \cos{k} + 2 J_{2} m_{0} \cos{k} - J_{2} m_{z} \cos{2k} - J_{2} m_{3} + h \bigr) \\
\gamma_{k} &=& 2 m_{2} J_{2} \sin{k}\\
\delta_{k} &=& 2 m_{1} J_{2} \sin{k}
\end{eqnarray}
The dispersion $\epsilon_{a,k}$ is plotted in Fig.~{\ref{fig:dis}}. The FM side being ordered, we get dispersion minimum at $k = 0$ for a particular value of $J_{2}$. That value of $J_{2}$ decides the boundary of the FM phase. For example, the dispersion is plotted for $J_{1} = -1.0$ and exactly for $J_{2} = 0.0$ line, the dispersion touched $\epsilon_{k} = 0$ at $k=0$. For other $J_{2}$ values, the energy minima does not touch zero. 

The average number of quasi-particle are defined as $ \big< \fhat^{\dagger}_{k} \fhat^{}_{k} \big> = \ntilde_{a,k} $ and $ \big< \fhat^{\dagger}_{- k} \fhat^{}_{- k} \big>  = \ntilde_{b,k} $. The mean-field parameters are 
\begin{eqnarray}
m_{0} &=& {\frac{1}{L}} {\sum_{k}} \frac{ ( \alpha_{k} \sin{k} + \beta_{k} \cos{k} ) }{\omega_{k}}  \bigl( \ntilde_{a,k} + \ntilde_{b,k} - 1 \bigr) {\label{m0}}\\
m_{1} &=& {\frac{1}{L}} {\sum_{k}} { \sin{k}} \bigl( \ntilde_{a,k} - \ntilde_{b,k}  \bigr) \\
m_{2} &=& {\frac{1}{L}} {\sum_{k}} \sin{k} \frac{\gamma_{k} }{\omega_{k}}  \bigl( \ntilde_{a,k} + \ntilde_{b,k} - 1 \bigr) \\
m_{3} &=& {\frac{1}{L}} {\sum_{k}} \frac{ ( \alpha_{k} \sin{2 k} + \beta_{k} \cos{2 k} ) }{\omega_{k}}   \bigl( \ntilde_{a,k} + \ntilde_{b,k} - 1 \bigr) \\
m_{z} &=& {\frac{1}{L}} {\sum_{k}} {\frac{ \beta_{k}}{\omega_{k}}} \big( {\ntilde_{a,k} + \ntilde_{b,k} - 1} \big) {\label{mz}}
\end{eqnarray}


\subsubsection{Phase diagram}

We self-consistently solve the mean-field parameters in the ground state of  Eq.~{\eqref{eqn:hdiag}. The average number of quasi-particles obey Fermi-Dirac statistics and expectation values are taken as thermal average at a very low temperature. We find the dispersion minima.
\begin{figure}[htbp] 
   \centering
   \includegraphics[width=0.44\textwidth]{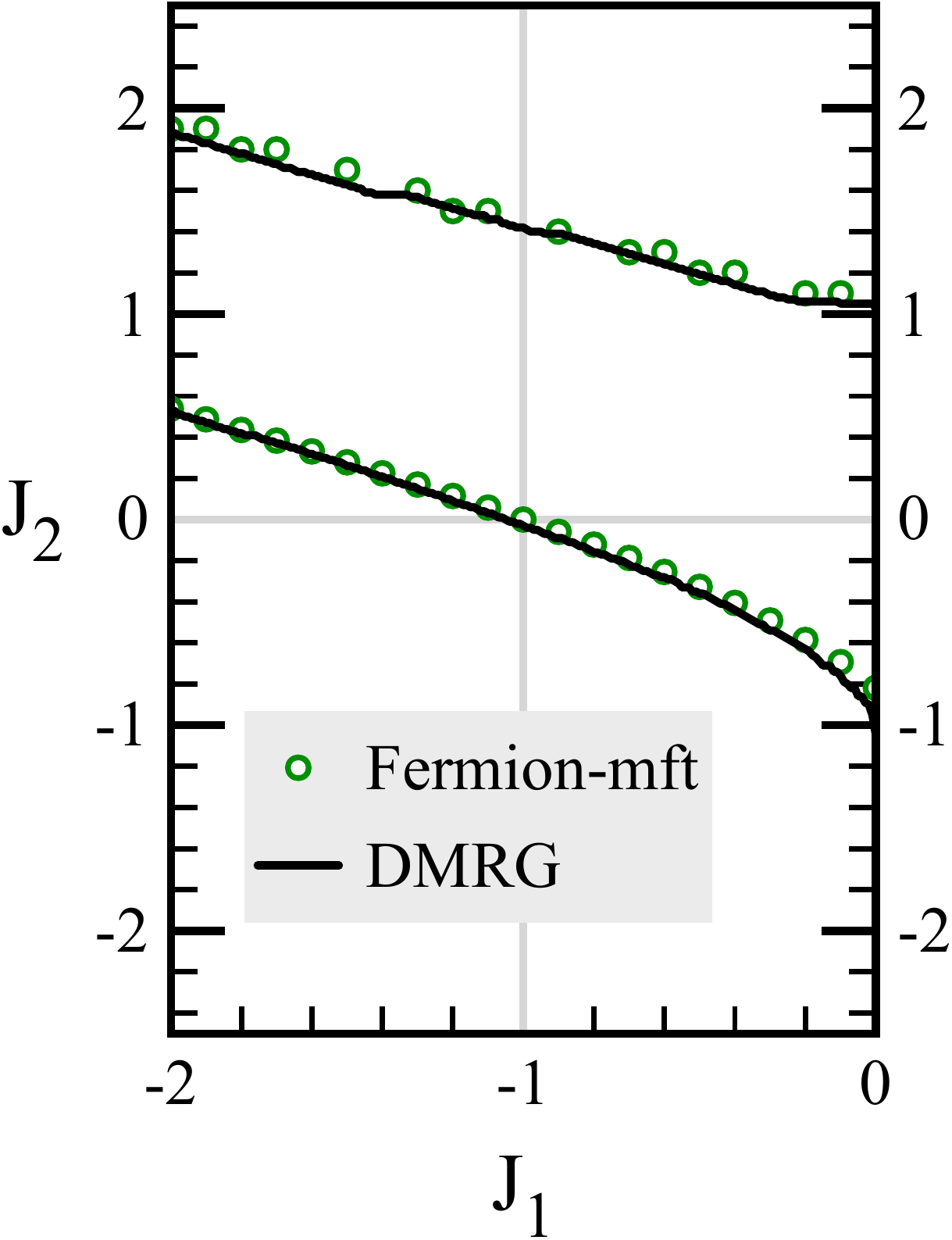} 
   \caption{Quantum phase diagram of Eq.~({\ref{h-majorana}}) on $J_{1}$-$J_{2}$ plane. Solid line from DMRG results and points are from simple fermionic mean-field theory method.}
   \label{fig:fmft-fm}
\end{figure}
 Self-consistent equation for $J_{1}$ as a function of $J_{2}$, $h$, $m_{0}$, $m_{3}$ and $m_{z}$ is 
\begin{eqnarray}
J_{1} = - h - J_{2} ( 2 m_{0} - m_{3} - m_{z} ) {\label{eq:j1j2-k0}}
\end{eqnarray}
Eq.~{\eqref{eq:j1j2-k0}} is calculated when the dispersion minimum is set at $k=0$ in the ferromagnetic side. To get FM boundary of the phase diagram, we choose some initial value of $J_{1}$. Put some value of $J_{2}$, calculate $m_{0}$, $m_{3}$ and $m_{z}$ using Eq.~{\eqref{m0}}-{\eqref{mz}}. From Eq.~{\eqref{eq:j1j2-k0}} we recalculate new $J_{1}$. If it does not match with initial $J_{1}$, then we replace new $J_{1}$, and thus for a particular $J_{2}$ we self-consistently solve for $J_{1}$. We take a new $J_{2}$ value and self-consistently find a new saturated $J_{1}$ from Eq.~{\eqref{eq:j1j2-k0}}. Thus we get the lower phase boundary of FM phase which matches well with DMRG and CMFT calculations done previously. See Fig.~{\ref{fig:fmft-fm}}. The upper phase boundary is crucial here. Dispersion nature shows that for $J_{1} = -1.0$ and $J_{2}$ in double-staggered side, the minima happens to be at $k=\pm \pi/2$. Making $J_{2}$ larger and larger in the positive direction does not make the dispersion to become zero at $k=\pm\pi/2$. Rather, there is always some finite energy gap at those $k$ points. The gap does not closes to zero anywhere between $-\pi < k < \pi$. As the phase is double-staggered, one expects the dispersion becomes gapless at $k=\pm\pi$. 

We proceed to find the upper phase diagram by this understanding that, in DS-AFM side, the nearest neighbor is frustrated but next-nearest neighbor interaction is always satisfied because of the doubled staggered ($ + + - -++-- \dots$) arrangement of spins. The $\langle \sigma^{x}_{i} \sigma^{x}_{i+2}\rangle$ term will be a negative number for any $i^{th}$ site. So $m_{3} = \langle i \psi_l \phi_{l+2} \rangle = \langle \sigma^{x}_{i} \sigma^{z}_{i+1}\sigma^{x}_{i+2} \rangle$ will be positive number (because $h=1$, and $\sigma^{z}$ expectation is negative) in the DS-AFM side and will change its sign when it crosses that upper boundary. We calculate $m_{3}$ as a function of $J_{2}$  for different $J_{1}$ value and collect the points to get the upper part of the phase diagram, which also matches, surprisingly, with DMRG data.


\subsubsection{Edge modes}

As discussed above the nn-QI open chain has two Majorana modes at the end of it. One can see this by looking at the wavefunctions corresponding to the zero eigenvalues in the eigenvalue spectrum. They will decay very rapidly into the bulk. The wavefunctions corresponding to the non-zero eigenvalues will be bound-state wave function and they have finite probability inside the chain. The eigenvalue spectrum of the Hamiltonian (\ref{eq:j1j2majo}) for $100$ sites are shown in the inset of Fig.~{\ref{fig:2majo}}. As we have written the fermionic problem in the Nambu basis, the Hamiltonian is of linear dimension $2 L$ for $L$ number of sites. Thus there are $2 L$ number of eigenvalues and an eigenvector is a $2 L \times 1$ dimensional column vector. Two zero energy eigenvalues ($100^{th}$ and $101^{st}$) are shown in the inset. Wavefunctions corresponding to these states are plotted, which decays rapidly from boundary. For example $\Psi (R)$ is constructed as $$\Psi(i) = \psi(i,R)+\psi(i+L,R) \qquad i = 1, 2, \dots L$$where $\psi(i,R)$ contains $R^{th}$ eigenvector's first $L$ elements and $\psi(i+L,R)$ is for the next $L$ elements. $R = 100$ and $101$ are understood as the Majorana zero energy modes. Here we set $J_{1} = -4.0$, $J_{2} = 0.0$ and $h = 1.0$ to be sure we are inside FM. The eigenspectrum is sorted in ascending order. The wavefunction corresponding to the first non-zero eigenvalue ($102^{nd}$) is a bound state having a finite probability in the bulk.
\begin{figure}[ht] 
   \centering
	\includegraphics[width=0.7\textwidth]{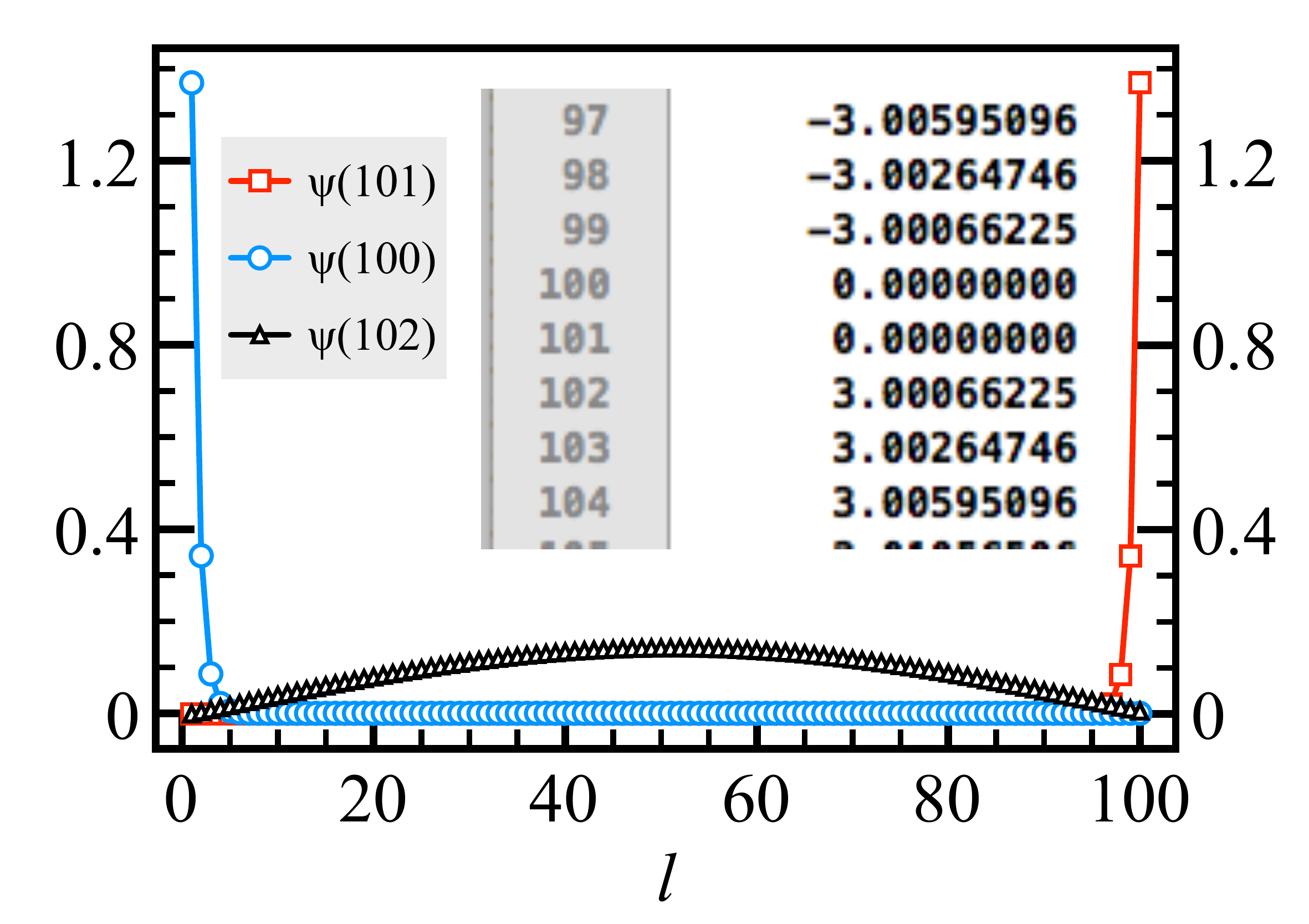} 
	\caption{The eigenvalue spectrum and the wavefunctions from uniform mean-field theory. $J_{1} = -4.0, J_{2} = 0.0, h = 1.0$}
	\label{fig:2majo}
\end{figure}

In the DS-AFM side we do not get any such zero eigenvalues or wavefunction. That is the reason we do a more systematic local mean-field theory, described in the next Subsection~{\ref{subsec:local-mft}}.

\subsection{{\label{subsec:local-mft}}Local mean-field theory}

We have fermionic mean field (mf) theory which is based on the idea that in ferromagnetic side, the spin configuration could be all $\uparrow$ or $\downarrow$ according to the sign of external field $h$. We take  $m_0 (= \left< i \psi_l \phi_{l+1} \right>$) and  $m_3 (= \left< i \psi_l \phi_{l+2} \right>$)  uniform throughout all the nearest neighbor and next nearest neighbor bonds respectively. This is justified as we got one part of the phase boundary which agrees with the DMRG calculation and cluster mean-field theory result. But consider the case of frustrated side where $J_2$ is +ve and effect of $J_2$ is dominant. The ordered phase is four-fold degenerate and we have $++--++--++-- \dots$  arrangement of spins (called this phase DS-AFM). In this case we can still do a mean-field theory and take the case where nearest neighbor interaction is not uniform (thus frustrated) (its alternatively +ve and -ve on each bond) but next nearest neighbor is always interacting anti-ferromagnetically (unfrustrated). See Fig.~{\ref{fig:ds-model}}.
%
%
\begin{figure}[htbp] 
\centering
\includegraphics[width=0.9\textwidth]{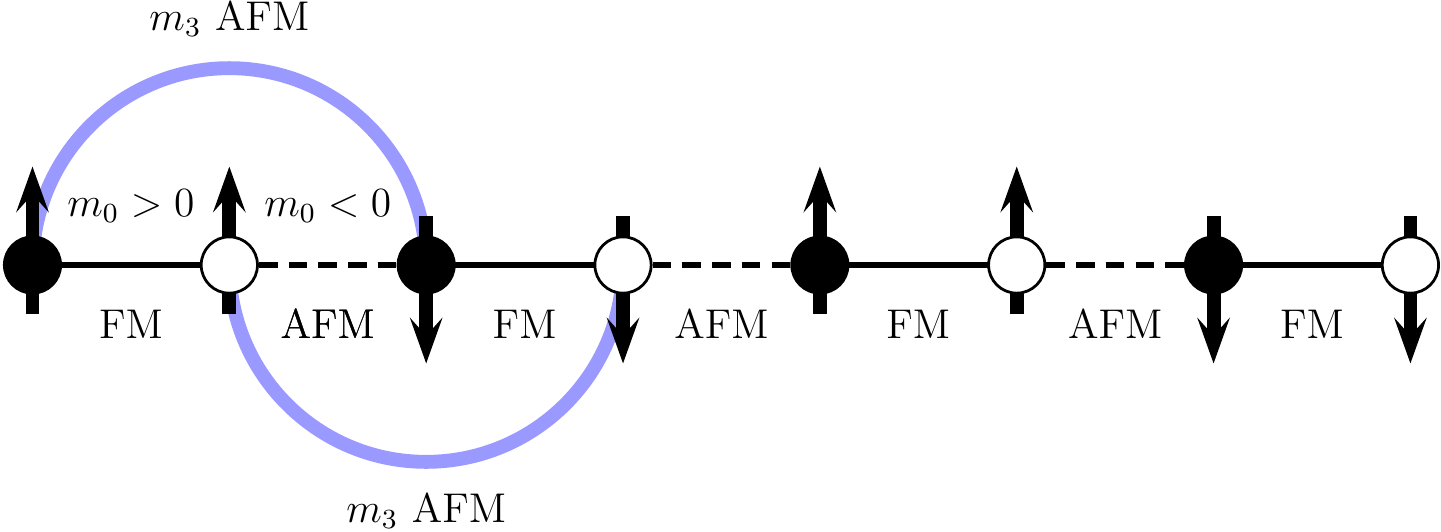} 
\caption{ Diagram of $J_{1}$-$J_{2}$ QI model in the DS-AFM phase. Here, it is physically correct to take uniform $m_3$, $m_z$ but non-uniform $m_0$.}
\label{fig:ds-model}
\end{figure}
So we take a site dependent mean-field theory where we start from Eq (\ref{h-majorana}) and define the mean-field parameters as
\begin{equation}
\left.
\begin{aligned}
& A_{l} =  \left< i\psi_{l} \phi_{l+1} \right>, \forall {\mbox{ {\it l} = 1,2 \dots (L-1)}} \\
& B_{l} =  \left< {i \psi_{l} \phi_{l+2}} \right>, \forall {\mbox{ {\it l}  = 1,2 \dots (L-2)}} \\
& C_{l} = \left< i \psi_{l+1} \phi_{l+1} \right>, \forall {\mbox{ {\it l}  = 1,2 \dots L}} 
\end{aligned}
\right\}{\label{eq:alblcl}}
\end{equation}
and the mean-field Hamiltonian becomes (remembering $\phi_l = \chat_l + \chat^{\dagger}_l$ and $i \psi_l = \chat^{\dagger}_l -\chat_l$)
\begin{eqnarray}
&& H = {\sum^{L-1}_{l=1}}  J_{1} i \psihat_{l} \phihat_{l+1} + {\sum^{L-2}_{l=1}} J_{2} \biggl[ A_l  i \psihat_{l+1} \phihat_{l+2} + A_{l+1} i \psi_{l} \phi_{l+1} \cr && - B_l i \psihat_{l+1} \phihat_{l+1}  - C_{l+1} i \psihat_{l}\phihat_{l+2} \biggr] \cr && + h {\sum^{L}_{l=1}} i \psihat_{l} \phihat_{l} + J_{2} {\sum_{l=1}^{L-2}} (B_{l} C_{l+1} - A_{l} A_{l+1} ) \\
&& = {{(J_{1} + J_{2} A_{2})}{}} {i \psi_{1} \phi_{2}} +  {{(J_{1} + J_{2} A_{L-2})}{}} {i \psi_{L-1} \phi_{L}} \cr && + {\sum^{L-2}_{l=2}} {\left( {J_1 + J_2 A_{l+1} +J_{2} A_{l-1} } \right)} {i \psi_{l}\phi_{l+1}} \cr && - {\sum^{L-2}_{l=1}}{{ J_2 C_{l+1}}} {i \psi_{l} \phi_{l+2}} + {\sum^{L-1}_{l=2}} (h - J_2 B_{l-1}) {i \psi_{l} \phi_{l}} + h {i \psi_{1} \phi_{1}} +  h {i \psi_{L} \phi_{L}} \cr && + {\sum^{L-2}_{l=1}} J_2 ( B_l C_{l+1} - A_l A_{l+1} ){\label{hr-abc}}
\end{eqnarray}
Here all the mean-field parameters from Eq.~({\ref{eq:alblcl}}) are site dependent and thus the method called ``local mean-field theory". The Eq.~(\ref{hr-abc}) is in bilinear form in fermionic operator and we can write this Hamiltonian in Nambu basis. For $L$ sites the Hamiltonian matrix will be of dimension is $2L \times 2L$. The mean-field parameters are one-dimensional arrays. For $L$ sites $A_l$, $B_l$ and $C_l$ has $L-1$, $L-2$ and $L$ values respectively. We have to solve for each element in these array self-consistently. The method is as follows. Take some initial values of all the $A_l $\rq{}s, $B_l$\rq{}s and $C_l$\rq{}s and find the ground state wave vector of the Hamiltonian. This ground-state will define the values of new $A_l$, $B_l$ and $C_l$ by calculating the expectation value from Eq.~{\eqref{eq:alblcl}}.



\subsubsection{Edge modes}

\begin{figure}[ht] 
   \centering
	\includegraphics[width=0.76\textwidth]{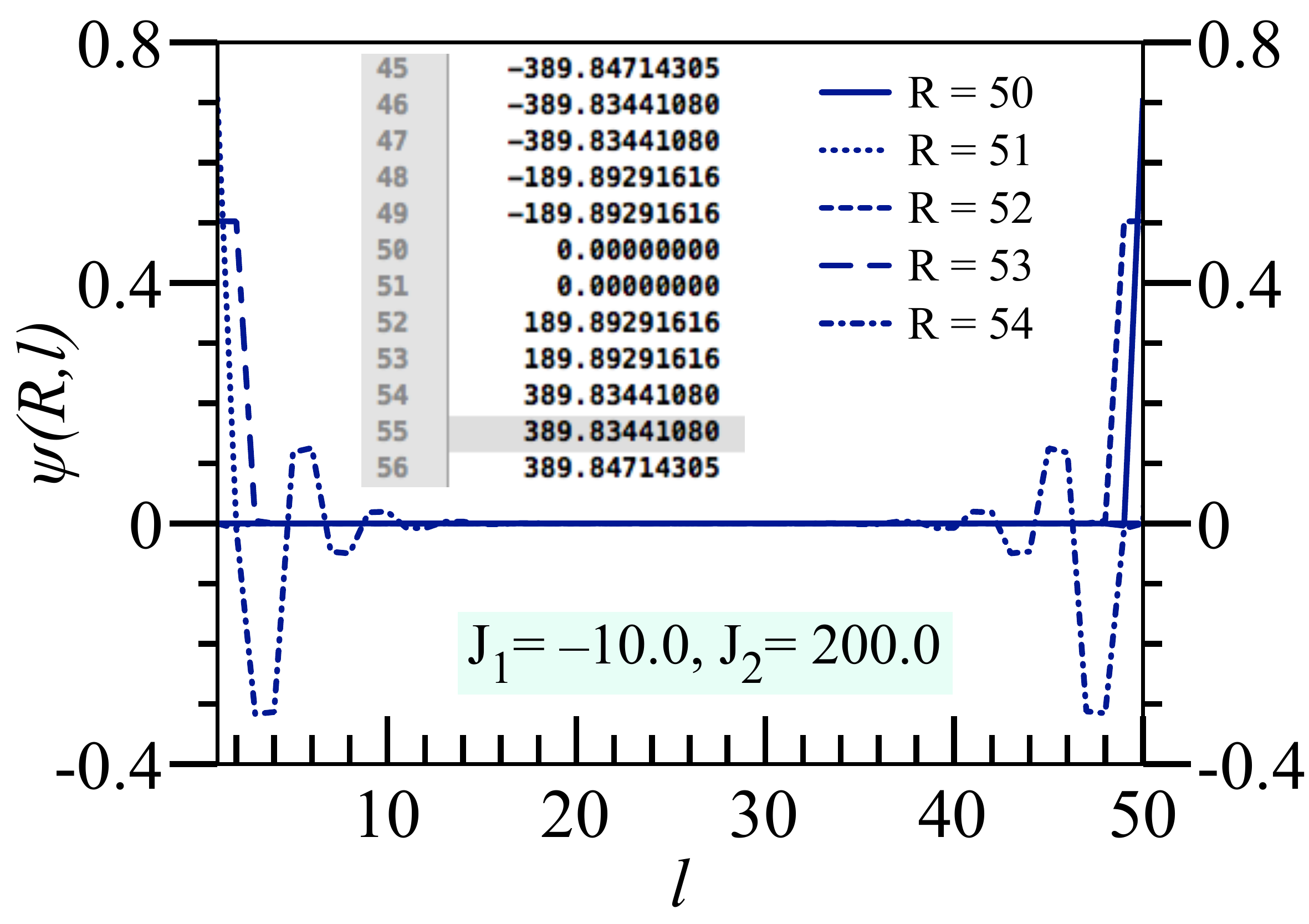} 
	\caption{The real space wave functions corresponding to different eigenvalues are plotted for the DS-AFM  phase. $R$ is eigenvalue index.}
	\label{fig:4majo}
\end{figure}

The eigenvalue spectrum of Hamiltonian from Eq.~({\ref{hr-abc}}) are discussed in the DS-AFM region. In Nambu basis there are $2L$ eigenvalues. $L^{th}$ and $(L+1)^{th}$ eigenvalues are exactly zero. Their wave vectors show decay from the boundary, and thus they are identified as Majorana zero modes or edge modes.. The next excited states ($(L+2)^{th}$ and $(L+3)^{th}$) have non-zero eigenvalue (see Fig.~{\ref{fig:4majo}}) and are degenerate. Their wavefunctions also decay rapidly from boundary like it was for the case of zero energy states. We understand them as Majorana-like edge modes although their eigenvalues are non-zero, and well separated from the rest of the eigenvalue spectrum. From $(L+4)^{th}$ and so on, the eigenvalues are increased continuously and their wave functions are not localized at boundary, as they are not Majorana modes. These eigenvalues are again well separated from the $(L+2)^{th}$ and $(L+3)^{th}$ eigenvalues as we can see from Fig.~{\ref{fig:4majo}} inset.


\section{\label{sec:sum} Summary}

\begin{figure}[htp] 
   \centering
   \includegraphics[width=0.6\textwidth]{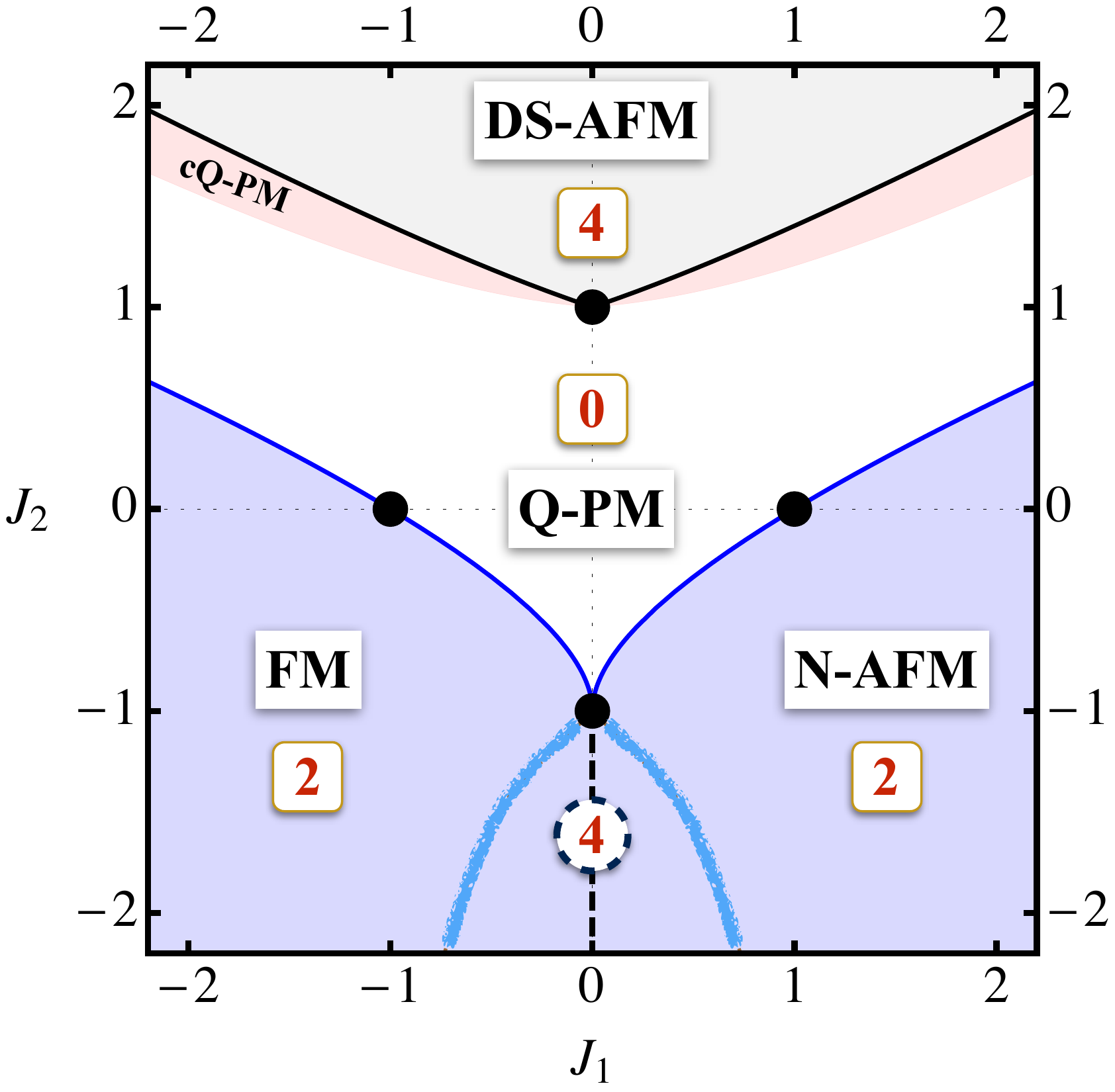} 
   \caption{The quantum phase diagram of the $J_{1}$-$J_{2}$ Ising chain in a transverse field ($h=1$), characterised in terms of the number of edge modes in each phase.}
   \label{fig:summary}
\end{figure}

Our basic motivation behind this work was to study the effect of competing Ising interactions on the occurrence of the edge-modes in a quantum Ising chain. This we have done by numerically investigating the ground state properties of the one-dimensional $J_{1}$-$J_{2}$ Ising model in a transverse field, that is, the $\Hhat$ of Eq.~(\ref{eq:model}). By doing systematic DMRG and CMFT calculations on $\Hhat$, we have generated its quantum phase diagram, and ascertained the presence of the edge modes in the ordered phases. The outcome of this effort is summarised in Fig.~\ref{fig:summary}.

There are two ordered regions in the ground state of $\Hhat$, one of which shows twofold degenerate ferromagnetic (FM) or N\'eel antiferromagnetic order (N-AFM) for negative or positive values of $J_{1}$, respectively.  The other phase, dominated by $J_{2}$, exhibits fourfold degenerate double-staggered antiferromagnetic (DS-AFM) order. In between lies a quantum disordered region consisting mostly of a gapped paramagnetic phase (Q-PM), and a small gapless critical phase (cQ-PM) adjacent to the DS-AFM phase. There are four exact quantum critical points, $(J_{1},J_{2})=(\pm 1,0)$ and $(0,\pm 1)$, shown as filled black circles in Fig.~\ref{fig:summary}. The dashed, $J_{1}=0$ and $J_{2}<-1$, black line is the exact level-crossing line between the FM and N-AFM phases. The rest of the phase boundaries are generated numerically. These phase boundaries in Fig.~\ref{fig:summary} have been smoothened by empirically fitting the data of Fig.~\ref{fig:QPD} to $\sqrt{1+a|x|+bx^{2}}$ for the DS-AFM and cQ-PM phase boundaries, and $-1+ c|x|+(1-c)\sqrt{|x|}$ for the FM phase boundary. These forms are guided by the data itself, and are constrained to pass through four exact critical points, and to become linear for large $x$. One can further improve it, but here we keep it simple.

We characterise these ordered phases in terms of the edges modes that occur therein. We compute different end-to-end spin-spin correlations on the chains with open boundaries, and compare them with the $8^{th}$, $5^{th}$ and $2^{nd}$ power of the order parameter. Sufficiently inside the ordered phase, if an end-to-end correlation goes as the $8^{th}$ power of the order parameter, then it indicates the occurrence of two edge modes. If it goes as the $2^{nd}$ power, then it is clearly a bulk behaviour, and implies no edge modes. The $5^{th}$ power behaviour indicates that one of the two concerned sites supports a Majorana-like edge mode, while the other behaves as bulk. These empirical rules are inspired by the exact $8^{th}$ power behaviour for the end-to-end  correlation in the nearest-neighbor QI chain, and are born out well by our numerical calculations. 

Thus, we come to conclude that the DS-AFM phase supports four edge modes (indicated by a 4 inside the little box in Fig.~\ref{fig:summary}). There are no edge-modes in the quantum paramagnetic phases (Q-PM as well as cQ-PM). The FM and N-AFM phases support two edge modes for most parts. However, very near the dashed $J_{1}=0$ line that has four edge modes (indicated by a 4 inside the circle), our calculations seem to suggest that the FM and N-AFM phases may also realise four edge modes. This possible cross-over from two to four edge modes is denoted by two thick hazy light-blue lines surrounding the $J_{1}=0$ dashed line, and is roughly estimated by $\log{\rho^{x}_{2,L-1}}/\log{p}$ going above $7$ as $J_{2}$ grows more and more negative for a small $J_{1}$. Notably, the number of edge modes is also same as the degeneracy of the ordered phase. It roughly makes sense, because a pair of Majorana modes leads to twofold degeneracy, therefore, two pairs of these edge modes would cause fourfold degeneracy. In view of this, a more careful analysis needs to be done to check whether the cross-over from two to four edge modes in the FM/N-AFM phase indeed occurs.


We also did a fermionic mean-field calculation that looks okay well inside the FM/N-AFM phase, because the decoupling of the $J_{2}$ interaction generates $J_{1}$ like terms. But more care is needed for discussing the DS-AFM phase, and regions close to $J_{1}=0$. Overall there seems to be a broad agreement between DMRG and fermionic MFT calculations. But we are looking for alternative ways to study the corresponding fermion model to see the Majorana edge modes in the $J_{1}$-$J_{2}$ QI chain.



\clearpage
\bibliographystyle{unsrt}
\bibliography{chapters/ref-all}


%% file: chapters/falicov-kimball-model.tex
 \def\f{\hat{f}}
 \def\fdag{\hat{f}^\dagger}

 \def\xhat{\hat{x}}
 \def\yhat{\hat{y}}
\def\phihat{\hat{\phi}}
\def\psihat{\hat{\psi}}
\def\a{\hat{a}}
\def\adag{\hat{a}^\dagger}
\def\b{\hat{b}}
\def\bdag{\hat{b}^\dagger}

\chapter{Some studies on Falicov-Kimball model}

\begin{center}
\parbox{0.8\textwidth}{
\footnotesize
{\bf {\small About this chapter}}
\\[12pt]
In this chapter, we study the half-filled Falicov-Kimball model on square lattice. A canonical representation for electron~[Physical Review B, {\bf 77}, 205115 (2008)] has been used to map the problem in terms of spinless fermion and Pauli matrices. In strong correlation limit (onsite repulsion, $U \gg t$, the hopping) the problem leads to the classical Ising model. We are interested in studying the effects  of quantum charge fluctuations (for finite $U/t$) on the Ising behaviour. 
\\[15pt]
}
\end{center}

\minitoc

\section{Introduction}
Falicov-Kimball (FK) {\index{Falicov-Kimball model}}model was introduced by L. M. Falicov and J. C. Kimball~{\cite{fkmodel}} to study metal-insulator transitions in mixed valence compounds of rare earth and transition metal oxides. The model is also very effective to study interaction between localized f-electrons or classical or heavy particles and itinerant d-electrons or quantum or light-weight particles. There have been lots of numerical studies on this model using Monte Carlo simulation~\cite{zonda,maska}, dynamical mean-field theory~\cite{freericks}, exact diagonalization~\cite{farkasovsky-iop-2002,farkasovsky-prb-1995}. The main theme of the studies have been phase diagram at and away from half-filling, presence of disorder~\cite{fm-model-disorder}, temperature dependence of chemical potentials of localized and itinerant objects~\cite{musial}, effects of correlated hopping on the ground state properties of FK model~{\cite{umesh-2010}} etc.


In its simplest form, it is a model in which the electrons of one spin state ($\uparrow$ or $\downarrow$) hop with amplitude $t$, while the others with opposite spin state don't hop. In addition to this spin-dependent hopping, it also has a local Hubbard interaction, $U$. In the limit $U/t\gg1$, at half-filling, this model behaves as the classical Ising model. However, for any finite $U$, the quantum mechanical charge fluctuations are always present. Therefore, we like to think of the Falicov-Kimball model also as a quantum Ising problem with `quantum mechanical charge-fluctuations', unlike the spin fluctuations due to the transverse field in the standard definition of the quantum Ising problem. Our objective is to understand the effect of charge fluctuations on the standard Ising thermodynamics and other properties. As it is an unfinished calculation, we present only some results from the Monte Carlo simulations on square lattice, and note the deviations from the known results for the classical Ising model.

\section{Falicov-Kimball model}

The Hubbard model Hamiltonian is written as 
\begin{equation}
H = - {\sum_{<i,j>}} {\sum_{\sigma = \uparrow , \downarrow}} \frac{t_{ij}}{2}  \left(  f_{i,{\sigma}}^{\dagger} f_{j,\sigma} + h.c  \right) + U {\sum_i} {\left(n_{i,{\uparrow}} - {1\over 2} \right)}{\left(n_{i{\downarrow}} - {1\over 2} \right)}
\end{equation}
here $t_{ij}$ is hopping amplitude of itinerant electron between nearest-neighbor sites $i$ {\&} $j$. $f_{i,{\sigma}}$ is annihilation operator of these electrons. $U$ is onsite repulsion energy. $n_{i,\uparrow}$ and $n_{i,\downarrow}$ are number of $\uparrow$ electrons and $\downarrow$ electrons on site $i$. We are at half-filling. That means number of particle on each site is exactly 1 (minimum being $0$ and maximum being $2$ for real electron). The geometry we have taken is a bipartite square lattice. Let's consider $ t_{\uparrow} = 0$ and $t_{\downarrow} \neq 0 $ i.e., only $\downarrow$ type of electrons are moving but $\uparrow$ type of electrons are heavy and are not allowed to hop. The Hamiltonian can be written as 
\begin{equation}
H_{FK} = - \frac{t}{2} {\sum_{r,\delta = \pm \hat{x} , \pm \hat{y}}} \left(  f_{r,{\downarrow}}^{\dagger} f_{r + \delta, {\downarrow}} + h.c  \right) + U {\sum_r} {\left(n_{r,{\uparrow}} - {1\over 2} \right)}{\left(n_{r{\downarrow}} - {1\over 2} \right)}{\label{eq:fk}}
\end{equation}
Eq.~{\eqref{eq:fk}} is called Falicov-Kimball model.
We use the canonical representation of electron~\cite{bk-canonical}
\begin{equation}
\hat{f_{\uparrow}^{\dagger}} = \hat{\phi} {\sigma^{+}}
\end{equation}
 \begin{equation}
\hat{f_{\downarrow}^{\dagger}} = i \frac{\hat{\psi}}{2} -  \frac{\hat{\phi}}{2}{\sigma^{z}}
\end{equation}
where $\phi = \a + \adag$ and $i \psi = \adag - \a$, and $\a$ are spinless fermions, and $\sigma$'s are Pauli spin matrices. Some useful identities 
\begin{eqnarray}
n_{\uparrow} = \fdag_\uparrow \f_\uparrow  &=& {1 \over 2} + \frac{\sigma^{z}}{2}\\
n_{\downarrow} = \fdag_\downarrow \f_\downarrow &=& {1 \over 2} + ({1 \over 2} - n ){\sigma^{z}}
\end{eqnarray}
$n = \adag \a$ is number of spinless fermions. On a bipartite square lattice the above mentioned representation will look like 
\begin{equation}
f_{a_{\downarrow}}^{\dagger} = \frac{i {\psihat_{a}}}{2} - \frac{{\phihat_{a}}{\sigma^{z}}}{2} 
\end{equation}
\begin{equation}
f_{b_{\downarrow}}^{\dagger} =  \frac{\phihat_{b}}{2} -  \frac{i{\psihat_{b}}{\tau^{z}}}{2}
\end{equation}
where we have taken $\a$ type and $\b$ type of spinless fermions and $\sigma$ and $\tau$ type of spin matrices on two neighboring sites. We can change the number of particle by changing  the chemical potential. Using the constraint of total number of electrons ($N_{e} = {{\sum}_{r=1}^{L}}{[ 1 + ( 1 - n_{r} ){\sigma}_{r}^{z}]}$) we get the final Hamiltonian of the form
\begin{equation}
H = H_{FK} - {\mu} N_{e}
\end{equation}
\begin{equation}
H = H_{FK} - {\mu}{{\sum}_{r=1}^{L}}{[ 1 + ( 1 - n_{r} ){\sigma}_{r}^{z}]}
\end{equation}
The problem becomes a coupled problem of spinless fermions and Pauli spin matrices. The bipartite property of the square lattice is making  the Hamiltonian look like 
\begin{equation}
H = H_{U} + H_{t}
\end{equation}
where the two parts $ H_{U}$ and $H_{t}$ are given by
\begin{eqnarray}
H_{U} &=&  {\mu} \left( N_{e} - L \right) - {\frac{\mu}{2}}{\sum}_{r \in A}{\sigma_{r}^{z}}  - {\frac{\mu}{2}}{\sum}_{r \in B}{\tau_{r}^{z}}  \cr &&
+ {\sum}_{r \in A} \left( \a_{r}^{\dagger}\a_{r} 
 - \a_{r} \a_{r}^{\dagger} \right) \left( {\frac{\mu {\sigma_{r}^{z}}}{2}} - \frac{U}{4} \right)  \cr &&
  + {\sum}_{r \in B} \left( b_{r}^{\dagger}b_{r} - b_{r}b_{r}^{\dagger} \right) \left( {\frac{\mu {\tau_{r}^{z}}}{2}} - \frac{U}{4} \right) 
\end{eqnarray}

And 

\begin{eqnarray}
H_{t} &=& -{\frac{t}{2}} {\sum_{r \in A}} {\sum_{\delta \pm \xhat,\yhat}} \big[ \left( { \adag_{r} \bdag_{r+\delta} } + h.c  \right) \left( { 1 - {\sigma_{r}^{z}} {\tau_{r+\delta}^{z}}} \right) \cr && + \left( {\adag_{r} \b_{r+\delta}} + h.c \right) \left({1 + {\sigma_{r}^{z}}{\tau_{r+\delta}^{z}}}\right)  \big]
\end{eqnarray}

So we can now write the Hamiltonian on Nambu basis of spinless fermions. And the problem becomes classical Ising model on 2D. We solve this Hamiltonian to find the eigenvalues $\epsilon_{i}$. We use Monte Carlo (MC) simulations to study it. We benchmark our calculations by first checking the simples cases of atomic ($t=0$) and band ($U =0$) limits of the FK model. Then we compute the staggered magnetization as a function of temperature and as a function of onsite repulsion $U$ with the help of MC simulation. The MC steps are discussed below. 

\section{Monte Carlo Simulation}
Here are the simple steps we followed during Monte Carlo simulation for this problem

\begin{enumerate}
\item Fix the values of $U$, $t$, $\mu$ and Temperature $T$ 
\item We choose some initial configuration for spins $\sigma$.
\item{\label{step:new-config}} Calculate $\sum_{i}{\sigma_i}$ and $\sum_{i}{(-1)^{x+y}{\sigma_i}}$ \footnote{ x and y are the Cartesian co-ordinate of the $i^{th}$ site of the square lattice structure. }. We then write the Hamiltonian $H$. The Hamiltonian is written on Nambu basis.  Then we calculate the energy eigenvalues and eigenvectors of the Hamiltonian.
\item  Calculate the free energy $F[{\{}\sigma{\}}]$
\item{\label{step:calculate-f}} The formula is
\begin{equation}
F = E_{0} - T {\sum_{i}}{ \log [ 2 \cosh( \beta \epsilon_{i} ) ] }
\end{equation}
\begin{equation}
E_{0} = -\mu L - {\frac{\mu}{2}}{\sum_{i}}\sigma_{i}
\end{equation}
\item Let's call this free energy to be $F_{initial}$. Now we choose a particular site and flip it to get a new $\sigma$ configuration and calculate the new free energy as done from Step~{\ref{step:new-config}} to Step~{\ref{step:calculate-f}}. Let's call this free energy to be $F_{final}$.
\item We check if the final free energy is is minimized. If so, then we accept the flip and try a different $\sigma$ configuration. And if not we reject the flip with the probability $e^{-\beta \Delta F}$ where $ \Delta F = (F_{final} - F_{initial} ) $ and (remember this is not real spin flip. We are just changing the value of $\sigma$ from +1 to -1).
\item Thus we reach an thermodynamic stable $\sigma$ configuration and minimum free energy for that  configuration. 
\item This is one Monte Carlo step.
\item We calculate some physical quantity such as magnetization etc.
\item We repeat same steps for $10^5$ times or so and averages over these many number of values to get the final stable value for a physical quantity
\end{enumerate}

\section{Results}
\textbf{Verifying the Atomic limit and Band limits:}

\subsection*{Atomic limit ($t = 0$) }
\begin{figure}[htp] 
   \centering
   \includegraphics[width=0.65\textwidth]{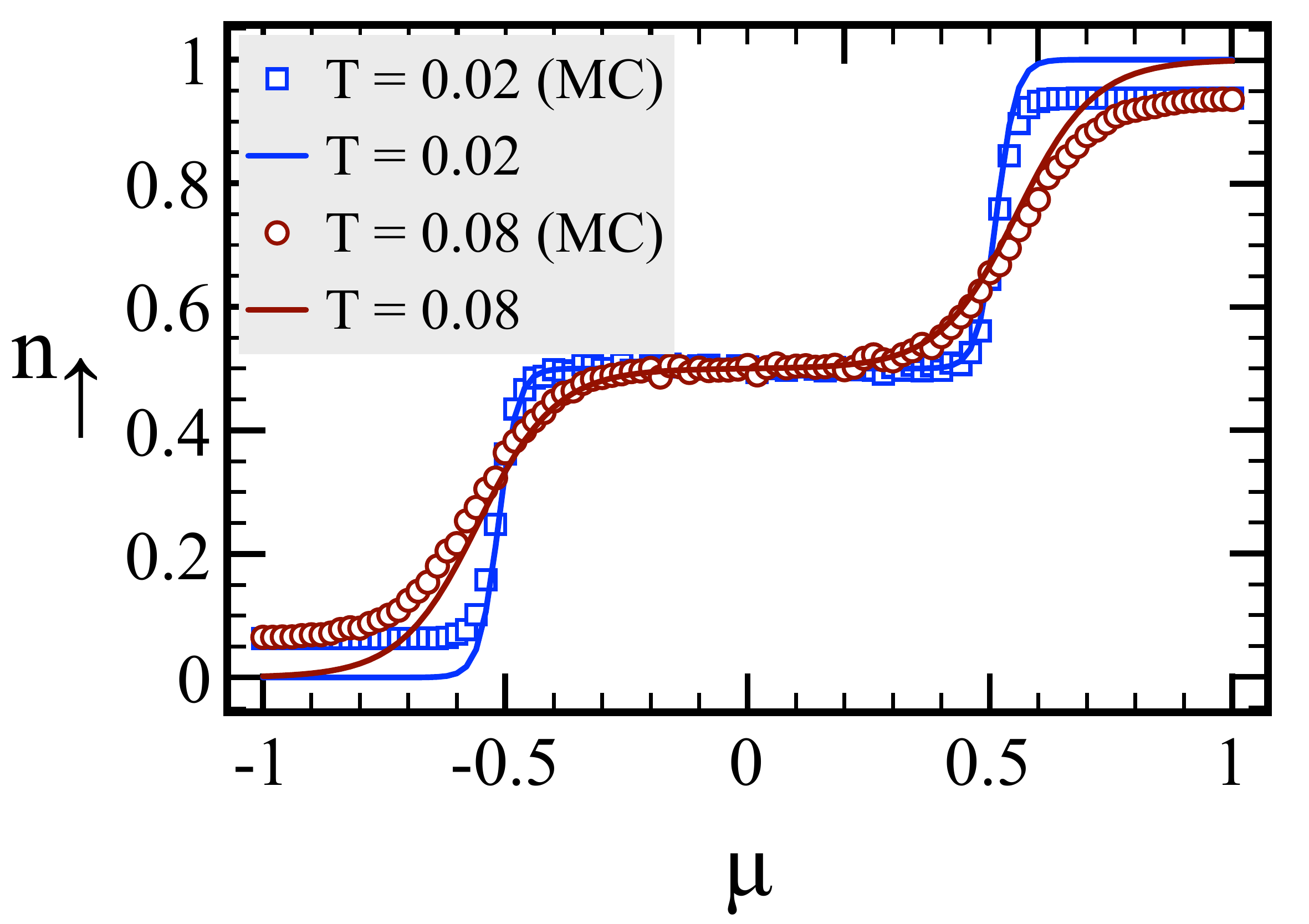} 
   \caption{This figure is $n_{\uparrow}$ Vs $\mu$ plot (for $U=1.0$) for different temperature in atomic limit. Both from Monte Carlo simulation (data points) and exact results (solid lines) have been plotted}
   \label{fig:atomic}
\end{figure}
In the atomic limit, $t=0$ and $U\neq 0$. This is obviously an insulating state. We get localized states. 

\begin{equation}
H = U {\sum_i} {\left(n_{i,{\uparrow}} - {1\over 2} \right)}{\left(n_{i{\downarrow}} - {1\over 2} \right)}  - {\mu}N
\end{equation}
And so $\langle {n_{\uparrow}}\rangle$ will be given as $\frac{Tr \left( n_{\uparrow} \exp[-\beta H] \right)}{Tr\left( \exp[-\beta H] \right)} $ and so for the $\langle {n_{\downarrow}}\rangle$. The final expressions are given as

\begin{equation}
n_{\downarrow} = n_{\uparrow} =  \frac{ {e^{\beta U/4 + \beta \mu}} + {e^{-\beta U/4 + 2 \beta \mu}} }{ {e^{-\beta U/4}} + {2 e^{\beta U/4 + \beta \mu}} + {e^{-\beta U/4 + 2 \beta \mu}} }
\end{equation}
So we can plot $n_{\uparrow}$ or $n_{\downarrow}$ as  a function of $\mu$. Now how we get the same result for $n_{\uparrow}$ from Monte Carlo simulation result is written below. As we know the expression for $n_{\uparrow}$ is given by $n_{\uparrow} = {1 \over 2} + {{\sigma}^{z} \over 2}$ and $n_{\downarrow} = {1 \over 2} + ({1 \over 2} - n) {\sigma^{z}}$  we calculate the total value of the spins on the lattice after the completion of one Monte Carlo step. And then take average. This is for single value of $\mu$ with $U, T, t $ all fixed. See Fig.~{\ref{fig:atomic}} Notably, at $\mu=0$, $n_{\uparrow}=0.5$, as it correctly should be at half-filling. Far away from it, $n_{\uparrow}$ either tends to 0 or 1.


\subsection*{Band limit ($U = 0$) }
\begin{figure}[htp] 
   \centering
   \includegraphics[width=0.65\textwidth]{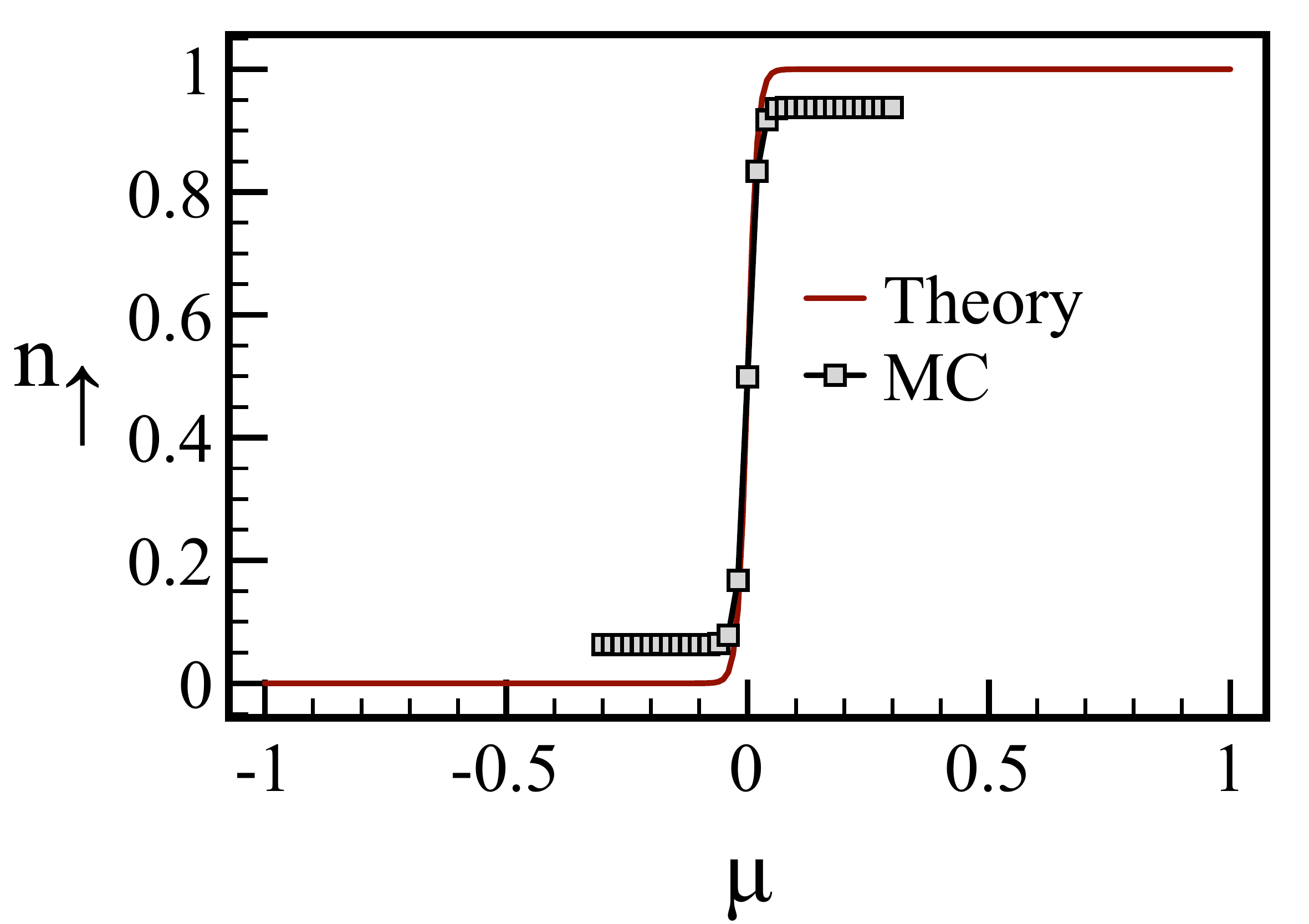} 
   \caption{This figure is $n_{\uparrow}$ vs $\mu$ plot for temperature $T=0.01$ in band Limit. Both from Monte Carlo simulation (data points) and exact results (solid line) have been plotted.}
   \label{fig:band}
\end{figure}

In the band limit the system acts like metal. There is no onsite repulsion term. i.e; $U = 0$. The hopping parameter $t$ is making the electrons to hop from one site to other. The Hamiltonian in this case is 
\begin{equation}
H = -t{\sum_{<i,j>}} \left(  f_{i,{\downarrow}}^{\dagger} f_{j, {\downarrow}} + h.c  \right)  - {\mu}N.
\end{equation}
We do the Fourier transform, and get the diagonal form.
\begin{equation}
H = \sum_{k} ( \epsilon_{k} - \mu ) n_{k \downarrow} - {\mu}{\sum_{k}} n_{k \uparrow}
\end{equation}
The expression for $n_{\uparrow}$ is given by
\begin{equation}
n_{\uparrow} = \frac{1}{e^{\beta \mu} + 1}
\end{equation}
We can again compare this result with MC simulation result by setting $U=0$. See Fig.~{\ref{fig:band}}, in which, at $\mu=0$, $n_{\uparrow}=0.5$, as it correctly should be, and away from it, $n_{\uparrow}$ approaches 0 or 1.

\subsection{Magnetization}

We now proceed to calculate the staggered-magnetization for this system by the Monte Carlo simulations for both $t$ and $U\neq 0$. The staggered magnetization as 
\begin{eqnarray}
M_{\sigma}^{staggered} &=& {\sum_{i}}{(-1)^{x+y}}{\sigma^{z}_{i}} \\
M_{spin}^{staggered} &=& {\sum_{i}}{(-1)^{x+y}}{\frac{{n_{i}}{\sigma^{z}_{i}}}{2}}
\end{eqnarray}

\begin{figure}[htbp]
\centering
\subfloat[Magnetization (staggered) as vs. U for different values of temperature , $t = 1.0$, $36$ sites, $10^4$ MC steps.]{
\label{kol}
\includegraphics[width=0.45\textwidth]{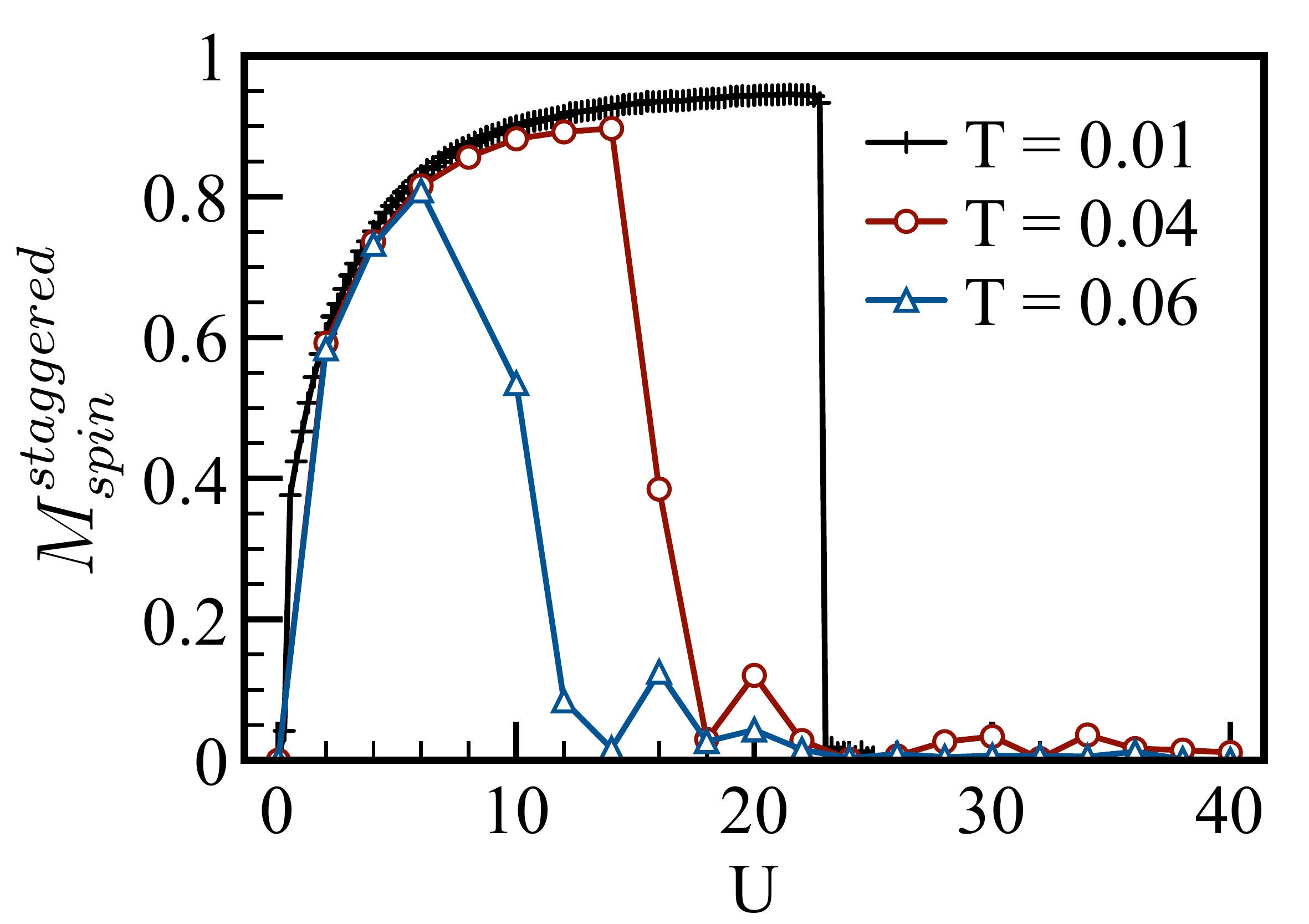} 
} \qquad
\subfloat[Staggered magnetization ($M_{\sigma}^{\rm staggered}$)�vs temperature  for U = 10, t=1.0, $\mu =0$]{
\label{fig:gap-vs-j2-j1-1p5}
\includegraphics[width=0.45\textwidth]{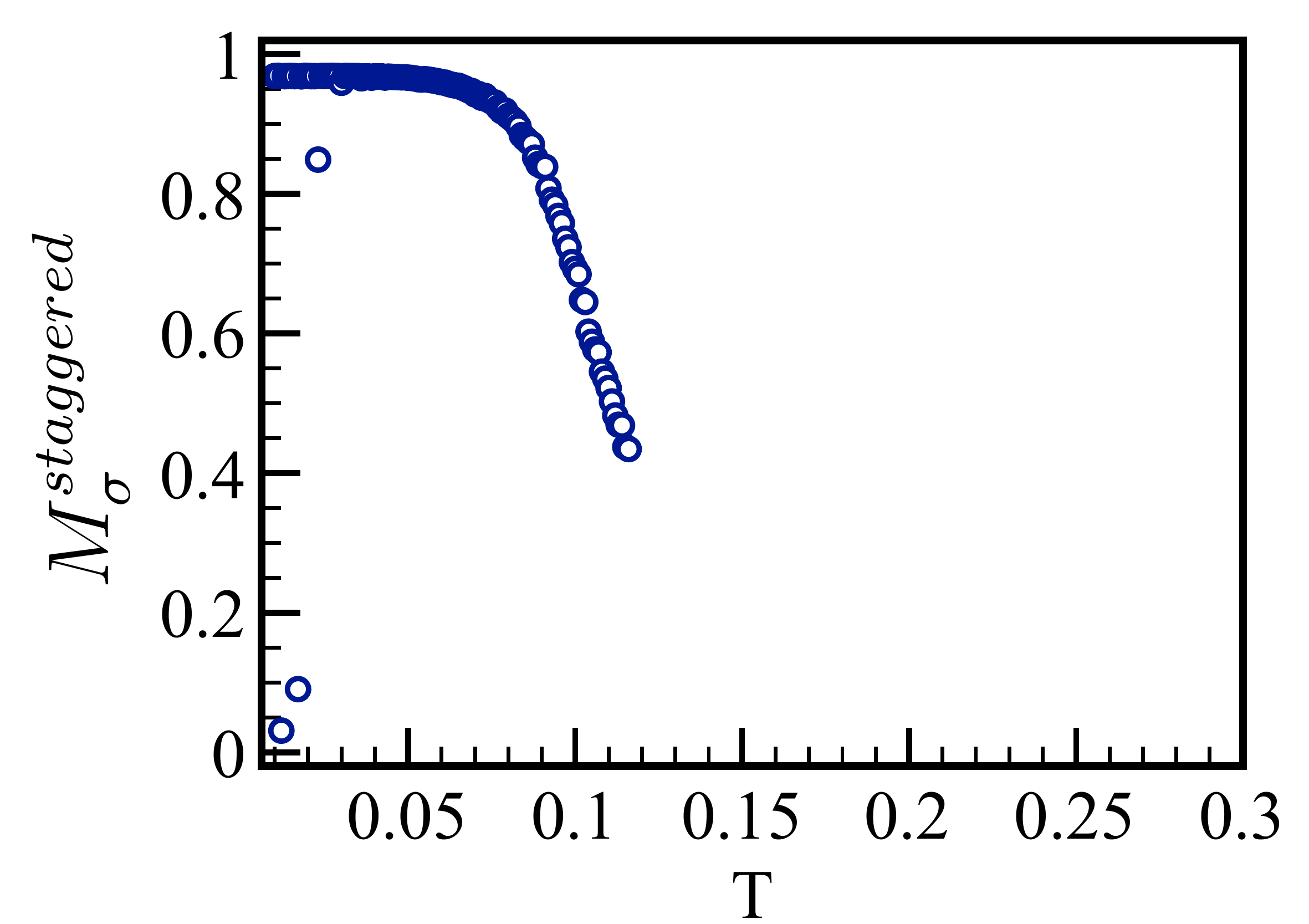} 
} 
\caption{Magnetization plot.}
\label{fig:mst-t}
\end{figure}

There are two notable features in the data presented above. For a fixed $T$, as we increase $U/t$, there is first order transition from an antiferromangetic ordered phase to a paramagnetic phase, as shown in Fig.~\ref{fig:mst-t}(a). Secondly, for a fixed $U/t$, the staggered magnetisation [in Fig.~\ref{fig:mst-t}(b)] decreases with increasing temperature almost linearly, as compared to power $1/8$ for the two-dimensional classical Ising model. This is all at half-filling. Clearly, the quantum mechanical charge fluctuations seem to be causing a deviation from the standard Ising behaviour. More calculations are in progress at this stage.

%

\section{Summary}

Here, we have described our motivations and preliminary studies of the Falicov-Kimball model. Using an interesting canonical representation~{\cite{bk-canonical}} for electrons, we have developed this problem into a problem of spin-less fermions and Pauli spin operators. In this formulation, the charge fluctuations and Ising interactions become more explicit. The Monte Carlo method, together with exact diagonalization of the bilinear fermion problem, is used to calculate physical magnetization for antiferromagnetic case at half filling. It is an ongoing work, through which, we look forward to address many different questions. For instance, why the magnetization seems to decrease with a different power from $1/8$?  Moreover, in line with the main theme of this thesis, our immediate interest is in investigating the Majorana edge modes in this model (that is, with charge fluctuations).


%


\clearpage
\bibliographystyle{unsrt}
\bibliography{chapters/ref-all}


%% file: thesis-main.bbl
\begin{thebibliography}{10}

\bibitem{fkmodel}
L.~M. Falicov and J.~C. Kimball.
\newblock Simple model for semiconductor-metal transitions:
  Sm${\mathrm{b}}_{6}$ and transition-metal oxides.
\newblock {\em Phys. Rev. Lett.}, 22:997--999, May 1969.

\bibitem{zonda}
Martin {\v Z}onda.
\newblock Phase transitions in the falicov-kimball model away from
  half-filling.
\newblock {\em Phase Transitions}, 85(1-2):96--105, 2012.

\bibitem{maska}
Maciej~M. Ma\ifmmode~\acute{s}\else \'{s}\fi{}ka and Katarzyna Czajka.
\newblock Thermodynamics of the two-dimensional falicov-kimball model: A
  classical monte carlo study.
\newblock {\em Phys. Rev. B}, 74:035109, Jul 2006.

\bibitem{freericks}
J.~K. Freericks and V.~Zlatic.
\newblock Exact dynamical mean-field theory of the falicov-kimball model.
\newblock {\em Rev. Mod. Phys.}, 75:1333, October 2003.

\bibitem{farkasovsky-iop-2002}
Pavol Farkasovsk{\'y} and Nat{\'a}lia Hud{\'a}kov{\'a}.
\newblock Ground-state properties of the falicov-kimball model with correlated
  hopping in two dimensions.
\newblock {\em Journal of Physics: Condensed Matter}, 14(3):499, 2002.

\bibitem{farkasovsky-prb-1995}
Pavol Farka\ifmmode~\check{s}\else \v{s}\fi{}ovsk\'y.
\newblock Falicov-kimball model and the problem of valence and metal-insulator
  transitions.
\newblock {\em Phys. Rev. B}, 51:1507--1512, Jan 1995.

\bibitem{fm-model-disorder}
M.~A. Gusm\~ao.
\newblock Phase diagram of the anderson-falicov-kimball model at half filling.
\newblock {\em Phys. Rev. B}, 77:245116, Jun 2008.

\bibitem{musial}
G.~Musial, L.~Debski, and J.~Wojtkiewicz.
\newblock A monte carlo study of the falicov-kimball model in the perturbative
  regime.
\newblock {\em Low Temperature Physics}, 33(9):797, 2007.

\bibitem{umesh-2010}
Umesh~K Yadav, T~Maitra, Ishwar Singh, and A~Taraphder.
\newblock A ground state phase diagram of a spinless, extended falicov--kimball
  model on the triangular lattice.
\newblock {\em Journal of Physics: Condensed Matter}, 22(29):295602, 2010.

\bibitem{bk-canonical}
Brijesh Kumar.
\newblock Canonical representation for electrons and its application to the
  hubbard model.
\newblock {\em Phys. Rev. B}, 77:205115, May 2008.

\end{thebibliography}


\begin{thebibliography}{10}

\bibitem{quantum-simulation}
I.~M. Georgescu, S.~Ashhab, and Franco Nori.
\newblock Quantum simulation.
\newblock {\em Reviews of Modern Physics}, 86:153, March 2014.

\bibitem{BH.ColdAtom}
D.~Jaksch, C.~Bruder, J.~I. Cirac, C.~W. Gardiner, and P.~Zoller.
\newblock Cold bosonic atoms in optical lattices.
\newblock {\em Phys. Rev. Lett.}, 81:3108--3111, Oct 1998.

\bibitem{ColdAtoms}
I.~Bloch, J.~Dalibard, and W.~Zwerger.
\newblock {\em Rev. Mod. Phys.}, 80:885, 2008.

\bibitem{Tomadin.Fazio.10}
A.~Tomadin and R.~Fazio.
\newblock {\em J. Opt. Soc. Am. B}, 27(6):A130, 2010.

\bibitem{CircuitQED.Girvin}
R.~J. Schoelkopf and S.~M. Girvin.
\newblock {\em Nature}, 451:664, 2008.

\bibitem{Greentree.06}
A.~D. Greentree, C.~Tahan, J.~H. Cole, and L.~C.~L. Hollenberg.
\newblock {\em Nature Physics}, 2:856, 2006.

\bibitem{Hartmann}
M.~J. Hartmann, F.~G. S~L. Brandao, and M.~B. Plenio.
\newblock {\em Nat. Phys.}, 2:849, 2006.

\bibitem{pathria}
R.~K. Pathria.
\newblock {\em Statistical Mechanics}.
\newblock Butterworth Heinemann, second edition edition, 1996.

\bibitem{huang}
Kerson Huang.
\newblock {\em Statistical Mechanics}.
\newblock John Wiley and Sons, second edition edition, 1987.

\bibitem{deGennes}
P.~G. de~Gennes.
\newblock {\em Solid State Commun}, 1(132), 1963.

\bibitem{book.Bikas}
B.~K. Chakrabarti, A.~Dutta, and P.~Sen.
\newblock {\em Quantum Ising Phases and Transitions in Transverse Ising
  models}.
\newblock Springer-Verlag, Berlin, 1996.

\bibitem{randomQI.Fisher}
D.~S. Fisher.
\newblock {\em Phys. Rev. B}, 51:6411, 1995.

\bibitem{Moessner}
R.~Moessner, S.~L. Sondhi, and P.~Chandra.
\newblock {\em Phys. Rev. Lett.}, 84:4458, 2000.

\bibitem{QI.Holo.Girvin}
P.~B. Chakraborty, P.~Henelius, H.~Kj{\o}nsberg, A.~W. Sandvik, and S.~M.
  Girvin.
\newblock {\em Phys. Rev. B}, 70:144411, 2004.

\bibitem{Coldea}
R.~Coldea, D.~A. Tennant, E.~M. Wheeler, E.~Wawrzynska, D.~Prabhakaran,
  M.~Telling, K.~Habicht, P.~Smeibidl, and K.~Kiefer.
\newblock {\em Science}, 327:177, 2010.

\bibitem{bermudez-qi-expt}
A~Bermudez, J~Almeida, K~Ott, H~Kaufmann, S~Ulm, U~Poschinger, F~Schmidt-Kaler,
  A~Retzker, and M~B Plenio.
\newblock Quantum magnetism of spin-ladder compounds with trapped-ion crystals.
\newblock {\em New J. Phys.}, 14:093042, 2012.

\bibitem{Mila.Square}
Sandro Wenzel, Tommaso Coletta, Sergey~E. Korshunov, and Fr\'ed\'eric Mila.
\newblock Evidence for columnar order in the fully frustrated transverse field
  ising model on the square lattice.
\newblock {\em Phys. Rev. Lett.}, 109:187202, Nov 2012.

\bibitem{QPT.Sachdev}
Subir Sachdev.
\newblock {\em Quantum Phase Transitions}.
\newblock Cambridge University Press, 2011.

\bibitem{QIsing.Pfeuty}
P.~Pfeuty.
\newblock {\em Ann. Phys. (N.Y.)}, 57:79, 1970.

\bibitem{LSM}
E.~H. Lieb, T.~Schultz, and D.~C. Mattis.
\newblock {\em Ann. Phys. (N.Y.)}, 16:407, 1961.

\bibitem{marcel-race4majorana}
Marcel Franz.
\newblock Race for majorana fermions.
\newblock {\em Physics}, 3:24, Mar 2010.

\bibitem{Kitaev.FreeMajorana}
A.~Yu. Kitaev.
\newblock {\em Phys.-Usp.}, 44:131, 2001.

\bibitem{Nayak.RMP}
Chetan Nayak, Steven~H. Simon, Ady Stern, Michael Freedman, and Sankar
  Das~Sarma.
\newblock Non-abelian anyons and topological quantum computation.
\newblock {\em Rev. Mod. Phys.}, 80:1083--1159, Sep 2008.

\bibitem{Topo_QC_Kitaev}
A.~Kitaev and C.~Laumann.
\newblock Topological phases and quantum computation.
\newblock arXiv:0904.2771, 2009.

\bibitem{das-nature2012}
Anindya Das, Yuval Ronen, Yonatan Most, Yuval Oreg, Moty Heiblum, and Hadas
  Shtrikman.
\newblock Zero-bias peaks and splitting in an al-inas nanowire topological
  superconductor as a signature of majorana fermions.
\newblock {\em Nat Phys}, 8(12):887--895, 12 2012.

\bibitem{Mourik.Majorana}
V.~Mourik, K.~Zou, S.~M. Frolov, S.~R. Plissard, E.~P. A.~M. Bakkers, and L.~P.
  Kouwenhoven.
\newblock {\em Science}, 336:1003, 2012.

\bibitem{Wilczek.Majorana}
F.~Wilczek.
\newblock {\em Nat. Phys.}, 5:614, 2009.

\bibitem{note.Majorana}
A fermion is created (annihilated) by an operator $\chat^\dag$ ($\chat$), and
  it satisfies anti-commutation with other fermions. For a given $\chat^\dag$
  and $\chat$, we can define two Majorana fermions $\phihat = \chat+\chat^\dag$
  and $\psihat=i(\chat-\chat^\dag)$ such that $\phihat^\dag=\phihat$ and
  $\phihat^2=1$ (and likewise for $\psihat$). The Majorana operators $\phihat$
  and $\psihat$ anti-commute with each other, and with other fermion.

\bibitem{Scully.Zubairy}
M.~O. Scully and M.~S. Zubairy.
\newblock {\em Quantum Optics}.
\newblock Cambridge University Press, 1997.

\bibitem{Jaynes.Cummings}
E.~T. Jaynes and F.~W. Cummings.
\newblock {\em Proc. IEEE}, 51:89, 1963.

\bibitem{Littlewood}
M.~Aichhorn, M.~Hohenadler, C.~Tahan, and P.~B Littlewood.
\newblock {\em Phys. Rev. Lett.}, 100:216401, 2008.

\bibitem{Blatter}
S.~Schmidt and G.~Blatter.
\newblock {\em Phys. Rev. Lett.}, 103:086403, 2009.

\bibitem{QD.Photonic}
K.~Hennessy, A.~Badolato, M.~Winger, D.~Gerace, M.~Atat\ ̈ure, S.~Gulde, S.~F\
  ̈alt, E.~L. Hu, and A.~Imamo\v{g}lu.
\newblock {\em Nature (London)}, 445:896, 2007.

\bibitem{CircuitQED.Fink}
J.~M. Fink, R.~Bianchetti, M.~Baur, M.~G\"oppl, L.~Steffen, S.~Filipp, P.~J.
  Leek, A.~Blais, and A.~Wallraff.
\newblock {\em Phys. Rev. Lett.}, 103:083601, 2009.

\bibitem{Houck.Koch}
A.~A. Houck, H.~E. T\"ureci, and J.~Koch.
\newblock {\em Nat. Phys.}, 8:292, 2012.

\bibitem{bkumar.somenath}
Brijesh Kumar and Somenath Jalal.
\newblock Quantum ising dynamics and majorana-like edge modes in the rabi
  lattice model.
\newblock {\em Phys. Rev. A}, 88:011802(R), July 2013.

\bibitem{J1J2QI}
Somenath Jalal and Brijesh Kumar.
\newblock Edge modes in a frustrated quantum ising chain.
\newblock arXiv:1407.0201, 2014.

\bibitem{fkmodel}
L.~M. Falicov and J.~C. Kimball.
\newblock Simple model for semiconductor-metal transitions:
  Sm${\mathrm{b}}_{6}$ and transition-metal oxides.
\newblock {\em Phys. Rev. Lett.}, 22:997--999, May 1969.

\bibitem{srw-noack-prl1992}
S.~R. White and R.~M. Noack.
\newblock Real-space quantum renormalization group.
\newblock {\em Phys. Rev. Lett.}, 68(24):3487, June 1992.

\bibitem{srw-prl1992}
S.~R. White.
\newblock Density matrix formulation for quantum renormalization groups.
\newblock {\em Phys. Rev. Lett.}, 69(19), November 1992.

\bibitem{srw-prb1993}
Steven~R. White.
\newblock Density-matrix algorithms for quantum renormalization groups.
\newblock {\em Phys. Rev. B}, 48:10345--10356, Oct 1993.

\bibitem{Schollwock2005}
U.~Schollw\"ock.
\newblock {\em Rev. Mod. Phys.}, 77:259, 2005.

\bibitem{chiara}
Gabriele~De Chiara, Matteo Rizzi, Davide Rossini, and Simone Montangero.
\newblock Density matrix renormalization group for dummies.
\newblock {\em arXiv:cond-mat/0603842}, April 2009.

\bibitem{noack}
Reinhard~M. Noack and Salvatore~R. Manmana.
\newblock {\em arXiv:cond-mat/0510321}.

\bibitem{saad}
Youcef Saad.
\newblock {\em Numerical Methods for large Eigenvalue problem}.
\newblock Manchester University Press Series in Algorithms and Architecture for
  Advanced Scientific Computing. Fourth edition, 1991.

\bibitem{davidson}
Andreas Stathopoulos and Charlotte~F. Fischer.
\newblock A davidson program for finding a few selected extreme eigenpairs of a
  large, sparse, real, symmetric matrix.
\newblock {\em Computer Physics Communication}, 79(2):268--290, April 1994.

\bibitem{lapack}
E.~Anderson, Z.~Bai, C.~Bischof, S.~Blackford, J.~Demmel, J.~Dongarra,
  J.~Du~Croz, A.~Greenbaum, S.~Hammarling, A.~McKenney, and D.~Sorensen.
\newblock {\em LAPACK Users' Guide}.
\newblock Society for Industrial and Applied Mathematics, Philadelphia, PA,
  third edition, 1999.

\bibitem{sandvik}
A.~W. Sandvik.
\newblock Aip conf. proc.
\newblock volume 1297, page 135, http://dx.doi.org/10.1063/1.3518900, 2010.

\bibitem{lin-prb-ed}
H.~Q. Lin.
\newblock Exact diagonalization of quantum-spin models.
\newblock {\em Phys. Rev. B}, 42(10):6561, October 1990.

\bibitem{lauchli-ed-talk}
Andreas Lauchli.
\newblock {\em Introduction to Exact Diagonalization}.
\newblock Boulder Summer School, 2010.
\newblock (presentation note).

\bibitem{krauth}
Werner Krauth.
\newblock {\em Statistical Mechanics: Algorithm and Computation}.
\newblock Oxford Master Series in Statistical, Computational, and Theoretical
  Physics. Oxford University Press, 2006.

\bibitem{binder}
K.~Binder and D.~H. Heermann.
\newblock {\em Monte Carlo Simulation in Statistical Physics}.
\newblock Springer Series in Solid-State Sciences. Springer, fourth edition,
  2002.

\bibitem{assa-1988}
Daniel~P. Arovas and Assa Auerbach.
\newblock Functional integral theories of low-dimensional quantum heisenberg
  models.
\newblock {\em Phys. Rev. B}, 38(1):316, July 1988.

\bibitem{assa-book}
Assa Auerbach.
\newblock {\em Interacting electrons and quantum magnetism}.
\newblock Graduate Texts in Contemporary Physics. Springer, Technion, Israel
  Institute of Technology, 1997.

\bibitem{BH_Fisher}
M.~P.~A. Fisher, P.~B. Weichman, G.~Grinstein, and D.~S. Fisher.
\newblock {\em Phys. Rev. B}, 40:546, 1989.

\bibitem{BH_Sheshadri}
K.~Sheshadri, H.~R. Krishnamurthy, R.~Pandit, and T.~V. Ramakrishnan.
\newblock {\em Europhys. Lett.}, 22(257), 1993.

\bibitem{QIsing.deGennes}
P.~G. de~Gennes.
\newblock {\em Solid State Commun.}, 1:132, 1963.

\bibitem{QIsing.Katsura}
S.~Katsura.
\newblock {\em Phys. Rev.}, 127:1508, 1962.

\bibitem{Columbite1}
R.~Coldea, D.~A. Tennant, E.~M. Wheeler, E.~Wawrzynska, D.~Prabhakaran,
  M.~Telling, K.~Habicht, P.~Smeibidl, and K.~Kiefer.
\newblock {\em Science}, 327:177, 2010.

\bibitem{quench.QI}
K.~Sengupta, S.~Powell, and S.~Sachdev.
\newblock {\em Phys. Rev. A}, 69:053616, 2004.

\bibitem{Moessner.QI}
R.~Moessner and S.~Sondhi.
\newblock {\em Phys. Rev. B}, 63:224401, 2001.

\bibitem{RabiModel.Graham}
R.~Graham and M.~H\"ohnerbach.
\newblock {\em Z. Phys.B}, 57:233, 1984.

\bibitem{RabiModel.Reik}
H.~G. Reik, P.~Lais, M.~E. St\"utzle, and M.~Doucha.
\newblock {\em J. Phys. A: Math. Gen.}, 20:6327, 1987.

\bibitem{RabiModel.Braak}
D.~Braak.
\newblock {\em Phys. Rev. Lett.}, 107:100401, 2011.

\bibitem{RabiLattice.ZHeng}
H.~Zheng and Y.~Takada.
\newblock {\em Phys. Rev. A}, 84:043819, 2011.

\bibitem{srw-prb-1993}
Steven~R. White.
\newblock Density-matrix algorithms for quantum renormalization groups.
\newblock {\em Phys. Rev. B}, 48(14):10345--10356, Oct 1993.

\bibitem{RabiLattice.Schiro}
M.~Schir\'o, M.~Bordyuh, B.~\"Oztop, and H.~E. T\"ureci.
\newblock {\em Phys. Rev. Lett.}, 109:053601, 2012.

\bibitem{Solano}
D.~Ballester, G.~Romero, J.~J. Garc\'ia-Ripoll, F.~Deppe, and E.~Solano.
\newblock {\em Phys. Rev. X}, 2:021007, 2012.

\bibitem{Bogoliubov}
N.~N. Bogoliubov.
\newblock {\em Qauntum Statistical Mechanics}.
\newblock World Scientific, Singapore, 2000.

\bibitem{Sheshadri.93}
K.~Sheshadri, H.~R. Krishnamurthy, R.~Pandit, and T.~V. Ramakrishnan.
\newblock {\em Europhys. Lett.}, 22(257), 1993.

\bibitem{bk-ent}
Brijesh Kumar.
\newblock Emergent radiation in an atom--field system at twice resonance.
\newblock {\em J. Phys. A: Math. Theor.}, 42, 2009.

\bibitem{pfeuty}
Pierre Pfeuty.
\newblock The one-dimensional ising model with a transverse field.
\newblock {\em Annals of Physics}, 57:79--90, 1970.

\bibitem{Kitaev.QWire}
A~Yu Kitaev.
\newblock Unpaired majorana fermions in quantum wires.
\newblock {\em Physics-Uspekhi}, 44(10S):131, 2001.

\bibitem{jason-majorana}
Jason Alicea.
\newblock New directions in the pursuit of majorana fermions in solid state
  systems.
\newblock {\em Rep. Prog. Phys}, 75(7):076501, 2012.

\bibitem{QIj1j2.Barber}
M.~N. Barber and P.~M. Duxbury.
\newblock {\em J. Phys. A: Math. Gen.}, 14:L251, 1981.

\bibitem{ANNNI.Selke}
W.~Selke.
\newblock {\em Phys. Reports}, 170:213, 1988.

\bibitem{dirk}
Fabian Hassler and Dirk Schuricht.
\newblock Stongly interacting majorana modes in an array of josephson
  junctions.
\newblock {\em New J. Phys.}, 14(125018), December 2012.

\bibitem{MG}
C.~K. Majumdar and D.~K. Ghosh.
\newblock {\em J. Math. Phys.}, 10:1399, 1969.

\bibitem{QIj1j2.Bikas}
P.~Sen and B.~K. Chakrabarti.
\newblock {\em Phys. Rev. B}, 43:13559, 1991.

\bibitem{QIj1j2.Brazil}
P.~R.~C. Guimar$\tilde{\rm a}$es, J.~Plascak, F.~C.~S. Barreto, and
  J.~Florencio.
\newblock {\em Phys. Rev. B}, 66:064413, 2002.

\bibitem{QIj1j2.Subinay}
A.K. Chandra and S.~Dasgupta.
\newblock {\em Phys. Rev. E}, 75:021105, 2007.

\bibitem{QIj1j2.DMRG}
M.~Beccaria, M.~Campostrini, and A.~Feo.
\newblock {\em Phys. Rev. B}, 76:094410, 2007.

\bibitem{QIj1j2.Adam}
A.~Nagy.
\newblock {\em New J. Phys.}, 13:023015, 2011.

\bibitem{review.dutta}
A.~Dutta, U.~Divakaran, D.~Sen, B.~K. Chakrabarti, T.~F. Rosenbaum, and
  G.~Aeppli.
\newblock Quantum phase transitions in transverse field spin models: From
  statistical physics to quantum information.
\newblock arXiv:1012.0653, Dec 2010.

\bibitem{dmrg1}
Steven~R. White.
\newblock Density-matrix algorithms for quantum renormalization groups.
\newblock {\em Phys. Rev. B}, 48(14):10345--10356, Oct 1993.

\bibitem{dmrg2}
U.~Schollw\"ock.
\newblock {\em Rev. Mod. Phys.}, 77:259, 2005.

\bibitem{niu2012prb}
Yuezhen Niu, Suk~Bum Chung, Chen-Hsuan Hsu, Ipsita Mandal, S.~Raghu, and Sudip
  Chakravarty.
\newblock Majorana zero modes in a quantum ising chain with longer-ranged
  interactions.
\newblock {\em Phys. Rev. B}, 85:035110, Jan 2012.

\bibitem{diptiman-talk}
Diptiman Sen.
\newblock Majorana fermions.
\newblock Quantum Condensed Matter Journal Club Talk, IISc, Bangalore, May
  2012.

\bibitem{volovik}
G.~E. Volovik.
\newblock Fermion zero modes on vortices in chiral superconductors.
\newblock {\em JETP Letters}, 70(9):609, Nov 1999.

\bibitem{zonda}
Martin {\v Z}onda.
\newblock Phase transitions in the falicov-kimball model away from
  half-filling.
\newblock {\em Phase Transitions}, 85(1-2):96--105, 2012.

\bibitem{maska}
Maciej~M. Ma\ifmmode~\acute{s}\else \'{s}\fi{}ka and Katarzyna Czajka.
\newblock Thermodynamics of the two-dimensional falicov-kimball model: A
  classical monte carlo study.
\newblock {\em Phys. Rev. B}, 74:035109, Jul 2006.

\bibitem{freericks}
J.~K. Freericks and V.~Zlatic.
\newblock Exact dynamical mean-field theory of the falicov-kimball model.
\newblock {\em Rev. Mod. Phys.}, 75:1333, October 2003.

\bibitem{farkasovsky-iop-2002}
Pavol Farkasovsk{\'y} and Nat{\'a}lia Hud{\'a}kov{\'a}.
\newblock Ground-state properties of the falicov-kimball model with correlated
  hopping in two dimensions.
\newblock {\em Journal of Physics: Condensed Matter}, 14(3):499, 2002.

\bibitem{farkasovsky-prb-1995}
Pavol Farka\ifmmode~\check{s}\else \v{s}\fi{}ovsk\'y.
\newblock Falicov-kimball model and the problem of valence and metal-insulator
  transitions.
\newblock {\em Phys. Rev. B}, 51:1507--1512, Jan 1995.

\bibitem{fm-model-disorder}
M.~A. Gusm\~ao.
\newblock Phase diagram of the anderson-falicov-kimball model at half filling.
\newblock {\em Phys. Rev. B}, 77:245116, Jun 2008.

\bibitem{musial}
G.~Musial, L.~Debski, and J.~Wojtkiewicz.
\newblock A monte carlo study of the falicov-kimball model in the perturbative
  regime.
\newblock {\em Low Temperature Physics}, 33(9):797, 2007.

\bibitem{umesh-2010}
Umesh~K Yadav, T~Maitra, Ishwar Singh, and A~Taraphder.
\newblock A ground state phase diagram of a spinless, extended falicov--kimball
  model on the triangular lattice.
\newblock {\em Journal of Physics: Condensed Matter}, 22(29):295602, 2010.

\bibitem{bk-canonical}
Brijesh Kumar.
\newblock Canonical representation for electrons and its application to the
  hubbard model.
\newblock {\em Phys. Rev. B}, 77:205115, May 2008.

\end{thebibliography}


\begin{thebibliography}{10}

\bibitem{srw-noack-prl1992}
S.~R. White and R.~M. Noack.
\newblock Real-space quantum renormalization group.
\newblock {\em Phys. Rev. Lett.}, 68(24):3487, June 1992.

\bibitem{srw-prl1992}
S.~R. White.
\newblock Density matrix formulation for quantum renormalization groups.
\newblock {\em Phys. Rev. Lett.}, 69(19), November 1992.

\bibitem{srw-prb1993}
Steven~R. White.
\newblock Density-matrix algorithms for quantum renormalization groups.
\newblock {\em Phys. Rev. B}, 48:10345--10356, Oct 1993.

\bibitem{Schollwock2005}
U.~Schollw\"ock.
\newblock {\em Rev. Mod. Phys.}, 77:259, 2005.

\bibitem{chiara}
Gabriele~De Chiara, Matteo Rizzi, Davide Rossini, and Simone Montangero.
\newblock Density matrix renormalization group for dummies.
\newblock {\em arXiv:cond-mat/0603842}, April 2009.

\bibitem{noack}
Reinhard~M. Noack and Salvatore~R. Manmana.
\newblock {\em arXiv:cond-mat/0510321}.

\bibitem{saad}
Youcef Saad.
\newblock {\em Numerical Methods for large Eigenvalue problem}.
\newblock Manchester University Press Series in Algorithms and Architecture for
  Advanced Scientific Computing. Fourth edition, 1991.

\bibitem{davidson}
Andreas Stathopoulos and Charlotte~F. Fischer.
\newblock A davidson program for finding a few selected extreme eigenpairs of a
  large, sparse, real, symmetric matrix.
\newblock {\em Computer Physics Communication}, 79(2):268--290, April 1994.

\bibitem{lapack}
E.~Anderson, Z.~Bai, C.~Bischof, S.~Blackford, J.~Demmel, J.~Dongarra,
  J.~Du~Croz, A.~Greenbaum, S.~Hammarling, A.~McKenney, and D.~Sorensen.
\newblock {\em LAPACK Users' Guide}.
\newblock Society for Industrial and Applied Mathematics, Philadelphia, PA,
  third edition, 1999.

\bibitem{sandvik}
A.~W. Sandvik.
\newblock Aip conf. proc.
\newblock volume 1297, page 135, http://dx.doi.org/10.1063/1.3518900, 2010.

\bibitem{lin-prb-ed}
H.~Q. Lin.
\newblock Exact diagonalization of quantum-spin models.
\newblock {\em Phys. Rev. B}, 42(10):6561, October 1990.

\bibitem{lauchli-ed-talk}
Andreas Lauchli.
\newblock {\em Introduction to Exact Diagonalization}.
\newblock Boulder Summer School, 2010.
\newblock (presentation note).

\bibitem{krauth}
Werner Krauth.
\newblock {\em Statistical Mechanics: Algorithm and Computation}.
\newblock Oxford Master Series in Statistical, Computational, and Theoretical
  Physics. Oxford University Press, 2006.

\bibitem{binder}
K.~Binder and D.~H. Heermann.
\newblock {\em Monte Carlo Simulation in Statistical Physics}.
\newblock Springer Series in Solid-State Sciences. Springer, fourth edition,
  2002.

\bibitem{assa-1988}
Daniel~P. Arovas and Assa Auerbach.
\newblock Functional integral theories of low-dimensional quantum heisenberg
  models.
\newblock {\em Phys. Rev. B}, 38(1):316, July 1988.

\bibitem{assa-book}
Assa Auerbach.
\newblock {\em Interacting electrons and quantum magnetism}.
\newblock Graduate Texts in Contemporary Physics. Springer, Technion, Israel
  Institute of Technology, 1997.

\end{thebibliography}


\begin{thebibliography}{10}

\bibitem{deGennes}
P.~G. de~Gennes.
\newblock {\em Solid State Commun}, 1(132), 1963.

\bibitem{book.Bikas}
B.~K. Chakrabarti, A.~Dutta, and P.~Sen.
\newblock {\em Quantum Ising Phases and Transitions in Transverse Ising
  models}.
\newblock Springer-Verlag, Berlin, 1996.

\bibitem{randomQI.Fisher}
D.~S. Fisher.
\newblock {\em Phys. Rev. B}, 51:6411, 1995.

\bibitem{Moessner}
R.~Moessner, S.~L. Sondhi, and P.~Chandra.
\newblock {\em Phys. Rev. Lett.}, 84:4458, 2000.

\bibitem{QI.Holo.Girvin}
P.~B. Chakraborty, P.~Henelius, H.~Kj{\o}nsberg, A.~W. Sandvik, and S.~M.
  Girvin.
\newblock {\em Phys. Rev. B}, 70:144411, 2004.

\bibitem{Coldea}
R.~Coldea, D.~A. Tennant, E.~M. Wheeler, E.~Wawrzynska, D.~Prabhakaran,
  M.~Telling, K.~Habicht, P.~Smeibidl, and K.~Kiefer.
\newblock {\em Science}, 327:177, 2010.

\bibitem{bermudez-qi-expt}
A~Bermudez, J~Almeida, K~Ott, H~Kaufmann, S~Ulm, U~Poschinger, F~Schmidt-Kaler,
  A~Retzker, and M~B Plenio.
\newblock Quantum magnetism of spin-ladder compounds with trapped-ion crystals.
\newblock {\em New J. Phys.}, 14:093042, 2012.

\bibitem{Mila.Square}
Sandro Wenzel, Tommaso Coletta, Sergey~E. Korshunov, and Fr\'ed\'eric Mila.
\newblock Evidence for columnar order in the fully frustrated transverse field
  ising model on the square lattice.
\newblock {\em Phys. Rev. Lett.}, 109:187202, Nov 2012.

\bibitem{bkumar.somenath}
Brijesh Kumar and Somenath Jalal.
\newblock Quantum ising dynamics and majorana-like edge modes in the rabi
  lattice model.
\newblock {\em Phys. Rev. A}, 88:011802(R), July 2013.

\bibitem{pfeuty}
Pierre Pfeuty.
\newblock The one-dimensional ising model with a transverse field.
\newblock {\em Annals of Physics}, 57:79--90, 1970.

\bibitem{Kitaev.QWire}
A~Yu Kitaev.
\newblock Unpaired majorana fermions in quantum wires.
\newblock {\em Physics-Uspekhi}, 44(10S):131, 2001.

\bibitem{Nayak.RMP}
Chetan Nayak, Steven~H. Simon, Ady Stern, Michael Freedman, and Sankar
  Das~Sarma.
\newblock Non-abelian anyons and topological quantum computation.
\newblock {\em Rev. Mod. Phys.}, 80:1083--1159, Sep 2008.

\bibitem{jason-majorana}
Jason Alicea.
\newblock New directions in the pursuit of majorana fermions in solid state
  systems.
\newblock {\em Rep. Prog. Phys}, 75(7):076501, 2012.

\bibitem{QIj1j2.Barber}
M.~N. Barber and P.~M. Duxbury.
\newblock {\em J. Phys. A: Math. Gen.}, 14:L251, 1981.

\bibitem{ANNNI.Selke}
W.~Selke.
\newblock {\em Phys. Reports}, 170:213, 1988.

\bibitem{dirk}
Fabian Hassler and Dirk Schuricht.
\newblock Stongly interacting majorana modes in an array of josephson
  junctions.
\newblock {\em New J. Phys.}, 14(125018), December 2012.

\bibitem{MG}
C.~K. Majumdar and D.~K. Ghosh.
\newblock {\em J. Math. Phys.}, 10:1399, 1969.

\bibitem{QIj1j2.Bikas}
P.~Sen and B.~K. Chakrabarti.
\newblock {\em Phys. Rev. B}, 43:13559, 1991.

\bibitem{QIj1j2.Brazil}
P.~R.~C. Guimar$\tilde{\rm a}$es, J.~Plascak, F.~C.~S. Barreto, and
  J.~Florencio.
\newblock {\em Phys. Rev. B}, 66:064413, 2002.

\bibitem{QIj1j2.Subinay}
A.K. Chandra and S.~Dasgupta.
\newblock {\em Phys. Rev. E}, 75:021105, 2007.

\bibitem{QIj1j2.DMRG}
M.~Beccaria, M.~Campostrini, and A.~Feo.
\newblock {\em Phys. Rev. B}, 76:094410, 2007.

\bibitem{QIj1j2.Adam}
A.~Nagy.
\newblock {\em New J. Phys.}, 13:023015, 2011.

\bibitem{review.dutta}
A.~Dutta, U.~Divakaran, D.~Sen, B.~K. Chakrabarti, T.~F. Rosenbaum, and
  G.~Aeppli.
\newblock Quantum phase transitions in transverse field spin models: From
  statistical physics to quantum information.
\newblock arXiv:1012.0653, Dec 2010.

\bibitem{dmrg1}
Steven~R. White.
\newblock Density-matrix algorithms for quantum renormalization groups.
\newblock {\em Phys. Rev. B}, 48(14):10345--10356, Oct 1993.

\bibitem{dmrg2}
U.~Schollw\"ock.
\newblock {\em Rev. Mod. Phys.}, 77:259, 2005.

\bibitem{niu2012prb}
Yuezhen Niu, Suk~Bum Chung, Chen-Hsuan Hsu, Ipsita Mandal, S.~Raghu, and Sudip
  Chakravarty.
\newblock Majorana zero modes in a quantum ising chain with longer-ranged
  interactions.
\newblock {\em Phys. Rev. B}, 85:035110, Jan 2012.

\bibitem{diptiman-talk}
Diptiman Sen.
\newblock Majorana fermions.
\newblock Quantum Condensed Matter Journal Club Talk, IISc, Bangalore, May
  2012.

\bibitem{volovik}
G.~E. Volovik.
\newblock Fermion zero modes on vortices in chiral superconductors.
\newblock {\em JETP Letters}, 70(9):609, Nov 1999.

\bibitem{LSM}
E.~H. Lieb, T.~Schultz, and D.~C. Mattis.
\newblock {\em Ann. Phys. (N.Y.)}, 16:407, 1961.

\end{thebibliography}


\begin{thebibliography}{10}

\bibitem{Tomadin.Fazio.10}
A.~Tomadin and R.~Fazio.
\newblock {\em J. Opt. Soc. Am. B}, 27(6):A130, 2010.

\bibitem{Hartmann}
M.~J. Hartmann, F.~G. S~L. Brandao, and M.~B. Plenio.
\newblock {\em Nat. Phys.}, 2:849, 2006.

\bibitem{Greentree.06}
A.~D. Greentree, C.~Tahan, J.~H. Cole, and L.~C.~L. Hollenberg.
\newblock {\em Nature Physics}, 2:856, 2006.

\bibitem{Littlewood}
M.~Aichhorn, M.~Hohenadler, C.~Tahan, and P.~B Littlewood.
\newblock {\em Phys. Rev. Lett.}, 100:216401, 2008.

\bibitem{Blatter}
S.~Schmidt and G.~Blatter.
\newblock {\em Phys. Rev. Lett.}, 103:086403, 2009.

\bibitem{QD.Photonic}
K.~Hennessy, A.~Badolato, M.~Winger, D.~Gerace, M.~Atat\ ̈ure, S.~Gulde, S.~F\
  ̈alt, E.~L. Hu, and A.~Imamo\v{g}lu.
\newblock {\em Nature (London)}, 445:896, 2007.

\bibitem{CircuitQED.Girvin}
R.~J. Schoelkopf and S.~M. Girvin.
\newblock {\em Nature}, 451:664, 2008.

\bibitem{CircuitQED.Fink}
J.~M. Fink, R.~Bianchetti, M.~Baur, M.~G\"oppl, L.~Steffen, S.~Filipp, P.~J.
  Leek, A.~Blais, and A.~Wallraff.
\newblock {\em Phys. Rev. Lett.}, 103:083601, 2009.

\bibitem{Houck.Koch}
A.~A. Houck, H.~E. T\"ureci, and J.~Koch.
\newblock {\em Nat. Phys.}, 8:292, 2012.

\bibitem{ColdAtoms}
I.~Bloch, J.~Dalibard, and W.~Zwerger.
\newblock {\em Rev. Mod. Phys.}, 80:885, 2008.

\bibitem{BH_Fisher}
M.~P.~A. Fisher, P.~B. Weichman, G.~Grinstein, and D.~S. Fisher.
\newblock {\em Phys. Rev. B}, 40:546, 1989.

\bibitem{BH_Sheshadri}
K.~Sheshadri, H.~R. Krishnamurthy, R.~Pandit, and T.~V. Ramakrishnan.
\newblock {\em Europhys. Lett.}, 22(257), 1993.

\bibitem{QIsing.deGennes}
P.~G. de~Gennes.
\newblock {\em Solid State Commun.}, 1:132, 1963.

\bibitem{QIsing.Katsura}
S.~Katsura.
\newblock {\em Phys. Rev.}, 127:1508, 1962.

\bibitem{QIsing.Pfeuty}
P.~Pfeuty.
\newblock {\em Ann. Phys. (N.Y.)}, 57:79, 1970.

\bibitem{Columbite1}
R.~Coldea, D.~A. Tennant, E.~M. Wheeler, E.~Wawrzynska, D.~Prabhakaran,
  M.~Telling, K.~Habicht, P.~Smeibidl, and K.~Kiefer.
\newblock {\em Science}, 327:177, 2010.

\bibitem{quench.QI}
K.~Sengupta, S.~Powell, and S.~Sachdev.
\newblock {\em Phys. Rev. A}, 69:053616, 2004.

\bibitem{Moessner.QI}
R.~Moessner and S.~Sondhi.
\newblock {\em Phys. Rev. B}, 63:224401, 2001.

\bibitem{QPT.Sachdev}
Subir Sachdev.
\newblock {\em Quantum Phase Transitions}.
\newblock Cambridge University Press, 2011.

\bibitem{LSM}
E.~H. Lieb, T.~Schultz, and D.~C. Mattis.
\newblock {\em Ann. Phys. (N.Y.)}, 16:407, 1961.

\bibitem{Kitaev.FreeMajorana}
A.~Yu. Kitaev.
\newblock {\em Phys.-Usp.}, 44:131, 2001.

\bibitem{Wilczek.Majorana}
F.~Wilczek.
\newblock {\em Nat. Phys.}, 5:614, 2009.

\bibitem{note.Majorana}
A fermion is created (annihilated) by an operator $\chat^\dag$ ($\chat$), and
  it satisfies anti-commutation with other fermions. For a given $\chat^\dag$
  and $\chat$, we can define two Majorana fermions $\phihat = \chat+\chat^\dag$
  and $\psihat=i(\chat-\chat^\dag)$ such that $\phihat^\dag=\phihat$ and
  $\phihat^2=1$ (and likewise for $\psihat$). The Majorana operators $\phihat$
  and $\psihat$ anti-commute with each other, and with other fermion.

\bibitem{Nayak.RMP}
Chetan Nayak, Steven~H. Simon, Ady Stern, Michael Freedman, and Sankar
  Das~Sarma.
\newblock Non-abelian anyons and topological quantum computation.
\newblock {\em Rev. Mod. Phys.}, 80:1083--1159, Sep 2008.

\bibitem{Topo_QC_Kitaev}
A.~Kitaev and C.~Laumann.
\newblock Topological phases and quantum computation.
\newblock arXiv:0904.2771, 2009.

\bibitem{Mourik.Majorana}
V.~Mourik, K.~Zou, S.~M. Frolov, S.~R. Plissard, E.~P. A.~M. Bakkers, and L.~P.
  Kouwenhoven.
\newblock {\em Science}, 336:1003, 2012.

\bibitem{bkumar.somenath}
Brijesh Kumar and Somenath Jalal.
\newblock Quantum ising dynamics and majorana-like edge modes in the rabi
  lattice model.
\newblock {\em Phys. Rev. A}, 88:011802(R), July 2013.

\bibitem{Scully.Zubairy}
M.~O. Scully and M.~S. Zubairy.
\newblock {\em Quantum Optics}.
\newblock Cambridge University Press, 1997.

\bibitem{RabiModel.Graham}
R.~Graham and M.~H\"ohnerbach.
\newblock {\em Z. Phys.B}, 57:233, 1984.

\bibitem{RabiModel.Reik}
H.~G. Reik, P.~Lais, M.~E. St\"utzle, and M.~Doucha.
\newblock {\em J. Phys. A: Math. Gen.}, 20:6327, 1987.

\bibitem{RabiModel.Braak}
D.~Braak.
\newblock {\em Phys. Rev. Lett.}, 107:100401, 2011.

\bibitem{RabiLattice.ZHeng}
H.~Zheng and Y.~Takada.
\newblock {\em Phys. Rev. A}, 84:043819, 2011.

\bibitem{srw-prb-1993}
Steven~R. White.
\newblock Density-matrix algorithms for quantum renormalization groups.
\newblock {\em Phys. Rev. B}, 48(14):10345--10356, Oct 1993.

\bibitem{Schollwock2005}
U.~Schollw\"ock.
\newblock {\em Rev. Mod. Phys.}, 77:259, 2005.

\bibitem{RabiLattice.Schiro}
M.~Schir\'o, M.~Bordyuh, B.~\"Oztop, and H.~E. T\"ureci.
\newblock {\em Phys. Rev. Lett.}, 109:053601, 2012.

\bibitem{Solano}
D.~Ballester, G.~Romero, J.~J. Garc\'ia-Ripoll, F.~Deppe, and E.~Solano.
\newblock {\em Phys. Rev. X}, 2:021007, 2012.

\bibitem{Bogoliubov}
N.~N. Bogoliubov.
\newblock {\em Qauntum Statistical Mechanics}.
\newblock World Scientific, Singapore, 2000.

\bibitem{Sheshadri.93}
K.~Sheshadri, H.~R. Krishnamurthy, R.~Pandit, and T.~V. Ramakrishnan.
\newblock {\em Europhys. Lett.}, 22(257), 1993.

\bibitem{bk-ent}
Brijesh Kumar.
\newblock Emergent radiation in an atom--field system at twice resonance.
\newblock {\em J. Phys. A: Math. Theor.}, 42, 2009.

\end{thebibliography}
